\documentclass[10pt]{article}
\usepackage{placeins}
\usepackage{graphicx}
\usepackage{amssymb}
\usepackage{fancyhdr}
\usepackage{xcolor}
\usepackage{url}
\usepackage{mathtools}
\usepackage{amsmath}
\usepackage[utf8]{inputenc}
\usepackage[english]{babel}
\usepackage{breakcites}
\usepackage{tabu}
\usepackage{mathrsfs}
\usepackage{dsfont}
\usepackage{booktabs}
\usepackage{multirow}
\usepackage{amsthm}
\usepackage{hyperref}
\usepackage{rotating}
\usepackage{graphicx}
\usepackage{lscape}
\usepackage{enumitem}
\usepackage{bbm}
\usepackage[]{interval}
\usepackage{setspace}
\usepackage{adjustbox}

\usepackage{geometry}
\intervalconfig{
  soft open fences,
  separator symbol =;,
}
\definecolor{cornellred}{rgb}{0.7, 0.11, 0.11}
\hypersetup{colorlinks,linkcolor={red},citecolor={cornellred},urlcolor={red}}

\newtheoremstyle{break}
  {\topsep}{\topsep}%
  {\itshape}{}%
  {\bfseries}{}%
  {\newline}{}%
  
\theoremstyle{break}
\newtheorem{assumption}{A}
\newtheorem{theorem}{Theorem}
\newtheorem{example}{Example}
\newtheorem{definition}{Definition}
\newtheorem{proposition}{Proposition}
\newtheorem{rem}{Remark}
\newtheorem{lemma}{Lemma}

\DeclareMathOperator{\argmax}{argmax}

\DeclareMathOperator{\spargel}{sp}
\DeclareMathOperator{\cspargel}{\overline{\spargel}}

\DeclareMathOperator{\E}{\mathbb E}

\DeclareMathOperator{\proj}{proj}

\DeclareMathOperator{\tr}{tr}
\DeclareMathOperator{\esssup}{ess\,sup}

\DeclareMathOperator{\vect}{vec}

\DeclareMathOperator{\ulim}{\underline{\lim}}

\newcommand\utimes{\mathbin{\ooalign{$\cup$\cr%
   \hfil\raise0.42ex\hbox{$\scriptscriptstyle\times$}\hfil\cr}}}
\newcommand\bigutimes{\mathop{\ooalign{$\bigcup$\cr%
   \hfil\raise0.36ex\hbox{$\scriptscriptstyle\boldsymbol{\times}$}\hfil\cr}}}

\makeatletter
\renewenvironment{proof}[1][\proofname]{%
  \par\pushQED{\qed}\normalfont%
  \topsep6\p@\@plus6\p@\relax
  \trivlist\item[\hskip\labelsep\bfseries#1\@addpunct{.}]%
  \ignorespaces
}{%
  \popQED\endtrivlist\@endpefalse
}
\makeatother

\DeclarePairedDelimiter\abs{\lvert}{\rvert}%
\DeclarePairedDelimiter\norm{\lVert}{\rVert}%

\makeatletter
\let\oldabs\abs
\def\abs{\@ifstar{\oldabs}{\oldabs*}}
\let\oldnorm\norm
\def\norm{\@ifstar{\oldnorm}{\oldnorm*}}
\makeatother

\usepackage[round]{natbib}

\makeatletter
\newcommand{\bianca}{\renewcommand\NAT@open{[}\renewcommand\NAT@close{]}}
\makeatother

\usepackage{stackengine}
\newcommand\ubar[1]{\stackunder[1.2pt]{$#1$}{\rule{.8ex}{.075ex}}}

\usepackage{xcolor}
\hypersetup{
    colorlinks,
    linkcolor={red!50!black},
    citecolor={blue!50!black},
    urlcolor={blue!80!black}
}

\usepackage{comment}
\excludecomment{ext}\excludecomment{comment}

\usepackage{tabu}

\newcolumntype{L}[1]{>{\raggedright\let\newline\\\arraybackslash\hspace{0pt}}m{#1}}
\newcolumntype{C}[1]{>{\centering\let\newline\\\arraybackslash\hspace{0pt}}m{#1}}
\newcolumntype{R}[1]{>{\raggedleft\let\newline\\\arraybackslash\hspace{0pt}}m{#1}}

\DeclareMathAlphabet\mathbfcal{OMS}{cmsy}{b}{n}

\newcommand{\beq}{\begin{equation}}
\newcommand{\eeq}{\end{equation}}
\newcommand{\bea}{\begin{eqnarray}}
\newcommand{\eea}{\end{eqnarray}}
\newcommand{\ba}{\begin{array}}
\newcommand{\ea}{\end{array}}
\newcommand{\bit}{\begin{itemize}}
\newcommand{\eit}{\end{itemize}}
\newcommand{\ben}{\begin{enumerate}} 
\newcommand{\een}{\end{enumerate}}
\newcommand{\bpm}{\begin{pmatrix}}
\newcommand{\epm}{\end{pmatrix}}
\newcommand{\bbm}{\begin{bmatrix}}
\newcommand{\ebm}{\end{bmatrix}}

\renewcommand{\l}{\left}
\renewcommand{\r}{\right}

\newcommand{\nn}{\nonumber}
\newcommand{\wh}{\widehat}

\newtheoremstyle{break}
  {\topsep}{\topsep}%
  {\itshape}{}%
  {\bfseries}{}%
  {\newline}{}%

\begin{document}
\onehalfspacing
%
\title{The Canonical Decomposition of Factor Models:\\
Weak Factors are Everywhere}
%
\author{Philipp Gersing\footnote{Department of Statistics and Operations Research, University of Vienna, \url{philipp.gersing@univie.ac.at}} \hskip .4cm
Matteo Barigozzi\footnote{Department of Economics, Universit\`a di Bologna} \hskip .4cm
Christoph Rust\footnote{Vienna University of Business and Economics} \hskip .4cm
Manfred Deistler\footnote{Department of Statistics, Vienna University of Technology, Institute for Advanced Studies Vienna}
}
\maketitle
%
%
%
%
%
%
\begin{abstract}

We derive a novel canonical decomposition of factor models encompassing both the static factor model---where factors are loaded only contemporaneously---and the Generalised Dynamic Factor Model---where factors are loaded with lags. This decomposition features a new term: the weak common component, defined as the difference between the dynamic and static common components. It is driven by (possibly infinitely many) non-pervasive weak factors which belong to the dynamically common space. Through theoretical and empirical examples---both on U.S. macroeconomic indicators and global financial volatilities---we show that, in general, the weak common component shall not be neglected. Furthermore, we show that, by accounting for the presence of weak common components, we are likely to obtain Impulse Response Functions with more plausible shapes than those obtained from purely static approaches.  
In addition, we provide consistent estimators for all terms of the canonical decomposition and for the weak factors.

%
%
\end{abstract}
\section{Introduction}
%
%
%
%
Consider a high-dimensional time series as a realisation of an underlying double indexed (zero-mean, covariance stationary) process $(y_{it}: i \in \mathbb N, t \in \mathbb Z)=: (y_{it})$. There are two approaches to factor models in the high-dimensional time series literature. On the one hand, there is the static approach or the approximate static factor decomposition, firstly introduced by \cite{chamberlain1983funds}, \cite{chamberlain1983arbitrage}, and then extended and studied in detail by \citet{stock2002forecasting, stock2002macroeconomic}, \citet{bai2002determining}, and \citet{bai2003inferential}, among many others. The static decomposition reads
\begin{align}
    y_{it} = C_{it} + e_{it} = \Lambda_i F_t + e_{it} ,   \label{eq: static factor model rep} 
\end{align}
where $(F_t)$ is a vector process of latent pervasive factors of fixed small dimension $r$, which are loaded statically by the loadings $\Lambda_i$ (an $r$-dimensional row vector) into $C_{it}$. We shall call $C_{it}$ the ``static common component''. The factors are pervasive in the sense that all $r$ eigenvalues of $\Lambda^{n'}\Lambda^n$ diverge as $n \to \infty$, with $\Lambda^n := (\Lambda_1' \cdots \Lambda_{n}')'$. The ``static idiosyncratic component'', $(e_{it})$, is contemporaneously orthogonal to $(F_t)$ and assumed to be weakly correlated within the cross-section. This can formally be expressed by the assumption that the largest eigenvalue of the covariance matrix $\Gamma_e^n := \E[e_t^n e_t^{n'}]$ stays bounded as $n \to \infty$, with $e_t^n = (e_{1t} \cdots e_{nt})'$. 

On the other hand, there is the dynamic approach, or the dynamic factor decomposition, also referred to as the {Generalized Dynamic Factor model} (GDFM), firstly introduced by \cite{forni2000generalized} and \citet{forni2001generalized}:
\begin{align}
    y_{it} = \chi_{it} + \xi_{it} = 
    \sum_{j = 0}^\infty K_i(j) \varepsilon_{t-j} + \xi_{it}, \label{eq: GDFM rep}
\end{align}
where the ``dynamic common component'' $(\chi_{it})$ is driven by a vector of latent factors $(\varepsilon_t)$ of fixed small dimension $q$, which, without loss of generality can always be assumed to be an orthonormal vector white noise process. The $K_i(j)$'s are $1\times q$ vectors  coefficients square-summable in $j$. The factors $(\varepsilon_t)$ are often called ``dynamic factors'' or even ``common  shocks'' and are dynamically pervasive in the sense that all $q$ eigenvalues of $k^{n\prime}(\theta) k^{n}(\theta)$ diverge as $n\to \infty$, for almost all frequencies $\theta \in [-\pi, \pi]$, where $k^{n}(\theta) = \l(k_1(\theta)' \cdots k_n(\theta)'\r)'$ with $k_i(\theta) = \sum_{j = 0}^\infty K_i(j)e^{-\iota\theta}$.  The ``dynamic idiosyncratic component'', $(\xi_{it})$, is uncorrelated with $(\varepsilon_t)$ at all leads and lags and assumed to be weakly correlated within the cross-section \textit{and over time}. This is formally expressed by the assumption that the largest eigenvalue of the spectral density $f_\xi^n (\theta)$ of the process $\xi_t^n:=(\xi_{1t}\cdots \xi_{nt})'$ is essentially bounded on the frequency band $\theta \in [-\pi, \pi]$ as $n\to \infty$. 

No unified framework that covers both models is currently available. The existing literature has largely treated the two approaches as equivalent \citep{bai2007determining,forni2009opening,doz2011two, stock2016dynamic}, and has therefore tended to ignore the fact that they actually imply conceptually distinct notions of what is considered ``common'' and what is deemed ``idiosyncratic''.

The main contribution of this paper is to derive a canonical representation of $(y_{it})$, encompassing both the static and dynamic factor decomposition, of the form (Theorem \ref{thm: relation of r-SFS and q-DFS}): 
\begin{align}
    y_{it} &= \lefteqn{\underbrace{\phantom{C_{it} + e_{it}^\chi}}_{\chi_{it}}}  C_{it} + \overbrace{e_{it}^\chi + \xi_{it}}^{e_{it}}. 
    \label{eq: 3fold decomp intro}
\end{align}
The term $e_{it}^\chi = \chi_{it} - C_{it} = e_{it} - \xi_{it}$ is what we call the \textit{weak common component}. It is the difference between the dynamic and the static common component or, equivalently, the static and the dynamic idiosyncratic component. Furthermore, we show that the weak common component is statically idiosyncratic and driven by (possibly infinitely many) non-pervasive factors, denoted as $F_t^w$, which live in the dynamically common space, i.e., the space spanned by the common shocks $(\varepsilon_{t})$. This means that we can also write the decomposition \eqref{eq: 3fold decomp intro} as (Theorem \ref{thm: weakfactors}):
\begin{align}
 y_{it} &= \underbrace{\Lambda_i F_t + \Lambda_i^w F_t^w}_{\chi_{it}} + \xi_{it}, \label{eq: 3fold decomp intro2}
\end{align}
with $e_{it}^\chi = \Lambda_i^w F_t^w$. Here the eigenvalues of $\Lambda^{w, n'} \Lambda^{w, n}$ stay bounded as $n\to \infty$. Consequently, the factors $F_t^w$ are ``weak'' in the same sense as in  \cite{onatski2012asymptotics}.

We then ask and answer two related questions: First, are representations \eqref{eq: 3fold decomp intro} and \eqref{eq: 3fold decomp intro2} theoretically and empirically relevant? 
And, second, if yes, what are the main  implications for estimation of factor models? 

The central insight of the decomposition \eqref{eq: 3fold decomp intro} is that $(\chi_{it})$ is obtained by linearly projecting $(y_{it})$ onto the space spanned by its dynamic aggregations, which strictly contains the space generated by its static aggregations used to recover $(C_{it})$. The presence of a weak common component and, thus, of weak factors, is then to be expected in every high-dimensional time series panel, although not necessarily in every individual time series. 
Hence, our claim that \textit{weak factors are everywhere}. 

To support our claim, we begin by presenting several theoretical examples, through which we show that a weak common component emerges naturally when dynamic factors are loaded with varying strengths at different lags. We then provide empirical evidence by examining a panel of U.S. monthly macroeconomic time series and a panel of global financial volatility indicators, showing that the weak common component is in fact important for many series. For instance, weak factors explain roughly 10\% of the variance of the Unemployment rate (UNRATE) and Consumer Price Inflation (CPIAUCSL). 

However, not every series is influenced by weak factors. For example, in our empirical application, Industrial Production (INDPRO) exhibits an almost negligible weak common component, accounting for less than 2\% of its variance. 
Furthermore, the weak common component disappears entirely for all series in the highly improbable situation where the panel of time series follows a vector white noise process. 

The potential relevance of the weak common component, has important implications for the empirical analysis of Impulse Response Functions (IRF) to the structural shocks driving the economy. The standard macroeconometric approach to IRF estimation through factor models consists in assuming that: the common shocks $(\varepsilon_t)$ in \eqref{eq: GDFM rep} span the same space as the structural shocks, and the static and dynamic representations coincide \citep{giannone2006does,stock2016dynamic}. Under these assumptions, any VAR approach based on the estimated static common components or on the static pervasive factors can be used to retrieve the IRFs \citep{bernanke2005measuring,bai2006confidence,bai2007determining,BoivinGiannoniMihov2009,forni2009opening,ForniGambetti2010,bai2015identification,stock2016dynamic,forni2025common,han2025global}. 
However, when the variables of interest contain weak, but non-negligible, common components, the IRFs obtained with this static approach are likely to be inconsistent and display implausible patterns. The reason is that, in such a setting, the common shocks also transmit through these weak common components, which are being ignored. This issue is demonstrated both via simulations and through two empirical applications: (a) revisiting the analysis of U.S. monetary policy shocks in \citet{forni2025common}, and (b) examining the IRFs of volatilities to a global market-wide shock.

The main message for estimation is then clear. 
On the one hand, static estimation via Principal Component Analysis (PCA) is straightforward  \citep{stock2002forecasting,bai2003inferential}, while the existing dynamic approaches rely on more complex estimators based on spectral or dynamic PCA \citep{forni2000generalized,forni2017dynamic}. 
On the other hand, in order to justify the use of a static approach one should first assess the presence of weak common components. If these are null or negligible, static PCA should be preferred, but, if these are non-negligible, a fully dynamic approach should be chosen, because it automatically accounts for their presence. To this end, we propose an estimator for the weak common component and prove its consistency with rates (Theorem \ref{thm: weakCC}). In so doing, we also provide new consistency rates for the estimated dynamic common component (Proposition \ref{prop1}), extending the results by \cite{forni2000generalized,forni2004generalized, forni2017dynamic} and \citet{barigozzi2024inferential}.

Finally, a solution to most of the above issues might seem to simply estimate the static common component using an overestimated number of factors in such a way to include not only all pervasive ones (strong and rate weak), but also any non pervasive weak factor.  However, such approach is likely to fail. Indeed, while we can consistently recover pervasive factors, i.e., corresponding to eigenvalues diverging at linear of sub-linear rates in $n$, by means of PCA or other related methods employing the sample eigenvectors \citep{lam2012factor,uematsu2022estimation,freyaldenhoven2022factor, bai2023approximate}, none of those procedures is valid for recovering the weak factors which correspond to non-diverging eigenvalues \citep{onatski2012asymptotics}. In this respect, we also provide a solution, since, given our decomposition \eqref{eq: 3fold decomp intro2}, it is clear that, if there is a finite number of weak factors, these can always be retrieved consistently once we estimate the weak common component (Theorem \ref{thm: weakfactest}). Once again, for this we have to estimate the canonical decomposition first. Other possible solutions to the estimation of weak factors are suggested by 
\citet{BarigozziCho2020} who introduce a trimmed PCA estimator that enforces incoherence by post-processing,
\citet{LiaoTongWangXiu2026} who rotate PCA estimated factors to strengthen the signal of the weak ones, and
\citet{lettau2020estimating, lettau2020factors} who propose a modified PCA estimator based on a covariance matrix with overweighted mean, thus strengthening the signal of such factors.

The rest of the paper is organised as follows. In Section \ref{sec: Structure Theory: Reconciling the Schools in One Model}, we briefly review the main existing representation results (Section \ref{subsec: Hilbert}),  then we introduce Theorems \ref{thm: relation of r-SFS and q-DFS} and \ref{thm: weakfactors}, which characterize the new canonical decomposition encompassing the static and dynamic approaches (Section \ref{subsec: canonical}), and, last, we discuss implications for IRFs (Section \ref{subsec: IRFs}).
In Section \ref{sec: estimation} we provide a complete proof of consistency, with rates, for the estimators of the dynamic and static common component (Propositions \ref{prop1} and \ref{prop3}), as well as for the weak common component and the weak factors (Theorems \ref{thm: weakCC} and \ref{thm: weakfactest}). 
In Sections \ref{sec: simulation experiments} and \ref{sec: empiric} we provide numerical studies on simulated, macroeconomic, and financial data. In Section \ref{sec:conc} we conclude. Proofs and additional examples, as well as additional numerical and empirical results are in the Appendix.
\paragraph{Basic notation.} For a complex vector $a$ with elements $a_j$, $j=1,\ldots p$, we denote as $a^*$ its adjunct (transposed complex conjugate) and we use the norm $\Vert v\Vert =\sqrt{\sum_{j=1}^p |v_j|^2}$. 
For a random vector $\cspargel(X)$ is the closure of the space of all linear combinations of the elements of $X$.
For two random vectors $Y$ and $X$ we let $\proj(Y|\cspargel(X))= \E[YX']\{\E[XX']\}^{-1}X$.
 The limit with respect to mean-square convergence is denoted as $\ulim$.
For a symmetric or Hermitian matrix $A$ we denote its $j$-th largest eigenvalue (which is always real) as $\mu_j(A)$. We use the norms $\Vert A\Vert = \sqrt{\mu_1(AA')}$ and $\Vert A\Vert_F=\sqrt{\text{tr}(AA')}$. 

%
%
%
%
%
%
%
\section{Representation theory}\label{sec: Structure Theory: Reconciling the Schools in One Model}
\subsection{The theory of stationary stochastic double sequences}\label{subsec: Hilbert}
\paragraph{Fundamental  definitions.}
Throughout, we consider stochastic double sequences, i.e., a family of random variables indexed in time and cross-section: $(y_{it}: i \in \mathbb N, t \in \mathbb Z) = (y_{it})$. Such a process can also be thought of as a nested sequence of multivariate stochastic processes: $\left(y_t^n: t \in \mathbb Z \right) = (y_t^n)$, where $y_t^n = (y_{1t}\cdots y_{nt})'$ and $y_t^{n+1} = (y_t^{n'}, y_{n+1, t})'$ for $n \in \mathbb N \cup \{\infty\}$. For $n = \infty$, we write $(y_t: t \in \mathbb Z)=(y_t)$. 
\begin{assumption}[Stationary Double Sequence]\label{A: stat}
 Let $\mathcal P = (\Omega, \mathcal A, P)$ be a probability space and $L_2(\mathcal P, \mathbb C)$ be the Hilbert space of square integrable complex-valued, zero-mean, random-variables defined on $\Omega$ equipped with the covariance inner product $\langle u, v\rangle =  \E[ u \bar v]$ for $u, v \in L_2(\mathcal P, \mathbb C)$. 
 For all $n\in \mathbb N$, the process $(y^n_t : t \in \mathbb Z)$ is real valued, weakly stationary with zero-mean and such that:
\begin{itemize}
    \item[(i)] $y_{it} \in L_2(\mathcal P, \mathbb C)$ for all $(i, t) \in \mathbb N \times \mathbb Z;$
\item [(ii)] it has existing spectral density $f_{y}^n(\theta)$ for $\theta \in [-\pi, \pi]$ defined as the $n\times n$ matrix:\footnote{Given our definition of spectral density, its inverse Fourier transform is $\E[y_t^n y_{t-\ell}^{n'}]:=\frac 1{2\pi}\int_{-\pi}^\pi f_{y}^n(\theta)\mathrm d\theta$.}
\[
f_{y}^n(\theta):=\sum_{\ell=-\infty}^{\infty} e^{-\iota \ell \theta} 
\E [y_t^n y_{t-\ell}^{n'}],\quad  \theta \in[-\pi,\pi].
\]
\end{itemize}
\end{assumption}

Hereafter, we let $$
\mathbb H(y):=\cspargel(y_{it}, i \in \mathbb N, t \in \mathbb Z),
$$
which is a Hilbert space and a subspace of $L_2(\mathcal P, \mathbb C)$.

Finally, since the typical data analyzed in economics applications is  real valued, we assume, henceforth, that $y_{it} \in L_2(\mathcal P, \mathbb C)$ takes only real values. Consequently the $n\times n$ covariance matrix $\Gamma_y^n:=\E[y_t^ny_t^{n'}]$ is also real valued. While the spectral density matrix $f_y^n(\theta)$ is always a complex Hermitian matrix, i.e., $f_y^n(\theta)=(f_y^n(\theta))^*$ for all $\theta\in[-\pi,\pi]$, hence it has real eigenvalues. Notice that the sequences of matrices
$(\Gamma_{y}^n : n\in\mathbb N)$
and
$(f_{y}^n(\theta) : n\in\mathbb N)$ are nested.

\paragraph{The Approximate Static Factor Model.}
 Let the covariance matrices of $(C_t^n)$ and of $(e_t^n)$ be $\Gamma_C^n:=\E[C_t^nC_t^{n'}]$ and $\Gamma_e^n=\E[e_t^ne_t^{n'}]$,  with eigenvalues $\mu_j(\Gamma_C^n)$ and $\mu_j(\Gamma_e^n)$, $j=1,\ldots, n$, respectively, sorted in decreasing order. Then, we suppose that $(y_{it})$ follows an approximate static factor model, as formulated by \citet{chamberlain1983funds} and \citet{chamberlain1983arbitrage}.

\begin{assumption}[$r$-Static Factor Structure]\label{A: r-SFS struct}
The process $(y_{it})$ can be represented as:
\[
y_{it} = C_{it} + e_{it} = \Lambda_i F_t+e_{it}, \quad i\in\mathbb N, \quad t\in\mathbb Z,
\]
where $ \Lambda_i$ and $F_t$ are $1\times r$ and $r\times 1$, respectively, with $r < \infty$ and independent of $n$, $\E [F_t]=0$, $\E [F_t F_t'] = I_r$ for all $t\in\mathbb Z$, $\E [F_t e_{it}] = 0$ for all $i \in \mathbb N$ and $t\in\mathbb Z$, and 
\begin{itemize}
    \item[(i)]  $\sup_{n\in\mathbb N} \mu_r(\Gamma^n_C) = \infty$;
    \item[(ii)]  $\sup_{n\in\mathbb N} \mu_1(\Gamma^n_e) < \infty$. 
\end{itemize}
\end{assumption}
We shall say that a double sequence for which A\ref{A: r-SFS struct} holds is an $r$-Static Factor Sequence ($r$-SFS). Alternatively, and equivalently to A\ref{A: r-SFS struct}, we may specify the pervasiveness of the factors via the loadings and we may define idiosyncraticness via bounding the cross-correlations of the $e_{it}$'s \citep[see, e.g.,][Assumptions B and C3]{bai2003inferential}. Orthonormality of $(F_t)$ can be assumed without loss of generality.

%
 Denote by $\widehat L_2^\infty(\Gamma_y)$ the set of all \textit{constant} infinite-dimensional complex vectors $\widehat c =(\widehat c_1\, \widehat c_2\cdots) \in \mathbb C^{1\times \infty}$, with $\widehat c^{\{n\}}=(\widehat c_1\cdots \widehat c_n)$ and such that $\lim_{n\to\infty} \widehat c^{\{n\}} \Gamma_y^n  {\left(\widehat c^{\{n\}}\right)}^* < \infty$. Denote also $\widehat L_2^\infty(I)$
 as the the set of all vectors with $\lim_{n\to\infty} \widehat c^{\{n\}} {(\widehat c^{\{n\}})}^* < \infty$. 
Then, consider a sequence of infinite-dimensional vectors $(\widehat c^{(k)}\in\widehat L_2^\infty(\Gamma_y) ,k\in\mathbb N)$. These generate scalar  valued cross-sectional weighted sums with mean-square limit: 
$  \ulim_{k\to\infty} \widehat c^{(k)}y_t = \ulim_{k\to\infty} \ulim_{n\to\infty} \sum_{i = 1}^n \widehat c_i^{(k)} y_{it}$, 
where $\widehat c_i^{(k)}$ is the $i$-th entry of $\widehat c^{(k)}$.
Due to our focus on real data only, hereafter, we can  consider only static averaging sequences living in $\mathbb R^{1\times \infty}$, without loss of generality.\footnote{A simple example for a static averaging sequence would be the cross-sectional average, with $\widehat c^{(k)}$ having its first $k$ elements equal to $1/k$ and all others being zero,
so that, $\widehat c^{(k)}\widehat c^{(k)\prime} = \sum_{i = 1}^k 1/k^2 = 1/k \to 0$ as $k\to \infty$. }
\begin{definition}[Static Averaging Sequence (SAS)]\label{def: SAS}
Let $\widehat c^{(k)} \in \widehat L_2^\infty(I) \cap \widehat L_2^\infty(\Gamma_y) \cap \mathbb R^{1\times \infty}$ for $k \in \mathbb N$. The sequence $\left(\widehat c^{(k)}:k \in \mathbb N\right)$ is called Static Averaging Sequence (SAS) if
$
    \lim_{k\to\infty} \norm{\widehat c^{(k)}}_{\widehat L_2^\infty(I)} := \lim_{k\to\infty} \widehat c^{(k)}\left(\widehat c^{(k)}\right)' =  0 .
$ We denote the set of all SAS of $(y_{it})$  as 
$\mathcal S(\Gamma_y) := \left \{ (\widehat c^{(k)}) : \widehat c^{(k)} \in \widehat L_2^\infty(I) \cap \widehat L_2^\infty(\Gamma_y) \cap \mathbb R^{1 \times \infty} : k \in \mathbb N \mbox{  and  } \lim_{k\to\infty} \norm{\widehat c^{(k)}}_{\widehat L_2^\infty(I)} = 0 \right \} .$
\end{definition}

We then consider the set of all random variables that can be written as the limit of a SAS.
\begin{definition}[Static Aggregation Space]\label{def: static aggregation space}
The space $\mathbb S_t(y) := \left\{ z_t : z_t = \ulim_{k\to\infty} \widehat c^{(k)} y_t, \ \mbox{where} \ \left(\widehat c^{(k)}\right) \in \mathcal S(\Gamma_y) \right\}$ is called Static Aggregation Space at time $t$. 
\end{definition}

The static aggregation space is a closed subspace of $\cspargel(y_{it},i \in \mathbb N)$ \citep[the proof is analogous to][Lemma 6]{forni2001generalized}.  
%
%
Importantly, the static aggregation space changes over $t \in \mathbb Z$ as it emerges from aggregation of the $y_{it}$'s, while holding $t$ fixed.
\begin{definition}[Statically Idiosyncratic]
A stochastic double sequence $(z_{it})$ is called statically idiosyncratic if $\ulim_{k\to\infty} \widehat c^{(k)} z_t = 0$ for all $(\widehat c^{(k)}) \in \mathcal S (\Gamma_z)$ and for all $t \in \mathbb Z$.
\end{definition}
This definition of a \textit{statically} idiosyncratic double sequence is implicitly contained in \citet{chamberlain1983funds}.
Note that
\citet{stock2002forecasting}, \cite{bai2002determining}, \citet{bai2003inferential} define idiosyncratic in a different way that involves also a limitation of time dependence. However, such condition is not necessary for the present Hilbert space theory on static factor sequences \citep[see Example 2.2.8 in][]{gersing2023reconciling}. 

A static idiosyncratic component can be equivalently defined in terms of eigenvalues of its covariance matrix \citep[the proof is analogous to][Theorem 1]{forni2001generalized}.
\begin{theorem}[Characterisation of Statically Idiosyncratic Components]\label{thm: charact stat idiosyncratic}
The following statements are equivalent:
\begin{itemize}
    \item[(i)] a stochastic double sequence $(z_{it})$ is statically idiosyncratic; 
    \item[(ii)] 
    $
     \sup_{n \in \mathbb N} \mu_1(\Gamma_z^n) < \infty.
     $
\end{itemize}
\end{theorem}

By Theorem \ref{thm: charact stat idiosyncratic}, under Assumption A\ref{A: r-SFS struct}, $(e_{it})$ is statically idiosyncratic, thus it does not survive static aggregation, while $(C_{it})$, being orthogonal to $(e_{it})$, does. This is formalized in the next
 key result characterizing a static factor model and its common component 
 (see \citealp{gersing2023reconciling}, Section 2.2.3, for a proof).


%
\begin{theorem}[\citet{chamberlain1983arbitrage} - Static Representation Theorem]
\label{thm: projection on the static aggregation space}
Under Assumption A\ref{A: stat}:
\begin{itemize}
    \item[1.] $(y_{it})$ satisfies Assumption A\ref{A: r-SFS struct}  if and only if $\sup_{n \in \mathbb N} \mu_r(\Gamma_y^n) = \infty$ and $\sup_{n \in \mathbb N} \mu_{r+1}(\Gamma_y^n) < \infty$; 
    %
%
\item[2.]  $r$, $C_{it}$, and $e_{it}$ are uniquely identified from $(y_{it})$;
\item[3.]  $
C_{it} = \proj(y_{it} \mid \mathbb S_t(y))
$   for all  $i \in \mathbb N$ and $t \in \mathbb Z$;
\item[4.] $\proj(e_{it} \mid \mathbb S_t(y))=0$ for all  $i \in \mathbb N$ and $t \in \mathbb Z$.
\end{itemize}
\end{theorem}
\noindent Part 1 is also proved in \citet[Theorem 4]{chamberlain1983arbitrage} and \citet[Proposition 1]{barigozzi2025dynamic} and it is analogous to the proof in \citet[Theorem 2]{forni2001generalized}. 
By this result we see that the existence of an eigengap in the eigenvalues of $\Gamma_y^n$ implies the existence of a static factor decomposition. So,
by studying the eigenvalues of $\Gamma_y^n$ we can determine the number of static factors $r$ (see, e.g., \citealp{bai2002determining,ahn2013eigenvalue,Onatski2010_REStat}). Because of parts 1 and 2 we might therefore say that the static factor decomposition in Assumption A\ref{A: r-SFS struct} is a representation rather than just a statistical model. 

From part 3, we see that each static common component  $C_{it}$ can be expressed in terms of a mean square-limit of a static average of the observed variables $(y_{it})$, i.e., a cross-sectional average holding $t$ fixed, which survives aggregation and lives in the static aggregation space. By computing these averages, we filter out the static idiosyncratic component from the observed process, because its cross-sectional correlation, while not entirely absent, is sufficiently weak to be averaged away (part 4).

\paragraph{The Generalized Dynamic Factor Model.}
 Let $f_\chi^n(\theta)$ and $f_\xi^n(\theta)$ be the spectral density matrices of $(\chi_t^n)$ and $(\xi_t^n)$, respectively, at frequency $\theta \in [-\pi, \pi]$, with eigenvalues $\mu_j(f_\chi^n(\theta))$ and $\mu_j(f_\xi^n(\theta))$, $j=1,\ldots, n$, sorted in decreasing order. Then, we suppose that $(y_{it})$ has also a Generalized Dynamic Factor Model (GDFM) representation, as introduced by \cite{forni2000generalized} and \citet{forni2001generalized}. 
\begin{assumption}[$q$-Dynamic Factor Structure]\label{A: q-DFS struct}
The process $(y_{it})$ can be represented as: 
\[
y_{it}=\chi_{it}+\xi_{it}= \sum_{j=0}^\infty K_i(j)\varepsilon_{t-j}+\xi_{it},\quad i\in\mathbb N, \quad t\in\mathbb Z,
\]
$ K_i(j)=(K_{i1}(j)\cdots K_{iq}(j))$ and $\varepsilon_t$ are $1\times q$ and $q\times 1$, respectively, with $q < \infty$ and independent of $n$, $\sum_{j=0}^\infty\sum_{k=1}^q \vert K_{ik}(j)\vert^2<\infty$,
$\E[ \varepsilon_t]=0$, $\E [\varepsilon_t\varepsilon_t']=I_q$, 
$\E [\varepsilon_t\varepsilon_s']=0$, for all $t,s\in\mathbb Z$ with $t\ne s$,
$\E[\varepsilon_t\xi_{is}]=0$, for all $i \in \mathbb N$ and $t,s\in\mathbb Z$, and
\begin{itemize}    
    \item[(i)] $\sup_{n\in\mathbb N} \mu_q\left(f_\chi^n(\theta)\right) = \infty$ a.e. on $[-\pi,\pi]$; 
    \item[(ii)] $\esssup_{\theta\in[-\pi,\pi]} \sup_{n\in\mathbb N} \mu_1(f_\xi^n(\theta))<  \infty$,  where ``$\esssup$'' denotes the essential supremum of a measurable function.\footnote{For any real function $f$, $\esssup(f)= \inf\{M : \mathcal L(y : f( y)> M)=0\}$, where $\mathcal L$ is the Lebesgue measure on $\mathbb R$.
    }
\end{itemize}
\end{assumption}
We call a double sequence for which A\ref{A: q-DFS struct} holds a $q$-Dynamic Factor Sequence ($q$-DFS) \citep[see][Definition 10]{forni2001generalized}. Note that the original formulation of the $q$-DFS is stated in terms of two-sided filters. The existence of the innovation form in Assumption A\ref{A: q-DFS struct} with one-sided filters is proved in the ARMA framework in \cite{forni2015dynamic} and in more general terms in \cite{gersing2026on}.


Let $L_2^\infty(f_y)$ be the complex linear space of all infinite row vectors of complex valued equivalence classes $c = (c_1(\cdot)\, c_2(\cdot) \cdots)$, such that for all $i \in \mathbb N$:
(1.) $c_i:[-\pi, \pi] \rightarrow \mathbb C$ is a measurable function; 
(2.) $\lim_{n\to\infty} \int_{-\pi}^{\pi} c^{ \{n \} }(\theta) f_y^n (\theta) \{c^{\{ n \}}(\theta) \}^* d\theta < \infty$; 
(3.) the space $L_2^\infty(f_y)$ is endowed with the inner product
$        
\langle c, d \rangle_{f_y} := \lim_{n\to\infty} (2\pi)^{-1}\int_{-\pi}^\pi c^{\{n\}}(\theta) f_y^n(\theta) \{d^{\{n\}}(\theta)\}^* d\theta,
$
 and the norm $\norm{c}_{L_2^\infty(f_y)} := \sqrt{\langle c, c\rangle_{f_y}}$; 
[4.] two vectors $c_1, c_2$, are equivalent if $\norm{c_1- c_2}_{L_2^\infty(f_y)} = 0$. 
Accordingly, we write $L_2^\infty(I)$, if $f_y^n$ is the identity matrix $I_n$ for all $n$.

\citet[][Lemma 1, 2]{forni2001generalized}, show that also for infinite dimensional stochastic processes $(y_t)$ there exists an 
isomorphism  $\Phi(\cdot)$ between the frequency domain $L_2^\infty(f_y)$ and the time domain $\mathbb H(y)$, preserving the inner product,
%
%
so that for processes in $\mathbb H(y)$ that are outputs of linear filters, we write
$\ubar{c}(L)y_t := \Phi^{-1}\left(c(\theta) e^{\iota \theta t}\right)$ (see also \citealp[Chapter 4]{brockwell2009time}, for the finite dimensional case). 
%
%

Consider \textit{sequences} of infinite-dimensional row vectors of functions:
$(c^{(k)}\in L_2^\infty(f_y): k \in \mathbb N).$
Through the isomorphism $\Phi(\cdot)$ each $c^{(k)}$ defines a linear filter. The scalar valued limit of a filter sequence $(c^{(k)})$ applied to the double-sequence $(y_{it})$, is denoted as the mean-square limit:
$
    \ulim_{k\to\infty} \ubar c^{(k)}(L)y_t = \ulim_{k\to\infty}\ulim_{n\to\infty} \sum_{i=1}^n \ubar c_i^{(k)}(L)y_{it},$
where $c_i^{(k)}$ is the $i$-th entry of the infinite row vector of functions $c^{(k)}$.

\begin{definition}[Dynamic Averaging Sequence (DAS)]\label{def: dynamic averaging sequence}
Let $c^{(k)} \in L_2^\infty(I) \cap L_2^\infty(f_y)$ for $k \in \mathbb N$. The sequence of filters $(c^{(k)} : k \in \mathbb N)$ is called Dynamic Averaging Sequence (DAS) if
$
    \lim_{k\to\infty} \norm{c^{(k)}}_{L_2^\infty (I)} := \lim_{k\to\infty} \int_{-\pi}^\pi c^{(k)}(\theta)\left(c^{(k)}(\theta)\right)^* d \theta =  0 .
$
We denote the set of all DAS of $(y_{it})$  as    $\mathcal D (f_y) := \left \{ (c^{(k)}) : c^{(k)} \in L_2^\infty(I) \cap L_2^\infty(f_y) : k \in \mathbb N \mbox{  and  } \lim_{k\to\infty} \norm{c^{(k)}}_{L_2^\infty(I)} = 0 \right \}. $
\end{definition}

If $(c^{(k)})$ is a DAS, the scalar valued output process $(z_t)$ is called \textit{dynamic aggregate}. Similarly to a SAS, a DAS is a sequence of vanishing weights, but this time we compute averages employing not only the cross-section and but also leads and lags of the observed variables. We then consider the set of all random variables that can be written as the limit of a DAS.
\begin{definition}[Dynamic Aggregation Space]\label{def :DynAggSpace}
The space $\mathbb D(y) := \left\{ z_t : z_t = \ulim_{k\to\infty}\ubar c^{(k)} (L) y_t \ \mbox{where} \ \left(c^{(k)}\right) \in \mathcal D(f_y)  \right\}$ is called Dynamic Aggregation Space.
\end{definition}
%
%
For a stationary double sequence $(y_{it})$, satisfying A\ref{A: stat}, the dynamic aggregation space $\mathbb D(y)$ is a closed subspace of the time domain $\mathbb H(y)$ \citep[see][Lemma 6]{forni2001generalized}. Note that in contrast to the static aggregation space $\mathbb S_t(y)$, the dynamic aggregation space $\mathbb D(y)$ does not depend on time.
\begin{definition}[Dynamically Idiosyncratic]
A stochastic double sequence $(z_{it})$ is called dynamically idiosyncratic if it satisfies A\ref{A: stat} and
$
    \ulim_{k\to\infty} \ubar c^{(k)}(L) z_t = 0$ for all  $(c^{(k)}) \in \mathcal D (f_z)$.
\end{definition}
%
We can then express dynamic idiosyncraticness equivalently in terms of spectral eigenvalues \citep[see][Theorem 1, for a proof]{forni2001generalized}.
\begin{theorem}[Characterisation of Dynamically Idiosyncratic Components]\label{thm: charact dyn idiosyncratic}
The following statements are equivalent:
\begin{itemize}
    \item[(i)] a stochastic double sequence $(z_{it})$ is dynamically idiosyncratic; 
    \item[(ii)] 
    $
        \esssup_{\theta\in[-\pi,\pi]} \, \sup_{n \in \mathbb N} \mu_1(f_z^n(\theta)) < \infty .
    $
\end{itemize}
\end{theorem}

Analogously to the static case, by Theorem \ref{thm: charact dyn idiosyncratic}, under Assumption A\ref{A: q-DFS struct}, $(\xi_{it})$ is dynamically idiosyncratic, thus it does not survive dynamic aggregation, while $(\chi_{it})$, being orthogonal at all leads and lags to $(\xi_{it}),$ does.
This is formalized in the next result (see \citealp{forni2001generalized}, Theorems 2 and 3, for a proof).

\begin{theorem}[\citet{forni2001generalized} - Dynamic Representation Theorem]\label{thm: projection on the dynamic aggregation space}
Under Assumptions A\ref{A: stat}:
\begin{itemize}
\item[1.] $(y_{it})$ satisfies Assumption A\ref{A: q-DFS struct} if and only if $\sup_{n\in \mathbb N} \mu_q\l(f_y^n(\theta)\r) = \infty$ a.e. on $[-\pi, \pi]$ and \linebreak $\esssup_{\theta\in[-\pi,\pi]} \, \sup_{n \in \mathbb N} \mu_{q+1}(f_y^n)(\theta) < \infty$;
\item[2.] $q$, $\chi_{it}$, and $\xi_{it}$ are uniquely identified from $(y_{it})$;
\item[3.] $
 \chi_{it} = \proj(y_{it} \mid \mathbb D(y))$  for all  $i \in \mathbb N$ and $t \in \mathbb Z$;
 \item[4.] $\proj(\xi_{it} \mid \mathbb D(y))=0$  for all  $i \in \mathbb N$ and $t \in \mathbb Z$.
 \end{itemize}
\end{theorem}

From part 1 it follows that the existence of an eigengap in the spectral eigenvalues implies the existence of a dynamic factor decomposition. 
So by looking at the eigenvalues of $f_y^n$ we can determine the number of dynamic factors $q$ (see, e.g., \citealp{hallin2007determining,onatski2009testing}). Because of parts 1 and 2 we might therefore say that also the dynamic factor
decomposition in Assumption A\ref{A: q-DFS struct} is a representation rather than just a statistical model.

Part 3, which is also proved in \citet[Theorem ??]{hallin2013factor}, implies that each dynamic common component $\chi_{it}$ can be expressed in terms of a mean-square limit of a dynamic average of the observed variables $(y_{it})$, which survives aggregation and lives in the dynamic aggregation space. By computing dynamic averages,
we filter out the dynamic idiosyncratic component from the observed process (part 4).

\subsection{The canonical decomposition of factor models}\label{subsec: canonical}
Starting from the results from the previous sections, we can prove our first main result.
\begin{theorem}[The Canonical Decomposition of Factor Models]\label{thm: relation of r-SFS and q-DFS}
Under Assumption A\ref{A: stat}, the following holds.
\ben
\item[1.] For all $t \in \mathbb Z$, $\mathbb S_t(y) \subset \mathbb D(y)$, except if $(y_{it})$ is dynamically idiosyncratic, then $\mathbb S_t(y) = \mathbb D(y) = \l\{0\r\}$.
\item[2.] If also Assumptions A\ref{A: r-SFS struct} and A\ref{A: q-DFS struct} hold, then 
$(y_{it})$ admits the decomposition: 
    \begin{align}
        y_{it} = C_{it} + e_{it}^\chi + \xi_{it}, \label{eq: gen decomp}
    \end{align}
where $\chi_{it} = C_{it} + e_{it}^\chi$ and $e_{it} = e_{it}^\chi + \xi_{it}$, with 
$\E[e_{it}^\chi \xi_{js}]=0$,
$\E[C_{it} \xi_{js}]=0$, and
$\E[C_{it} e_{jt}^\chi]=0$
for all $i, j\in\mathbb N$ and  $s, t\in\mathbb Z$. Furthermore $(e_{it}^\chi)$ is the static idiosyncratic component of $(\chi_{it})$, i.e., is statically idiosyncratic.

%
\item[3.] The decomposition in \eqref{eq: gen decomp} is unique.
\een
\end{theorem}

\begin{proof} 
Every SAS is a DAS with zero coefficients at all leads and lags. It follows that every static aggregate is a dynamic aggregate and therefore $\mathbb S_t(y) \subset \mathbb D(y)$ for all $t \in \mathbb Z$. The second statement is trivial. 
This proves part 1.

Now, for all $t\in\mathbb Z$, consider the orthogonal decomposition
\begin{equation}
\chi_{it}=\proj\left(\chi_{it}\mid \mathbb S_t(y)\right)+e_{it}^\chi.\label{eq:decomp chi}
\end{equation}
Then, since
\begin{align}
    C_{it} &=  \proj\left( y_{it} \mid \mathbb S_t(y)\right) \qquad \mbox{(by  Theorem \ref{thm: projection on the static aggregation space})}\nonumber \\
    &= \proj\left( \chi_{it} \mid \mathbb S_t(y)\right) +  \proj\left(\xi_{it} \mid \mathbb S_t(y)\right) \qquad \mbox{(by Assumption A\ref{A: q-DFS struct})}\nonumber\\
    &= \proj\left( \chi_{it} \mid \mathbb S_t(y)\right) \qquad \mbox{(by part 1 and Theorem \ref{thm: projection on the dynamic aggregation space}) },\label{eq:chi on S}
\end{align}
%
we have 
\begin{equation}
    e_{it}^\chi = \chi_{it} - C_{it}= y_{it}-\xi_{it}-C_{it}=e_{it}-\xi_{it},\label{eq:decomp_echi}
\end{equation}
because of \eqref{eq:decomp chi} and Assumptions A\ref{A: r-SFS struct} and A\ref{A: q-DFS struct}.
Moreover, since $e_{it}^\chi \in \mathbb D(y)$ by \eqref{eq:decomp chi}, then
\begin{itemize}
\item [(a)] $\E[e_{it}^\chi  \xi_{js}]=0$ for all $i,j\in\mathbb N$ and $s, t\in\mathbb Z$, because of Assumption 
A\ref{A: q-DFS struct};

\item [(b)] $C_{it}\in \mathbb D(y)$ by \eqref{eq:decomp_echi}, and so $\E[C_{it} \xi_{js}]=0$ for all $i,j\in\mathbb N$ and $s, t\in\mathbb Z$, because of Assumption 
A\ref{A: q-DFS struct}.
\end{itemize}
Then, from \eqref{eq:decomp_echi} we also have $\E[C_{it} e_{jt}^\chi]=\E[C_{it} e_{jt}]-\E[C_{it} \xi_{jt}]=0$, for all $i,j\in\mathbb N$ and $t\in\mathbb Z$, by Assumption A\ref{A: r-SFS struct} and (b).
Since $\Gamma_e^n = \Gamma_{e^\chi}^n + \Gamma_{\xi}^n$ it follows that $\sup_{n\in\mathbb N} \mu_1\l(\Gamma_{e}^\chi\r) \leq \sup_{n\in \mathbb N}\mu_1\l(\Gamma_e^n\r) < \infty$, by Weyl's inequality \citep[Theorem 1]{merikoski2004inequalities} and Assumption A\ref{A: r-SFS struct}, so $(e_{it}^\chi)$ is statically idiosyncratic. 

The uniqueness of the decomposition follows directly from the uniqueness of the static and dynamic factor decompositions (Theorems \ref{thm: projection on the static aggregation space} and \ref{thm: projection on the dynamic aggregation space}) or more generally from the uniqueness of the orthogonal projection theorem \citep[Proposition 1.11]{deistler2022modelle}. This completes the proof.
\end{proof}
%
%
%
%
%
%

Hereafter, we call $(e_{it}^\chi)$ in (\ref{eq: gen decomp})  the {\it weak common component}. We now introduce a series of remarks aimed at explaining the meaning of Theorem \ref{thm: relation of r-SFS and q-DFS}.

\begin{rem}[Interpreting the weak common component]
\upshape{
On the one hand, $(e_{it}^\chi)$ is the residual term from the projection of the dynamic common component on the static aggregation space $\mathbb S_t(y)$. As such, it is the static idiosyncratic component of the dynamic common component, hence, it vanishes under static averaging and it lives in the dynamic aggregation space $\mathbb D(y)$. On the other hand, $(e_{it}^\chi)$ is also the projection of the static idiosyncratic component on the dynamic aggregation space, i.e. $e_{it}^\chi = \proj(e_{it} \mid \mathbb D(y))$ or that part of the static idiosyncratic component which is dynamically common to $(y_{it})$.  }
\end{rem}

\begin{rem}
[Explained variance of the dynamic and static common components]
\upshape{The dynamic common component is the projection of $(y_{it})$ onto the Hilbert space $\mathbb D(y)$ (Theorem \ref{thm: projection on the dynamic aggregation space}). Now, since, by Theorem \ref{thm: relation of r-SFS and q-DFS}, $\mathbb S_t(y)\subset \mathbb D(y)$, it follows that, in general, the dynamic common component explains a larger part of the variation of the observables $(y_{it})$ with respect to the static common component. Moreover,   $\mathbb D(y)$ contains \textit{all} static aggregation spaces, i.e., $\cspargel\left(\bigcup_{t \in \mathbb Z} \mathbb S_t (y) \right) = \cspargel(C_{it}: i \in \mathbb N, t \in \mathbb Z)\subseteq \mathbb D(y)$.}
%
%
\end{rem}


\begin{rem}[No serial correlation implies no weak common component]\label{rem:whitenoise}
    \upshape{
    Let $(y_{it})$ be such that $\E[y_{it} y_{jt-h}]=0$ for all $i,j\in\mathbb N$ and all $h\in\mathbb Z\setminus\{0\}$ so that $(y_t^n)$ is a white noise process for all $n\in\mathbb N$.  This is a case where there are no weak common components.
    To see this, first note that $f_y^n(\theta)=\Gamma_y^n$ for all $\theta\in[-\pi,\pi]$ and all $n\in\mathbb N$,
    so the spectral density and the covariance matrix have the same eigenvalues and eigenvectors and it must be that $q=r$. Let $V_n(\theta)$ and $\Pi_n$ be the $q\times n$ matrices having as rows the normalized eigenvectors of $f_y^n(\theta)$ and $\Gamma_y^n$, respectively, corresponding to their $q$-largest eigenvalues. 
    Then, $V_n(\theta)=\Pi_n$ for all $\theta\in[-\pi,\pi]$ and all $n\in\mathbb N$.
    It follows that, for all $i\in\mathbb N$,
    \begin{align}
    \chi_{it}&
    =\ulim_{n\to\infty} s_i'\ubar V_n'(L)\ubar V_n(L) y_t^n\nn\\
    &=\ulim_{n\to\infty} s_i'
    \l\{\frac 1{2\pi}\sum_{k=-\infty}^\infty \l(\int_{-\pi}^\pi V_n(\theta) e^{ik\theta} \mathrm d\theta\r) L^k\r\}'
\l\{\frac 1{2\pi}\sum_{k=-\infty}^\infty \l(\int_{-\pi}^\pi V_n(\theta) e^{ik\theta} \mathrm d\theta\r) L^k\r\} y_t^n\nn\\
&=\ulim_{n\to\infty} s_i'
    \l\{\frac 1{2\pi}\Pi_n \sum_{k=-\infty}^\infty \l(\int_{-\pi}^\pi  e^{ik\theta} \mathrm d\theta\r) L^k\r\}'
\l\{\frac 1{2\pi} \Pi_n \sum_{k=-\infty}^\infty \l(\int_{-\pi}^\pi  e^{ik\theta} \mathrm d\theta\r) L^k\r\} y_t^n\nn\\
    &= \ulim_{n\to\infty} s_i'\Pi_n'\Pi_n y_t^n = C_{it},\nn
    \end{align}
    where $s_i$ is the $i$th element of the canonical basis of $\mathbb R^{n\times 1}$, the first equality comes from 
    \citet[Theorem 5]{forni2001generalized}, the last follows from Assumption A\ref{A: r-SFS struct} and Theorem \ref{thm: charact stat idiosyncratic}, and we recall that $\int_{-\pi}^\pi  e^{ik\theta} \mathrm d\theta=0$ for all $k\in\mathbb Z\setminus\{0\}$.
  In practice, it is extremely unlikely that a large panel of time series is made of serially uncorrelated elements. 
    }
\end{rem}

\begin{rem}[Correlation between $C_{it}$ and $e_{it}$ at leads and lags is possible]
\label{rem:BAING}
\upshape{
From the point-of view of the static factor model (\ref{eq: static factor model rep}),  assuming uncorrelatedness across all leads and lags between $(C_{it})$ and $(e_{it})$ \citep[see e.g.][]{forni2005generalized,forni2009opening,FL2024hallin} or even independence \citep[see e.g.][]{bai2006confidence,doz2011two,anderson2022linear,forni2025common},  would be unduly restrictive as it would imply $e_{it}^\chi=0$ for all $i\in\mathbb N$. 
All we must require is just {contemporaneous} orthogonality, i.e., $
\E[C_{it}e_{it}]=0$ for all $i\in\mathbb N$ and $t\in\mathbb Z$, which, in serially correlated time-series panels, does not imply anything about the correlation at lags and leads. This is also enough for consistent estimation of $C_{it}$ via static PCA (see Proposition \ref{prop3}). In this respect, notice also that \citet{bai2002determining} and \citet{bai2003inferential}, among many others, assume only the ``weak-dependence'' condition:
$\sup_{n,T\in\mathbb N} \max_{i=1,\ldots, n}
 \E[T^{-1}(\sum_{t=1}^T C_{it} e_{it})^2] \le M<\infty$, which it still implies 
{contemporaneous} orthogonality, while allowing for dependence but only at the level of fourth-order moments. 

 }
\end{rem}

\begin{rem}
[The weak common component  does not need to be dynamically idiosyncratic]\label{rem: weak CC not dyn idio}
\upshape{
Although Theorem \ref{thm: relation of r-SFS and q-DFS} shows that the weak common component, $(e_{it}^\chi)$ is always statically idiosyncratic, i.e., $\sup_{n\in \mathbb N}\mu_1\l(\Gamma_{e^\chi}^n\r) < \infty$, it is not necessarily dynamically idiosyncratic. In other words, even though the cross-sectional correlation of $(e_{it}^\chi)$ is weak, the serial correlation can by so strong that the largest spectral eigenvalue diverges, i.e., $\sup_{n\in \mathbb N} \mu_1\l(f_{e^\chi}^n(\theta)\r)  = \infty$ a.e. on $[-\pi, \pi]$, as illustrated in Example \ref{exmp: 1-DFS but 0-SFS}, where $r=0$ but $q>0$. Consequently, the typical assumption that $q \leq r$, i.e., more static factors than common shocks, might not hold \citep{forni2005generalized,forni2009opening,forni2025common,stock2016dynamic}. 
}
\end{rem}

%
%
%
\begin{example}[A dynamic common component without static common component: $q=1$ and $r=0$]\label{exmp: 1-DFS but 0-SFS}
\upshape{Let $(\varepsilon_t)$ be a scalar white noise process with unit variance. Consider a dynamic common component of the form
$
    \chi_{it} = \varepsilon_{t-i+1}.
    $
Then, the spectral density of $(\chi_{it})$ is 
\begin{align*}
f_\chi (\theta) = \begin{pmatrix}
    1                  & e^{\iota \theta} & e^{2\iota \theta} & \cdots \\
    e^{-\iota\theta}   & 1                & e^{\iota \theta} & \cdots   \\
    e^{-2\iota \theta} &  e^{-\iota\theta}                &  1          & \cdots\\
      \vdots           &      \vdots            &       \vdots       & \ddots  
\end{pmatrix} ,
\end{align*}
so the first row of $f_\chi^n$ equals the $k$-th row of $f_\chi^n$ times $e^{\iota k\theta}$. Therefore, $f_\chi^n$ has rank one a.e. on $[-\pi, \pi]$ with $\mu_1(f_\chi^n(\theta)) = \tr (f_{\chi}^n(\theta)) = n$. 
By Theorem \ref{thm: projection on the dynamic aggregation space} it follows that $q=1$.
Moreover, 
$\Gamma_\chi^n = I_n$ for all $n\in\mathbb N$, and so the largest eigenvalue of $\Gamma_\chi^n$ is bounded for all $n$. Thus, by Theorem \ref{thm: projection on the static aggregation space}, $r=0$, and, therefore, 
$C_{it}=0$ and  $\chi_{it} = e_{it}^\chi$, for all  $i\in \mathbb N$, $t \in \mathbb Z$,
by the canonical decomposition \eqref{eq: gen decomp}. This means that the weak common component is dynamically common and contains all predictive power. Indeed,
 since $\chi_{2,t+1} = \varepsilon_{t+1 - 2 + 1} = \varepsilon_{t} = \chi_{1t}$ and $\chi_{3, t+1} = \varepsilon_{t+1-3 +1} = \varepsilon_{t-1} = \chi_{2t}$ and so, then, for any $i\geq 2$, can perfectly predict $\chi_{i,t+1}$ through $\chi_{i-1,t}$. However, if we were to consider just a static aggregate of $(\chi_{it})$, then all predictive power would be lost.
 }
\end{example}

%
%
%
%
%
%
%
%
%
%
%

Let us now introduce the following definition.

\begin{definition}[Static Factors]\label{def: static factors}
We call \textit{static factor} any basis coordinate of $\cspargel (\chi_{it}: i \in \mathbb N)$, i.e., any basis coordinate that drives the dynamic common component. 
\end{definition}

From Theorem \ref{thm: relation of r-SFS and q-DFS} we know that the weak common component is statically idiosyncratic. Hence, the basis coordinates of $\cspargel\l(e_{it}^\chi : i \in \mathbb N\r)$, which, according to Definition \ref{def: static factors}, are static factors, must be loaded non-pervasively. This is formalized in the next result, which gives an alternative expression for the canonical decomposition \eqref{eq: gen decomp}, this time formulated in terms of pervasive and weak static factors. 
\begin{theorem}[Canonical Decomposition in terms of Pervasive and Weak Static Factors]\label{thm: weakfactors}
     Under Assumptions A\ref{A: stat}, A\ref{A: r-SFS struct}, and A\ref{A: q-DFS struct}, for any fixed $n\in\mathbb N$, we can write the canonical decomposition \eqref{eq: gen decomp}, in vector notation, as:
\begin{align}
    y_t^n &= C_t^n + e_t^{\chi, n} + \xi_t^n = \Lambda^n F_t + \Lambda^{w, n} F_t^{w, n}  + \xi_t^n 
    %
    , \label{eq: canonical rep weak and strong factors}
\end{align}
which implies $\chi_t^n=\Lambda^n F_t + \Lambda^{w, n} F_t^{w, n}$, 
where:
\begin{itemize}
    \item [(i)] $\Lambda^n$ is $n\times r$
    such that  $\sup_{n\in\mathbb N}\mu_r(\Lambda^{n'}\Lambda^n) = \infty$, $F_t$ are $r$  pervasive static factors with $\E[F_t F_t'] = I_r$, so $C_t^n= \Lambda^n F_t$; 
\item [(ii)] $\Lambda^{w, n}$ is $n\times r_w(n)$ with $r_w(n) \leq n$ and such that
$\sup_{n\in \mathbb N} \mu_1(\Lambda^{w, n\, \prime} \Lambda^{w, n}) < \infty$, 
$F_t^{w, n}$ are $r_w(n)$ weak static factors with 
$\E[F_t^{w, n} {F_t^{w, n\, \prime}}] = I_{r_w(n)}$;

\item [(iii)] $\E[F_t^{w, n}F_t'] = 0$ for all $t\in \mathbb Z$;

\item [(iv)] $\cspargel(F_t^{w,n})\subseteq \mathbb D(y)$.
\end{itemize}

\end{theorem}

\begin{proof} Part (i) follows directly from Assumption A\ref{A: r-SFS struct} with $\Lambda^n=(\Lambda_1'\cdots \Lambda_n')'$, and from Theorem \ref{thm: projection on the static aggregation space}. Indeed, $\sup_{n\in\mathbb N}\mu_r(\Lambda^{n'}\Lambda^n)=\sup_{n\in\mathbb N}\mu_r(\Gamma_C^n)=\infty$.

For parts (ii) and (iii), 
to obtain an orthonormal basis $F_t^{w, n}$, we use the Gram-Schmidt orthogonalisation procedure and orthogonalise with respect to $F_t$. Choose the first $i$ in order for which $\chi_{it} - \proj(\chi_{it} \mid \cspargel(F_t)) \neq 0$, set this to $i_1$. Set $v_{1t} = \chi_{{i_1}, t} - \proj(\chi_{i_1, t} \mid \cspargel(F_t))$ and set $F_{1t}^w = \norm{v_{1t}}^{-1} v_{1t}$. Let $i_2 > i_1$ be the next $i$ in order such that $\chi_{it} - \proj(\chi_{it} \mid \cspargel(F_t, F_{1t}^w)) \neq 0$ and set $v_{2t} = \chi_{i_2,t} - \proj(\chi_{i_2, t} \mid \cspargel(F_t, F_{1t}^w))$ and $F_{2t}^{w} = \norm{v_{2t}}^{-1} v_{2t}$. In this way we obtain indices $i_1, i_2, \ldots, i_{r_w(n)}$ with $ r_{w}(n) \leq n$ together with $F_t^{w,n} = (F_{1t}^w \cdots F_{r_w(n), t}^w)'$ which, by construction, is such that $\E[F_t^{w,n}{F_t^{w,n\,\prime}}]=I_{r_w(n)}$ and also $\E[F_{it}^{w,n}F_{jt}]=0$ for all $i,j=1,\ldots, n$ and $t\in\mathbb Z$.

Furthermore, because $F_t$ and $F_t^{w,n}$ are orthogonal, then $\Lambda^{w,n} F_t^{w, n}$ are the residuals from the projection of $\chi_t^n$ on $\mathbb S_t(y)$. And since  $e_t^{\chi,n}$ is statically idiosyncratic, by 
Theorem \ref{thm: relation of r-SFS and q-DFS}, then,
$\mathbb S_t(y)= \mathbb S_t(\chi)$. Therefore, $e_t^{\chi,n}=\Lambda^{w,n} F_t^{w, n}$ and
$
\sup_{n \in \mathbb N}\mu_1 (\Lambda^{w, n\,\prime} \Lambda^{w, n}) =\sup_{n\in \mathbb N}\mu_1\l(\Gamma_{e^\chi}^n\r) < \infty,
$
by Theorem \ref{thm: charact stat idiosyncratic}.

For part (iv), $\cspargel(F_t^{w,n})=\cspargel (e_{t}^{\chi,n}) \subseteq\cspargel (\chi_{t}^n)
\subseteq\cspargel (\chi_{it}: i \in \mathbb N, t\in\mathbb Z) = \mathbb D(y)$, 
by parts (i) and (ii), Theorems \ref{thm: projection on the dynamic aggregation space} and \ref{thm: relation of r-SFS and q-DFS}, and Definition \ref{def :DynAggSpace}.
\end{proof}

We introduce the following terminology to distinguish between two types of static factors.

\begin{definition}[Pervasive and Weak Static Factors]\label{def:pervasive and weak}
For any fixed $n\in\mathbb N$,
let 
\[
r_\chi(n):=r+r_w(n).
\]
By Assumption A\ref{A: r-SFS struct}, $r$ is independent of $n$. Moreover, let
$r_w:=\sup_{n\in\mathbb N} r_w(n)$ and
\beq
r_\chi:=\sup_{n\in\mathbb N} r_\chi(n)
= r+r_w.
\nonumber
\eeq
From Theorem \ref{thm: weakfactors} $r_\chi(n)$ is the total number of static factors driving $\chi_t^n$, so, by Definition \ref{def: static factors},
$r_\chi(n)=\dim\cspargel(\chi_{it}, i=1,\ldots, n)$ and
$\Gamma_\chi^n:=\E[\chi_t^n\chi_t^{n'}]$ has rank $r_\chi(n)$. According to the representation in \eqref{eq: canonical rep weak and strong factors}, the $r_\chi(n)$ static factors are then distinguished  between:
\ben
 \item [(i)] {\it pervasive static factors} which are the $r$ coordinates of $(F_t)$, associated to those eigenvalues of $\Gamma_\chi^n$ which diverge as $n\to\infty$; these include both strong factors, i.e., when the eigenvalues are diverging linearly at rate $n$, as well as rate-weak factors, i.e., when the eigenvalues are diverging at rate $n^\alpha$, with $\alpha\in(0,1)$;
 \item [(ii)] {\it weak (non-pervasive) static factors} which are the $r_w(n)$ coordinates of $(F_t^{w,n})$, associated to those eigenvalues of the  $\Gamma_\chi^n$ which are uniformly bounded for all $n\in\mathbb N$.
\een
\end{definition}

We now introduce some remarks on the meaning and nature of weak factors.

\begin{rem}[Weak factors are everywhere]
\upshape{
A non-null weak common component is to be expected in every high-dimensional panel of time series. Indeed, this is the case for all those series for which the dynamic common component has a larger variance than the static one. Now, by Theorem \ref{thm: weakfactors} the presence of a weak common component is equivalent to the presence of weak factors. 
It follows that although weak factors are not pervasive so they are not to be expected in each series, they are to be expected in every high-dimensional panel of time series meaning that either few series load them with relatively large loading coefficients (in the case of sparse loadings) or that all series load them very weakly. 
This is what we mean by claiming that ``Weak Factors are Everywhere''. Although weakly influential for the whole panel excluding them beforehand would lead to a mis-specification that can be substantial when looking at individual series (see, e.g., the empirical application in Section \ref{sec: empiric}). 
}
\end{rem}

\begin{rem}[Recovering the weak factors in the dynamic common space]
\label{rem: weakF}
\upshape{For any fixed $n\in\mathbb N$, consider the factor model 
$y_t^n = \Phi^{w,n} G_t^w + z_t^n$ with $G_t^w$ being $k$-dimensional with $k$ finite and independent of $n$, $\Phi^{w,n}$ being $n\times k$ such that
$\sup_{n \in \mathbb N}\mu_1 ({\Phi^{w, n'}}\Phi^{w, n} )<\infty$, and $z_t^n$ being statically idiosyncratic, i.e., $\sup_{n\in\mathbb N}\mu_1(\E[z_t^nz_t^{n'}])<\infty$. This model, is studied by \cite{onatski2012asymptotics}, is a static factor model driven only by a finite number of weak factors, $G_t^{w}$, associated to $k$ non-diverging eigenvalues of $\Gamma_y^n$. Similarly to $(G_t^{w})$ also the weak factors $(F_t^{w,n})$ in representation \eqref{eq: canonical rep weak and strong factors} in Theorem \ref{thm: weakfactors} are associated to non-diverging eigenvalues, but of $\Gamma_\chi^n$ rather than $\Gamma_y^n$. Indeed, the $(F_t^{w,n})$ live in the dynamic common space of $(y_{it})$.
This distinction between $(G_t^w)$ and $(F_t^{w,n})$ is fundamental for their estimation. 

On the one hand, PCA cannot produce consistent estimates of $G_t^{w}$ \citep[see][Theorem 1, for a proof]{onatski2012asymptotics}. Intuitively, the reason is that PCA works as long as we can disentangle the static common component from the static idiosyncratic one, and when the latter is cross-correlated, this separation is only feasible as $n \to \infty$. However, because the factors are weak the information they convey does not grow with $n$, so this usual ``blessing of dimensionality'' argument no longer applies. 

On the other hand, we can consistently estimate the weak factors $F_t^{w,n}$ via PCA applied to the estimated weak common component---provided they are a finite number and an identification scheme is imposed.
The intuition for this result is simply that in this case $F_t^{w,n}$ is a finite dimensional basis of $\cspargel(e_{it}^\chi, i\in\mathbb N)$ and, in turn, $e_{it}^\chi$ can always be consistently recovered as the  difference between the estimated static and dynamic common components, i.e., by exploiting the dynamic information. We refer to Section \ref{sec: estimation} for details on the estimation procedure. 
Clearly, if the dimension of $F_t^{w,n}$ grows with $n$, then $\cspargel(e_{it}^\chi, i\in\mathbb N)$ has an infinite dimensional basis which cannot be estimated using a fixed number of weak common components. And, obviously, we cannot estimate any weak factors driving the dynamic idiosyncratic component either, for the same reason that we could not estimate $G_t^w$. 

}
\end{rem}

\begin{rem}[There can be infinite static pervasive and/or weak factors]
\label{rem:infinite weak}
\upshape{In general, the total number of static factors does not have to be fixed. On the one hand, the number of weak factors, $r_w(n)$, may increase with $n$, i.e., by adding new variables in (\ref{eq: canonical rep weak and strong factors}) new weak factors appear. On the other hand, we could also have $r=\infty$, so that the static representation in Assumption A\ref{A: r-SFS struct} does not hold for $(y_{it})$, but we can still have a dynamic representation as in Assumption A\ref{A: q-DFS struct}.
Both these cases imply that  $r_\chi=r+r_w=\infty$ and are illustrated in Examples \ref{exmp: infinitely many weak factors} and \ref{exmp: infinite dimensional CCC-space}, respectively.
It follows that, in general, the rank of the dynamic common component covariance matrix, $\Gamma_\chi^n$, is not necessarily fixed and independent of $n$, as instead often assumed in the literature \citep{forni2009opening, doz2011two,stock2016dynamic}. As a consequence, $(\chi_{it})$ cannot be estimated using standard static PCA and must instead be obtained through alternative estimation methods \citep{forni2000generalized, forni2017dynamic}, which are discussed in Section \ref{sec: estimation}.
}
\end{rem}

\begin{example}[Infinitely many weak factors but one statically pervasive factor: $r_w = \infty$ and $r = 1$]
\label{exmp: infinitely many weak factors}
\upshape{
Let $(\varepsilon_t)$ be scalar white noise and $\alpha_i \sim iidU(-0.9,0.9)$. Define the dynamic common component by
$
    \chi_{it} = \lambda_{i0} \varepsilon_t + \lambda_{i1} (1 - \alpha_i L)^{-1} \varepsilon_{t-1} , 
$
with $\lim_{n\to\infty}\sum_{i=1}^n \lambda_{i0}^2 = \infty$, while $\lim_{n\to\infty}\sum_{i=1}^n \lambda_{i1}^2 < \infty$ and $\lambda_{i1} \neq 0$ for all $i\in\mathbb N$. 
Clearly, $q=1$, and $C_{it}=\lambda_{i0} \varepsilon_t$, i.e., $\dim\mathbb S_t(y)=r=1$. Then, $e_{it}^\chi= \lambda_{i1} (1 - \alpha_i L)^{-1} \varepsilon_{t-1}$ and $\dim \cspargel(e_{it}^\chi: i \in \mathbb N)= \sup_{n\in\mathbb N} r_w(n) =r_w= \infty$, i.e., we have infinitely many weak  factors.
}
\end{example}
\begin{example}[Infinitely many statically pervasive factors: $r = \infty$]\label{exmp: infinite dimensional CCC-space}
\upshape{Let $(\varepsilon_t)$ be scalar white noise and $\alpha_i \sim iidU(-0.9,0.9)$.  Define a dynamic common component by
%
%
%
%

%
\begin{align}
\begin{pmatrix}
\chi_{1t}\\ \hline
  \chi_{2t} \\
  \chi_{3t}\\ \hline
 \chi_{4t}\\
 \chi_{5t} \\
 \chi_{6t} \\ \hline  \vdots 
\end{pmatrix} 
= 
\begin{pmatrix}
(1-\alpha_1 L)^{-1} \\ \hline 
(1-\alpha_1 L)^{-1}  \\
(1-\alpha_2 L)^{-1}  \\ \hline
(1-\alpha_1 L)^{-1}  \\
(1-\alpha_2 L)^{-1} \\
(1-\alpha_3 L)^{-1}  \\ \hline  \vdots 
\end{pmatrix} \varepsilon_t=:\begin{pmatrix}
z_{1t}\\ \hline
z_{1t} \\
z_{2t}\\ \hline
z_{1t}\\
z_{2t} \\
z_{3t} \\ \hline  \vdots 
\end{pmatrix}.\nn
\end{align}
Clearly, $q=1$, but,
%
since each $z_{it}$ is replicated infinite times,
we can produce infinitely many linearly independent static aggregates 
$F_{it}=z_{it}$, $i\in\mathbb N$, by computing the cross-sectional average over the sub-sequence $(\chi_{jt})$ selecting only $z_{it}$. Consequently $\dim\mathbb S_t(y)=r=\infty$.
%
%
}
\end{example}




We then discuss the conditions under which the static and dynamic common components coincide (see Appendix \ref{app:th2} for a proof).
\begin{proposition}[Relationship between the Dynamic and the Static Approach]\label{thm: finite dimensional CCC}
Under Assumptions A\ref{A: stat} and A\ref{A: q-DFS struct} the following holds.
\ben
\item [1.] If $ r_\chi < \infty$, then $(y_{it})$ satisfies Assumption A\ref{A: r-SFS struct} with $0\leq r\leq r_\chi$.

%
%
%


\item [2.] Assumption A\ref{A: r-SFS struct} holds with $C_{it} = \chi_{it}$, for all $i \in \mathbb N$, $t \in \mathbb Z$ if and only if $r_\chi < \infty$ and $r_\chi =r$.
\een
\end{proposition}
%



\begin{rem}[The dynamic factor model with finite dimensional dynamic common component]
\label{rem: stacking}
\upshape{
 Proposition \ref{thm: finite dimensional CCC}.1 shows that 
 if $(y_{it})$ admits a GDFM representation as in Assumption A\ref{A: q-DFS struct}, but with finite dimensional dynamic common component, i.e., with $r_\chi < \infty$, then it admits also a static factor representation as in Assumption A\ref{A: r-SFS struct} with $r\le r_\chi$ static pervasive factors. To further illustrate this, let us consider a prototypical  dynamic factor model satisfying Proposition \ref{thm: finite dimensional CCC}.1:
%
\begin{align}
 y_{it} &= \chi_{it}+\xi_{it},\qquad
 \chi_{it}=\ubar \lambda_i(L) f_t
=  \lambda_{i0} f_{t}+\ldots+ \lambda_{ip} f_{t-p}, 
 \label{eq: SW example dyn fac}\\
 \ubar a (L) f_t &= (1 - a_1 L - \ldots - a_{p_f} L^{p_f}) f_t = b \varepsilon_t, \label{eq: SW example dyn fac2}
\end{align}
where $(f_t)$ is a $q$-dimensional process, $\ubar a (L)$ is a stable $q\times q$ polynomial filter,
$(\varepsilon_t)$ is a $q$-dimensional orthonormal white noise, and $b$ is $q\times q$ with full-rank. 
%
 Let $\lambda_j^n=(\lambda_{1j}'\cdots \lambda_{nj}')'$, $j=1,\ldots, p$, which are  matrices of dimensions $n\times q$. Henceforth, we also assume that: 
\beq
\sup_{n \in \mathbb N}\, \max_{j = 0,\ldots, p} \mu_q ({\lambda_j^n}' \lambda_j^n)= \infty.\label{eq:divergingloadings}
\eeq
Furthermore, it is assumed that $(\xi_{it})$ is dynamically idiosyncratic and orthogonal to $(\varepsilon_t)$ at all leads and lags.
In Appendix \ref{app:example gdfm} we prove that, under the above conditions, $(y_{it})$ satisfies Assumption A\ref{A: q-DFS struct}, hence it admits a GDFM representation. Clearly, $(y_{it})$ satisfies  Assumption A\ref{A: stat} too. From \eqref{eq: SW example dyn fac} we see that, for all $n\in\mathbb N$,
the dynamic common component can be written as: 
\begin{equation}
    \chi_t^n=
        \left(
        \lambda_0^n \cdots \lambda_p^n
        \right)
    \left(
    \begin{array}{c}
         f_t  \\
          \vdots\\
          f_{t-p}
    \end{array}
    \right)
    =:L^n x_t, \; \text{say}.\label{eq:stacked}
\end{equation}
By Definition \ref{def:pervasive and weak}, the number of static factors driving $\chi_t^n$ is
$r_\chi(n)=\dim\cspargel(\chi_{it}, i=1,\ldots, n) =\dim\cspargel(x_{t})= q(p+1)$, which is finite and independent of $n$, thus, $r_\chi=q(p+1)<\infty$. By Proposition \ref{thm: finite dimensional CCC}.1, it follows that $(y_{it})$ satisfies also Assumption A\ref{A: r-SFS struct} with $r\le r_\chi$ pervasive static factors. 
In particular, if $r<r_\chi$, from  \eqref{eq:stacked} we can derive the canonical representation \eqref{eq: canonical rep weak and strong factors} in Theorem \ref{thm: weakfactors}, where $x_t$ is written as the linear combination of $r<r_\chi$ pervasive factors plus $r_w=r_\chi-r>0$ weak factors (see Appendix \ref{app:example weak} for details).
Whereas, if $r=r_\chi$, by Proposition \ref{thm: finite dimensional CCC}.2, we have that $\chi_{it}=C_{it}$ for all $i\in\mathbb N$ and $t\in\mathbb Z$ and \eqref{eq:stacked} is a static factor model obtained by stacking the dynamic factors and their lags \citep[see e.g.][]{forni2005generalized,forni2009opening, bai2007determining,stock2016dynamic}. In Example \ref{exemp: stacking} we show that assuming $r=r_\chi$ can be quite restrictive.
}
 \end{rem}

 \begin{example} [Static pervasive factors as ``stacked''  lags of dynamic factors can be restrictive] \label{exemp: stacking}
 \upshape{
 In model \eqref{eq: SW example dyn fac}-\eqref{eq: SW example dyn fac2} it is not unreasonable to assume that the effect of the dynamic factors $(f_t)$ on $(y_{it})$ decreases with the lags so that the columns of $\lambda_{j}^n$ ``taper off'', i.e., there exists some lag $\tilde p$ such that 
 \begin{align}
& \sup_{n \in \mathbb N}\mu_q({\lambda_j^{n'}\lambda_j^n})= \infty,\;     \text{ for $1 \leq j \leq \tilde p-1$},\label{eq:taper1} \\
  & \sup_{n \in \mathbb N}\mu_1({\lambda_j^{n'}\lambda_j^n})
  < \infty,\;     \text{ for $\tilde p \leq j \leq p$},\label{eq:taper2}
 \end{align}
An economic justification for such conditions is given in Example \ref{exmp: IRF}. Note that, because of \eqref{eq:taper1}, condition \eqref{eq:divergingloadings} still holds, and $(y_{it})$ still admits a GDFM representation.
Then, letting $l_j^n$, $j = 1, \ldots, r_\chi$, be the columns of $L^n$ in \eqref{eq:stacked}, there exists a $j^*$ such that $\sup_{n\in \mathbb N}\mu_{1}(l_{j^*}^{n'} l_{j^*}^n) = \sup_{n \in \mathbb N}\norm{l_{j^*}^n}^2< \infty$. And, since
\begin{align}
    \sup_{n \in \mathbb N}\mu_{r_\chi} ({L^n}' L^n)
    =\sup_{n\in\mathbb N}\min_{\substack{w\in\mathbb R^{1\times r_\chi} \\ ww'=1}} w {L^n}' L^n w'
    \le \sup_{n\in\mathbb N} s_{j^*}{L^n}' L^n s_{j^*}'
    &\leq \sup_{n \in \mathbb N} {l_{j^*}^n}' l_{j^*}^n
   = \sup_{n \in \mathbb N}\norm{l_{j^*}^n}^2 
    < \infty, \nonumber 
    \end{align}
    where $s_{j^*}$ is the $j^*$th element of the canonical basis of $\mathbb R^{1\times r_{\chi}}$.
    Then, by the multiplicative Weyl's inequality \citep[Theorem 7]{merikoski2004inequalities},
    \begin{align}
    \sup_{n \in \mathbb N}\mu_{r_\chi} (\Gamma_\chi^n) &= \sup_{n \in \mathbb N}\mu_{r_\chi}(\Gamma_x L^{n'} L^n) \leq \sup_{n \in \mathbb N} \mu_1(\Gamma_x) \mu_{r_\chi} (L^{n'}L^{n})  < \infty, \label{eq: taper off SW}
\end{align}
where $\Gamma_x:=\E[x_t x_t']$ with $\mu_1(\Gamma_x)<\infty$, because $(x_t)$ is finite dimensional and stationary by Assumption A\ref{A: stat}. Now if $r=r_\chi$, then, by Proposition \ref{thm: finite dimensional CCC}.2, we would have $C_{it}=\chi_{it}$ for all $i\in\mathbb N$ and $t\in\mathbb Z$, but \eqref{eq: taper off SW} implies $\sup_{n\in\mathbb N} \mu_{r}(\Gamma_C^n)= \sup_{n\in\mathbb N} \mu_{r_\chi}(\Gamma_\chi^n)<\infty$, which is a contradiction. Hence, it must be that $r<r_\chi$ and there is at least one non-zero weak common component.
}
 \end{example}

\subsection{Implications for impulse response analysis}\label{subsec: IRFs}
    Empirical macroeconomists are often interested in the impulse response functions (IRFs) of the observed variables $(y_{it})$ to some structural shocks of the economy. 
    
    According to the the GDFM, given in Assumption A\ref{A: q-DFS struct}, the $q$ common shocks $(\varepsilon_t)$ driving $(\chi_{it})$ generate all major co-movements of the observables, so it is reasonable to assume that they span the same space as the structural shocks, while $(\xi_{it})$ contains just idiosyncratic dynamics or measurement errors (\citealp{giannone2006does,stock2016dynamic}). Denote the structural shocks by $(\eta_t)$ and suppose we focus on the structural IRFs of the first $q$ observables with dynamic common components $(\chi_t^q)$. Then, by Assumption A\ref{A: q-DFS struct}, we get:
\begin{align}\label{eq:trueIRF}
\chi_t^q &= 
K^{q}(0)\varepsilon_t+\sum_{j=1}^\infty K^q(j) \varepsilon_{t-j}
= \ubar k^{\chi,q}(L)\l\{K^{q}(0)\r\}^{-1}d^{-1} d
K^{q}(0)\varepsilon_t
=: 
\ubar k^{\chi,q,*}(L)\eta_t,
\end{align}
where $d$ is an invertible $q\times q$ matrix, which depends on the chosen identification scheme, and such that $\eta_t :=  d K^q(0) 
\varepsilon_t$.\footnote{Note that $(K^q(0) 
\varepsilon_t)$ is the process of innovations defining the Wold representation of $(\chi_t^q)$.} 
We call $k^{\chi,q,*}(L)$ the structural or true IRFs.
    Since, by Theorem \ref{thm: relation of r-SFS and q-DFS}, $\chi_t^q=C_t^q+e_t^{\chi,q}$, it is clear that
   the true IRFs are given by the effect of the structural both via the static common component and via the weak common component. A consistent estimator of the true IRFs is given in \citet{forni2017dynamic}.
    
The typical approach to retrieve structural IRFs in a data-rich environment, is a static one based on the assumption that the structural shocks have an effect on the observables only through the static common component \citep{bernanke2005measuring,bai2006confidence,bai2007determining,BoivinGiannoniMihov2009,forni2009opening,stock2016dynamic,forni2025common}.
     It follows that, we can always derive the structural representation:
    \beq\label{eq:staticIRF}
    C_t^q
    =:\ubar k^{C,q,*}(L)\eta_t.
    \eeq
    We call $\ubar k^{C,q,*}(L)$ static IRFs. There are various ways to estimate these IRFs, the most recent one being the Common Component VAR (CC-VAR) approach proposed by \citet{forni2025common}, which is shown to nest both the
    Factor Augmented VAR \citep{bernanke2005measuring} and the Dynamic Factor Model  \citep{forni2009opening}. A necessary condition for identification of the common shocks is $r\ge q$, which, despite the peculiar case in Example \ref{exmp: 1-DFS but 0-SFS}, is a reasonable assumption often supported by the data \citep{dagostoni2012comparing}.\footnote{We are here ruling out the case $r=\infty$, which, 
     as shown in Example \ref{exmp: infinite dimensional CCC-space}, implies that there is no static representation and then the static IRFs would be undefined.}
  
From \eqref{eq:trueIRF} and \eqref{eq:staticIRF} it is clear that the true and static IRFs coincide only if $C_{t}^q=\chi_{t}^q$, or equivalently, $e_{t}^{\chi,q}=0$, for all $t\in\mathbb Z$, and for all $q$ series of interest.
    Note that even if we are interested in the IRF of just one observable say $i=1$, the static IRFs do not coincide with the true ones unless all $q$ variables used for identification have zero weak common component. This is because the matrix $d$ in \eqref{eq:trueIRF} mixes all rows of the IRFs.

    However, in general, the presence of at least some non-zero weak common components is reasonable and likely to be a common feature of large panels of time series reasonable, and, thus, it should not be ruled out a priori (see Example \ref{exemp: stacking} and the empirical results in Section \ref{sec: empiric}). Thus, in general, the static IRFs do not coincide with the true ones, and we should always estimate the IRFs starting from the GDFM representation instead (see Example \ref{exmp: infinitely many weak factors}
    for an infinite number of weak factors and Example \ref{exmp: IRF} for a finite number).

The issue discussed so-far is not only related to the choice of the factor representation, but it has important implications for estimation too, for we do not know a priori which series have a non-zero weak common component. 
On the one hand, we could estimate the true IRFs in \eqref{eq:trueIRF} via the dynamic approach of \citet{forni2017dynamic}, which, being based on the GDFM representation in Assumption A\ref{A: q-DFS struct}, is always valid. 
On the other hand, say we opt for the static approach and estimate the static IRFs in \eqref{eq:staticIRF} as in \citet{forni2025common}. But, in order to justify the use of the latter as estimators of the true IRFs, we should first estimate all the weak common components and establish that these are null for all relevant series used for identification.

Finally, one might argue adopting a static estimation approach based on more than $r$ static pervasive factors might be a possible solution, as the additional factors make up for the weak common component. However, this estimation procedure is non-consistent, for we are essentially trying to estimate the weak factors via static PCA (see the discussion in Remark \ref{rem: weakF} for details). 

\begin{example}[Static IRFs might not coincide with true IRFs]
\label{exmp: IRF}
\upshape{
Consider again model \eqref{eq: SW example dyn fac}-\eqref{eq: SW example dyn fac2}, when $q=1$, $p=1$, and $p_f=1$. In this case, the common and the structural shock, having both unit variance, coincide, up to a sign, which can be easily fixed, i.e., w can assume $\varepsilon_t=u_t$ for all $t\in\mathbb Z$. Then, for a given a $n$-dimensional panel of series, $(y_{t}^n)$, the true IRFs are: 
\beq\label{eq: DGP1 IRF model dynamicex5}
\ubar k_i^\chi(L) 
= (\lambda_{i0}+\lambda_{i1}L)(1-a L)^{-1} b, \quad i=1,\ldots, n.
\eeq
Let us now 
assume that 
$\lim_{n\to\infty}\sum_{i=1}^n \lambda_{i0}^2\to \infty$, while there exists a fixed $\bar n<n$ such that $\lambda_{i1}=0$ for $\bar n+1\le i\le n$, so that $\lim_{n\to\infty}\sum_{i=1}^n \lambda_{i1}^2=\lim_{n\to\infty}\sum_{i=1}^{\bar n}\lambda_{i1}^2<\infty$. Then, conditions \eqref{eq:taper1}-\eqref{eq:taper2} in Example \ref{exemp: stacking} are satisfied for $\tilde p=1$.
 In other words, after lag-1 the shocks have stronger (in absolute value) effects on the first $\bar n$ variables with respect to the remaining ones. For  example, \citet{furlanetto2025estimating} find that demand shocks have a negative effect on output and unemployment, but no effect on labor productivity.
 As seen from Example \ref{exemp: stacking} and the stacked representation \eqref{eq:stacked}, we have $r_\chi = q(p+1)=2$. Moreover, as shown in Appendix \ref{subsec: weak factor sim example details1} the canonical decomposition \eqref{eq: gen decomp} in Theorem \ref{thm: relation of r-SFS and q-DFS} for this model gives $C_{it}=(\lambda_{i0}+\lambda_{i1}a) f_t$
 and $e_{it}^\chi=\lambda_{i1}(f_{t-1}-a f_t)$, so $r=1$ and $r_w=1$, i.e., there is one pervasive static factor, $f_t$, loaded by all series and one weak factor, $(f_{t-1}-a f_t)$, loaded only by the first $\bar n$ series, which, then, have a non-zero weak common component. 

Now consider the static approach. Then, since the only static pervasive factor, $f_t$, follows the VAR in \eqref{eq: SW example dyn fac2}, the static IRFs are:
\[
\ubar k_i^C(L)= (\lambda_{i0}+\lambda_{i1}a)(1-aL)^{-1}b, \quad i=1,\ldots,n.
\]
Clearly, $\ubar k_i^C(L)\ne \ubar k_i^\chi(L)$ for $1\le i\le \bar n$. So, by starting from the static common component $C_{it}$, we cannot recover the correct IRFs for the first $\bar n$ series. Furthermore, notice that, by construction, all $n$ static IRFs are proportional, since the only truly dynamic component is given by the VAR for $(f_t)$ in \eqref{eq: SW example dyn fac2}. In other words, the static approach enforces all common components to have the same persistence in their reaction to the common shocks, whereas in the dynamic case the weak common component allows for some series to have an heterogeneous decay in the IRFs. This constraint is likely to produce unreasonable shapes for static IRFs (see, for example, the empirical application in Section \ref{sec:emp_vol}).}  

\end{example}

\section{Estimation of the weak common component and factors}\label{sec: estimation}
%
%

%

In this section, we study estimation of the canonical decomposition as given in Theorems \ref{thm: relation of r-SFS and q-DFS} and \ref{thm: weakfactors}.
Throughout, we assume to observe $n$ time series of length $T$, i.e., the realizations
$(y_{it} : i=1,\ldots, n,\, t=1,\ldots, T)$ of the $n$-dimensional process $(y_t^n : t\in\mathbb Z)$ satisfying Assumption A\ref{A: stat}. We assume to work with pre-centered data. If data is not pre-centered then all the following applies to the centered realizations $(y_{it}-T^{-1}\sum_{t=1}^T y_{it} : i=1,\ldots, n,\, t=1,\ldots, T)$.

We propose to proceed as follows.
\begin{itemize}
     \item[I.] Estimate the dynamic common component applied on $y_t^n$, thus giving $\widehat \chi_t^n$ with elements $\widehat \chi_{it}$, $i=1,\ldots, n$, $t=1,\ldots, T$,  as in (a) \citet{forni2000generalized} via dynamic PCA or in (b) \citet{forni2017dynamic}. Both approaches require to estimate the number of dynamic factors $q$, e.g., as in \cite{hallin2007determining}.
     %
     \item[II.] Estimate the static common component via static PCA applied on  $\widehat \chi_t^n$, thus giving $\widehat C_t^n$ with elements $\widehat C_{it}$, $i=1,\ldots, n$, $t=1,\ldots, T$. This step requires estimating the number of static pervasive factors $r$, e.g., as in \citet{bai2002determining}. 
  \item[III.] Estimate the weak common component, given in \eqref{eq: gen decomp} in Theorem \ref{thm: relation of r-SFS and q-DFS}, as
$
\wh e_{t}^{\,\chi,n} = \wh{\chi}_{t}^n- \wh{C}_{t}^n,
$
with elements $\widehat e_{it}^{\,\chi}$, $i=1,\ldots, n$, $t=1,\ldots, T$. 

\item [IV.] Estimate the weak factors, given in \eqref{eq: canonical rep weak and strong factors} in Theorem \ref{thm: weakfactors}, via static PCA applied on $\wh e_{t}^{\,\chi,n}$. This step requires the number of weak factors to be finite and it can be retrieved by standard rank tests on the sample covariance of $\wh e_{t}^{\,\chi,n}$.
\end{itemize}

We now describe in details all estimation steps (Section \ref{subsec: estimation in practice}), and then we prove consistency of the proposed estimators (Section \ref{sec:cons}).

\subsection{Estimation in practice}\label{subsec: estimation in practice}
\paragraph{Part I - Estimation of the dynamic common component.}
We can proceed in two ways. First, following the approach by \citet{forni2000generalized} we have these steps.
\ben
\item [I.a.i] Estimate the spectral density matrix of $(y_t^n)$ as:
\beq\label{eq:bartlett}
\widehat{f}_y^n(\theta_h):=\frac 1{2\pi}\sum_{\ell=-T+1}^{T-1} \kappa\left(\frac{\ell}{\mathcal B_T}\right)e^{-\iota \ell \theta_h}\left(\frac 1{T}\sum_{\vert \ell\vert+1}^T y_t^n y_{t-\vert \ell\vert }^{n'}\right),\quad  \theta_h = \frac{\pi h}{\mathcal B_T}, \quad \vert h\vert\le  \mathcal B_T,
\eeq
where $\kappa(\cdot)$ is a pre-specified kernel and $\mathcal B_T$ is the associated bandwidth such that $\mathcal B_T<T$.

\item [I.a.ii] Let $\widehat W_n(\theta_h)$ be the $q\times n$ matrix having as rows the normalized eigenvectors of $\widehat{f}_y^n(\theta_h)$ corresponding to the $q$ largest eigenvalues then let
\[
\widehat d_n(\theta_h) := \widehat W_n^* (\theta_h)\widehat W_n(\theta_h),\quad  \theta_h = \frac{\pi h}{\mathcal B_T}, \quad \vert h\vert\le \mathcal B_T,
\]
and
\[
\widehat D_n(\ell):= \frac {2\pi}{2B_T+1}\sum_{ h=-\mathcal B_T}^{\mathcal B_T} e^{\iota \ell \theta_h}\widehat d_n(\theta_h), \quad \vert\ell\vert\le \mathcal B_T.
\]
Then, for any $t=\mathcal M_T+1,\ldots, T-\mathcal M_T-1$, let
\[
\widehat{\chi}^n_t:=\sum_{\ell=-\mathcal M_T}^{\mathcal M_T} \widehat D_n(\ell) y_{t-\ell}^n,
\]
%
for some integer $\mathcal M_T<T$ and such that $\mathcal M_T\le \mathcal B_T$. Let $\widehat{\chi}_{it}$ be the $i$th component of $\widehat{\chi}^n_t$.
\een

\noindent Alternatively, second, following the approach by \citet{forni2017dynamic} we have the following steps.
\ben
\item  [I.b.i] Estimate the spectral density matrix $\widehat{f}_y^n(\theta_h)$ as in \eqref{eq:bartlett} in step I.a.i above.
\item [I.b.ii] Collect the $q$ largest eigenvalues ${\mu}_j(\wh f_y^n(\theta_h))$, $j=1,\ldots,q$, of $\widehat{f}_y^n(\theta_h)$ in decreasing order into the $q\times q$ diagonal matrix $\widehat M_n(\theta_h)$, then
estimate the spectral density matrix of $(\chi_t^n)$ as
\beq\label{eq:sigmachiest}
\widehat{f}_\chi^n(\theta_h):=\widehat W_n^*(\theta_h)\widehat M_n(\theta_h)\widehat W_n(\theta_h),\quad  \theta_h = \frac{\pi h}{\mathcal B_T}, \quad \vert h\vert\le \mathcal B_T,
\eeq
where $\widehat W_n(\theta_h)$ is the $q\times n$ matrix having as rows the normalized eigenvectors of $\widehat{f}_y^n(\theta_h)$ corresponding to the $q$ largest eigenvalues.
\item [I.b.iii] Estimate the lag-$k$ autocovariances by discrete inverse Fourier transform:
\[
\widehat \Gamma_\chi^n(\ell) := \frac {2\pi}{2B_T+1}\sum_{h=- \mathcal B_T}^{\mathcal B_T} e^{\iota \ell \theta_h} \widehat{f}_\chi^n(\theta_h), \qquad \vert \ell\vert\le \mathcal B_T.
\]
\item [I.b.iv] For simplicity of notation let $m=n/(q+1)$ be an integer. Consider the $m$ consecutive sub-vectors $y_t^{[h]}$, $h=1,\ldots, m$, of $y_t^n$, each of dimension $(q + 1)$. For each sub-vector $y_t^{[h]}$, we estimate, via the Yule-Walker a VAR$(p_h)$, $h=1,\ldots, m$, using $\widehat \Gamma_\chi^n(\ell)$, $\ell=0,\ldots, p_h$ with $p_h<\mathcal B_T$. This yields, for the $h$th sub-vector, an estimated autoregressive filter $\widehat {\underline a}^{[h]}(L)$ of dimension $(q+1)\times (q+1)$. Let $\widehat {\underline a}_n(L)$ the $n\times n$ block diagonal autoregressive filter, of order $\bar p=\max_{h=1,\ldots ,m} p_h$, with diagonal blocks $\widehat {\underline a}^{[h]}(L)$, $h=1,\ldots, m$.\footnote{This step can be repeated by considering various permutations of the $n$ cross-sectional items and then by averaging the results over such permutations (see \citealp{forni2017dynamic}, for a theoretical justification of such approach).}

\item [I.b.v] Let $\widehat \psi_t^n:= \widehat {\underline a}_n(L)y_t^n$, with sample covariance matrix $\widehat{\Gamma}^n_{\widehat\psi}:=T^{-1}\sum_{t=\bar p+1}^T \widehat \psi_t^n\widehat \psi_t^{n'}$ and let  $\widehat Q_n$ be the $q\times n$ matrix having as rows the normalized eigenvector of $\widehat \Gamma_{\widehat \psi}^n$ corresponding to its $q$ largest eigenvalues.
Let also $\underline{\wh b}_n(L):=\{\widehat {\underline a}_n(L) \}^{-1}$, with coefficients $(\widehat B_{n}(\ell) : \ell\in\mathbb Z^+)$.
Then, let
\[
\widehat{\chi}^n_t:=  \sum_{\ell=0}^{\mathcal K_T}\widehat B_{n}(\ell)  \widehat Q_n'\widehat Q_n\widehat \psi_{t-\ell}^n,
\]
for some integer $\mathcal K_T<T-\mathcal B_T$ (note that $\bar p <\mathcal B_T$ so $\mathcal K_T<T-\bar p$). Let $\widehat{\chi}_{it}$ be the $i$th component of $\widehat{\chi}^n_t$. 
\een

\paragraph{Part II - Estimation of the static common component.} We can proceed in two equivalent ways. Either using the eigenvectors of the sample covariance of $(y_t^n)$.
\ben
\item [II.a] 
Let $\widehat \Pi_n$ be the $r\times n$ matrix having as rows the normalized eigenvectors of $\widehat \Gamma_y^n:=T^{-1}\sum_{t=1}^T y_t^n y_t^{n'}$ corresponding to its $r$ largest eigenvalues. Then, let
$\widehat C_t^n := \widehat \Pi_n'\widehat \Pi_n\widehat{\chi}^n_t.
$ and let $\widehat{C}_{it}$ be the $i$th component of $\widehat{C}^n_t$.
\een

\noindent Or using the eigenvectors of the sample covariance of $(\chi_t^n)$ obtained by integrating the estimated spectral density.
\ben
\item [II.b]
Let $\widehat P_n$ be the $r\times n$ matrix having as rows the normalized eigenvectors of 
\beq\label{eq:GammahatChiSI}
\widehat \Gamma_\chi^n := \frac {2\pi}{2\mathcal B_T+1}\sum_{h=- \mathcal B_T}^{\mathcal B_T}  \widehat{f}_\chi^n(\theta_h), 
\eeq
corresponding to its $r$ largest eigenvalues and where $\widehat{f}_\chi^n(\theta_h)$ is computed as in \eqref{eq:sigmachiest} in step I.b.ii above.
Then,
let 
$\widehat C_t^n := \widehat P_n'\widehat P_n\widehat{\chi}^n_t$ and let $\widehat{C}_{it}$ be the $i$th component of $\widehat{C}^n_t$.
\een


\paragraph{Part III - Estimation of the weak common component.}
Given estimators of the dynamic and static common components the weak common component is estimated as:
$\widehat e_{it}^{\,\chi}=\widehat \chi_{it}-\widehat C_{it}$. Given that there are two ways to compute $\wh{\chi}_{it}$ and two ways to compute $\wh C_{it}$ there are four possible estimators of the weak common component. Notice that by construction the sample covariance of the static and the weak common component is zero, thus we ensure that they are orthogonal, as requested by Theorem \ref{thm: relation of r-SFS and q-DFS}.
Moreover, if we follow step II.b to compute $\wh{C}_{it}$ then the sample variance of $\wh\chi_{it}$ is always greater or equal than the sample variance of  $\wh C_{it}$.

\paragraph{Part IV - Estimation of the weak factors.} Consider the estimator of the covariance matrix of the weak common component: 
\beq\label{eq:GammahatWCCSI}
\widehat\Gamma^{ n}_{e^\chi}:= \widehat\Gamma^{ n}_\chi - \widehat P_n'\widehat P_n \widehat\Gamma^{ n}_\chi\widehat P_n'\widehat P_n,
\eeq
where $\widehat\Gamma^{n}_\chi$ is obtained by inverse Fourier transform of $\wh{f}_\chi^n(\theta)$, as defined in \eqref{eq:GammahatChiSI}, and $\widehat P_n$ is the $r\times n$ matrix of normalized eigenvectors corresponding to its $r$ largest eigenvalues. 

We then rearrange the panel in such a way that the $n$ time series are ordered according to the variance contribution of their estimated weak common component (in decreasing order). Then, for a given finite number of weak factors $r_w$, let $\bar n$ be such that $r_w \leq \bar n$, and let $\widehat\Gamma^{\bar n}_{e^\chi}$ be the $\bar n\times \bar n$ top-left sub-matrix of the reordered version of $\widehat\Gamma^{n}_{e^\chi}$  given in \eqref{eq:GammahatWCCSI}. Let also $\widehat {\mathcal W}_{\bar n}$ be the  $r_w\times \bar n$ matrix having as rows the normalized eigenvectors of $\widehat\Gamma^{\bar n}_{e^\chi}$ corresponding its $r_w$ largest eigenvalues collected into the $r_w\times r_w$ diagonal matrix $\widehat {\mathcal M}$. The estimated weak factors are given by:
\[
\widehat F_t^{w}:=(\widehat {\mathcal M})^{-1/2}\, \widehat {\mathcal W}_{\bar n}\, \widehat e_t^{\,\chi,\bar n},
\]
where $\widehat e_t^{\,\chi,\bar n}$ is any of the four estimators defined in part III. Finally, the number of weak factors $r_w$ can be determined by by means of any standard rank estimation procedure applied to $\widehat\Gamma^{\bar n}_{e^\chi}$. In Section \ref{sec: simulation experiments}, we consider an eigenvalue ratio criterion as in \citet{ahn2013eigenvalue}.

\subsection{Consistency}\label{sec:cons}
In this section, we provide the first thorough analysis of consistency, including convergence rates, for estimating the dynamic common component, both in the framework of \citet{forni2000generalized} and that of \citet{forni2017dynamic}. Existing results are either only partial or not directly comparable across studies, and thus cannot be straightforwardly applied to our current setting. Regarding part I.a we remark that \citet{forni2000generalized}  provide no consistency rates, while those given in \citet{forni2004generalized}  for the same estimator are incomplete. For part I.b, consistency results are available in \citet{forni2017dynamic}, and \citet{barigozzi2024fnets,barigozzi2024inferential}. However, these works make use of similar, but not identical, assumptions and are based on a series of results on consistency of the estimated spectral density which are not comparable with each other. 

The static PCA approach in part II requires a new proof too. Indeed, the na\"ive approach, consisting estimating $C_{it}$ via standard PCA on $y_{it}$, does not ensure orthogonality. 
Instead, we must use the estimated dynamic common component $\wh{\chi}_t^n$ obtained in part I, and thus we must also account for the related estimation error. Therefore, we cannot directly apply the results in \citet{bai2003inferential}.

In order to prove consistency of the proposed estimators, we need to make some additional assumptions. First, we strengthen Assumption A\ref{A: q-DFS struct} to the following assumption taken from \citet{forni2017dynamic} and \citet{barigozzi2024fnets,barigozzi2024inferential}.
\begin{assumption}\label{ass:newGDFM}
\ben 
\item[(i)]  The dynamic common component can be written as
\[
\chi_{it} =\underline k_i(L)\varepsilon_t = \sum_{j=0}^\infty K_i(j) \varepsilon_{t-j}=\sum_{j=0}^\infty\sum_{\ell=1}^q K_{i,\ell}(j) \varepsilon_{\ell,t-j}, \quad i\in\mathbb N,\quad t\in\mathbb Z,
\]
where
\ben
\item $(\varepsilon_t  : t \in \mathbb Z)$ is an i.i.d. $q$-dimensional process, with $\E [\varepsilon_t]=0$ and  $\E [\varepsilon_t\varepsilon_t']=I_q$ for all $t\in\mathbb Z$;
\item the coefficients of $\underline k_i(L) $ which are
$(K_{i,\ell}(j)  : i\in\mathbb N,\, \ell =1,\ldots, q, \, j\in\mathbb Z^+)$ are such that 
$\vert K_{i,\ell}(j)\vert \le A_{i,\ell}^\chi (\rho^\chi)^{j}$ with
$\sup_{i\in\mathbb N}\sum_{\ell=1}^q A_{i,\ell}^\chi \le A^\chi<\infty$ and $\rho^\chi\in(0,1)$;
\item  $0<q<\infty$ independent of $n$.
\een
\item[(ii)]  The dynamic idiosyncratic component can be written as
\[
\xi_{it} =\underline c_i(L)\eta_t = \sum_{j=0}^\infty C_i(j) \eta_{t-j}= \sum_{j=0}^\infty\sum_{\ell=1}^\infty C_{i,\ell}(j) \eta_{\ell,t-j}, \quad i\in\mathbb N,\quad t\in\mathbb Z,
\]
where
\ben
\item $(\eta_t  : t \in \mathbb Z)$ is an i.i.d. $\infty$-dimensional process, with $\E [\eta_t]=0$ and  $\E [\eta_t\eta_t']=I_\infty$ for all $t\in\mathbb Z$;
\item the coefficients of $ \underline c_i(L) $ which are
$(C_{i,\ell}(j)  : i,\ell\in\mathbb N,\, j\in\mathbb Z^+)$ are such that 
$\vert C_{i,\ell}(j)\vert \le A_{i,\ell}^\xi (\rho^\xi)^{j}$ with
$\sup_{i\in\mathbb N}\sum_{\ell=1}^\infty A_{i,\ell}^\xi \le A^\xi<\infty$, 
$\sup_{\ell\in\mathbb N}\sum_{i=1}^q A_{i,\ell}^\xi \le A^\xi<\infty$ 
and $\rho^\xi\in(0,1)$. 
\een
\item [(iii)] $\E [\varepsilon_t\eta_{\ell s}]=0$ for all $\ell \in\mathbb N$ and $t,s\in\mathbb Z$.
\item [(iv)] $\max\{\max_{j=1,\ldots, q} \E [\vert \varepsilon_{jt} \vert^\nu],\sup_{\ell \in\mathbb N} \E[\vert \eta_{\ell t}\vert^\nu] \}\le \bar A$ for some $\nu \ge 4$.
\een
\end{assumption}

Notice that (i) and (ii) imply the behavior of eigenvalues assumed in Assumption A\ref{A: q-DFS struct}. For the common component this is obvious. For the idiosyncratic component we refer to \citet[Proposition 1]{forni2017dynamic}. 
In principle independence of innovations could be relaxed to allow for martingale difference processes (see, e.g., \citealp{barigozzi2024fnets}, Assumption 2.4).
Part (iii) is redundant since we already assumed common and idiosyncratic to be uncorrelated at all leads and lags in Assumption A\ref{A: q-DFS struct}, but for convenience it is repeated here so that, hereafter, Assumption A\ref{ass:newGDFM} replaces entirely Assumption A\ref{A: q-DFS struct}. Part (iv) is natural and it is necessary for estimation of second moments.

We then assume linear divergence of eigenvalues.
\begin{assumption} \label{ass:diveval} \ben \item [] $\!\!\!\! \!\!\!\!$ There exist functions: $\theta\mapsto \alpha_j^\chi(\theta)$, $j=1,\ldots, q$, and $\theta\mapsto \beta_j^\chi(\theta)$, $j=0,\ldots,q- 1$ such that  
\[
0<\alpha_j(\theta) \le \lim_{n\to\infty}\frac{\mu_j\left(f_\chi^n(\theta)\right) }{n}\le \beta_{j-1}^\chi(\theta)<\infty, \quad j= 1,\ldots, q, \quad \theta\in[-\pi,\pi].
\]
\een
\end{assumption}

Linear divergence is reasonable and assumed for convenience, we refer to the comments in \citet{barigozzi2025dynamic}, for a possible justification. The results in this section can also be derived for rate-weak dynamic factors, so when assuming a sub-linear divergence rate. The proofs are almost identical, although the consistency rates would be slower, so for simplicity we omit this case. 

Then, we state standard assumptions for the kernel and its related bandwidth in \eqref{eq:bartlett}, as well as for the truncation levels in parts I.a and I.b.
\begin{assumption} \label{ass:kernel}
\ben
\item [(i)] The kernel function $\kappa : [-1, 1] \to\mathbb R^+$ is symmetric and bounded and such that:\linebreak 
(a) $\kappa(0) = 1$; 
(b) $\vert\kappa(u)-1\vert = O(\vert u\vert^\vartheta)$ as $u\to 0$, for some $\vartheta>0$;
(c) $\int_{-1}^1 \kappa(u)\mathrm du\le \mathcal K_1<\infty$;
(d) $\sum_{h=-\infty}^\infty \sum_{h_2: \vert h_1-h_2\vert\le 1} \vert\kappa(h_1 u)-\kappa(h_2 u)\vert\le \mathcal K_2<\infty$ as $u\to 0$.

\item [(ii)]  
$\mathcal B_T= c_1T^{b_1}$, with $c_1\in(0,\infty)$ and $b_1\in(0,1)$, for all $T\in\mathbb N$;
\item [(iii)]   $\mathcal M_T= c_2\log T$, with $c_2\in(0,\infty)$.
\item [(iv)]   $\mathcal K_T= c_3\log T$, with $c_3\in(0,\infty)$.
%
\een
\end{assumption}

For estimation of $\chi_{it}$ in part I we need one more assumption for each considered method. Specifically, for part I.a only, we assume.

\begin{assumption}\label{eq:decay}\ben \item [] $\!\!\!\! \!\!\!\!\!\!\!\!$
Let $V_n(\theta_h)=V_n({f}_\chi^n(\theta_h))$ be the $q\times n$ matrix having as rows the normalized eigenvectors of ${f}_\chi^n(\theta_h)$ corresponding to its $q$ non-zero eigenvalues, and let $g_{n}(\theta) = V_n^* (\theta)V_n(\theta)$,   $\theta\in[-\pi,\pi]$, having as Fourier coefficients
$
G_{n}(\ell)= \int_{-\pi}^\pi e^{\iota \ell \theta} g_{n}(\theta)$,  $\ell\in\mathbb Z$.
Then, 
there exist $\mathcal C,\varphi\in(0,\infty)$ independent of $\ell$, such that: 
$
\Vert G_{n}(\ell) \Vert  \le\mathcal C \Vert G_{n}(0) \Vert (1+\varphi)^{-\vert \ell\vert}.
$
\een
\end{assumption}

Although this assumption might seem quite technical, it is, in fact, just requiring summability of the coefficient of the filter $\underline g_n(L)=\sum_{\ell=-\infty}^{\infty} G_n(\ell) L^\ell$, which is the population counterpart of the dynamic PCA filter $\wh {\underline d}_n(L) = \sum_{\ell=-\infty}^\infty \wh D_n(\ell) L^\ell$ defined in step I.a.ii in part I.a. Notice that by definition $\Vert G_{n}(0) \Vert\le 2\pi$ for all $n\in\mathbb N$.

Whereas, for part I.b only, we assume.

\begin{assumption}\label{ass:ARGDFM}\ben \item [] $\!\!\!\! \!\!\!\!\!\!\!\!$
For all $n\in\mathbb N$ the dynamic common component can be written as
\[
\underline a_n(L) \chi_t^n = \phi_t^n, \qquad \phi_t^n= R_n \varepsilon_t, \qquad t\in\mathbb Z,
\]
where $n=m(q+1)$ for some integer $1<m<n$, and
\een
\ben
\item [(i)]   $\underline a_n(L) $ is block diagonal with $q+1\times q+1$ dimensional diagonal blocks $\underline a^{[h]}(L)$, such that
\[
\underline a^{[h]}(z) = I_{q+1}-\sum_{j=1}^{p_h} A^{[h]}(j)z^j, \quad z\in\mathbb C,\quad p_h<\mathcal B_T,\quad h=1,\ldots,m, 
\]
and $\det(\underline a^{[h]}(z))\neq 0$ for all $|z|\le 1$;
\item [(ii)] let $\chi_t^{(h)}=(\chi_{(q+1)(h-1)+i,t} : i=1,\ldots, q+1)$ and $\Gamma^{(h)}_\chi=\E[\chi_{t-h}^{(h)}\chi_t^{(h)'}]$, then 
 the matrix
\[
C_\chi^{(h)} =
\l(
\begin{array}{cccc}
\Gamma_\chi^{(h)}&\Gamma_\chi^{(h)}(1)&\ldots &\Gamma_\chi^{(h)}(-p_h+1)\\
\vdots&\vdots&\ddots&\vdots\\
\Gamma_\chi^{(h)}(p_h-1)&\Gamma_\chi^{(h)}(p_h-2)&\ldots &\Gamma_\chi^{(h)}\\
\end{array}
\r),\quad h=1,\ldots,m, 
\]
is such that $\min_{h=1,\ldots, m} \det(C_\chi^{(h)})>0$;
\item [(iii)] $(\varepsilon_t : t\in\mathbb Z)$ is defined in Assumption A\ref{ass:newGDFM};
\item [(iv)] $R_n$ is $r\times q$ with $\lim_{n\to\infty}n^{-1}R_n'R_n=\Sigma_R$ which is finite and positive definite.
\een
\end{assumption}

Essentially, here we assume the existence of a block-diagonal autoregressive representation for the dynamic common component. This standard in the literature and it is also assumed in \citet{forni2017dynamic}, who, in turn, make this assumption based on the representation results for processes with singular rational spectral density by \citet{anderson2008generalized}  and \citet{forni2015dynamic} to which we refer for further details. 

For estimation of $C_{it}$ in part II we make the following assumption.

\begin{assumption} \label{ass:divevalC}\ben 
\item [(i)]
 The static common component can be written as 
\[
C_{it} =\Lambda_i F_t, \quad i\in\mathbb N,\quad t\in\mathbb Z,
\]
where
\ben
\item [(a)] $\Lambda_n$ is $n\times r$ with $\lim_{n\to\infty}n^{-1}\Lambda_n'\Lambda_n=\Sigma_\Lambda$ which is finite and positive definite, and there exists as $C_\Lambda\in(0,\infty)$, independent of $n$, such that $\max_{i=1,\ldots, n}\sup_{n\in\mathbb N}\Vert\Lambda_i\Vert\le C_\Lambda$;
\item [(b)] $\E [F_t F_{t}']=I_r$ for all $t\in\mathbb Z$;
\item [(c)] $0<r<\infty$ independent of $n$.
\een
\item [(ii)]The static idiosyncratic component is such that there exists a $C_e\in(0,\infty)$, independent of $n$, such that $\sup_{n\in\mathbb N} \mu_1(\Gamma^n_e) < C_e$;
\item [(iii)] $\E [F_t e_{it}] = 0$ for all $i \in \mathbb N$ and $t\in\mathbb Z$.
\een
\end{assumption}

As shown in our proofs, this assumption contains the minimal set of conditions needed to prove consistency of static PCA. Although parts of this assumption are already in Assumption  A\ref{A: r-SFS struct}, they are repeated here for convenience, so that, hereafter, Assumption A\ref{ass:divevalC} replaces entirely Assumption A\ref{A: r-SFS struct}. 

Consistency of part I follows (see Appendix \ref{prop1proof} for a proof). 

\begin{proposition}[Consistency of Dynamic Common Component  $\wh{\chi}_t^n$]\label{prop1}
\ben
\item[] \mbox{} \linebreak (I.a) If $\wh{\chi}_t^n$ is estimated as in part I.a, then, under Assumptions A\ref{A: stat}, A\ref{ass:newGDFM}, A\ref{ass:diveval}, A\ref{ass:kernel}, A\ref{eq:decay},
as $n,T\to\infty$, 
\ben
\item[] (i) for any fixed $t=\mathcal M_T+1,\ldots, T-\mathcal M_T-1$, 
 $$
\min\l(\frac{T^{1-2/\nu}}{\mathcal B_T},\sqrt{\frac{T}{\mathcal B_T\log \mathcal B_T}},\mathcal B_T^{\vartheta},n \r) \frac{\l\Vert \wh{\chi}_t^n -\chi_t^n\r\Vert}{\log T \sqrt n}=\mathcal O_P(1);
 $$
 \item[] (ii) for any fixed $t=\mathcal M_T+1,\ldots, T-\mathcal M_T-1$ and any given $i=1,\ldots, n$,
 $$
 \min\l(\frac{T^{1-2/\nu}}{\mathcal B_T},\sqrt{\frac{T}{\mathcal B_T\log \mathcal B_T}},\mathcal B_T^{\vartheta},\sqrt n\r) \frac{ \l\vert \wh{\chi}_{it} -\chi_{it}\r\vert}{\log T}=\mathcal O_P(1);
 $$
 \een
\item [] (I.b) if $\wh{\chi}_t^n$ is estimated as in part I.b, then, under Assumptions A\ref{A: stat}, A\ref{ass:newGDFM}, A\ref{ass:diveval}, A\ref{ass:kernel}, A\ref{ass:ARGDFM},
as $n,T\to\infty$, 
\ben
\item[] (iii) for any fixed $t=\mathcal K_T+1,\ldots, T$, 
$$
\min\l(
\frac{T^{1-2/\nu}}{n^{2/\nu}\log^3 n}
,\sqrt{\frac T{\log n}}, \frac{T^{1-2/\nu}}{\mathcal B_T},\sqrt{\frac{T}{\mathcal B_T\log \mathcal B_T}},\mathcal B_T^{\vartheta},\sqrt n \r) \frac{\l\Vert \wh{\chi}_t^n -\chi_t^n\r\Vert}{\log T \sqrt n}=\mathcal O_P(1);
$$
 \item[] (iv) for any fixed $t=\mathcal K_T+1,\ldots, T$ and any given $i=1,\ldots, n$,
  $$
\min\l(
\frac{T^{1-2/\nu}}{n^{2/\nu}\log^3 n}
,\sqrt{\frac T{\log n}}, \frac{T^{1-2/\nu}}{\mathcal B_T},\sqrt{\frac{T}{\mathcal B_T\log \mathcal B_T}},\mathcal B_T^{\vartheta},\sqrt n \r)  \frac{ \l\vert \wh{\chi}_{it} -\chi_{it}\r\vert}{\log T}=\mathcal O_P(1);
 $$
 \een
\een
where $\nu\ge 4$ is defined in Assumption A\ref{ass:newGDFM} and  $\vartheta>0$ is defined in Assumption A\ref{ass:kernel}.
\end{proposition}

The following remarks give an intuition of the rates.

\begin{rem}[Rates for the \citet{forni2000generalized} two-sided estimator of the dynamic common component]\label{rem:FHLR}
\upshape{By construction the method by \citet{forni2000generalized} is two-sided thus, as shown in the proof of part I.a, it can give consistent estimates of $\chi_{it}$ only for $t$ in the central part of the sample, provided we choose a truncation level $\mathcal M_T=[\log T]$ in agreement with Assumption \ref{ass:kernel}. The first term in the convergence rates is always dominated by the second one, since $\nu\ge 4$. Typically, we use the Bartlett kernel to estimate the spectral density, for which $\vartheta=1$, and this implies that the optimal bandwidth, making the second and third term asymptotically equivalent, is $\mathcal B_T=[T^{1/3}]$.
Clearly, we also need $n\to\infty$ to achieve consistency since the GDFM is identified only asymptotically (see Theorem \ref{thm: projection on the dynamic aggregation space}).}    
\end{rem}

\begin{rem}[Rates for the \citet{forni2017dynamic} ones-sided estimator  of the dynamic common component]
\label{rem:FHLZ}
\upshape{
The method by \citet{forni2017dynamic} is multi-step but one-sided. Still, because the dynamic common component is estimated by inverting its estimated autoregressive representation, we need to specify a truncation lag $\mathcal K_T$ in the past, which should be large enough to provide a good approximation of the true infinite MA representation, but not too large to affect consistency since we need to estimate as many coefficients as $\mathcal K_T$. A good trade-off is obtained if we choose  $\mathcal K_T=[\log T]$ in agreement with Assumption \ref{ass:kernel}. With respect to the proof of part I.a, the convergence rates have now two additional terms which depend on $n$ and $T$. These are due to the necessity of deriving uniform bounds over all entries for the estimation error of the autocovariance matrices.
To gain more insight, assume that $n=\mathcal O(T^\gamma)$ for some $\gamma\in(0,\infty)$. Then, the first term, we clearly need $\nu>4$ to have consistency. This is stronger than what needed for part I.a, and, for $\nu$ large enough, this term gets also dominated by the fourth one. The second term in the rates is always dominated by the fourth one, since $\nu\ge 4$ and obviously $T> T/\mathcal B_T$. The third term is always dominated by the fourth since $\nu\ge 4$.
When using the Bartlett kernel to estimate the spectral density, for which $\vartheta=1$, the optimal bandwidth, making the fourth and fifth term asymptotically equivalent,  is still $\mathcal B_T=[T^{1/3}]$. Moreover, we still obviously need $n\to\infty$ to identify the common component and achieve consistency.
}
\end{rem}

\begin{rem}[Dynamic estimation of reduced form IRFs]\label{rem:IRFhat}
    \upshape{Under Assumption A\ref{ass:ARGDFM} the reduced form IRFs in \eqref{eq:trueIRF} for series $i$ can be written as:
$\ubar k_i^{\chi}(L)= s_i'\l\{\ubar a(L)\r\}^{-1}R_n$, $i=1,\ldots, n$, where $s_i$ is the $i$th element of the canonical basis in $\mathbb R^{n\times 1}$. From the estimation procedure described in part I.b.v, we immediately have the estimator of these reduced form IRFs: $\widehat{\ubar k}^{\chi}_i(L) := s_i' \underline{\wh b}_n(L)\sqrt n\widehat Q_n'$, where the scaling by $\sqrt n$ is required by Assumption A\ref{ass:ARGDFM}.
Consistency of this estimator follows directly from the same results used to prove Proposition \ref{prop1} part I.b.
}
\end{rem}

We then prove consistency of parts II.a and II.b (see Appendix \ref{prop3proof} for a proof).

\begin{proposition}[Consistency of  Static Common Component $\wh{C}_t^n$]
\label{prop3}
\ben
\item[] \mbox{} \linebreak (I.a - II.a and II.b)
 If we compute $\wh{\chi}_t^n$ as in part I.a and $\wh C_t^n$ as in part II.a or II.b, then,  under Assumptions A\ref{A: stat}, A\ref{ass:newGDFM}, A\ref{ass:diveval}, A\ref{ass:kernel}, A\ref{eq:decay}, A\ref{ass:divevalC},
as $n,T\to\infty$,
\ben
\item[] (i) for any fixed $t=\mathcal M_T+1,\ldots, T-\mathcal M_T-1$,
 $$
 \min\l(\frac{T^{1-2/\nu}}{\mathcal B_T},\sqrt{\frac{T}{\mathcal B_T\log \mathcal B_T}},\mathcal B_T^{\vartheta},n \r) 
\frac{\l\Vert \wh{C}_t^n -C_t^n\r\Vert}{\log T\sqrt n}=\mathcal O_P(1);
 $$
\item[] (ii) for any fixed $t=\mathcal M_T+1,\ldots, T-\mathcal M_T-1$, and for any fixed $i=1,\ldots, n$,
  $$
 \min\l(\frac{T^{1-2/\nu}}{\mathcal B_T},\sqrt{\frac{T}{\mathcal B_T\log \mathcal B_T}},\mathcal B_T^{\vartheta},\sqrt n \r) 
\frac{\l\vert \wh{C}_{it} -C_{it} \r\vert}{\log T}=\mathcal O_P(1);
 $$
 \een
\item[] (I.b - II.a and II.b)
 if we compute $\wh{\chi}_t^n$ as in part I.b and $\wh C_t^n$ as in part II.a or II.b, then, under Assumptions A\ref{A: stat}, A\ref{ass:newGDFM}, A\ref{ass:diveval}, A\ref{ass:kernel}, A\ref{ass:ARGDFM}, A\ref{ass:divevalC}, as $n,T\to\infty$,
 \ben
\item[] (iii)  for any fixed $t=\mathcal K_T+1,\ldots, T$,
$$
\min\l(
\frac{T^{1-2/\nu}}{n^{2/\nu}\log^3 n}
,\sqrt{\frac T{\log n}}, \frac{T^{1-2/\nu}}{\mathcal B_T},\sqrt{\frac{T}{\mathcal B_T\log \mathcal B_T}},\mathcal B_T^{\vartheta},\sqrt n \r)\frac{\l\Vert \wh{C}_t^n -C_t^n\r\Vert}{\log T\sqrt n}=\mathcal O_P(1),
$$
\item [] (iv)  for any fixed $t=\mathcal K_T+1,\ldots, T$, and for any fixed $i=1,\ldots, n$,
$$
\min\l(
\frac{T^{1-2/\nu}}{n^{2/\nu}\log^3 n}
,\sqrt{\frac T{\log n}}, \frac{T^{1-2/\nu}}{\mathcal B_T},\sqrt{\frac{T}{\mathcal B_T\log \mathcal B_T}},\mathcal B_T^{\vartheta},\sqrt n \r)\frac{\l\vert \wh{C}_{it} -C_{it}\r\vert}{\log T}=\mathcal O_P(1);
$$
 \een
 \een
 where $\nu\ge 4$ is defined in Assumption A\ref{ass:newGDFM} and  $\vartheta>0$ is defined in Assumption A\ref{ass:kernel}.
\end{proposition}

\begin{rem}
[Rates for the estimator of the static common component]
\upshape{
As clearly seen in the proof, the rates derived in this proposition are dominated by the estimation error of the estimated dynamic common component given in Proposition \ref{prop1}, which dominate the PCA rate $\min(n,\sqrt T)$ that we would have if we applied PCA on $\chi_t^n$ rather than on $\wh{\chi}_t^n$.}
\end{rem}

By combining Propositions \ref{prop1} and  \ref{prop3}, we have consistency of the estimated weak common component, which we state under some additional requirements that allow us to simplify the rates (see Appendix \ref{weakCCproof} for a proof).

\begin{theorem}
[Consistency of Weak Common Component $\widehat e_{t}^{\,\chi,n}$]\label{thm: weakCC}
If Assumptions A\ref{A: stat}, A\ref{ass:newGDFM}, A\ref{ass:diveval}, A\ref{ass:kernel}, A\ref{ass:divevalC} hold and we compute $\wh{C}_{it}$ as in parts II.a or II.b, then, under either of the following set of additional conditions:
\ben
\item []  (I.a-II-III) we compute $\wh{\chi}_t^n$ as in part I.a using the Bartlett kernel, Assumption A\ref{eq:decay} holds, and we consider any 
fixed $t=\mathcal M_T+1,\ldots, T-\mathcal M_T-1$ and $i=1,\ldots, n$;
\item [] (I.b-II-III) we compute $\wh{\chi}_t^n$ as in part I.b using the Bartlett kernel, $n=\mathcal O(T^\gamma)$ for some $\gamma\in(0,\infty)$, Assumptions A\ref{ass:newGDFM} holds with $\nu>6$, Assumption A\ref{ass:ARGDFM} holds,  and we consider any 
fixed $t=\mathcal K_T+1,\ldots, T$ and $i=1,\ldots, n$;
\een
as $n,T\to\infty$:
\[
\min\l( \sqrt{\frac{T}{\mathcal B_T\log \mathcal B_T}},\mathcal B_T,\sqrt n \r) \frac{\vert \wh{e}_{it}^{\,\chi}-e_{it}^\chi \vert}{\log T} = \mathcal O_P(1).
\] 
\end{theorem}

\begin{rem}[Rates for the estimator of the weak common component]
\upshape{The simplfying conditions of relatively light tails, $\nu\ge 6$, and $n=T^\gamma$, allow us to derive simple rates made of three component. First and second, the classical variance term $\sqrt {\mathcal B_T\log \mathcal B_T/T}$ and bias term $1/\mathcal B_T$ due to estimation of the spectral density in part I, which dominate over the usual $\sqrt T$-consistency of the sample covariance matrix which enters part II. Third, the term $1/\sqrt{n}$, required for disentangling the population dynamic and static common components from their idiosyncratic counterparts via cross-sectional aggregation.}
\end{rem}

To estimate the weak factors, we first have to ensure they are a finite number and identified. This is formalized by means of the following assumption.

\begin{assumption}\label{A weakfactor ID}
\ben \item [] $\!\!\!\! \!\!\!\!\!\!\!\!$
For any fixed $n\in\mathbb N$, the $r_w(n)$-dimensional vector of weak factors $F_t^{w,n}$ defined in Theorem \ref{thm: weakfactors} is such that:
\een
\ben
\item [(i)]
    $r_w(n)$ is independent of $n$, so that $r_w=r_w(n)$ and, moreover,  $0<r_w<\infty$;
\item [(ii)] let the $r_w$-dimensional vector of weak factors be denoted as $F_t^{w}$, then, there exists a fixed $\bar n\in\mathbb N$ independent of $n$ with $r_w\le \bar n<n$ and such that,
for all $t\in\mathbb Z$,
$$
F_t^w := (\mathcal M)^{-1/2}\, \mathcal W_{\bar n}\, e_t^{\chi,\bar n},
$$ 
where $e_t^{\chi,\bar n}=(e_{1t}^\chi\cdots e_{\bar n t}^\chi)'$, 
$\mathcal W_{\bar n}$ is the $r_w\times \bar n$ matrix having as rows the normalized eigenvectors of $\Gamma_{e^\chi}^{\bar n}:=\E[e_t^{\chi,\bar n}e_t^{\chi,\bar n\prime}]$ and $\mathcal M$ is the $r_w\times r_w$ diagonal matrix containing the corresponding $r_w$ eigenvalues sorted in decreasing order;
\item [(iii)] $\mu_j(\Gamma_{e^\chi}^{\bar n})>\mu_{j+1}(\Gamma_{e^\chi}^{\bar n})$ for $j=1,\ldots, r_w-1$.
\een
\end{assumption}

In part (i) we restrict to the case of a finite and positive number of weak factors, so their number, $r_w$, does not depend on $n$ and we adopt the simpler notation $F_t^w$. The case of infinite weak factors is briefly discussed in Remark \ref{rem: rwinfty}. In part (ii)
we identify the weak factors as the $r_w$-largest normalized principal components obtained from the population covariance matrix of the first $\bar n$ weak common components. This implies that weak factors are orthonormal, $\E[F_t^wF_t^{w'}]=I_{r_w}$, and it can be done without loss of generality, as, first, it is a common identifying assumption in the literature  \citep[see, e.g.,][]{forni2009opening}, and, second, the weak common components can always be ordered according to their variance, in such a way that, the first $\bar n\ge r_w$ are enough to recover the $r_w$ weak factors. This is because, under part (i) and Theorem \ref{thm: weakfactors}, $\Gamma_{e^\chi}^{\bar n}$ has exactly rank $r_w$.
Part (iii) requires distinct eigenvalues which is also standard in the literature and allows us to identify the eigenvectors.

We then prove consistency of the estimated weak factors (see Appendix \ref{weakfactproof} for a proof).

\begin{theorem}[Consistency of Weak Factors $\widehat F_t^w$]\label{thm: weakfactest}
If Assumptions A\ref{A: stat}, A\ref{ass:newGDFM}, A\ref{ass:diveval}, A\ref{ass:kernel}, A\ref{ass:divevalC}, A\ref{A weakfactor ID} hold and we compute $\wh{C}_{it}$ as in parts II.a or II.b, then, under either of the following set of additional conditions:
\ben
\item []  (I.a-II-III-IV) we compute $\wh{\chi}_t^n$ as in part I.a using the Bartlett kernel, Assumption A\ref{eq:decay} holds, and we consider any 
fixed $t=\mathcal M_T+1,\ldots, T-\mathcal M_T-1$;
\item [] (I.b-II-III-IV) we compute $\wh{\chi}_t^n$ as in part I.b using the Bartlett kernel, $n=\mathcal O(T^\gamma)$ for some $\gamma\in(0,\infty)$, Assumptions A\ref{ass:newGDFM} holds with $\nu>6$, Assumption A\ref{ass:ARGDFM} holds,  and we consider any 
fixed \linebreak $t=\mathcal K_T+1,\ldots, T$;
\een
as $n,T\to\infty$:
$$
\min\l(
    \sqrt{\frac{T}{\mathcal B_T\log \mathcal B_T}}, \mathcal B_T, \sqrt n\r
    ) \frac{\Vert\wh{F}_t^w-\mathcal S F_t^w \Vert}{\log T} = \mathcal O_P(1),
$$
where $\mathcal S$ is a $r_w\times r_w$ diagonal matrix with entries $\pm 1$.
\end{theorem}

\begin{rem}[The number of weak factors]
    \upshape{In order to compute our estimator of the weak factors we need to determine their number too. In practice, assuming these are a finite number, we can proceed as follows. After sorting the estimated weak common components according to their variance  (in decreasing order), we know that, if we choose $\bar n$ large enough, then, ${\Gamma}_{e^\chi}^{\bar n}$ has reduced rank $r_w$. Consider the sample covariance $\wh{\Gamma}_{e^\chi}^{\bar n}$ obtained from the top-left sub-matrix of the reordered version of $\wh{\Gamma}_{e^\chi}^n$ given in \eqref{eq:GammahatWCCSI}. Then, we can estimate $r_w$ by applying  to $\wh{\Gamma}_{e^\chi}^{\bar n}$ any standard method for determining the number of static factors, but in a fixed $n$ context and with no idiosyncratic component.
%
%
    In Section \ref{subsec: sim consist weak factors}, we show on simulated data that an eigenvalue-ratio criterion seems to work well  \citep{ahn2013eigenvalue}. In practice, as far as the choice of $\bar n$ is concerned, we can always pick a quite large $\bar n$ analogously to the choice of the maximum number of factors we are willing to allow for. 
 Given that in the proof of Theorem \ref{thm: weakfactest} we prove consistency of $\wh{\Gamma}_{e^\chi}^{\bar n}$, consistency of any estimator of $r_w$ built from such matrix would follow directly by the similar arguments as those used in the above mentioned works.}
\end{rem}

\begin{rem}[The case of an infinite number of weak factors]
\label{rem: rwinfty}
    \upshape{If the number of weak factors grows with $n$, then, Assumption A\ref{A weakfactor ID} does not hold, and we cannot retrieve them consistently for we would need an infinite dimensional panel. However, if $r_w(n)=n$ as it would be reasonable to assume, then, $\dim\cspargel(e_{t}^{\chi,n})=n$ for all $n\in\mathbb N$, and, therefore, for any given $n$-dimensional panel, we can always approximate the $n$-dimensional vector of weak factors $F_t^{w,n}$ by means of an $n$-dimensional orthonormal basis built starting from the vector of estimated weak common components $\wh{e}_t^{\,\chi,n}$, e.g., by the Gram-Schmidt orthogonalisation procedure.
    }
\end{rem}

\section{Simulation Experiments}\label{sec: simulation experiments}

In the following sections we assess in finite samples the impact of correctly accounting for the canonical decomposition in Theorems \ref{thm: relation of r-SFS and q-DFS} and \ref{thm: weakfactors} when estimating a factor decomposition and its implied IRFs.
To this end, we simulate data according to the GDFM in Assumption A\ref{A: q-DFS struct}: $y_{it}=\chi_{it}+\xi_{it}$, and we specify different data generating processes (DGPs)  for the dynamic common component $(\chi_{it})$, all based on Remark \ref{rem: stacking} where $r_\chi<\infty$ so that we have only a finite number of pervasive and weak static factors. Instead, for the dynamic idiosyncratic component, $(\xi_{it})$, we always consider the following model allowing for cross-sectional and serial correlations:
\begin{align}
    \xi_{it} = \alpha_i \xi_{i, t-1} + \varepsilon^\xi_{it}, \nn
\end{align}
with $(\varepsilon_{it}^\xi)$, $i=1,\ldots, n$, having elements $\varepsilon_{it}^\xi \sim iidN(0, 1)$ and independent of the common shocks $(\varepsilon_t)$ at all leads and lags. Moreover, $\E[\varepsilon_{it}^\xi \varepsilon_{jt}^\xi]= \tau^{\abs{i-j}}$, $i, j = 1, \ldots, n$, if $\abs{i-j}\leq 10$ and $\E[\varepsilon_{it}^\xi \varepsilon_{jt}^\xi] = 0$ otherwise, and $\alpha_i  \sim iidU(0, \delta)$.
Throughout, we set $\tau = 0.5$  and $\delta = 0.5$, and we rescale the idiosyncratic component in such a way that the signal-to-noise ratio is 1:1.

\subsection{Non-consistency of static PCA vs. consistency of dynamic PCA}\label{subsec: sim nonconsist}
In this section, we show that if the dynamic common component is partly driven by a non-trivial weak common component, static PCA  is non-consistent irrespective of the number of static factors considered, but dynamic PCA is consistent. To this end we consider the following DGP for the common component.

\noindent
\textbf{DGP1} The dynamic common component is generated as:
\begin{align}
    \chi_{it} &= \lambda_{i0} f_t + \lambda_{i1} f_{t-1}, \label{eq: innocent factor model} \\
     f_t &= a f_{t-1} + \sigma \varepsilon_t , 
     \label{eq: dynamics of ft DGP1}
\end{align}
with  $(f_t)$ scalar, $a = 0.8$, $\varepsilon_t\sim iidN(0,1)$, and  $\sigma^2 :=1-a^2$ so that $\E[f_t^2]=1$. Moreover,
 we set 
\begin{align}
\lambda_{i0}=\l\{
\begin{array}{ccc}
0 &\text{if}& 1\le i\le 10,\\
1 &\text{if}& 11\le i\le 20,\\
\mathcal N(1,1)&\text{if}& i\ge 21,
\end{array}
\r.
\quad
\lambda_{i1}=\l\{
\begin{array}{ccc}
1 &\text{if}& 1\le i\le 10,\\
0 &\text{if}&  i\ge 11.\\
\end{array}
\r.\label{eq:DG1load}
\end{align}
This is the same setup considered in Example \ref{exmp: IRF}.
  From Appendix \ref{app:example gdfm} we know that $q=1$, and from its canonical representation derived in Appendix \ref{subsec: weak factor sim example details1}
  we see that there are one static pervasive factor loaded by all series, and one static weak factor loaded only by the first 10 series. So $r=1$, $r_w=1$, and, therefore, $(\chi_{it})$ has covariance $\Gamma_\chi^n$ of rank $r_\chi=2$, but with only the largest eigenvalue diverging as $n$ grows.
In particular, the first unit, $i = 1$, decomposes as
\begin{align}
    \chi_{1t} &= a f_t + (f_{t-1} - a f_t) = f_{t-1}, \quad
    C_{1t} =   a f_t , \quad e_{1t}^\chi= f_{t-1} - a f_t,\label{eq: DGP1 C1}
\end{align}
so it has both a non-zero static and weak common component, which are mutually orthogonal by construction.

We then consider two ways of estimating $\chi_{1t}$. Either we use static PCA \citep{stock2002forecasting} with various numbers of static factors $r$, thus assuming incorrectly that $e_{it}^\chi=0$ for all $i=1,\ldots,n$, and, possibly, overestimating the number of factors according to the stacked representation \eqref{eq:stacked}. Or we use dynamic PCA \citep{forni2000generalized} with $q=1$ dynamic factors. Henceforth, we use $\mathcal B_T = \lfloor 0.75 \sqrt{T} \rfloor$ using the lag window estimator for the spectral density with a Bartlett kernel. 

We generate and estimate the model  $B = 500$ times and,  
at each replication $j = 1, \ldots, B$ we consider the two estimators of $\chi_{1t}$, generically denoted as $\widehat \chi_{1t}^{[j]}$, and we consider the Mean-Squared Error (MSE): 
$$
\text{MSE} = \frac 1B \sum_{j = 1}^B \frac 1T\sum_{t = 1}^T \l(\widehat \chi_{1t}^{[j]} - \chi_{1t}^{[j]}\r)^2,
$$ 
and report results for different $(n, T)$ combinations.
\begin{figure}
    \centering
	\begin{tabular}{cc}
    \includegraphics[width=.4\textwidth]{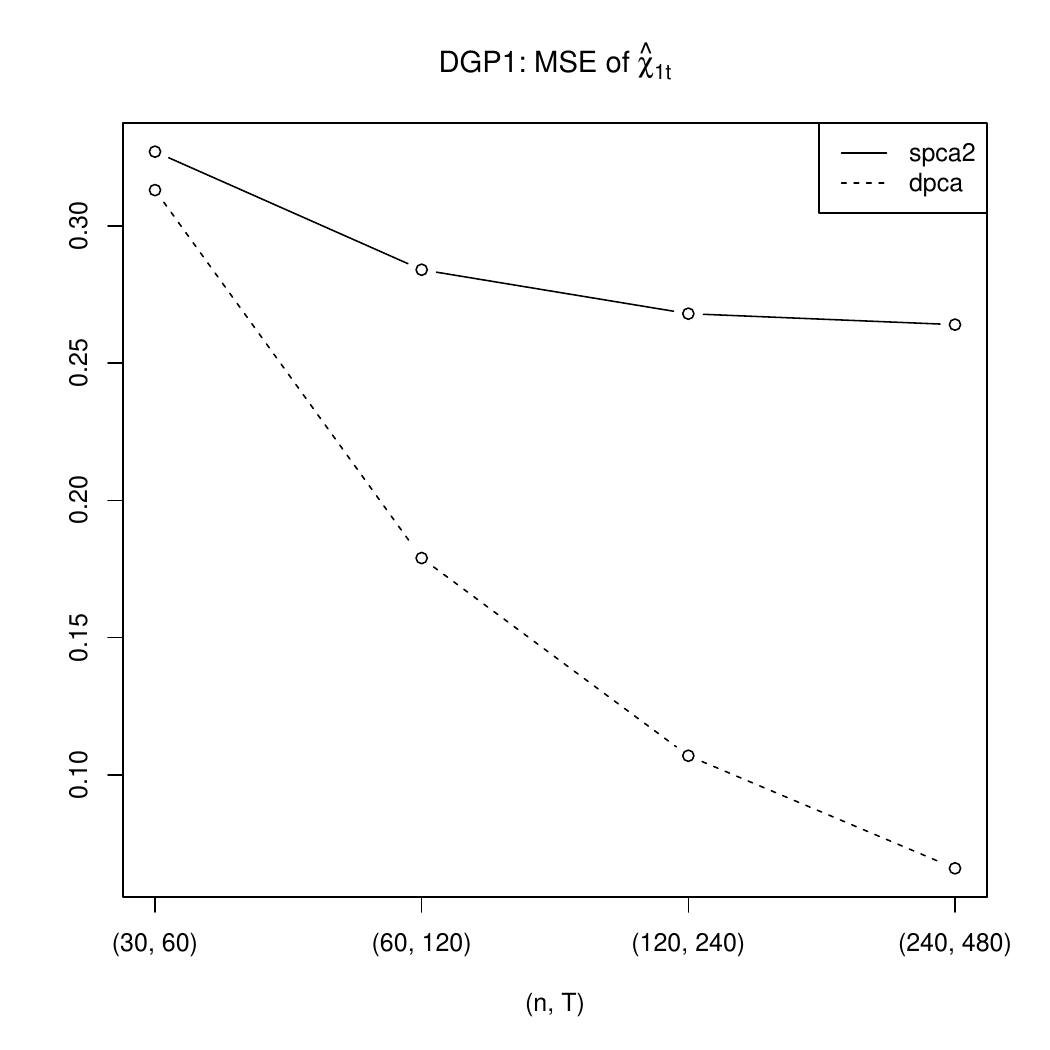}&
    \includegraphics[width=.4\textwidth]{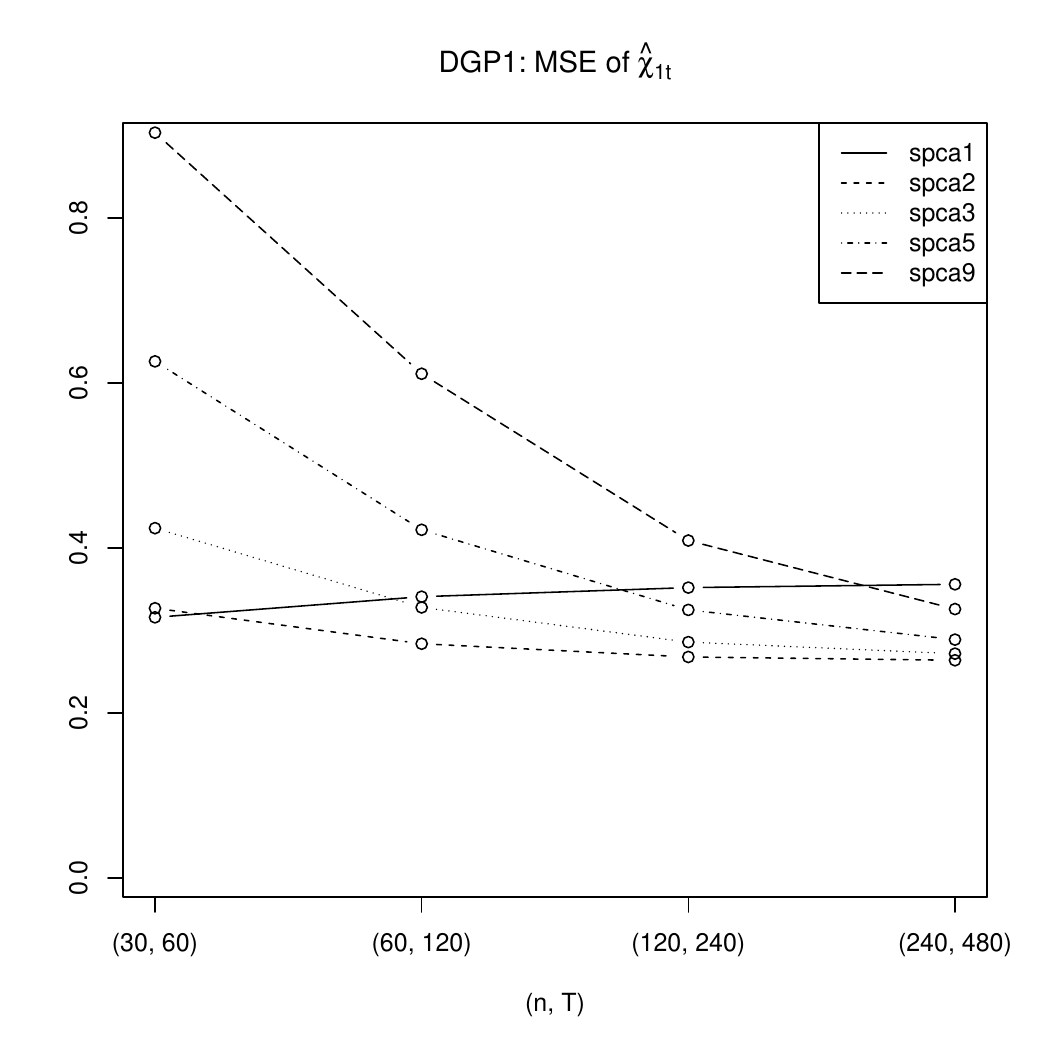}
    \end{tabular}
    \caption{ 
    \footnotesize DGP1. 
    Mean Squared Error of $\widehat \chi_{1t}$ over 500 replications for $(\delta, \tau) = (0.5, 0.5)$. \texttt{spcar}: estimation with static PCA with \texttt{r} $=1,2,3,5,9$, \texttt{dpca}: estimation by dynamic PCA with $q = 1$.}
    \label{fig: inconsistency of weak factors}
\end{figure}
The left plot in Figure \ref{fig: inconsistency of weak factors} shows, that, on the one hand, the static PCA with $r = 2$ (\texttt{spca2}), which is the total number of pervasive and weak static factors, is not consistent for $\chi_{1t}$, indeed, the MSE is not approaching zero as $n$ and $T$ becomes larger.  On the other hand dynamic PCA (\texttt{dpca}) is consistent as the MSE is monotonically decreasing with increasing $(n,T)$. Furthermore, the right plot of Figure \ref{fig: inconsistency of weak factors} shows that we cannot recover $\chi_{1t}$ by static PCA, when setting $r=1$ (\texttt{spca1}) or 
even if we increase the number of factors $r$, i.e., for $r = 3, 5, 9$.  This is because the second factor, being weak, cannot be estimated consistently (see Remark \ref{rem: weakF} and \citealp{onatski2012asymptotics}). It is, indeed, \textit{too weak} for the whole panel, though it is important individually, as it explains a large part of the variation for the first ten units. 

Analogous results are obtained for different sample size-paths with $n>T$ or $n = T$ (see Figures \ref{fig: inconsistency of weak factors T < n, 0505}-\ref{fig: inconsistency of weak factors T=n, 0505} in Appendix \ref{app: simres}). Notably, for $n > T$, we still obtain consistency for dynamic PCA, although, as expected from Proposition \ref{prop1}, $MSE$ decrease is slower.
\subsection{Consistency of the proposed estimators of the canonical decomposition}\label{subsec: sim consist candecomp}

We study the behavior of our proposed estimators for the canonical decomposition, as described in Section \ref{subsec: estimation in practice}, to $\textbf{DGP1}$ again for $i = 1$. 
We generate and estimate the model  $B = 500$ times and,  at each replication $j=1,\ldots, B$, we compare estimators $\wh\chi_{1t}^{type,[j]}$ with $type=$  I.a, I.b and $\widehat C_{1t}^{type,[j]}$, $\widehat e_{1t}^{\chi, type,[j]} = \widehat \chi_{1t}^{type,[j]} - \widehat C_{1t}^{type,[j]}$ with $type=$ I.a-II.a, I.a-II.b, I.b-II.a, and I.b-II.b. For $ t \in \mathcal T = \{\mathcal B_{T} + 1, \ldots, T - \mathcal B_T\}$, we evaluate:
\begin{align}
  \text{MSE} =\frac 1B \sum_{j=1}^B \frac{1}{T - 2 \mathcal B_T}\sum_{t\in \mathcal T} \l(\widehat \chi_{1t}^{type,[j]} - \chi_{1t}^{[j]}\r)^2, \nn
\end{align}
with analogous definitions for $\widehat C_{1t}^{type,[j]}$ and $\widehat e_{1t}^{\chi, type,[j]}$. The results for different $(n, T)$ combinations are shown in Table \ref{tbl: MSE canon decomp 0505}. For all the proposed estimators converge the MSE decreases with increasing $n$ and $T$. The estimator I.b-II.b yields mostly the best performance for all components across all cases. In particular $\wh\chi_{1t}^{Ib}$ built using the one-sided approach by \citet{forni2017dynamic}, while $\widehat C_{1t}^{IIb}$ is obtained via PCA on the covariance matrix obtained by  integration of the estimated spectral density of the dynamic common component. This approach ensures that the estimators of $\widehat C_{1t}^{IIb}$ and $\widehat e_{1t}^{\chi, Ib,IIb}$ have zero sample covariance, and the sum of their sample variances is equal to the sample variance of $\wh\chi_{1t}^{Ib}$.
Analogous results are obtained for different sample size-paths with $n>T$ or $n = T$ (see Tables \ref{tbl: MSE canon decomp 005 ngT} and \ref{tbl: MSE canon decomp 005  neqT} in Appendix \ref{app: simres}). %
%
%
\begin{table}[ht]
\centering
\begin{tabular}{l|llll}
\toprule
\multicolumn{5}{c}{\textbf{MSE, Canonical Decomposition Estimates}} \\
\midrule
\hline
$(n,T)=$ & (30,60) & (60,120) & (120,240) & (240,480) \\ 
\\[-1.8ex]\hline 
\hline \\[-1.8ex]

$\widehat{\chi}^{Ia}_{1t}$ 
& 0.318 (0.197) & 0.202 (0.122) & 0.11 (0.06) & 0.066 (0.033) \\[0.2em]

$\widehat{\chi}^{Ib}_{1t}$ 
& 0.429 (1.384) & 0.12 (0.086) & 0.061 (0.038) & 0.034 (0.021) \\[0.2em]

\hline \\[-0.8em]

$\widehat{C}^{Ia, IIa}_{1t}$ 
& 0.182 (0.134) & 0.097 (0.077) & 0.046 (0.038) & 0.023 (0.02) \\[0.2em]

$\widehat{C}^{Ia, IIb}_{1t}$ 
& 0.18 (0.134) & 0.099 (0.08) & 0.047 (0.038) & 0.023 (0.02) \\[0.2em]

$\widehat{C}^{Ib, IIa}_{1t}$ 
& 0.161 (0.178) & 0.044 (0.033) & 0.019 (0.017) & 0.008 (0.008) \\[0.2em]

$\widehat{C}^{Ib, IIb}_{1t}$ 
& 0.16 (0.175) & 0.046 (0.037) & 0.019 (0.017) & 0.008 (0.008) \\[0.2em]


\hline \\[-0.8em]

$\widehat{e}^{\chi, Ia, IIa}_{1t}$ 
& 0.196 (0.069) & 0.122 (0.038) & 0.071 (0.023) & 0.049 (0.014) \\[0.2em]

$\widehat{e}^{\chi, Ia, IIb}_{1t}$ 
& 0.188 (0.066) & 0.12 (0.037) & 0.071 (0.022) & 0.05 (0.014) \\[0.2em]

$\widehat{e}^{\chi, Ib, IIa}_{1t}$ 
& 0.359 (1.109) & 0.123 (0.083) & 0.06 (0.037) & 0.034 (0.022) \\[0.2em]

$\widehat{e}^{\chi, Ib, IIb}_{1t}$ 
& 0.348 (1.122) & 0.112 (0.077) & 0.056 (0.035) & 0.032 (0.021) \\[0.2em]



\hline \hline
\end{tabular}\\[3pt]

\caption{
\footnotesize 
Mean Squared Error and standard deviation (in parentheses) evaluated over $B = 500$ replications.
}
\label{tbl: MSE canon decomp 0505}
\end{table}

\subsection{Consistent estimation of weak factors}\label{subsec: sim consist weak factors}
Next, we consider estimation of weak factors under the assumption that they belong to the dynamic common space (see Remark \ref{rem: weakF}). 
To this end we consider the following DGP for the common component.

\noindent
\textbf{DGP2} The dynamic common component is generated as:
\begin{align}
     \chi_{it}  &= \lambda_{i0} f_t + \lambda_{i1} f_{t-1} + \lambda_{i2} f_{t-2} ,\nonumber\\
    f_t &= a_1 f_{t-1} + a_2 f_{t-2} + \sigma \varepsilon_t, 
    \label{eq: dynamics of ft AR2} 
\end{align}
with  $(f_t)$ scalar, $a_1 = 0.5$ and $a_2 = 0.3$, $\varepsilon_t\sim iidN(0,1)$, and  $\sigma^2 = 1-a_2^2 - \frac{a_1^2(1+a_2)}{1-a_2}$  so that $\E[f_t^2]=1$. Moreover,
we set 
\begin{align}
\lambda_{i0}=\l\{
\begin{array}{ccc}
0 &\text{if}& 1\le i\le 10,\\
1 &\text{if}& 11\le i\le 20,\\
\mathcal N(1,1)&\text{if}& i\ge 21,
\end{array}
\r.
\,
\lambda_{i1}=\l\{
\begin{array}{ccc}
1 &\text{if}& 1\le i\le 5,\\
0 &\text{if}&  i\ge 6,\\
\end{array}
\r.
\,
\lambda_{i2}=\l\{
\begin{array}{ccc}
0 &\text{if}& 1\le i\le 5,\\
0.5 &\text{if}& 6\le i\le 10,\\
0&\text{if}& i\ge 11.
\end{array}
\r.\nn
\end{align}
From Appendix \ref{app:example gdfm} we know that $q=1$, and from Appendix \ref{subsec: weak factor sim example details2} we see that there are one pervasive static factor and two weak static factors, i.e., $r=1$ and $r_w=2$, so that $(\chi_{it})$ has covariance $\Gamma_\chi^n$ of rank $r_\chi=3$, but with only the largest eigenvalue diverging as $n$ grows. In Appendix \ref{subsec: weak factor sim example details2} we also show how to generate the 2-dimensional vector of weak factors $F_t^w$ as the first two normalised population principal components of $(e_t^{\chi,\bar n})$, so that they satisfy Assumption A\ref{A weakfactor ID}, and how to fix their sign.
Here we choose $\bar n=10$.




We generate and estimate the model  $B = 500$ times and,  at each replication $j=1,\ldots, B$, we obtain the estimator $\wh F_t^{w,type,[j]}$, as described in Section \ref{subsec: estimation in practice} and with $type=$ I.a-II.a, I.a-II.b, I.b-II.a, and I.b-II.b, depending on the chosen estimator $\wh{e}_t^{\,\chi,\bar n}$.
For $ t \in \mathcal T = \{\mathcal B_{T} + 1, \ldots, T - \mathcal B_T\}$ we evaluate:
\begin{align*}
    \text{MSE} = \frac 1B\sum_{j=1}^B \frac{1}{T - 2 \mathcal B_T} \sum_{t \in \mathcal T} \norm{\widehat F_t^{w,type,[j]} - F_t^{w,[j]}}^2.
\end{align*}
Results are shown in Table \ref{tbl: MSE Fw 0505}  for different $(n, T)$ combinations.. In all cases the MSE converges to zero. Interestingly, convergence kicks in later (therefore we extended the sample size) for methods based on estimation of $\chi_{it}$ as in part I.b \citep{forni2017dynamic}. This is likely to be due to the instability of inverting a polynomial of a VAR(2) process in the last step of estimation.
\begin{table}[ht]
\centering
\begin{tabular}{l|lllll}
\toprule
\multicolumn{6}{c}{\textbf{MSE, Estimation of weak factors}} \\
\midrule
\hline
$(n,T)$ & (30,60) & (60,120) & (120,240) & (240,480) & (480,900) \\
\\[-1.8ex]\hline
\hline \\[-1.8ex]
$\widehat F_{t}^{w, Ia, IIa}$ & 0.61 (0.349)  & 0.309 (0.123) & 0.168 (0.041) & 0.123 (0.025) & 0.094 (0.016) \\[0.2em]
$\widehat F_{t}^{w, Ia, IIb}$ & 0.595 (0.339) & 0.304 (0.116) & 0.168 (0.04)  & 0.123 (0.025) & 0.094 (0.015) \\[0.2em]
$\widehat F_{t}^{w, Ib, IIa}$ & 3.571 (5.128) & 1.16 (2.096)  & 0.465 (0.399) & 0.297 (0.239) & 0.183 (0.143) \\[0.2em]
$\widehat F_{t}^{w, Ib, IIb}$ & 3.617 (5.18)  & 1.126 (2.105) & 0.431 (0.383) & 0.279 (0.23)  & 0.174 (0.138) \\[0.2em]
\hline 
\toprule
\multicolumn{6}{c}{\textbf{Share of correct estimates $\widehat r_w = 2$}} \\
\midrule
\hline
$\widehat r_w$             & 0.026  & 0.18   & 0.502    & 0.89   & 0.996  \\[0.2em]
\hline \hline
\end{tabular}
\caption{
\footnotesize
Mean Squared Error for $\widehat F_t^w$ and standard deviation (in parentheses) and share of correct estimates of $r_w$ evaluated over $B = 500$ replications.
}
\label{tbl: MSE Fw 0505}
\end{table}

For estimating $r_w$ we need to choose heuristically $\bar n \ge r_w$. A natural procedure for choosing $\bar n$ is to sort the variables according to the variance contribution by the weak common component (see Figures \ref{fig: share of var r8} and \ref{fig: share of var volatility r1 q1} obtained from real data) and to stop at $\bar n$ such that the estimated variance contribution of the weak common component becomes negligible. Having chosen $\bar n$ we estimate $r_w$ by means of an eigenvalue-ratio criterion as in \cite{ahn2013eigenvalue}, i.e., 
$\widehat r_w = \argmax_{1 \leq j \leq\bar n - 1} \mu_j(\widehat \Gamma_{e^\chi}^{\bar n}) / \mu_{j+1}(\widehat \Gamma_{e^\chi}^{\bar n})$. The results at the bottom of Table \ref{tbl: MSE Fw 0505} indicate that $\widehat r_w$ consistently identifies the true number of weak factors. 


       %
       %
       %
       %
       %
       %
       %

\subsection{Non-consistency of static IRFs as estimators of true IRFs}\label{subsec: sim est of IRF}

In light of the discussion in Section \ref{subsec: IRFs}, we show, by means of two exercises, that estimating the true dynamic IRFs using a static approach produces non-consistent estimates.


First consider again DGP1 where $r=q=1$. The generic expression of true IRFs is given in \eqref{eq: DGP1 IRF model dynamicex5} in Example \ref{exmp: IRF}. In particular, from \eqref{eq:DG1load}, for $\chi_{1t}$
we get:
    %
    \begin{align}
    \ubar k_1^\chi(L) 
    = L(1-aL)^{-1}\sigma = \sum_{j=0}^\infty a^j \sigma L^{j+1}=
    0 L^0 + \sum_{j = 1}^\infty a^{j-1}\sigma L^j. \label{eq: DGP1 IRF model dynamic} 
    %
    %
\end{align}
If instead we focus only on  the static common component, thus we ignore the existence of the weak common component, the static IRFs are obtained from \eqref{eq: DGP1 C1} and the AR(1) model of the factors as:
    \begin{align}
     \ubar k_1^C(L) &=a (1-aL)^{-1}\sigma= \sum_{j = 0}^\infty a^{j+1}\sigma L^j. \label{eq: DGP1 IRF Model Static}
\end{align}
Consequently $\ubar k_1^\chi(L) \neq \ubar k_1^C(L)$ and no estimator of $\ubar k_1^C(L)$ can be consistent for $\ubar k_1^\chi (L)$. 

By using the estimation approach of  \cite{forni2017dynamic},
we immediately have an estimate of the true IRFs (see Remark \ref{rem:IRFhat} for details).
Moreover, following the CC-VAR by \citet{forni2025common}, we estimate the static IRFs from an AR(1) model for 
 $\widehat C_{1t}$, which, in turn, is obtained via static PCA either when using the correct number of factors, $r = 1$, or when over-specifying the number of factors by choosing $r = 2$. Denote such estimator as $\widehat{\ubar k}_1^C(L)$.

\begin{figure}\label{Fig:IRFs}
    \centering
	\begin{tabular}{cc}
    \includegraphics[width=.4\textwidth]{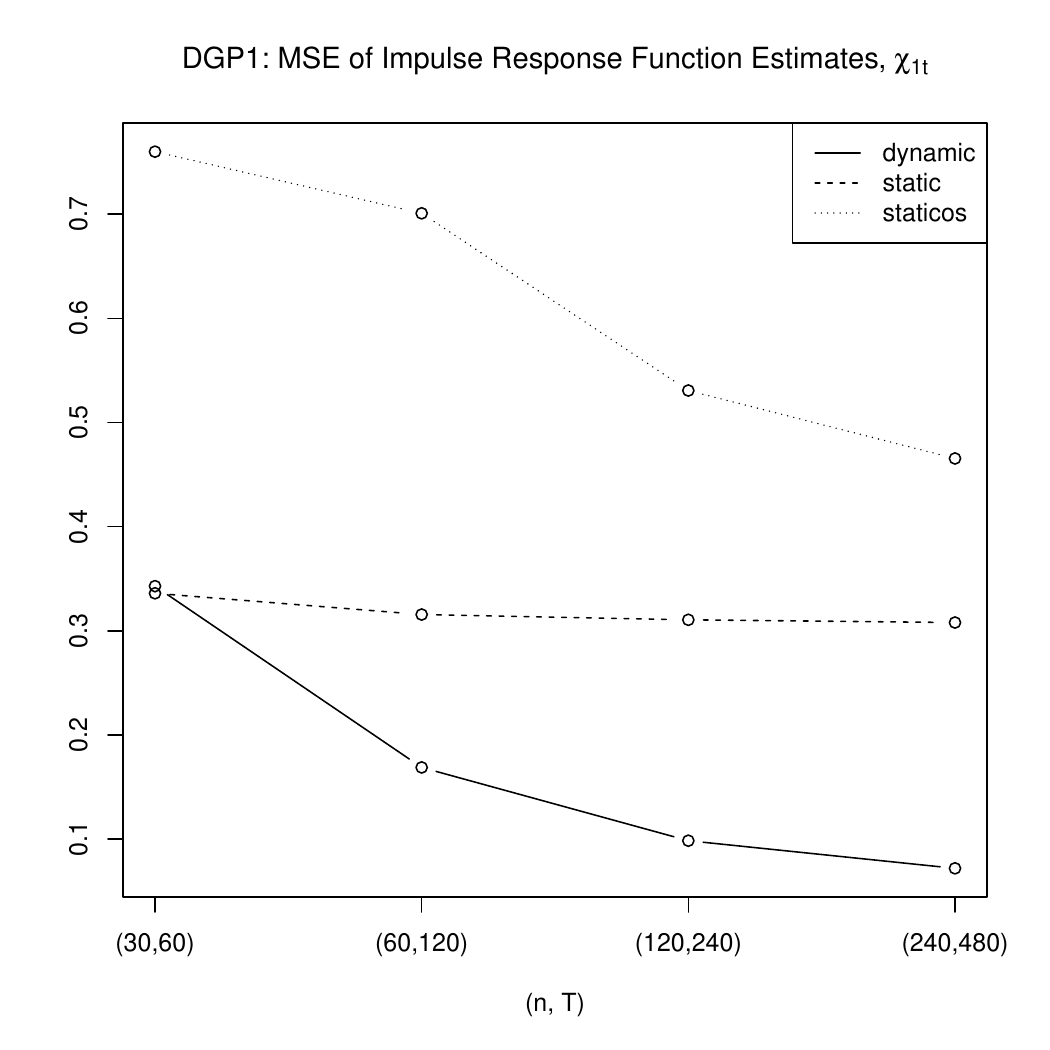}&
    \includegraphics[width=.4\textwidth]{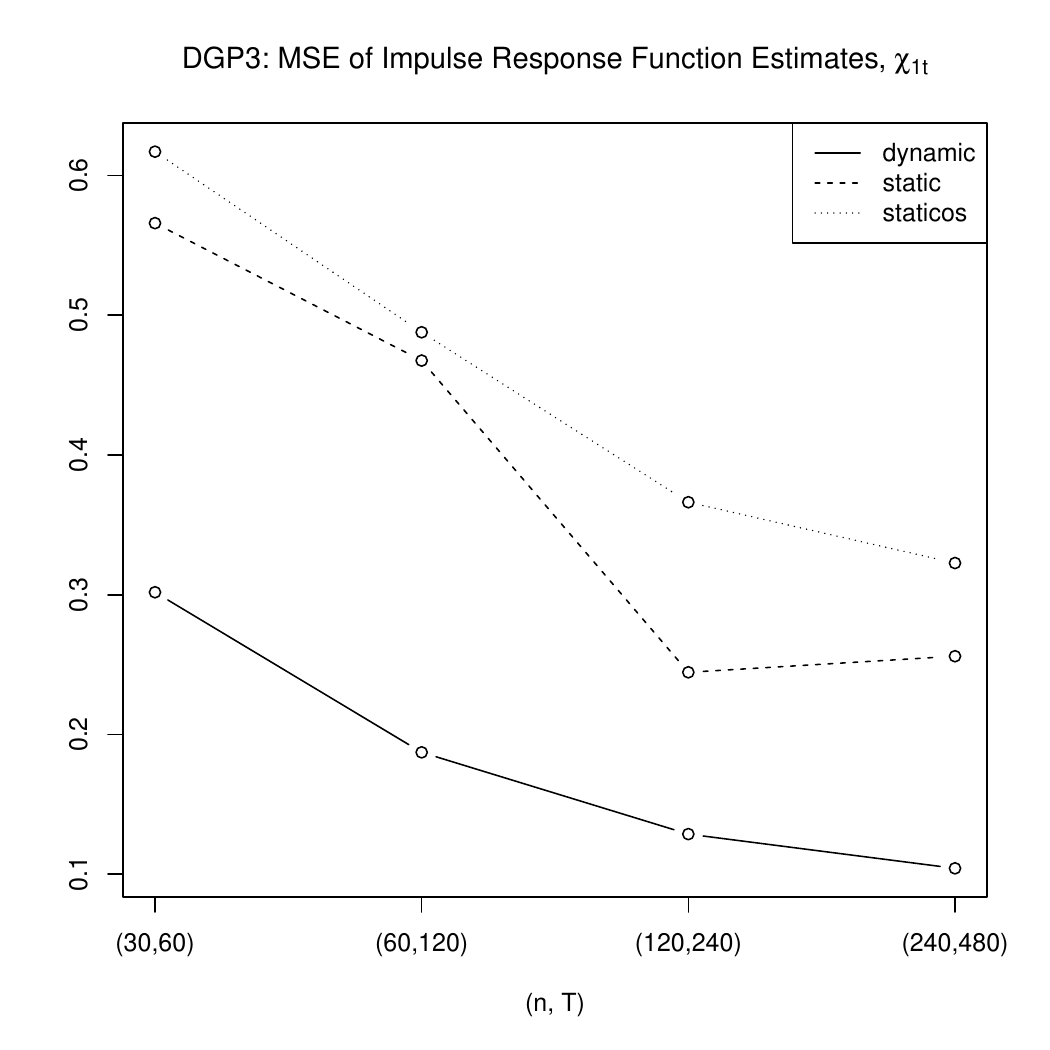}
    \end{tabular}
        \caption{ 
    \footnotesize  
   Mean Squared Errors of estimates of first 11 impulse response coefficients over 500 replications. Left: DGP1. Right: DGP3. \texttt{static}: Estimation of IRFs via static PCA plus VAR on $\wh C_{1t}$ with $r=1$ (for DGP1) or $(\wh C_{1t}\; \wh{C}_{11t})'$ with $r=2$ (for DGP3) as in \citet{forni2025common}.
   \texttt{dynamic}: Estimation of IRFs with $q=1$ as in \cite{forni2017dynamic}. \texttt{staticos}: Estimation of IRF as in the  \texttt{static} case but with over-specified number of factors, $r=2$ (for DGP1) or $r=3$ (for DGP3). }
\end{figure}

We generate and estimate the model  $B = 500$ times. Let $k^{\chi} := \l( K_1^{\chi}(0), \ldots, K_1^{\chi}(10)\r)' \in \mathbb R^{11}$ be the vector of the first 11 coefficients of the true IRF given in \eqref{eq: DGP3 IRF Model Dynamic}, which is fixed for all replications. Then, at each replication $j=1,\ldots, B$,  let $\widehat k^{[j]}$ be an estimate thereof obtained either from the dynamic estimate $\widehat {\ubar k}_1^{\chi,[j]}(L)$ or from the static one $\widehat {\ubar k}_1^{C,[j]}(L)$. Then, we compute 
\begin{align*}
\text{MSE}= \frac{1}{B}\sum_{j = 1}^B \norm{\widehat k^{[j]} - k^{\chi}}^2.
\end{align*}
The left panel of Figure \ref{Fig:IRFs} shows the results. The static approach (\texttt{static}) does not seem to be consistent, not even when over-estimating the number of factors  (\texttt{staticos}). The dynamic approach (\texttt{dynamic}) is instead consistent as the MSE is monotonically decreasing with increasing $(n,T)$.

In a second exercise we consider the following DGP where $r>q$.

\noindent
\textbf{DGP3} The dynamic common component is generated as:
\begin{align}
    \chi_{it} = \lambda_{i0} f_t + \lambda_{i1} f_{t-1} + \lambda_{i2} f_{t-2}, \nn
\end{align}
with $(f_t)$ scalar and following the same AR(2) as in \eqref{eq: dynamics of ft AR2}. Moreover, we set 
\begin{align}
\lambda_{i0}=\l\{
\begin{array}{ccc}
0 &\text{if}& 1\le i\le 10,\\
1 &\text{if}& 11\le i\le 20,\\
\mathcal N(1,1)&\text{if}& i\ge 21,
\end{array}
\r.
\,
\lambda_{i1}=\l\{
\begin{array}{ccc}
0 &\text{if}& 1\le i\le 20,\\
\mathcal N(0.5,0.5)&\text{if}& i\ge 21,
\end{array}
\r.
\,
\lambda_{i2}=\l\{
\begin{array}{ccc}
1 &\text{if}& 1\le i\le 10,\\
0 &\text{if}& i\ge 11.\\
\end{array}
\r.\nn
\end{align}
Using once again the results in Appendix \ref{app:example gdfm}, we see that $q = 1$, and from Appendix \ref{subsec: weak factor sim example details3} we see that there are two pervasive static factors and one weak static factor, i.e., $r=2$ and $r_w=1$, so that $(\chi_{it})$ has still covariance $\Gamma_\chi^n$ of rank $r_\chi=3$, but now with  the two largest eigenvalues diverging as $n$ grows.

Now, since $\chi_{1t}=f_{t-2}$, letting $\ubar k^f(L):=(1-a_1L-a_2L^2)^{-1}$, from \eqref{eq: dynamics of ft AR2} we see that the true IRFs are:
    %
    %
\begin{align}
    \ubar k_1^\chi (L) &= L^2\ubar k^f(L) \sigma = 0 L^0 + 0 L^1 + \sum_{j = 2}^\infty K^f(j-2) L^j \sigma. \label{eq: DGP3 IRF Model Dynamic} 
\end{align}
If instead we focus only on the static common component, then the IRFs can be obtained from the AR(2) representation of $f_t$. Indeed, from Appendix \ref{subsec: weak factor sim example details3},
$C_{1t}=\gamma_f(2)f_t+\gamma_f(1)f_{t-1}$, with $\gamma_f(h):=\E[f_tf_{t-h}]$, $h=1,2$,
%
    which implies 
    \begin{align}
    \ubar k_1^C(L) &= \l\{\gamma_f(2)  + L \gamma_f(1)\r\} \ubar k^f (L) \sigma 
    = \gamma_f(2) K^f(0) L^0+
    \sum_{j = 1}^\infty \l[\gamma_f(2) K^f(j) + \gamma_f(1) K^f(j-1)\r] \sigma L^j. \label{eq: DGP3 IRF Model Static}
\end{align}
Once again, $\ubar k_1^\chi(L) \neq \ubar k_1^C(L)$ and no estimator of $\ubar k_1^C(L)$ can be consistent for $\ubar k_1^\chi (L)$. 

As for DGP1 the true IRFs can be estimated as in \citet{forni2017dynamic}. While,  by following the CC-VAR by \citet{forni2015dynamic}, and since $q=1$, the static IRFs can be estimated by fitting any 2-dimensional VAR containing $\wh C_{1t}$ and any other $\wh C_{kt}$, $k>1$, provided they have different IRFs. Here we choose $k=11$. Again we consider static PCA implements either when using the true number of factors $r=2$, or when over-estimating their number and using $r=3$.

Results computed over $B=500$ replications are in the right panel of Figure \ref{Fig:IRFs} and confirm those obtained for DGP1. The static approach cannot recover the true IRFs neither when $r=2$ (\texttt{static}) nor when $r=3$ (\texttt{staticos}). The dynamic approach (\texttt{dynamic}) is instead consistent. 
%

\section{Empirical Applications}\label{sec: empiric}

\subsection{Macroeconomic indicators}
We analyze the FRED-MD dataset, which comprises monthly observations of U.S. macroeconomic time series. The data is described in \citet{mccracken2016fred}, who also provide details on the transformations necessary to achieve stationarity. With respect to those transformations we make three exceptions: the Consumer Price Index for All Urban Consumers (CPIAUCSL) is log-differenced, while the Civilian Unemployment Rate (UNRATE) and the 1-year Treasury Bills (GS1) are kept in levels. These choices do not affect the results of this section and ensure comparability with the analysis of \citet{forni2025common}, which we revisit below. Few series are removed due to missing values and outliers are removed using standard methods.\footnote{We remove univariate outliers when they exceed ten times the interquartile range, and multivariate outliers (entire rows of the panel) when detected using the isolation forest algorithm implemented in the R package \texttt{isotree} \citep{cortes2025package}. After removing outliers, we apply spline methods together with the EM algorithm with $r=8$ factors proposed by \cite{stock2002forecasting} to interpolate the remaining missing values.}

\begin{figure}[h!]
    \centering
    \includegraphics[width=.8\textwidth]{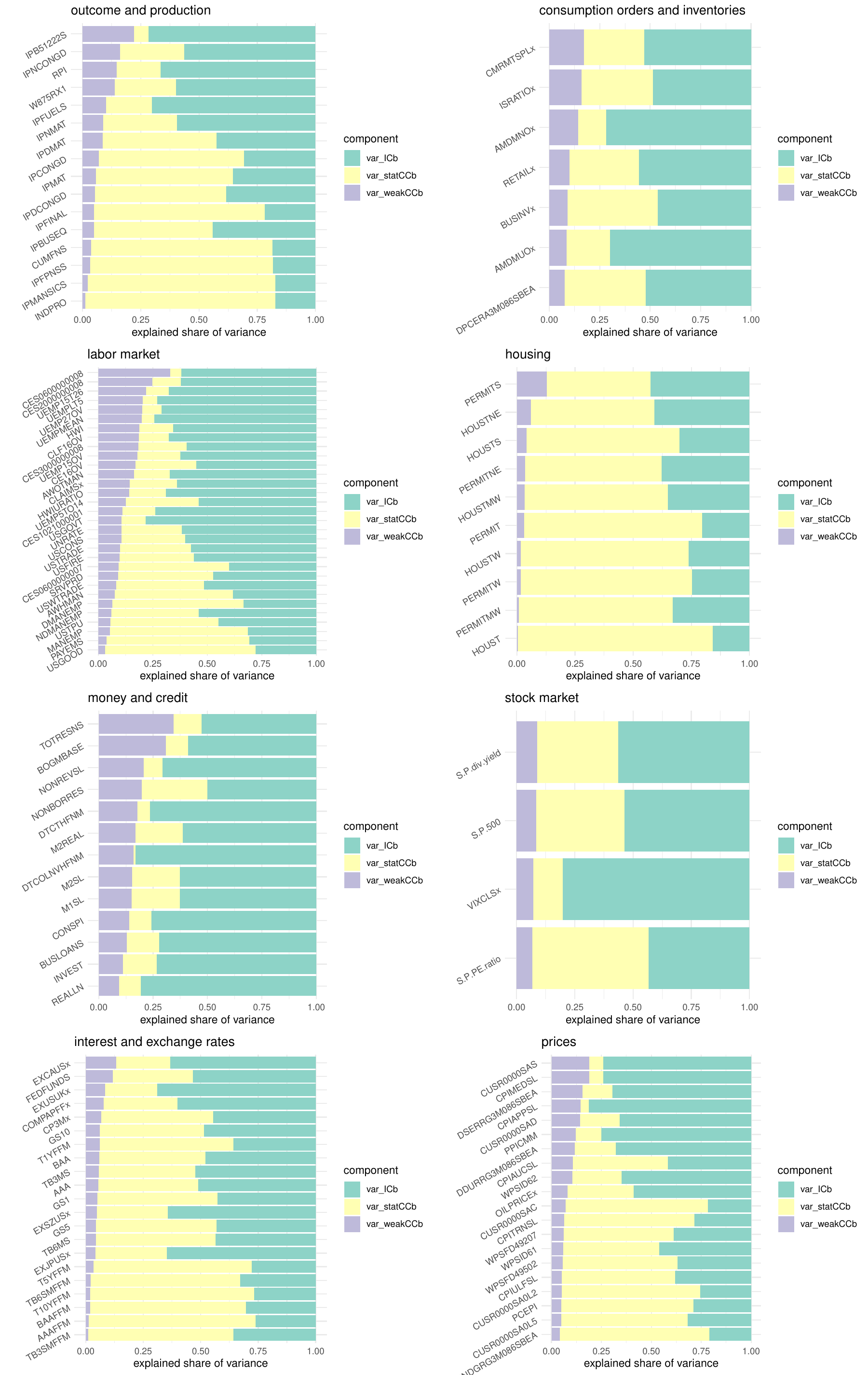}
    \caption{\footnotesize Share of variance explained by each component per variable with $q = 4$ and $r=8$. Estimates are obtained by using part II.b when estimating $C_{it}$. Here \texttt{var\_statCCb} $= EV_i^C$ (given in \eqref{eq:EVCa}),  \texttt{var\_weakCCb} $=EV_i^{e^\chi}$ (given in \eqref{eq:EVWC}).
     Last, \texttt{var\_ICb} is the variance explained by the dynamic idiosyncratic component, which is given by $EV_i^\xi=1-EV_i^C-EV_i^{e^\chi}$.}
    \label{fig: share of var r8}
\end{figure}

The final dataset contains $n=123$ series and $T=798$ monthly observations spanning 1959:01 to 2025:08, thereby including both the Global Financial Crisis and the COVID-19 period. Notably, several months during the 2008 financial crisis and the COVID-19 crisis are classified as multivariate outliers. An implicit list of the variables included in the dataset is provided in Figure \ref{fig: share of var r8}. 

\paragraph{Estimating the canonical decomposition.}
We estimate the canonical decomposition (\ref{eq: 3fold decomp intro}) using the procedures from Section \ref{sec: estimation} applied to the stationarity transformed centered and standardised data. All estimates are computed as in Section \ref{sec: simulation experiments} using a bandwidth $\mathcal B_T = \lfloor 0.75 \sqrt{T} \rfloor = 21$ and a Bartlett kernel.

To start, we determine the number of dynamic and static factors. Using the method by \cite{hallin2007determining} yields $q = 4$ dynamic factors and using the method by \cite{bai2002determining} with the penalty tuning suggested by \cite{alessi2010improved} yields $r = 8$ static pervasive factors. 
It is important to notice that, given the results in \citet{bai2023approximate}, among the $r = 8$ considered static factors are included all possible rate-weak factors. Indeed, when looking for the number of stronger factors, by means of the method by \citet{freyaldenhoven2022factor} we find $r = 6$ factors corresponding to eigenvalues diverging at rate $\sqrt n$ or faster, meaning that the remaining two factors correspond to eigenvalues diverging at rate $n^\alpha$, with $\alpha\in (0,1/2)$, i.e., they are still pervasive but very weakly pervasive.
Moreover, our eigenvalue ratio criterion suggests the presence of $r_w=5$ weak factors.

As stated above, the advantage of our approach is that it yields three components which are orthogonal also in sample, meaning they have zero sample correlation. In order to assess the importance of each component for each variable, we compute the variance share as follows.
First, since we are working with standardized data, each $y_{it}$ has zero sample mean and unit sample variance. Then, the share of variance explained by the dynamic common component is computed as:
\beq\label{eq:EVD}
EV^\chi_i =[\wh{\Gamma}_\chi^n]_{ii}= \frac {2\pi}{2\mathcal B_T+1}\sum_{h=- \mathcal B_T}^{\mathcal B_T}  [\widehat{f}_\chi^n(\theta_h)]_{ii},
\eeq
where $\widehat{f}_\chi^n(\theta_h)$ is computed as in \eqref{eq:sigmachiest} in step I.b.ii above. Note that the quantity $EV^\chi_i$ is the same regardless whether we use the estimator in part I.a or I.b. The share of variance explained by the static common component is computed as:
\beq\label{eq:EVCa}
EV^C_i  = \l[\widehat P_n'\widehat P_n \wh{\Gamma}_\chi^n \widehat P_n'\widehat P_n\r]_{ii},
\eeq
with $\widehat P_n$  the $r\times n$ matrix having as rows the normalized eigenvectors of $\wh{\Gamma}_\chi^n$  which in turn is computed as in \eqref{eq:EVD}.
This, corresponds to the sample variance of $\wh C_{it}$ when computed as in part II.b. The share of variance explained by the weak common component is then:
\beq\label{eq:EVWC}
EV^{e^\chi}_i=EV^\chi_i -EV^C_i,
\eeq
which, by construction, is always a non-negative quantity.

The results are shown in Figure \ref{fig: share of var r8}. 
Overall we see that the weak common component accounts for a non-negligible part of the variation for most of the series. It explains more than 5\% of the total variation for 94 of the 123 series. The weak common component seems to play an important role in particular the labour market and the money and credit sector but also for the other sectors. 

The variance explained by the weak common component of a few noteworthy indicators is:  Civilians Unemployed - Less Than 5 Weeks (UEMPLT5) with $20.4\%$, Avg Hourly Earnings in the Goods-Producing sector (CES0600000008) with $32.9\%$, and Total Monetary Base (BOGMBASE) with $30.9\%$. The weak common component shares of the variables used below to estimate IRFs are: Civilian Unemployment Rate (UNRATE) with $10.7\%$, CPI
Inflation (monthly growth rate of CPIAUCSL) with $10.8\%$, Industrial Production growth rate (INDPRO) with $1.17\%$ and the 1-year Treasury Bills (GS1) with $5\%$.  Industrial production has a small weak common component share. This makes sense from a theoretical point of view. 
Indeed, Industrial Production is a contemporaneous aggregate (static idiosyncratic part vanishes under averaging) so it is mainly driven by the contemporaneously pervasive factors and it is likely to have small dynamic idiosyncratic component.\footnote{Results when computing $EV_i^C$ starting from the estimator in part II.a and when $r=6$ are in Figures \ref{fig: share_of_var_r8_method_a} and \ref{fig: share of var r6} in Appendix \ref{app:EVfig}, respectively. These are overall similar to those shown here.} 





Finally, 
we examine how the empirical results change if we increase $r$. For instance, one might argue that the large part of variation of the weak common component obtained above, in fact stems from an underspecification of the number of statically pervasive factors. Results for the case $r=12$ are presented in Figure \ref{fig: share of var r12} in Appendix \ref{app:EVfig}. Of course, the share of variance explained by weak common component is reduced but the overall picture remains similar. Even with $r= 12$, for instance 68 of the 123 series have a weak common component share larger than $5\%$.
This is consistent with the fact that we cannot consistently recover the weak common component via contemporaneous aggregation as, e.g., PCA (see Remark \ref{rem: weakF}).

\paragraph{IRFs.}
We then study the effect of monetary policy shocks by comparing the results obtained both with a static and a dynamic approach. 
In order to match the original empirical analysis by \citet{forni2025common}, the sample considered for this exercise is shorter than before and spans from 1979:01 to 2011:01.\footnote{This choice is due to limited availability of the instrument used as the proxy of monetary policy shocks.}
 
First, we estimate the static common components via PCA and the dynamic common components as in \citet{forni2017dynamic}. 
In agreement with the results in the previous section, we still have $r=8$ and $q=4$.

Second, following the choice by \citet{forni2025common}, we focus on the following four variables: Industrial Production growth rate (INDPRO), Unemployment rate (UNRATE), CPI Inflation (CPIAUCSL), 1-year Treasury Bills (GS1), and we estimate the following VARs, with lags selected via BIC:
\begin{enumerate}
    \item [(a)] using the static common components of the four main variables, so that $m=q=4$;
    \item [(b)] using the static common components of the four main variables plus the first four principal components, so that $m=r=8$;\footnote{Results for case (b) are unchanged if besides the four main variables we use just two principal components together with the static common components of Oil Price (OILPRICEx) and the Excess Bond Premium (EBP) available from \citet{Gilchrist2012credit}.}
    \item [(c)] using the dynamic common components of the four main variables, so that $m=q=4$.
\end{enumerate} 

Third, the identification of the monetary policy shock is based on the proxy VAR method \citep{Stock2018identification}, implemented using the \citet{Gertler2015monetary} instrument and with the policy rate given by the 1-year Treasury Bills included in the VARs. The effect of the shock is such that the impact on the 1-year Treasury Bills is 0.5.

\begin{figure}[h!]
    \makebox[\textwidth][c]{
    \begin{tabular}{@{}>{\centering\arraybackslash}m{5mm}@{}  >{\centering\arraybackslash}m{.22\textwidth}
                  >{\centering\arraybackslash}m{.22\textwidth}
                  >{\centering\arraybackslash}m{.22\textwidth}
                  >{\centering\arraybackslash}m{.22\textwidth}
                  @{}}
    &\scriptsize Static $m=q=4$ &\scriptsize Static $m=r=8$ & \scriptsize Dynamic $m=q=4$ &\scriptsize Difference\\
\raisebox{1\height}{\rotatebox[origin=c]{90}{\scriptsize Ind. Production} }
&
\includegraphics[width=\linewidth,trim={65 40 55 38},clip]{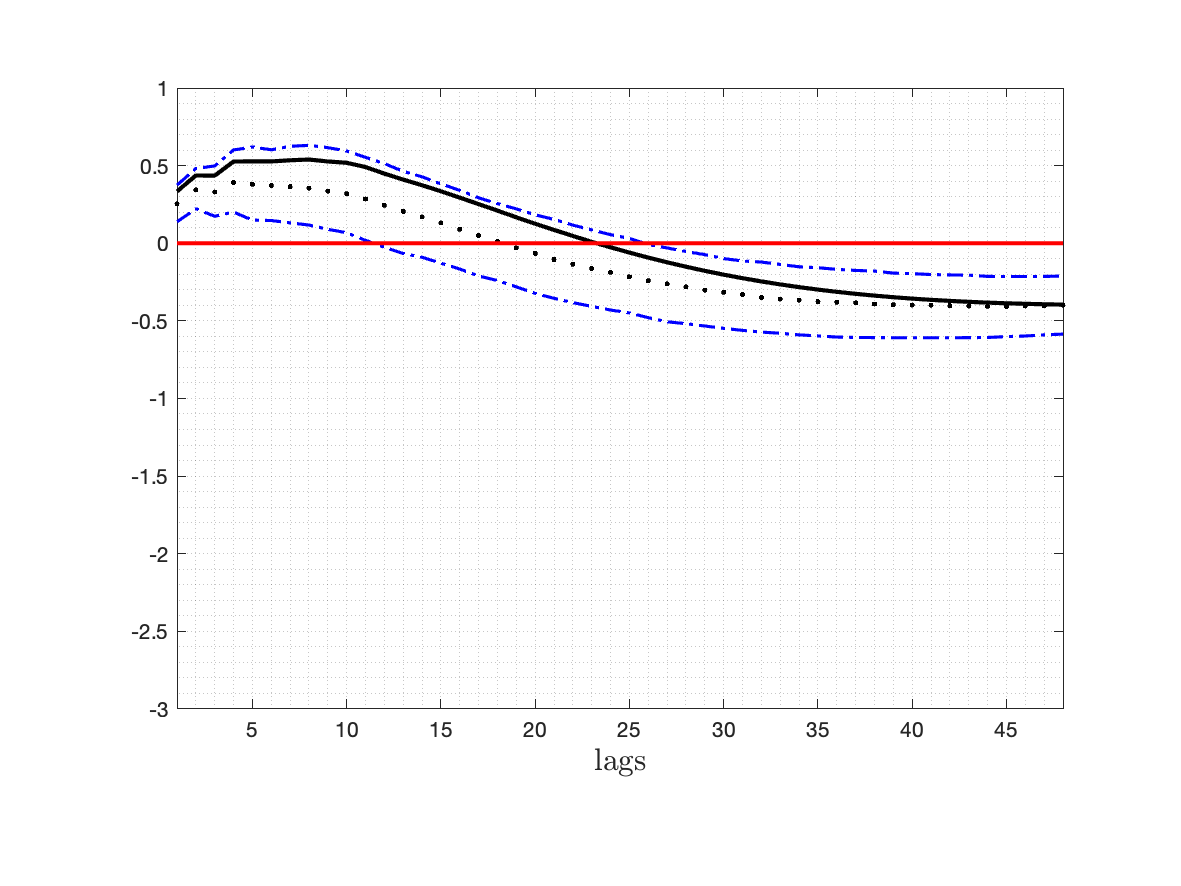} &
\includegraphics[width=\linewidth,trim={65 40 55 38},clip]{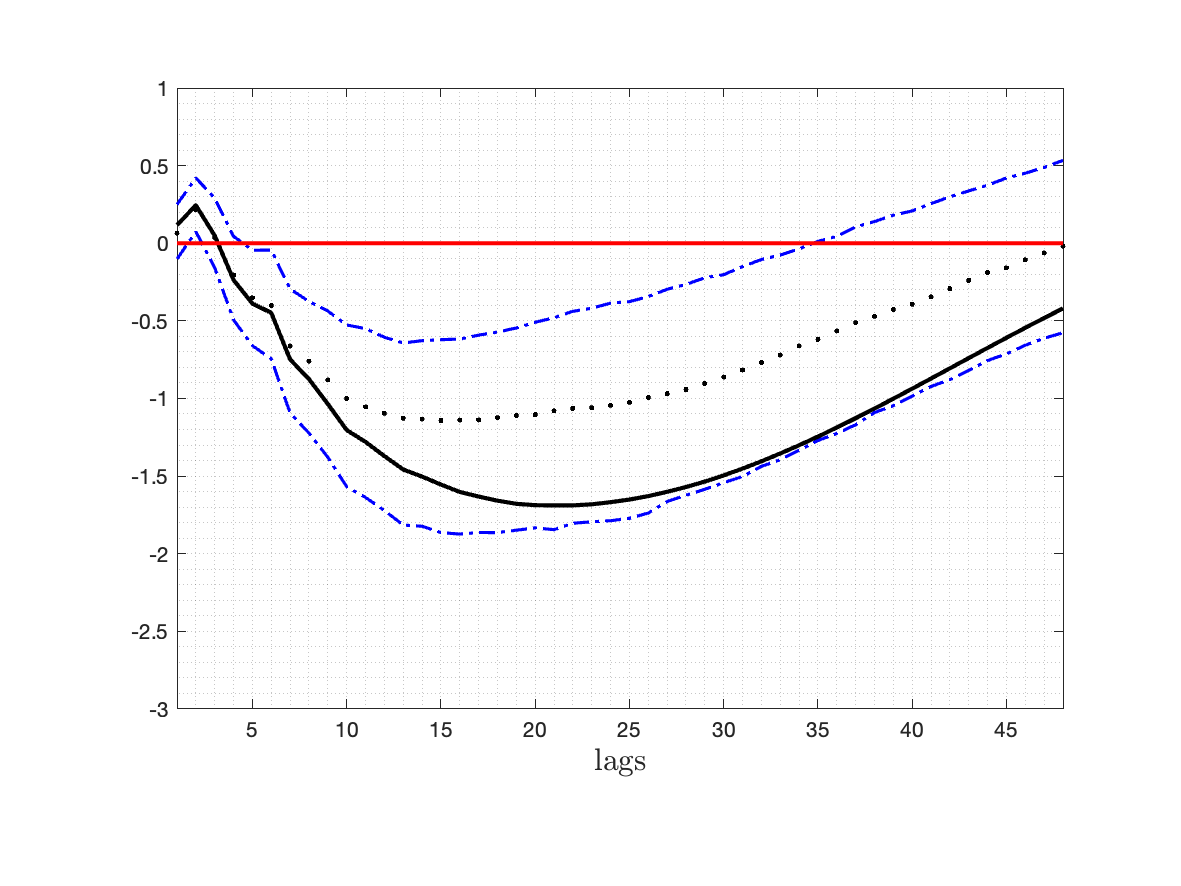} &
\includegraphics[width=\linewidth,trim={65 40 55 38},clip]{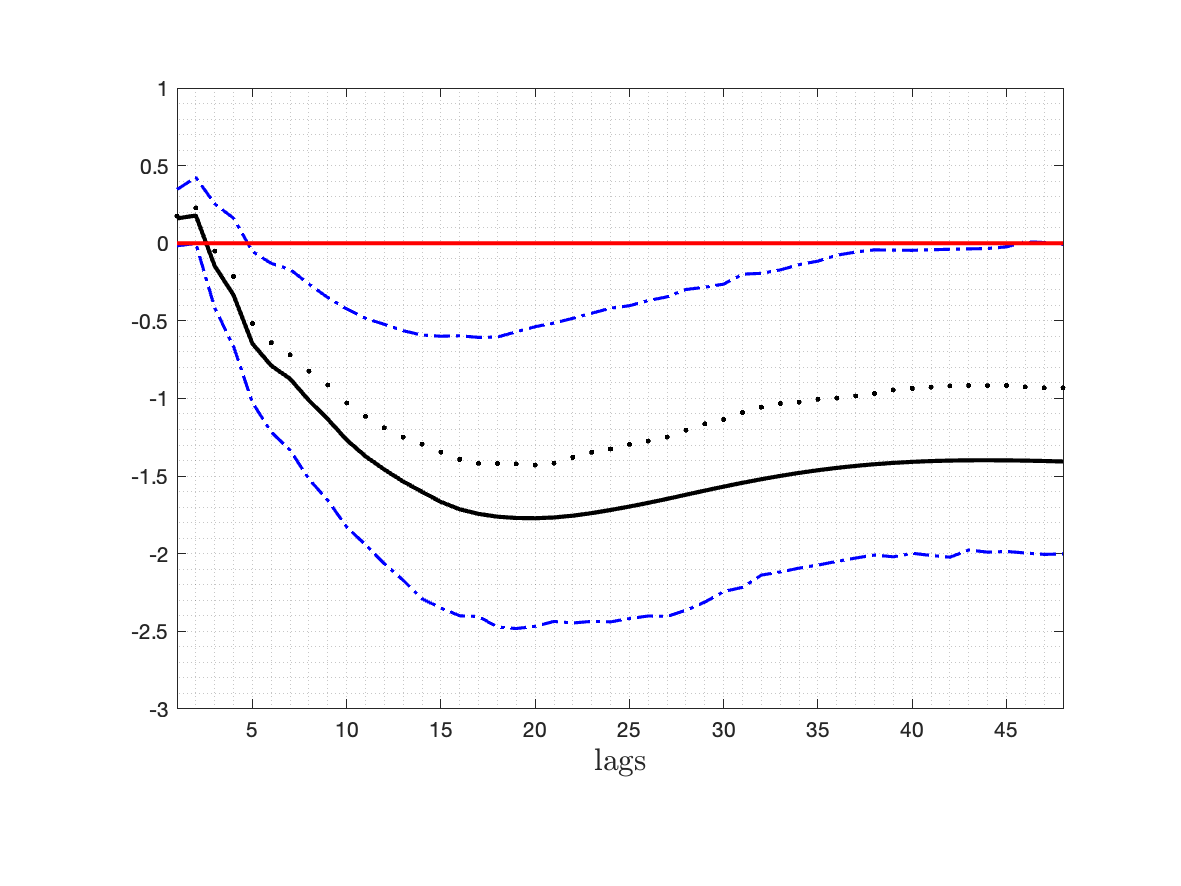} &
\includegraphics[width=\linewidth,trim={65 40 55 38},clip]{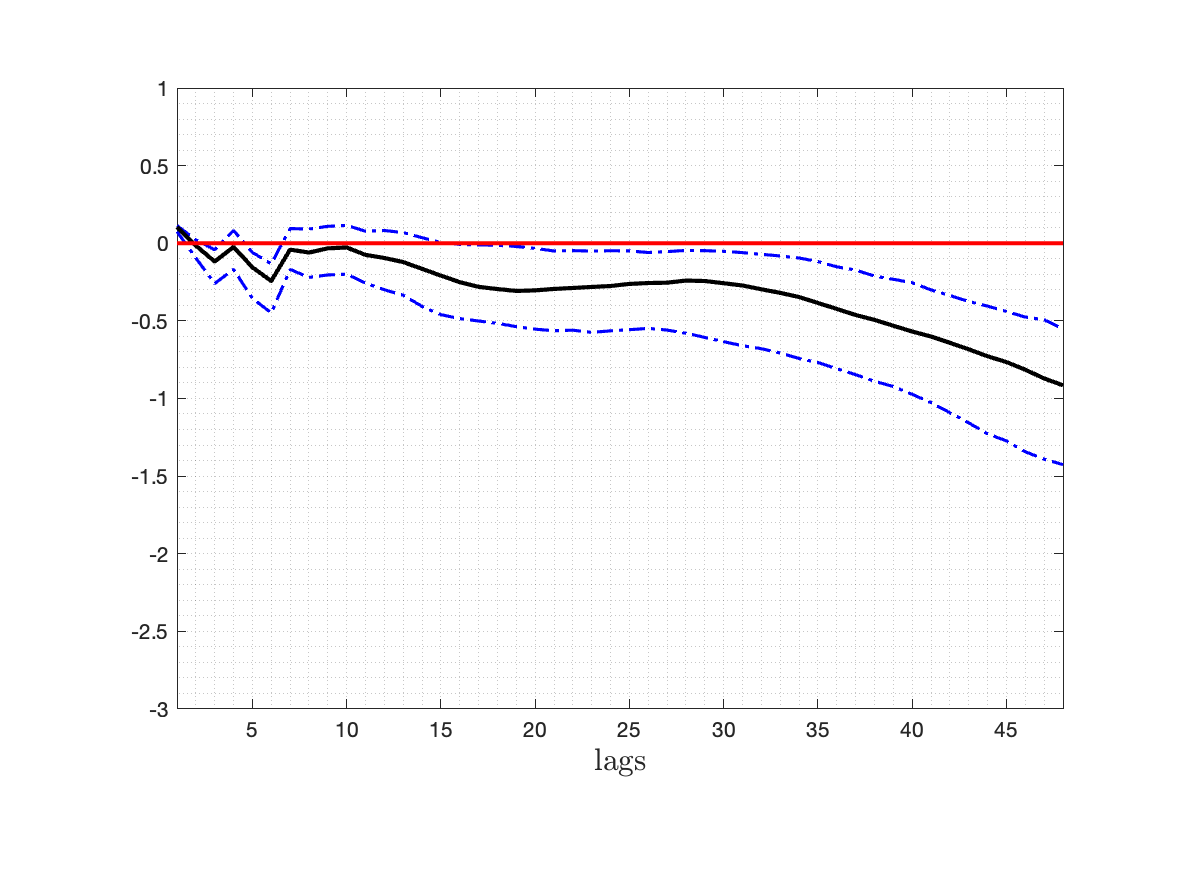} \\
\raisebox{1\height}{\rotatebox[origin=c]{90}{\scriptsize Unemployment} }
&
\includegraphics[width=\linewidth,trim={65 40 55 38},clip]{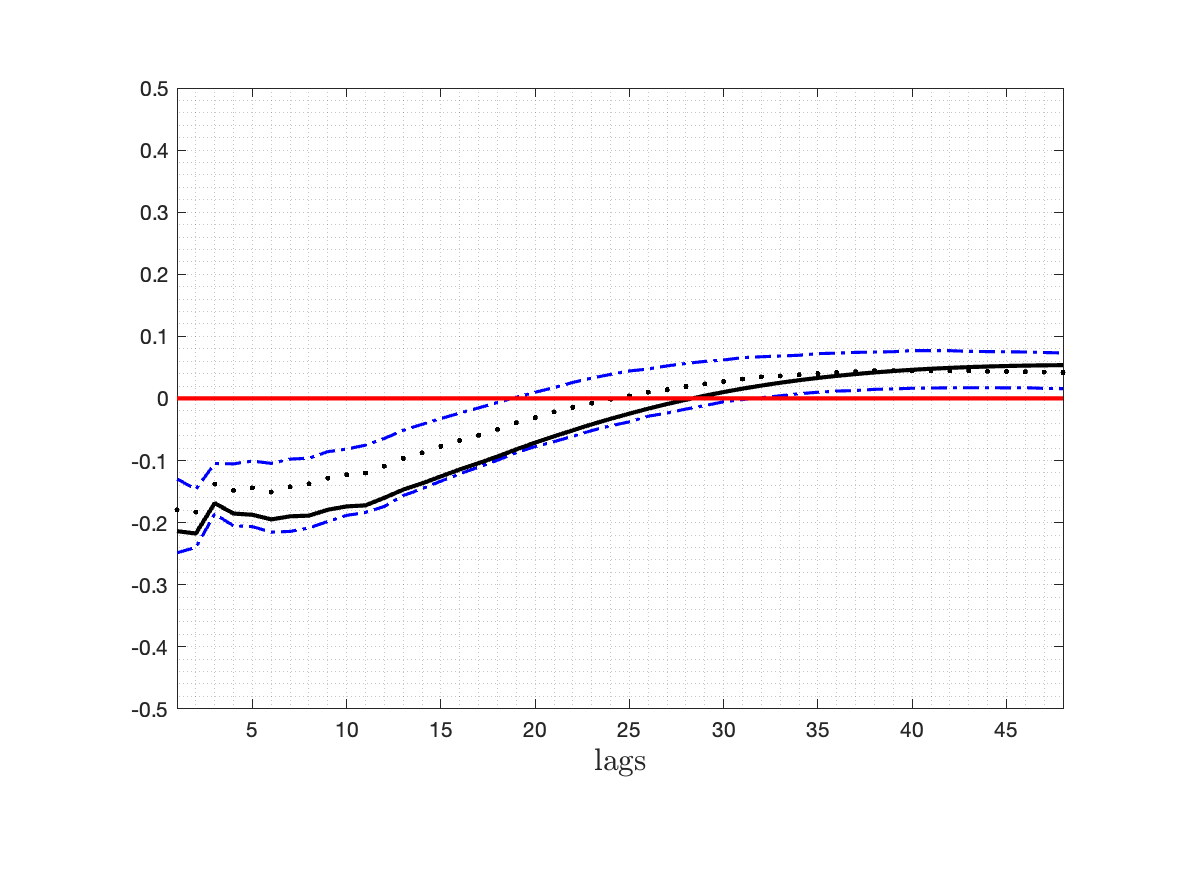} &
\includegraphics[width=\linewidth,trim={65 40 55 38},clip]{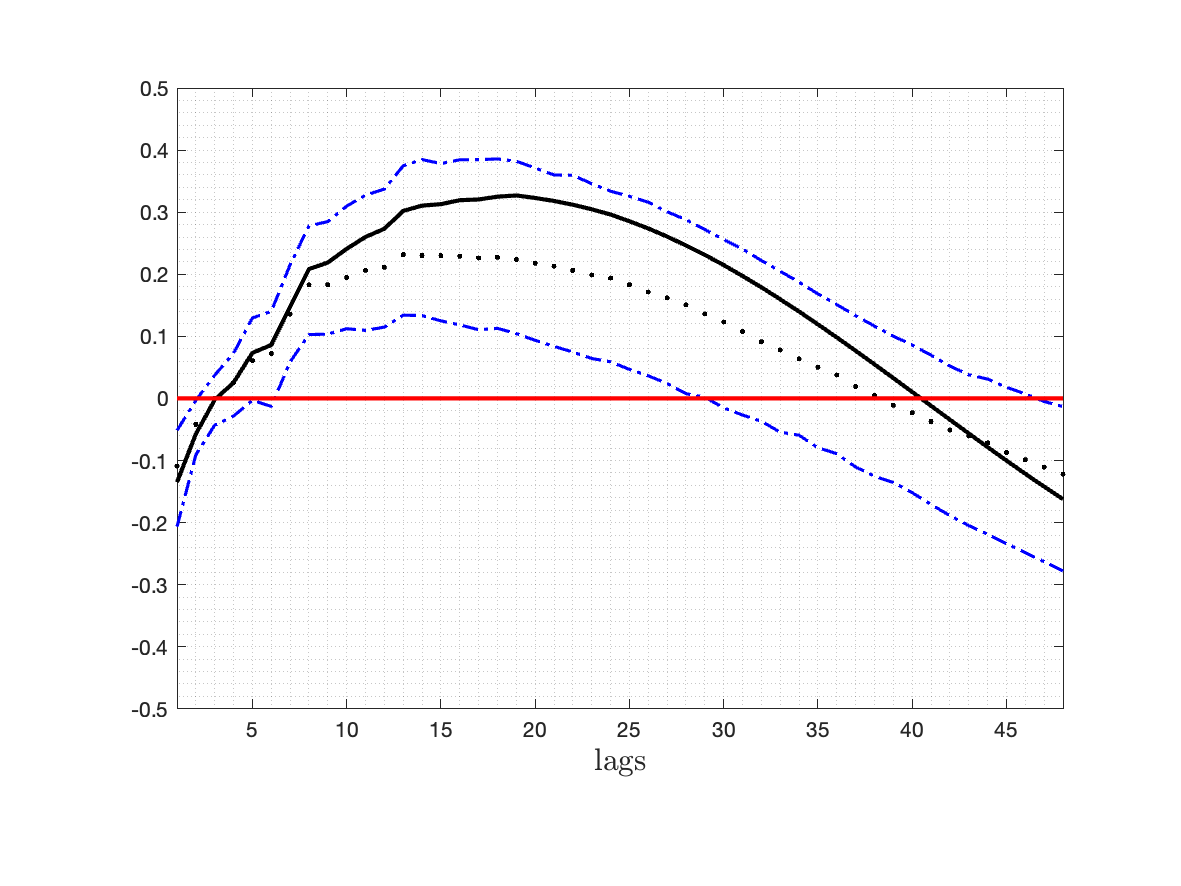} &
\includegraphics[width=\linewidth,trim={65 40 55 38},clip]{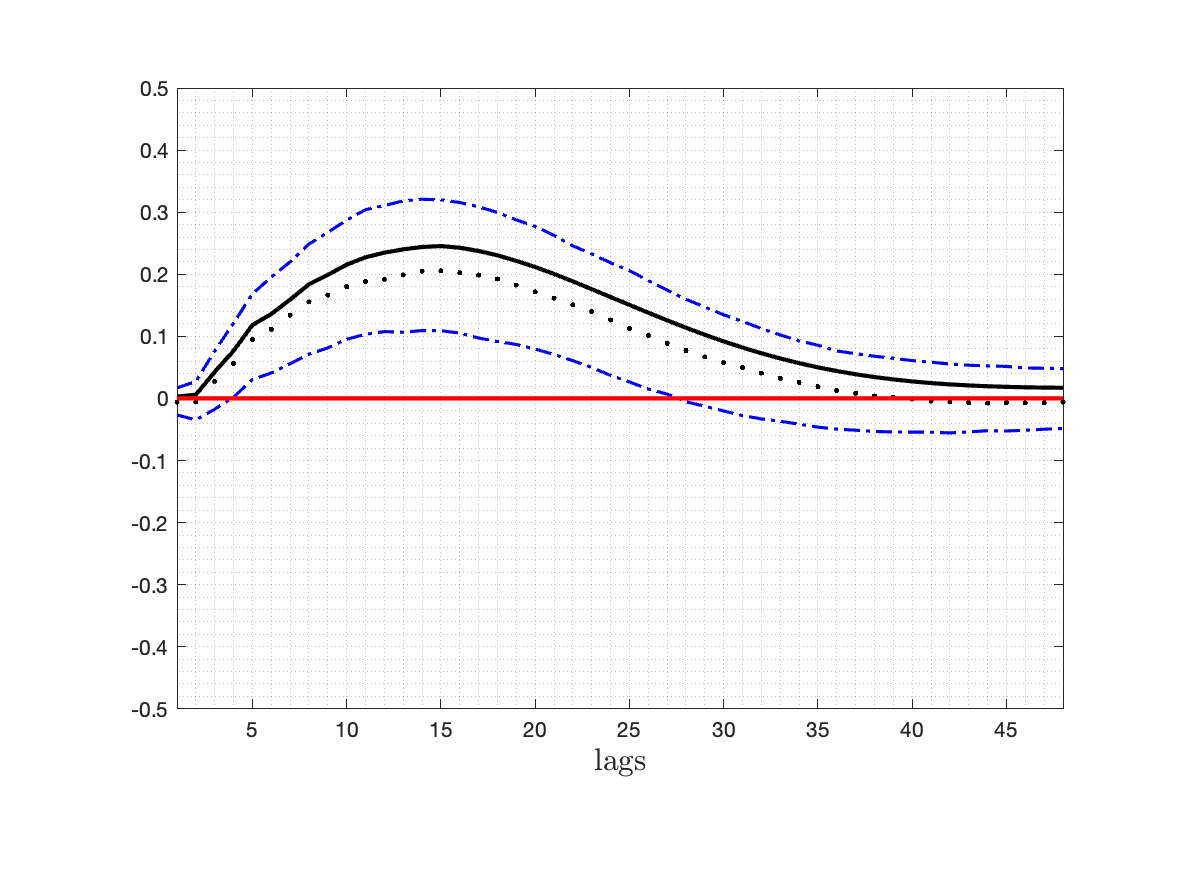} &
\includegraphics[width=\linewidth,trim={65 40 55 38},clip]{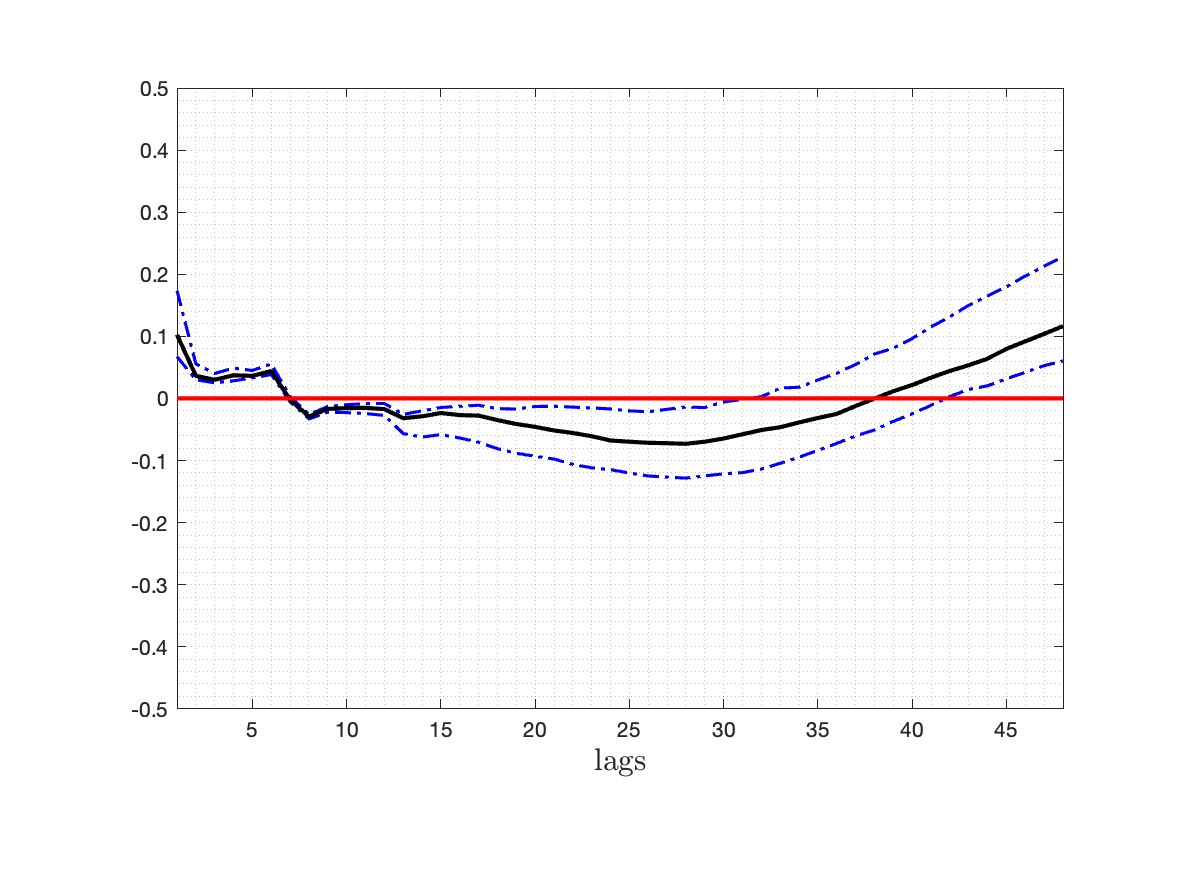} \\
\raisebox{1\height}{\rotatebox[origin=c]{90}{\scriptsize Prices} }
&
\includegraphics[width=\linewidth,trim={65 40 55 38},clip]{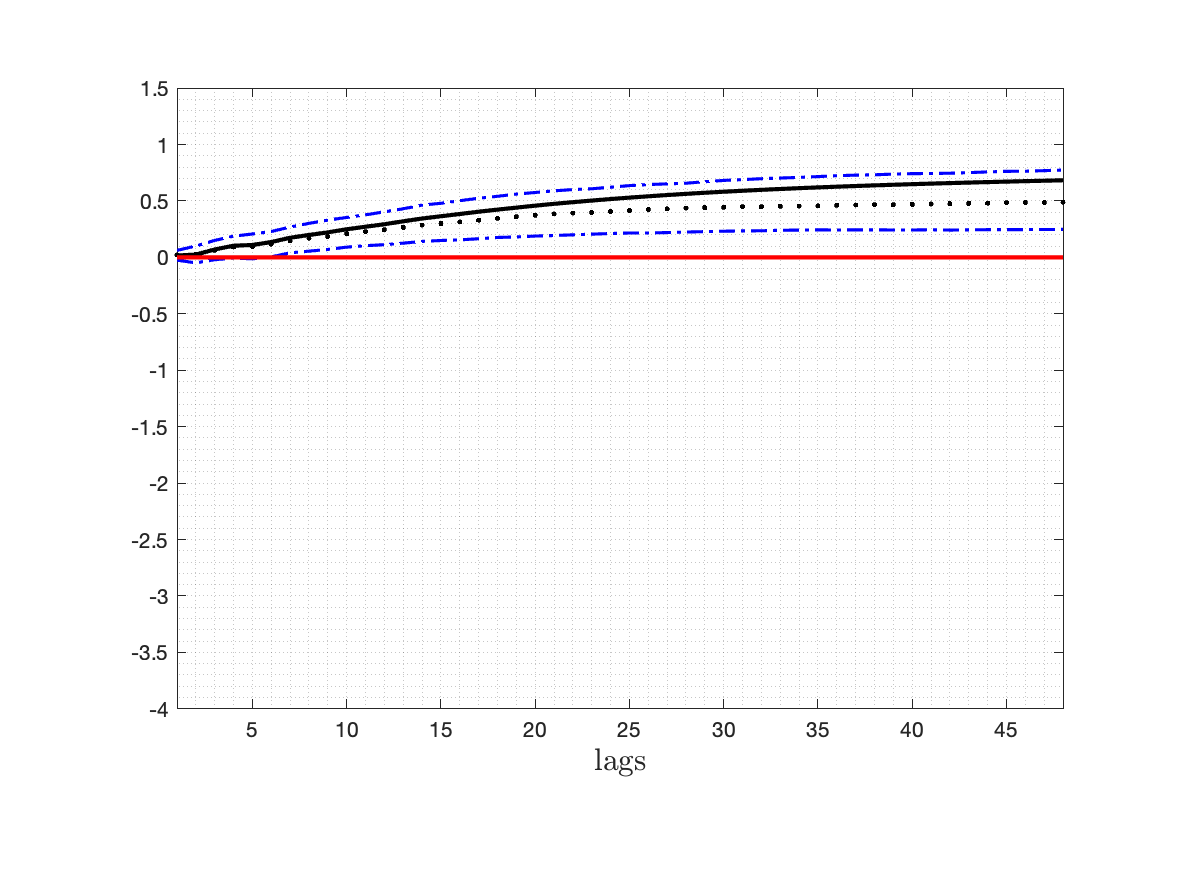} &
\includegraphics[width=\linewidth,trim={65 40 55 38},clip]{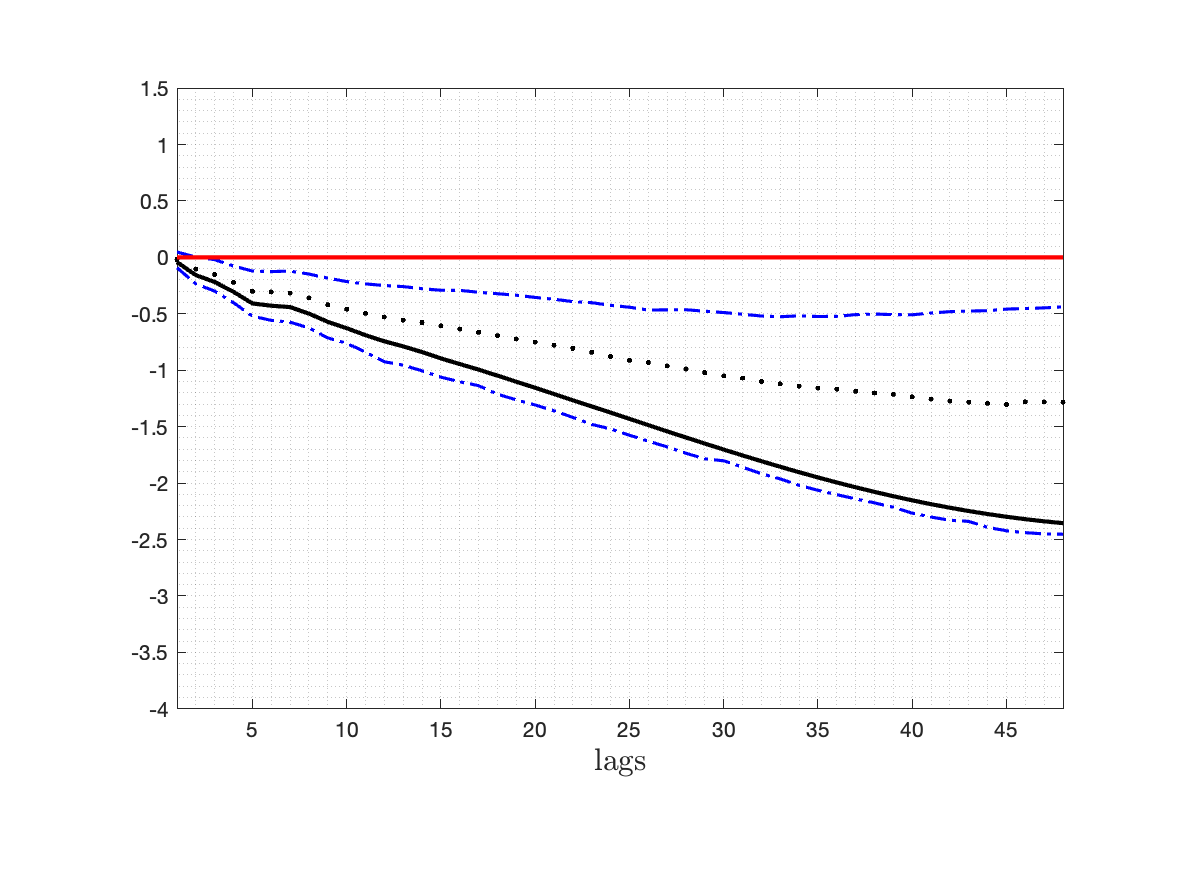} &
\includegraphics[width=\linewidth,trim={65 40 55 38},clip]{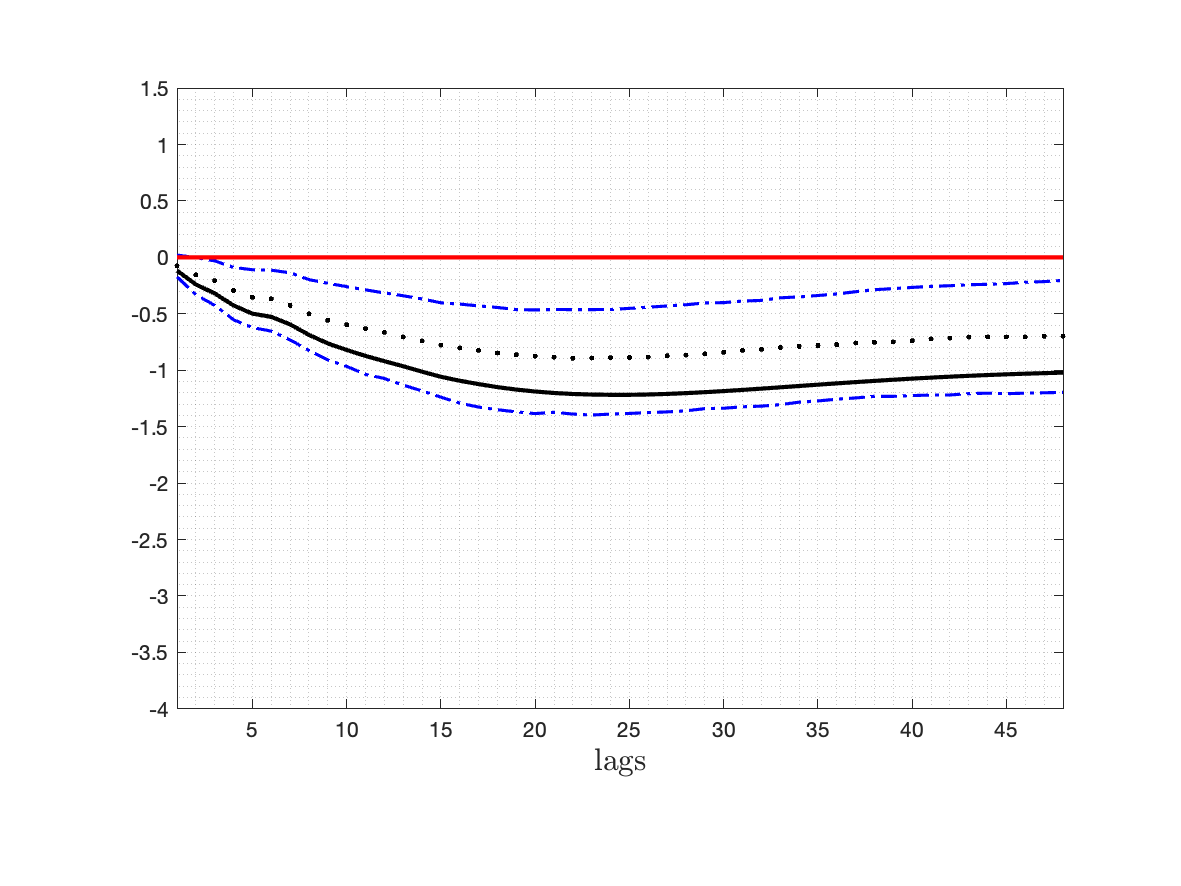} &
\includegraphics[width=\linewidth,trim={65 40 55 38},clip]{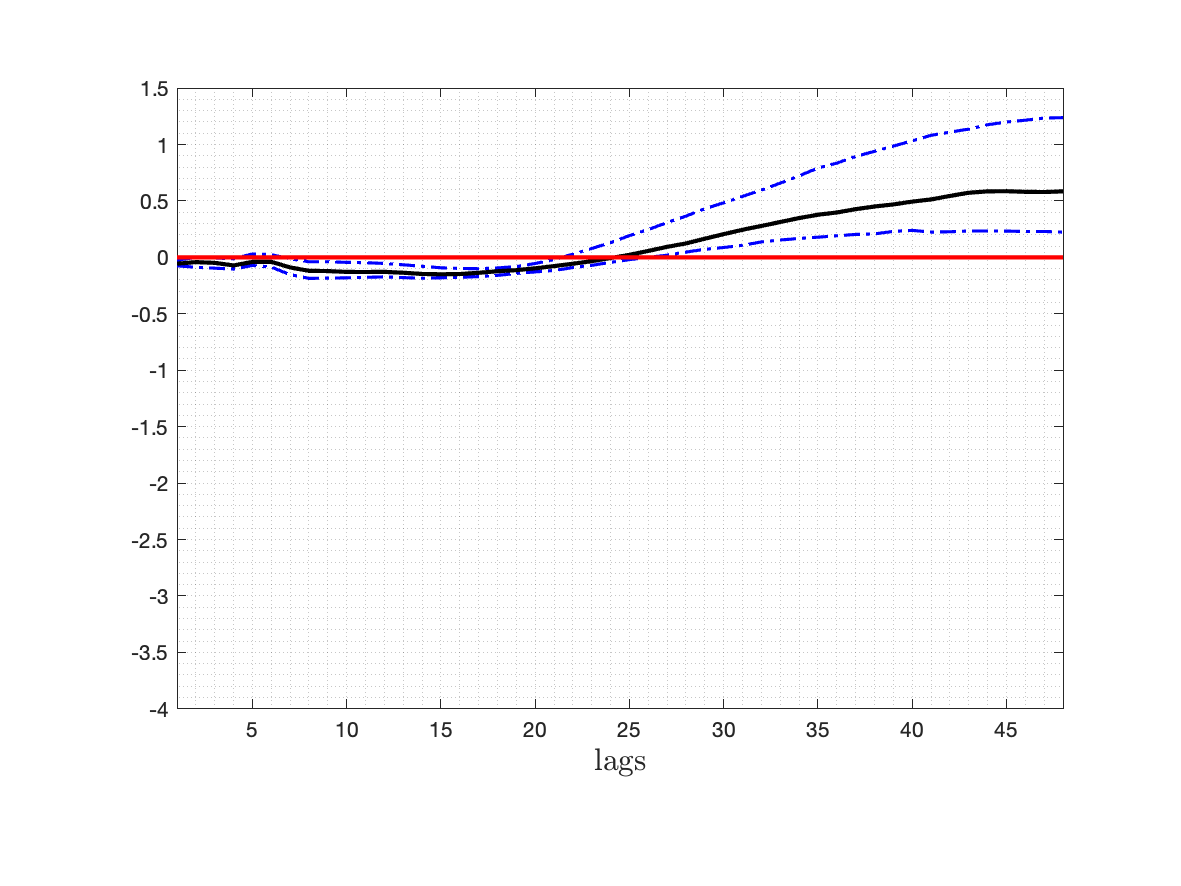} \\
\raisebox{1\height}{\rotatebox[origin=c]{90}{\scriptsize 1-yr T-bills} }
&
\includegraphics[width=\linewidth,trim={65 40 55 38},clip]{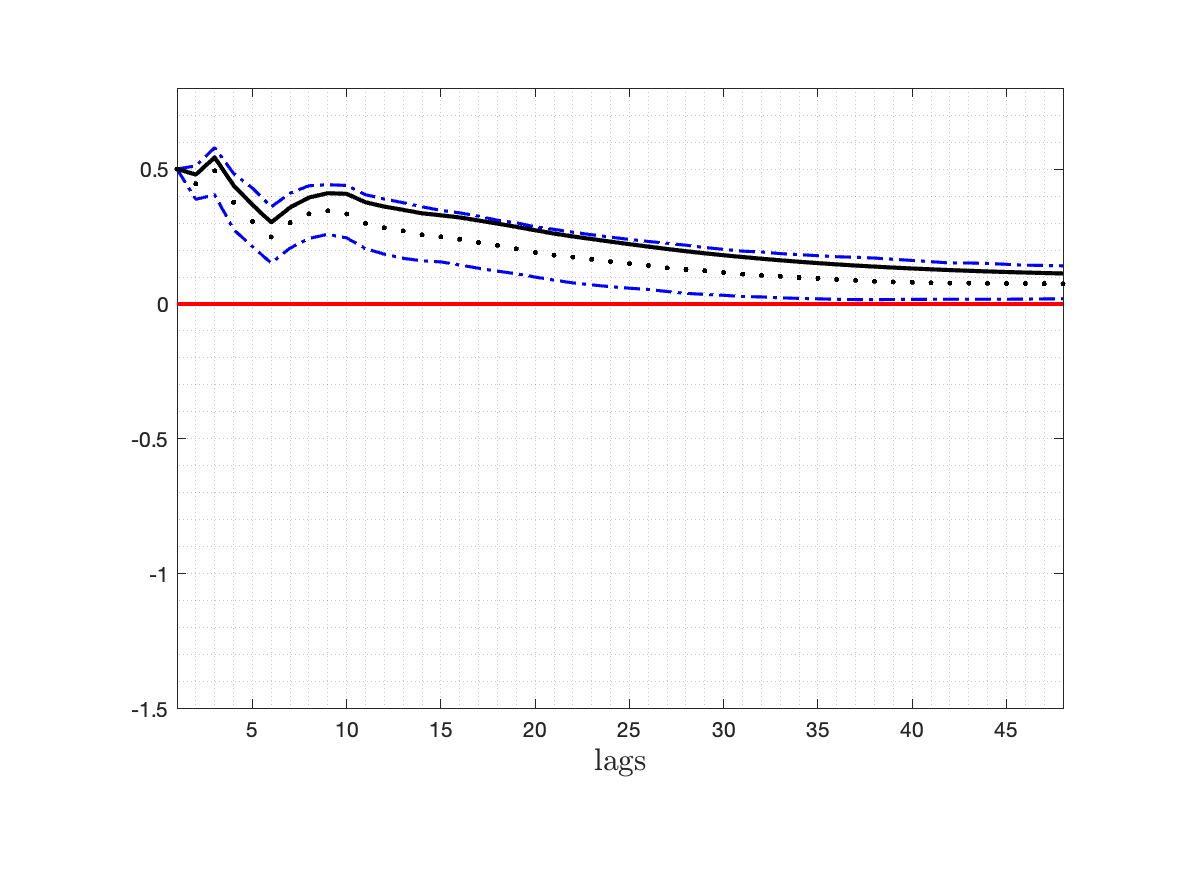} &
\includegraphics[width=\linewidth,trim={65 40 55 38},clip]{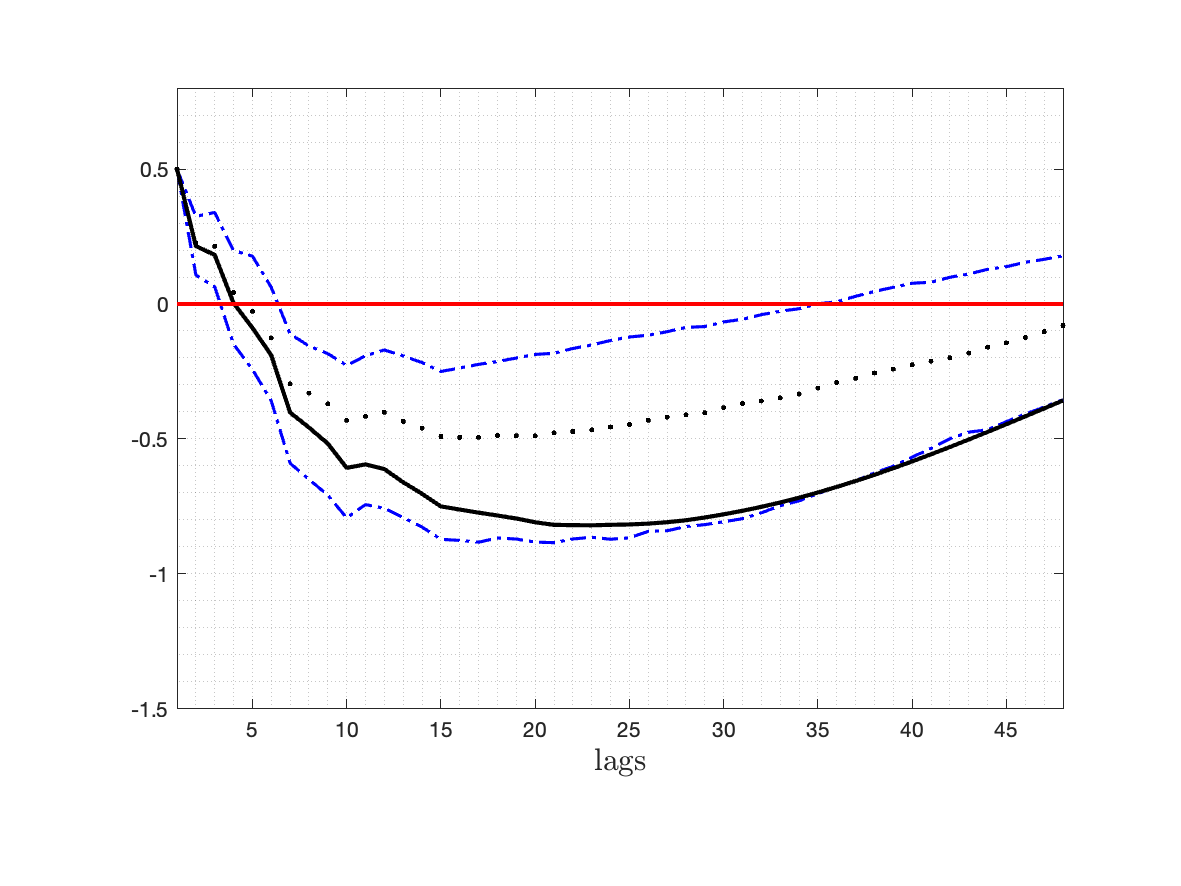} &
\includegraphics[width=\linewidth,trim={65 40 55 38},clip]{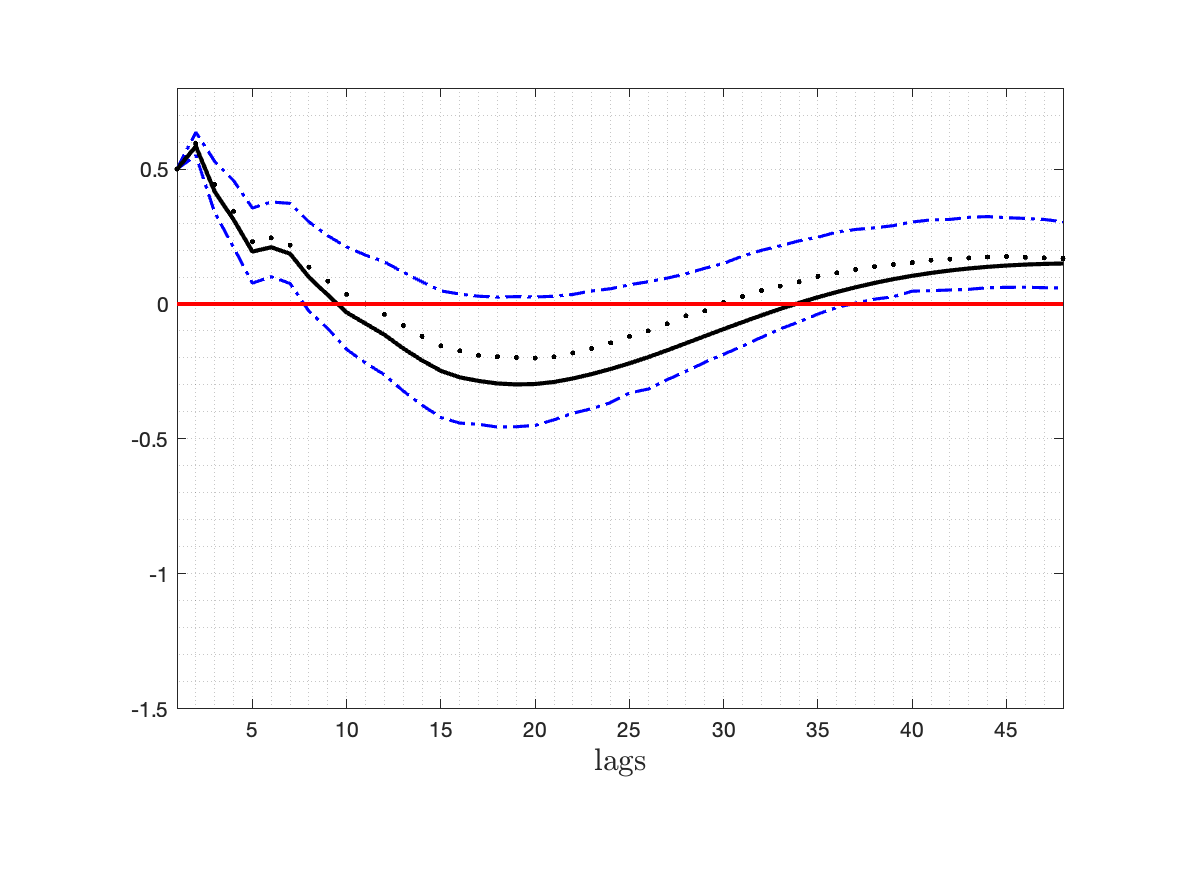} &
\includegraphics[width=\linewidth,trim={65 40 55 38},clip]{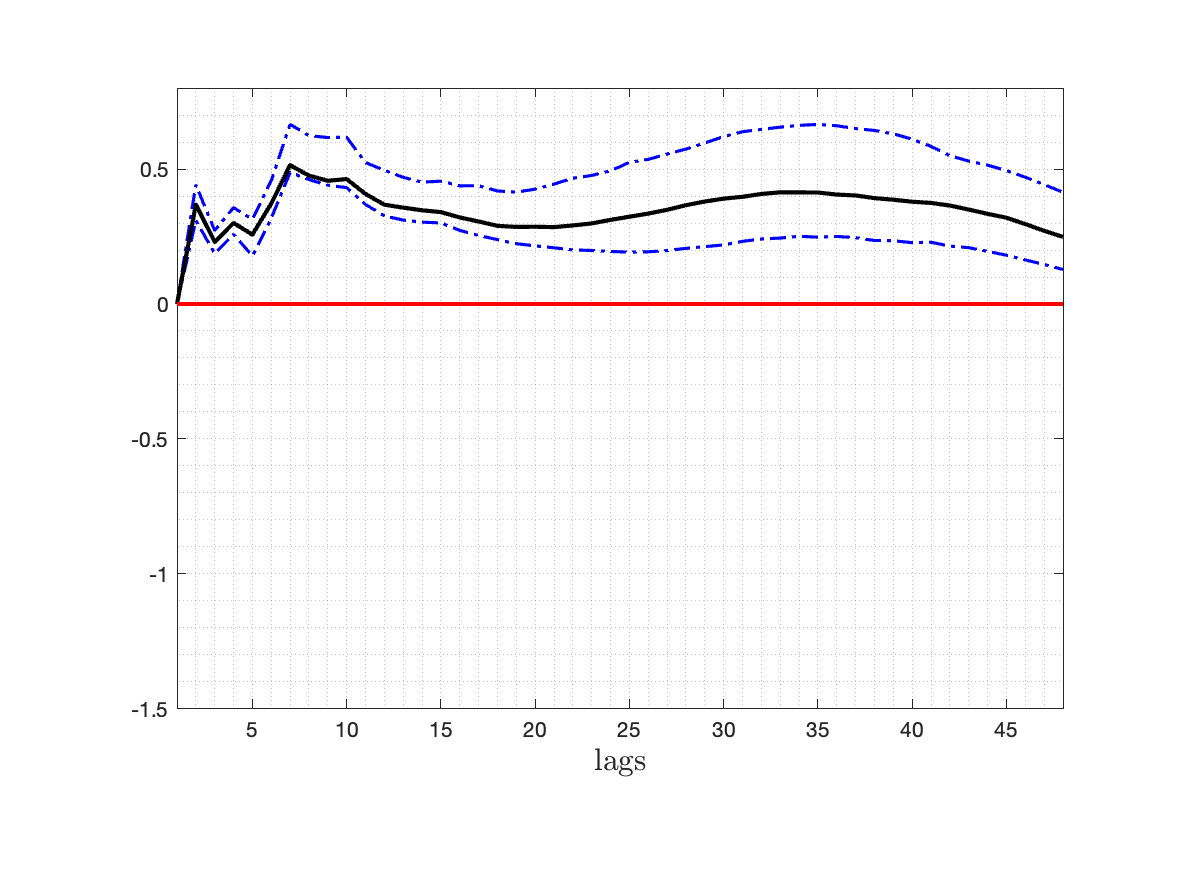} \\
\end{tabular}
    }
    \caption{\footnotesize IRFs to a monetary policy shock. First three columns---solid lines: CC-VAR on static common components with $m=q=4$, on static common components with $m=r=8$, on dynamic common components with $m=q=4$. Static common components estimated via PCA, dynamic common components estimated as in \citet{forni2017dynamic}.
    Dashed-dotted lines: upper and lower bounds of bootstrap 68\% confidence bands. Dotted lines: median bootstrap IRFs.\\ Last column---solid lines: difference between IRFs of third and second column. Dashed-dotted lines: upper and lower bounds of bootstrap 68\% confidence bands.}
    \label{fig: IRFs}
\end{figure}

In the first three columns of Figure \ref{fig: IRFs} we report the IRFs of the four main variables (solid lines) for the three considered approaches. In the last column we also report the difference between the dynamic IRFs (case (c)) and the static IRFs (case (b)). In all plots we also report 68\% confidence bands computed via 500 replications of the wild bootstrap procedure recommended by \citet[see also their replication files available online]{forni2025common}.
Since for factor analysis all variables are first transformed to stationarity, in order to report the effects of the shocks on the levels of the variables, the reported IRFs of Industrial Production and Prices are obtained by cumulating the estimated IRFs. 

If we just use $m=q$ (first column) variables the static approach does not produce reasonable IRFs---for example it is evident a ``price puzzle''. This is of no surprise, since, as explained by \citet{forni2025common} a necessary condition in their static model for correctly recovering the space of the shocks is to consider a singular VAR. Indeed, when using $m>q$ variables (second column) all IRFs have now a reasonable shape. However, we still notice an explosive IRF of Prices, a long run negative IRF of the Unemployement rate,  a long run positive IRF of Industrial Production, and a prolonged negative IRF of 1-year Treasury Bills.

If we estimate a VAR using the dynamic common components, we are including also the weak common components of the variables. This has two effects. First, as expected from our theory, it is enough to use $m=q$ (third column) dynamic common components to guarantee that we recover correctly the space of the shocks. Indeed, the IRFs are all reasonable, but different from those obtained by the static approach. In particular, the long run IRF of Industrial Production stabilizes to a negative level, the IRF of the Unemployment rate reverts to zero, the IRF of Prices remains negative but stabilizes, and the IRF of 1-year Treasury Bills is significant only in the very short run. All these differences are significant (fourth column) and hint at a better reconstruction of the IRFs, due to a non-zero weak common component of the considered variables, especially of Unemployment rate and CPI Inflation (see the previous section).


\subsection{Financial volatilities}
\label{sec:emp_vol}
We analyze a dataset of stock price volatilities based on daily observations from February 20, 2014, to June 14, 2023, yielding a  sample size of $T=2433$. The dataset comprises volatilities for a  set of $n=78$ global banking assets. The data are a sub-set of those used in \cite{krampe2024decomposing} to which we refer for details on the construction of the volatility proxies. Stock price data are sourced from Datastream.

\paragraph{Estimating the canonical decomposition.}
\begin{figure}[htbp]
    \centering
    \includegraphics[width=.8\textwidth]{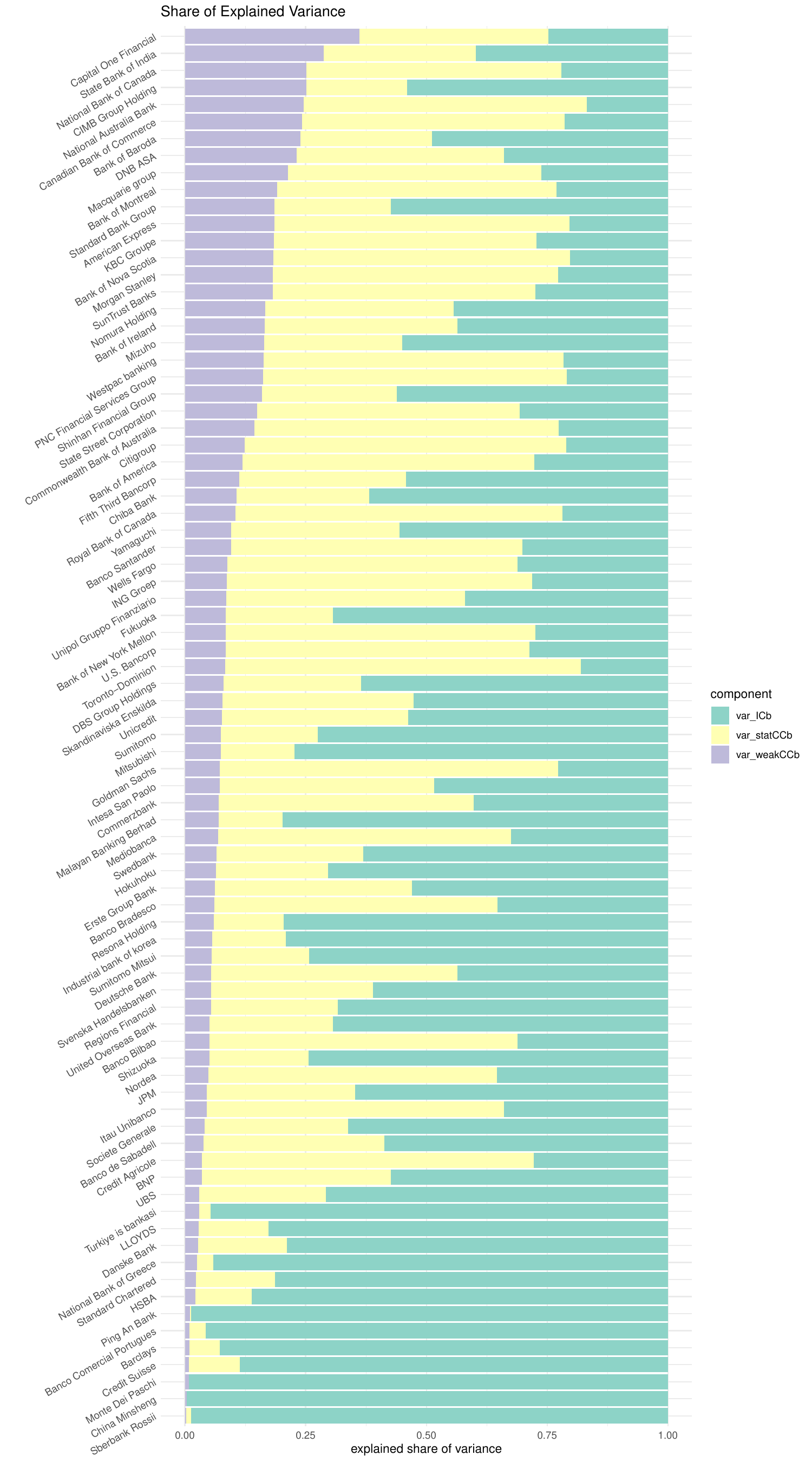}
    \caption{\footnotesize Share of variance explained by each component per variable with $q = 1$ and $r=1$. Estimates are obtained by using part II.b when estimating $C_{it}$. Here \texttt{var\_statCCb} $= EV_i^C$ (given in \eqref{eq:EVCa}),  \texttt{var\_weakCCb} $=EV_i^{e^\chi}$ (given in \eqref{eq:EVWC}). Last, \texttt{var\_ICb} is the variance explained by the dynamic idiosyncratic component, which is given by $EV_i^\xi=1-EV_i^C-EV_i^{e^\chi}$.}
    \label{fig: share of var volatility r1 q1}
\end{figure}

We estimate the canonical decomposition (\ref{eq: 3fold decomp intro}) using the procedures from Section \ref{sec: estimation} applied to the transformed centered and standardised data. All estimates are computed as in Section \ref{sec: simulation experiments} using a bandwidth $\mathcal B_T = \lfloor 0.75 \sqrt{T} \rfloor = 21$ and a Bartlett kernel.
In this case, we find evidence of $q = 1$ dynamic factor and $r=1$ static factor using the methods by \cite{hallin2007determining} and \cite{bai2002determining}, respectively. Moreover, our eigenvalue ratio criterion suggests the presence of $r_w=6$ weak factors.

Results are in Figure \ref{fig: share of var volatility r1 q1}. In total 61 out of 82 variables have a weak common component share larger than 5\%.

\paragraph{IRFs.}
In this section, we consider the static and dynamic IRFs to a single common shock, which can be seen as a market-wide shock. In this case $r=q$ and 
we can estimate the static IRFs by estimating a univariate AR model for all $n$ static common component, estimated via PCA.\footnote{This is equivalent to estimating an AR model on the the first estimated factor and then multiply its IRF times the estimated loadings.} Similarly, since $q=1$, we can estimate dynamic IRFs by fitting an AR model on all $n$ dynamic common component, estimated as in \citet{forni2017dynamic}. All AR orders are determined via BIC. 
Last, since $q=1$, identification requires just to impose a scale and a sign on the estimated reduced form IRFs. These are fixed by imposing that the common shock has a unit impact on all series.

\begin{figure}[htbp]
    \centering
    \begin{tabular}{cc}
    \scriptsize{Static}&\scriptsize{Dynamic}\\
    \includegraphics[width=.45\textwidth,trim={65 40 55 38},clip]{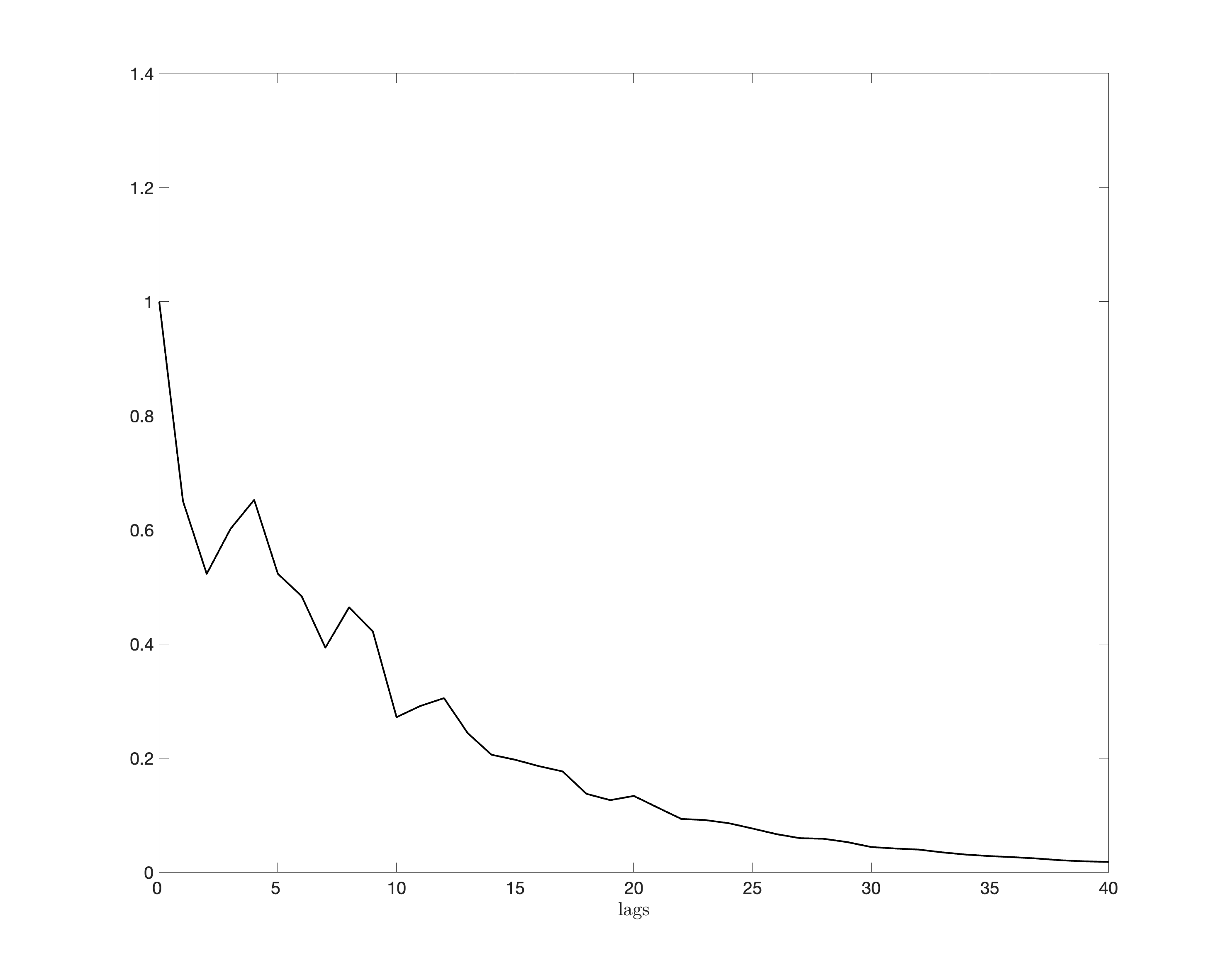}
    &
    \includegraphics[width=.45\textwidth,trim={65 40 55 38},clip]{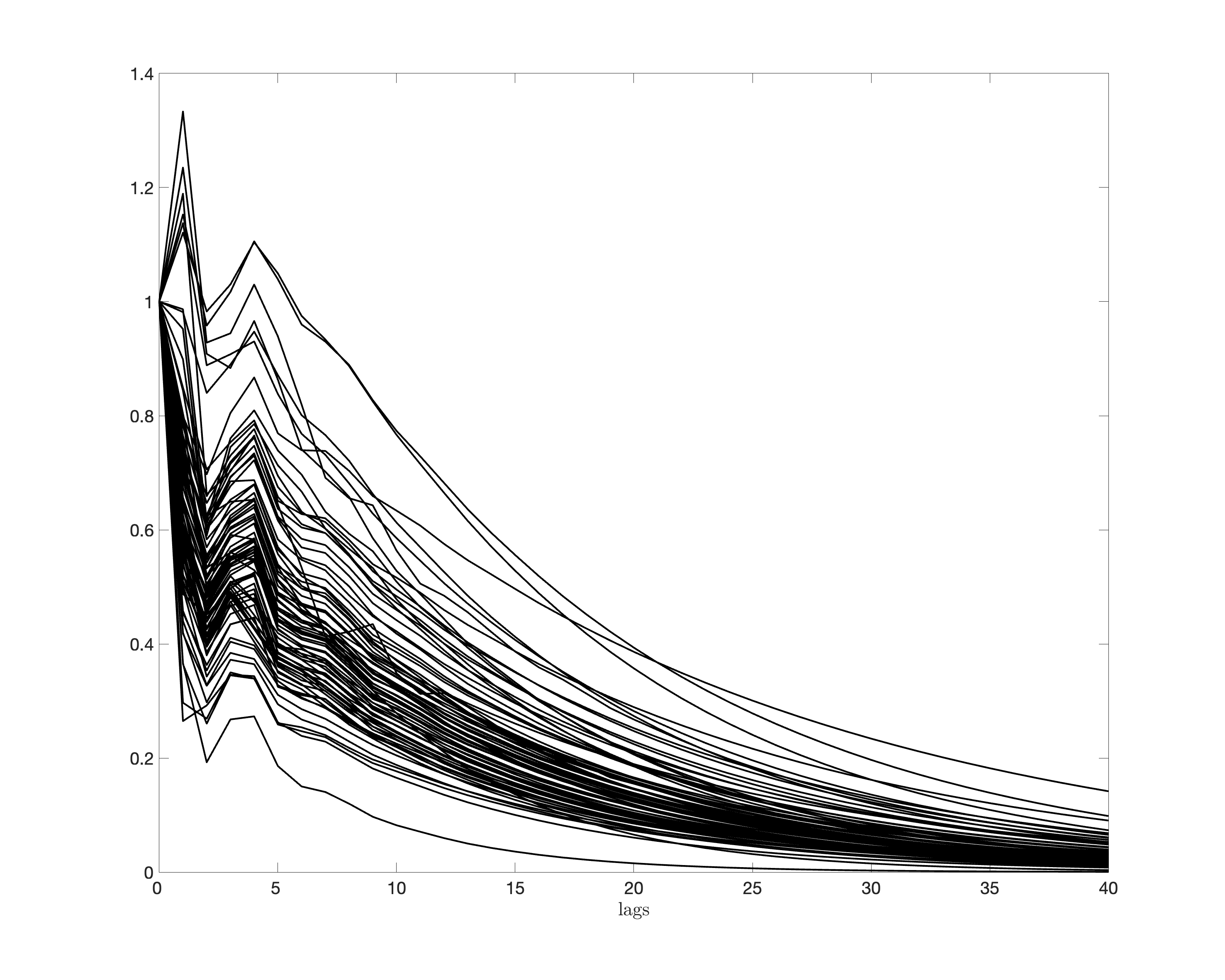}
    \end{tabular}

    \caption{\footnotesize IRFs to a market-wide shock. Left: AR on $n$ static common components estimated via PCA. Right: AR on $n$ dynamic common components estimated as in \citet{forni2017dynamic}.}
    \label{fig:VOLIRFs}
\end{figure}

Results are in Figure \ref{fig:VOLIRFs}. On the one hand, once identified, all static IRFs coincide by construction (left plot). This is a clear limitation of the static approach since it is unreasonable to think that all assets reacts in the same way to a market-wide shock. On the other hand, the dynamic IRFs (right plot) are heterogeneous both in the short and in the long run and seems to capture more clearly the well-documented long-memory of volatilities (see, e.g., \citealp{ergemen2023parametric}). Such heterogeneity, which is likely to be more realistic, is the result of including non-negligible weak common components in the estimation.

\section{Conclusions}\label{sec:conc}

The main implication of this paper is that empirical researchers should always consider estimating a GDFM first, to assess the presence of a weak common component. If there is evidence of such component then the dynamic approach seems to be better suited for the analysis. 


We document the presence of weak common components both in the standard panel of U.S. macroeconomic indicators and in a panel of global financial volatilities. These results have non-negligible implications for IRF analysis. In addition, previous work by \citet{forni2018dynamic} has already shown that the GDFM can deliver superior forecasts of CPI inflation compared with the static approach of \citet{stock2002forecasting}. 


Finally, in light of these conclusions and the findings of this paper, it is evident that there is a need for a GDFM estimator more efficient than the existing ones relying on spectral analysis. Developing such an estimator is the focus of our ongoing research.

\section*{Conflict of Interest Statement }
On behalf of all authors, the corresponding author states that there is no conflict of interest. 
\section*{Acknowledgements}
{{The authors would also like to thank Paul Eisenberg, Sylvia Frühwirth-Schnatter, Tobias Hartl and Dominik Liebl for helpful comments that lead to the improvement of the paper. The authors gratefully acknowledge financial support from the Austrian Central Bank under Anniversary Grant No. 18287 and the DOC-Fellowship of the Austrian Academy of Sciences (ÖAW).
}}

\bibliographystyle{apalike} 
\bibliography{references.bib}

\clearpage
\begin{center}
{\LARGE \bf Appendix}
\end{center}

\appendix
\renewcommand{\theequation}{\thesection.\arabic{equation}}

\section{Proof of Proposition \ref{thm: finite dimensional CCC}}\label{app:th2}

First, notice that, if $\dim \cspargel(\chi_{it}, i\in\mathbb N) = r_\chi<\infty$, then there exists some $r_\chi$ dimensional process $(z_t)$, together with $r_\chi$-dimensional row-loadings $L_i$ such that $\chi_{it} = L_i z_t$ for all $t\in \mathbb Z$ and $i\in \mathbb N$. 

For part 1, by Theorem \ref{thm: relation of r-SFS and q-DFS} we know that $\mathbb S_t(y) \subseteq \cspargel(\chi_{it}, i\in\mathbb N)$. Therefore, there exists $F_t$ a basis of $\mathbb S_t(y)$, which is a linear transformation of $z_t$ (the same transformation for all $t\in \mathbb Z$ by stationarity A\ref{A: stat}), of dimension $r$ such that $0\leq r \leq r_\chi<\infty$. Thus, by Assumption A\ref{A: q-DFS struct}, we can write
\begin{align}
    y_{it} = \chi_{it}+\xi_{it}= \proj(\chi_{it}\mid \mathbb S_t(y)) + v_{it} + \xi_{it} = \Lambda_i F_t + v_{it} + \xi_{it}, \ \mbox{say} \ \label{eq: intermed. 3fold decomp}
\end{align}
where $v_{it}$ is the residual from the projection of $\chi_{it}$ on $\mathbb S_t(y)$. Moreover, $\mathbb S_t(y) = \mathbb S_t(\chi)$ because $(e_{it}^\chi)$ is statically idiosyncratic by Theorem \ref{thm: relation of r-SFS and q-DFS}. Therefore, $(v_{it})$ vanishes under static aggregation since it lives in the complement of $\mathbb S_t(\chi)$, and it is statically idiosyncratic. So $\mu_1(\Gamma_v^n + \Gamma_\xi^n) \leq \mu_1 (\Gamma_v^n) + \mu_1(\Gamma_\xi^n) < \infty$ by Theorem \ref{thm: relation of r-SFS and q-DFS} from which it follows that a dynamically idiosyncratic component is also statically idiosyncratic. So also $v_{it} + \xi_{it} = e_{it}$ is statically idiosyncratic by Theorem \ref{thm: charact stat idiosyncratic}. Finally, 
let $C_{it} = \Lambda_i F_t$ in (\ref{eq: intermed. 3fold decomp}) and since $\mu_r(\Gamma_C^n)$ is monotonically non-decreasing, it either diverges with $n\to\infty$ or is bounded (and converges). Suppose $r > 0$ and $\sup_{n \in \mathbb N}\mu_r(\Gamma_C^n)<\infty$, so by Theorem \ref{thm: charact stat idiosyncratic}, $(C_{it})$ is statically idiosyncratic and therefore $C_{it}$ is orthogonal to $\mathbb S_t(\chi)$, but this is a contradiction, so it must be that $\sup_{n \in \mathbb N}\mu_r(\Gamma_C^n)=\infty$, which finishes the proof by Theorem \ref{thm: projection on the static aggregation space} The case $r=0$ is trivial.

For part 2, the proof is obvious from the decomposition in \eqref{eq: intermed. 3fold decomp}. Indeed, $v_{it}=0$ if and only if $C_{it}=\chi_{it}$ but also if and only if $r=r_\chi$. \hfill \rule{.7em}{.7em}
\section{On the stacked dynamic factor model}\label{Appendix sec: proof for exmp SW dyn fac}

\subsection{Existence of a GDFM representation}\label{app:example gdfm}
First, we prove that the model in \eqref{eq: SW example dyn fac}-\eqref{eq: SW example dyn fac2} satisfies Assumption A\ref{A: q-DFS struct}.

Since $f_{t}= \ubar{a}^{-1}(L) b \varepsilon_t$, the 
spectral density of $(\chi_{it})$ is:
$f_\chi^n(\theta) = \frac 1{2\pi} \lambda^n(\theta) a^{-1}(\theta) b$,
where $\Psi(\theta) := I_q - a_1 e^{-\iota \theta} - \cdots - a_{p_f} e^{-p_f \iota \theta}$. So, 
\begin{align}
2\pi\mu_q(f_\chi^n(\theta))&=
   \mu_q\left[a^{-1}(\theta)bb'a^{-1, *}(\theta) \ \lambda^{n, *}(\theta) \lambda^n(\theta)  \right]\nonumber\\ 
   &\geq  \mu_q\left[a^{-1}(\theta)bb'a^{-1, *}(\theta)\right] \mu_q\left[\lambda^{n, *}(\theta) \lambda^n(\theta)\right]\nonumber\\
   &\geq \mu_q\left[a^{-1, *}(\theta)a^{-1}(\theta)\right] \mu_q(bb') \mu_q\left[\lambda^{n, *}(\theta) \lambda^n(\theta)\right],
   \label{eq: weyl multiplic}
\end{align}
where we used twice the multiplicative Weyl's inequality \citep[Theorem 7]{merikoski2004inequalities}.
Now, $b$ has full-rank by assumption and, clearly, also $a(\theta)$ has full rank $q$ a.e. on $[-\pi, \pi]$, so  
\begin{equation}
\mu_q(bb')>0 \; \text{ and }\; 
\mu_q\left[a^{-1, *}(\theta)a^{-1}(\theta)\right]>0\; \text{ a.e. on } [-\pi,\pi].\label{eq:weyl3}
\end{equation}
Let us suppose for simplicity and without loss of generality  that $p = 1$. Then,
\begin{align}
    \mu_q\left[\lambda^{n, *}(\theta) \lambda^n(\theta)\right] &= \mu_q\left[(\lambda_1^n + \lambda_2^n e^{\iota \theta})' (\lambda_1^n + \lambda_2^n e^{-\iota \theta})\right] \nonumber\\ 
    &= \mu_q\left[\lambda_1^{n'} \lambda_1^n + \lambda_2^{n'} \lambda_2^n +  \lambda_1^{n'} \lambda_2^n e^{-\iota \theta} + \lambda_2^{n'} \lambda_1^n e^{\iota \theta} \right] \geq \max_{j = 0, 1} \mu_q \left[\lambda_j^{n'} \lambda_j^n\right], \label{eq: wyel sum}
\end{align}
where the last inequality holds since the expression on the LHS is a sum of Hermitian matrices \citep[see][Theorem 1]{merikoski2004inequalities}. From the assumed condition of divergent loadings in \eqref{eq:divergingloadings} and \eqref{eq: wyel sum} it follows that
$\sup_{n\in\mathbb N} \mu_q\left[\lambda^{n, *}(\theta) \lambda^n(\theta)\right] =\infty$, which, once substituted in \eqref{eq: weyl multiplic} and jointly with \eqref{eq:weyl3}, proves that $\sup_{n\in\mathbb N} \mu_q(f_\chi^n(\theta))=\infty$. Since we directly assume $(\xi_{it})$ to be idiosyncratic and orthogonal to $(\chi_{it})$ at all leads and lags, it follows that $(y_{it})$ satisfies Assumption \ref{A: q-DFS struct}, thus it admits a GDFM representation.

\subsection{Existence of the canonical representation}\label{app:example weak}
Next, let us investigate how to obtain the canonical representations
\eqref{eq: gen decomp} and 
\eqref{eq: canonical rep weak and strong factors} in Theorems \ref{thm: relation of r-SFS and q-DFS} and \ref{thm: weakfactors}, respectively, from the model in \eqref{eq: SW example dyn fac}-\eqref{eq: SW example dyn fac2}. First, we give the general treatment which is then applied to Example \ref{exmp: IRF} and DGP1, DGP2, and DGP3 in Section \ref{sec: simulation experiments}.

Recall the stacked representation in \eqref{eq:stacked}: $\chi_t^n= L^n x_t$ with $L^n:=(\lambda_0^n \cdots \lambda_p^n)$ and $x_t:=(f_t'\cdots f_{t-p}')'$.
Then, collect all loading columns $l_j^n$, $j=1,\ldots, r_\chi$, of $L^n$ for which $\sup_{n \in \mathbb N} \norm{l_j^n}^2 = \infty$ in $L_1^n$. They are associated with factors $(x_t^1)$ of dimension $r_1\times 1$. Collect the non-explosive columns of $L^n$, i.e., such that $\sup_{n \in \mathbb N} \norm{l_j^n}^2 < \infty$, in $L_2^n$. They are associated with factors $(x_t^2)$ of dimension $r_2 \times 1$. Since $x_t^1$ is not contemporaneously orthogonal to $x_t^2$, we project out $x_t^1$ from $x_t^2$. Let,
\begin{align}
    \widetilde x_t^2 &:= x_t^2 - \proj\left(x_t^2 \mid \cspargel(x_t^1)\right) = x_t^2 - \underbrace{\E [x_t^2 {x_t^1}'] 
    \l\{\E [x_t^1 {x_t^1}']\r\}^{-1}}_{\beta_{21}} x_t^1 = x_t^2 - \beta_{21} x_t^1.  \nonumber 
\end{align}
Then,
\begin{align}
    \chi_t^n &= 
     \begin{bmatrix} L_1^n & L_2^n \end{bmatrix} 
       \begin{pmatrix}
        x_t^1 \\
        x_t^2
    \end{pmatrix} =
    \begin{bmatrix} L_1^n & L_2^n \end{bmatrix} 
    \begin{pmatrix}
        I_{r_1} & 0 \\
        \beta_{21}  & I_{r_2}
    \end{pmatrix}
    \begin{pmatrix}
        I_{r_1} & 0 \\
        - \beta_{21}  & I_{r_2}
    \end{pmatrix} 
    \begin{pmatrix}
        x_t^1 \\
        x_t^2
    \end{pmatrix} = 
     \begin{bmatrix} L_1^n + L_2^n \beta_{21} & L_2^n \end{bmatrix} 
     \begin{pmatrix}
     x_t^1 \\
     \widetilde x_t^2
     \end{pmatrix} \nonumber \\
    &= \begin{bmatrix} \widetilde L_1^n & L_2^n \end{bmatrix} \begin{pmatrix}
        x_t^1 \\
        \widetilde x_t^2
    \end{pmatrix} 
    = \widetilde L_1^n x_t^1 + L_2^n \widetilde x_t^2.    \label{eq: decomp. on SW example} 
\end{align}
We shall then distinguish between three cases.

First, if $\sup_{n \in \mathbb N}\mu_{r_1}(L_1^{n'} L_1^n) = \infty$ also $\sup_{n \in \mathbb N} \mu_{r_1}(\widetilde L_1^{n'} \widetilde L_1^n) = \infty$ and the representation \eqref{eq: decomp. on SW example} is already encompassing both canonical representations with $r = r_1$, and 
$$
C_t^n:=
\underbrace
{\widetilde L_1^n 
\l\{
\E\l[x_t^1x_t^{1'}\r]
\r\}^{1/2}
}_{\Lambda^n}
\underbrace{
\l\{
\E\l[x_t^1x_t^{1'}\r]
\r\}
^{-1/2}x_t^1
}
_{F_t}
$$ 
and 
$$
e_t^{\chi, n}:=
\underbrace{
L_2^n 
\l\{\E\l[\widetilde x_t^2 \widetilde x_t^{2'}\r]\r\}^{1/2}
}_{\Lambda^{w,n}}
\underbrace{
\l\{\E\l[\widetilde x_t^2 \widetilde x_t^{2'}\r]\r\}^{-1/2}
\widetilde x_t^2}_{F_t^w},
$$ 
where the vector of static factors 
$(F_t' \, F_t^{w'})'$ is orthonormal and we 
note that the weak factors do not depend on $n$ in this case since $r_\chi<\infty$. 
Since $r_\chi=r_1+r_2=r+r_w$, we also have
 $r_w=r_2$.   

Second, suppose $r_1 < r$, then since $\sup_{n\in \mathbb N}\mu_{1} (L_2^{n'} L_2^n)< \infty$, i.e., $L_2^n \widetilde x_t^2$ is statically idiosyncratic, and this implies that $\mathbb S_t(y) \subseteq \cspargel (x_t^1)$, which, however, is a contradiction since $\dim \mathbb S_t(y) = r$. Consequently it must be that $r_1 \geq r$. 

Finally, suppose $r_1 > r$, i.e., $\sup_{n \in \mathbb N} \mu_{r_1}(\widetilde L_1^{n'} \widetilde L_1^n) < \infty$. This is possible if there are loading columns in $L_1^n$ which are highly collinear with each other. In this case we may firstly get an $r$-dimensional orthonormal basis from the factors in $x_t^1$, e.g., via the Gram-Schmidt orthogonalisation procedure, and then rotate the resulting loadings accordingly. Denote the resulting $r$-dimensional factors as $\bar x_t^1$ and the corresponding $n\times r$ matrix of loadings as ${\bar L}_1^n$, then
\begin{align}
C_t^n&:=
\underbrace
{\bar L_1^n 
\l\{
\E\l[\bar x_t^1 \bar x_t^{1'}\r]
\r\}^{1/2}
}_{\Lambda^n}
\underbrace{
\l\{
\E\l[\bar x_t^1 \bar x_t^{1'}\r]
\r\}
^{-1/2} \bar x_t^1
}
_{F_t},\nn
\end{align}
while $e_t^{\chi,n}$ is the same as in the case $r_1=r$.

\subsubsection{The canonical representation of Example \ref{exmp: IRF} and DGP1}\label{subsec: weak factor sim example details1}

From Example \ref{exmp: IRF} and DGP1 in Section \ref{subsec: sim nonconsist}, by projecting $f_t$ out from $f_{t-1}$, we get
\begin{align}
\chi_{t}^n&=  
\underbrace{\l(\lambda_{0}^n + \lambda_{1}^n a\r) f_t}_{C_t^n}
+ 
\underbrace{\lambda_{1}^n\l(f_{t-1} - a f_t\r)}_{e_t^{\chi,n}},\nn
\end{align}
where $\lambda_j^n:=(\lambda_{1j}\cdots \lambda_{nj})'$, $j=0,1$, which are $n\times 1$. This is representation \eqref{eq: gen decomp}. Furthermore, representation \eqref{eq: canonical rep weak and strong factors} is obtained by letting $\Lambda^n:=\lambda_{0}^n + \lambda_{1}^n a$,
$F_t:= f_t$, $\Lambda^{w,n}:= \lambda_{1}^n \sigma$, and $F_t^w:=\sigma^{-1}(f_{t-1} - a f_t)$,
so that, the vector of static factors 
$(F_t \, F_t^{w})'$ is orthonormal. Given our assumptions on the loadings, we see that $r=1$ and $r_w=1$.

\subsubsection{The canonical representation of  DGP2}\label{subsec: weak factor sim example details2}
From DGP2 in Section \ref{subsec: sim consist weak factors}, by projecting $f_t$ out from $(f_{t-1}\;f_{t-2})'$, we  get
\beq
\chi_t^n = 
\underbrace{\l\{
\lambda_0^n+
\begin{bmatrix}
\lambda_1^n & \lambda_2^n
\end{bmatrix}
\begin{pmatrix}
\gamma_f(1)\\
\gamma_f(2)
\end{pmatrix}
\r\} f_t }_{C_t^n}
+
\underbrace{\begin{bmatrix}
\lambda_1^n & \lambda_2^n
\end{bmatrix}
\l\{
\begin{pmatrix}
f_{t-1}\\
f_{t-2}
\end{pmatrix}
-\begin{pmatrix}
\gamma_f(1)\\
\gamma_f(2)
\end{pmatrix} f_t
\r\}
}_{e_t^{\chi,n}},\nn
\eeq
where $\lambda_j^n:=(\lambda_{1j}\cdots \lambda_{nj})'$, $j=0,1,2$, which are $n\times 1$, and $\gamma_f(h):=\E[f_t f_{t-h}']$, $h=\pm 1,\pm 2$. This is representation \eqref{eq: gen decomp}. Now, let
\begin{align*}
F_t&:= f_t \quad \text { and }\quad
\Lambda^n:= \lambda_0^n+
\begin{bmatrix}
\lambda_1^n & \lambda_2^n
\end{bmatrix}
\begin{pmatrix}
\gamma_f(1)\\
\gamma_f(2)
\end{pmatrix},
\nn\\
  \widetilde  F_t^w &: = \begin{pmatrix}
        f_{t-1} \\
        f_{t-2}
    \end{pmatrix} -  
 \begin{pmatrix}
  \gamma_f(1)\\ 
  \gamma_f(2)
  \end{pmatrix}
    f_t \quad \text { and }\quad
\widetilde\Lambda^{w,n}:=\begin{bmatrix}
\lambda_1^n & \lambda_2^n
\end{bmatrix}.
    \nn
    \end{align*}
Then, by construction $F_t$ and $\widetilde F_t^w$ are orthogonal. So, given our assumptions on the loadings, we see that $r=1$ and $r_w=2$. 
    
    In this case, we need orthonormal weak factors that satisfy Assumption A\ref{A weakfactor ID}. We can proceed as follows. Recalling that, by construction, $\E[f_t^2]=1$, we have
        \begin{align*}    %
    \Gamma_{\widetilde F^w} := \E\l[ \widetilde F_t^w \widetilde F_t^{w'}\r] &= \begin{pmatrix}
        1 & \gamma_f(1) \\
        \gamma_f(1) & 1
    \end{pmatrix} - 
    \begin{pmatrix}
\{\gamma_f(1)\}^2&\gamma_f(1)\gamma_f(2)\\ 
\gamma_f(1)\gamma_f(2)  &\{\gamma_f(2)\}^2
  \end{pmatrix}.
    \end{align*}
    So the population covariance matrix of the first $\bar n$ weak common components is:
    \[
    \Gamma_{e^\chi}^{\bar n} =\widetilde \Lambda^{w,\bar n} \Gamma_{\widetilde F^w}\widetilde\Lambda^{w,\bar n'}
    = \mathcal W_{\bar n} \mathcal M\mathcal W_{\bar n}',
    \]
    where $\mathcal W_{\bar n}$ is the $2\times \bar n$ matrix having as rows the normalized eigenvectors of $\Gamma_{e^\chi}^{\bar n}$ and $\mathcal M$ is the diagonal $2\times 2$ matrix containing the corresponding eigenvalues sorted in decreasing order. Then, we identify the orthonormal weak factors as the population normalized principal components of $(e_t^{\chi,\bar n})$: 
    \begin{align}
    F_t^w &:= (\mathcal M)^{-1/2}\mathcal W_{\bar n} e_t^{\chi,\bar n}=
    (\mathcal M)^{-1/2}\mathcal W_{\bar n}\begin{bmatrix}
\lambda_1^{\bar n} & \lambda_2^{\bar n}
\end{bmatrix}
\l\{
\begin{pmatrix}
f_{t-1}\\
f_{t-2}
\end{pmatrix}
-\begin{pmatrix}
\gamma_f(1)\\
\gamma_f(2)
\end{pmatrix} f_t
\r\},\nn\\
\Lambda^{w,\bar n} &:= \mathcal W_{\bar n}'(\mathcal M)^{1/2}.\nn
\end{align}
With these definitions and those of $F_t$ and $\Lambda^n$ above we obtain representation \eqref{eq: canonical rep weak and strong factors}.
Finally, to fix the sign, we can always choose the sign of eigenvectors in such a way that the $2\times 2$ top sub-matrix of $\mathcal W_{\bar n}'$ has positive entries along the main diagonal.

In practice, the above procedure requires computing the autocovariances of $(f_t)$, which follows the AR(2) model:
\begin{align}
    \underbrace{\begin{pmatrix}
        f_t \\
        f_{t-1} \\
        f_{t-2}
    \end{pmatrix}}_{x_t} = \underbrace{\begin{pmatrix}
        a_1 & a_2 & 0\\
        1   & 0 & 0 \\
        0   & 1 & 0
    \end{pmatrix}}_{A} \underbrace{\begin{pmatrix}
        f_{t-1} \\
        f_{t-2} \\
        f_{t-3}
    \end{pmatrix}}_{x_{t-1}}  + \underbrace{\begin{pmatrix}
        \sigma \\
        0 \\
        0
    \end{pmatrix}}_{B} \varepsilon_t. 
\end{align}
And, from the Lyapunov equations, we obtain:
\begin{align*}
    \Gamma_x &:= \E\l[x_t x_t'\r] = A \Gamma_x A' + BB', \\
    \vect \Gamma_x &= \l(I_{4} - A \otimes A\r)^{-1} \vect (BB').
\end{align*}
We then immediately have $\gamma_f(1)$ and $\gamma_f(2)$ as the second and third elements of $\vect \Gamma_x$.

\subsubsection{The canonical representation of  DGP3}\label{subsec: weak factor sim example details3}

From DGP3 in Section \ref{subsec: sim est of IRF}, by projecting $(f_t\; f_{t-1})'$ out from $f_{t-2}$, we  get:
\begin{align}
    \chi_t^n &= 
\underbrace{\l\{\begin{bmatrix}\lambda_0^n& \lambda_1^n\end{bmatrix} + \lambda_2^n
\begin{bmatrix}
\gamma_f(2)&\gamma_f(1)
\end{bmatrix}
\r\}
\begin{pmatrix}
        f_t \\ f_{t-1}
    \end{pmatrix} }_{C_t^n}
        + 
 \underbrace{   \lambda_2^n 
 \l\{
        f_{t-2} - \begin{bmatrix}
\gamma_f(2)&\gamma_f(1)
\end{bmatrix} \begin{pmatrix}
            f_t \\ 
            f_{t-1}
        \end{pmatrix}
    \r\}
    }_{e_t^{\chi,n}}, \nn
\end{align} 
where $\lambda_j^n:=(\lambda_{1j}\cdots \lambda_{nj})'$, $j=0,1,2$, which are $n\times 1$, and $\gamma_f(h):=\E[f_t f_{t-h}']$, $h=\pm 1,\pm 2$. This is representation \eqref{eq: gen decomp}. Now, let
\begin{align*}
\widetilde F_t&:= \begin{pmatrix}
f_t\\
f_{t-1}
\end{pmatrix}\quad \text { and }\quad
\widetilde\Lambda^n:= \begin{bmatrix}\lambda_0^n& \lambda_1^n\end{bmatrix} + \lambda_2^n
\begin{bmatrix}
\gamma_f(2)&\gamma_f(1)
\end{bmatrix},
\nn\\
  \widetilde  F_t^w &: = 
        f_{t-2} - \begin{bmatrix}
\gamma_f(2)&\gamma_f(1)
\end{bmatrix} \begin{pmatrix}
            f_t \\ 
            f_{t-1}
        \end{pmatrix} \quad \text { and }\quad
\widetilde\Lambda^{w,n}:= \lambda_2^n.
    \nn
    \end{align*}
Then, by construction $\widetilde F_t$ and $\widetilde F_t^w$ are orthogonal. So, given our assumptions on the loadings, we see that $r=2$ and $r_w=1$. To obtain representation \eqref{eq: canonical rep weak and strong factors} it is enough to normalized the factors, i.e., by defining:
\begin{align}
&F_t:= \l\{\E[\widetilde F_t\widetilde F_t']
\r\}^{-1/2} \widetilde F_t
\quad \text { and }\quad
\Lambda^n:= \widetilde\Lambda^n \l\{\E[\widetilde F_t\widetilde F_t']
\r\}^{1/2},\nn\\
&F_t^w:= \l\{\E[\widetilde F_t^w\widetilde F_t^{w'}]
\r\}^{-1/2} \widetilde F_t^w
\quad \text { and }\quad
\Lambda^{w,n} := \widetilde\Lambda^{w,n}\l\{\E[\widetilde F_t^w\widetilde F_t^{w'}]
\r\}^{1/2}.\nn
\end{align}

\section{Proofs for Section \ref{sec: estimation}}\label{app:proofprop}
\subsection{Preliminary lemmas}
\begin{lemma}\label{lem:WZ}
Under Assumptions A\ref{A: stat}, A\ref{ass:newGDFM}, A\ref{ass:kernel}, for all $n,T\in\mathbb N$, there exists a $C\in(0,\infty)$ independent of $n$ and $T$, such that
\[
\max_{1\le i,j\le n}\E
\l[\max_{\vert h\vert \le \mathcal B_T} \l\vert
[\widehat{f}_{y}^n(\theta_h)]_{ij}
-[{f}_{y}^n(\theta_h)]_{ij}
\r\vert^2
\r]\le C\alpha_T,
\]
where
\[
\alpha_T=\max\l(\frac{\mathcal B_T^2}{T^{2-4/\nu}}, \frac{\mathcal B_T\log \mathcal B_T}{T},\frac 1{\mathcal B_T^{2\vartheta}} \r),
\]
and $\nu\ge 4$ is defined in Assumption A\ref{ass:newGDFM}.
\end{lemma}

\begin{proof}
We refer to \citet[Proposition 6]{forni2017dynamic}  who derive the result from \citet{wu2018asymptotic} but without the first term in $\alpha_T$, and to \citet[Proposition 1]{barigozzi2024inferential} or \citet[Proposition F.9]{barigozzi2024fnets}
who derive the result from \citet{zhang2021convergence}. 
\end{proof}

\begin{lemma} \label{lem:Cxi}
Under Assumptions A\ref{A: stat}, A\ref{ass:newGDFM}, for all $n\in\mathbb N$ there exists a $C_\xi\in(0,\infty)$ independent of $n$ such that $\sup_{\theta\in[-\pi,\pi]} \mu_1(f_\xi^n(\theta))\le C_\xi.$
\end{lemma}

\begin{proof}See  \citet[Proposition 1]{forni2017dynamic}.
\end{proof}

\begin{lemma} \label{lem:spectra}
Under Assumptions A\ref{A: stat}, A\ref{ass:newGDFM}, A\ref{ass:diveval}, A\ref{ass:kernel}, for all $n,T\in\mathbb N$,
\ben
\item [(i)]
$ n^{-2}\E\l[
\max_{\vert h\vert \le \mathcal B_T}\l\Vert
\wh{f}^n_y(\theta_h)-f^n_y(\theta_h)
\r\Vert^2
\r]\le C\alpha_T$;
\item [(ii)]
$ \max_{1\le i\le n} n^{-1}\E\l[\max_{\vert h\vert \le \mathcal B_T}
\l\Vert
s_i' \l\{\wh{f}^n_y(\theta_h)-f^n_y(\theta_h)\r\}
\r\Vert^2
\r]
\le C\alpha_T$;
\item [(iii)] 
$ \max_{\vert h\vert \le \mathcal B_T}n^{-1}
\l\Vert
{f}^n_y(\theta_h)-{f}^n_\chi(\theta_h)
\r\Vert
\le C_\xi n^{-1};
$
\een
where $s_i$ is the $i$th element of the canonical basis of $\mathbb R^{n\times 1}$,  $C$ and $\alpha_T$ are defined in Lemma \ref{lem:WZ}, and $C_\xi$ is defined in Lemma \ref{lem:Cxi}.
Moreover, for all $n,T\in\mathbb N$ there exists $C_1\in(0,\infty)$ independent of $n$ and $T$, such that
\ben
\item [(iv)] $n^{-2}\E\l[\max_{\vert h\vert \le \mathcal B_T}
\l\Vert
\wh{f}^n_y(\theta_h)-f^n_\chi(\theta_h)
\r\Vert^2
\r]\le C_1\max(\alpha_T,n^{-2})$;
\item [(v)]
$\max_{1\le i\le n} n^{-1}\E\l[\max_{\vert h\vert \le \mathcal B_T}
\l\Vert
s_i' \l\{\wh{f}^n_y(\theta_h)-f^n_\chi(\theta_h)\r\}
\r\Vert^2
\r]
\le C_1\max(\alpha_T,n^{-1})$.
\een
\end{lemma}

\begin{proof}
For part (i):
\begin{align}
n^{-2}\E\l[\max_{\vert h\vert \le \mathcal B_T}
\l\Vert
\wh{f}^n_y(\theta_h)-f^n_y(\theta_h)
\r\Vert^2
\r]&\le
\sup_{n\in\mathbb N} n^{-2}\E\l[\max_{\vert h\vert \le \mathcal B_T}
\l\Vert
\wh{f}^n_y(\theta_h)-f^n_y(\theta_h)
\r\Vert_F^2
\r]\nonumber\\
&= 
\sup_{n\in\mathbb N} n^{-2}\sum_{i,j=1}^n
\E\l[ \max_{\vert h\vert \le \mathcal B_T}\l\vert
[\widehat{f}_{y}^n(\theta_h)]_{ij}
-[{f}_{y}^n(\theta_h)]_{ij}
\r\vert^2\r]\nonumber\\
&\le 
\sup_{n\in\mathbb N}\max_{1\le i,j\le n}
\E\l[ \max_{\vert h\vert \le \mathcal B_T} \l\vert
[\widehat{f}_{y}^n(\theta_h)]_{ij}
-[{f}_{y}^n(\theta_h)]_{ij}
\r\vert^2\r]\le C\alpha_T,\nonumber
\end{align}
because of Lemma \ref{lem:WZ}, with $C$ independent of $n$ and $T$.

For part (ii), just notice that
\[
\max_{1\le i\le n}n^{-1} \l\Vert
s_i' \l\{\wh{f}^n_y(\theta_h)-f^n_y(\theta_h)\r\}
\r\Vert^2 =\max_{1\le i\le n} n^{-1}\sum_{j=1}^n \l\vert
[\widehat{f}_{y}^n(\theta_h)]_{ij}
-[{f}_{y}^n(\theta_h)]_{ij}
\r\vert^2,
\]
then the proof is identical to part (i).

For part (iii),
\begin{align}
 \max_{\vert h\vert \le \mathcal B_T}n^{-1}
\l\Vert
{f}^n_y(\theta_h)-{f}^n_\chi(\theta_h)
\r\Vert
&=  \max_{\vert h\vert \le \mathcal B_T}n^{-1}
\l\Vert{f}^n_\xi(\theta_h)
\r\Vert\nonumber\\
&= \max_{\vert h\vert \le \mathcal B_T} n^{-1}\mu_1(f_\xi^n(\theta_h))\le n^{-1}C_\xi,\nonumber
\end{align}
because of Lemma \ref{lem:Cxi}, with $C_\xi$ independent of $n$.

For part (iv)
\begin{align}
n^{-2}\E\l[\max_{\vert h\vert \le \mathcal B_T}
\l\Vert
\wh{f}^n_y(\theta_h)-f^n_\chi(\theta_h)
\r\Vert^2
\r]\le&\,
n^{-2}\E\l[\max_{\vert h\vert \le \mathcal B_T}
\l\Vert
\wh{f}^n_y(\theta_h)-f^n_y(\theta_h)
\r\Vert^2
\r] \nonumber\\
&+
 \max_{\vert h\vert \le \mathcal B_T}n^{-2}
\l\Vert
{f}^n_y(\theta_h)-f^n_\chi(\theta_h)
\r\Vert^2\nonumber\\
\le&\, C\alpha_T +C_\xi^2 n^{-2}
\le \max(C,C_\xi^2)\max(\alpha_T ,n^{-2}),\nonumber
\end{align}
because of parts (i) and (iii),  with $C$ and $C_\xi$ independent of $n$ and $T$.

For part (v) the proof follows directly from parts (ii) and (iii) and using the same approach as in part (iv).
\end{proof}

\begin{lemma} \label{lem:weyl}
Under Assumptions A\ref{A: stat}, A\ref{ass:newGDFM}, A\ref{ass:diveval}, A\ref{ass:kernel}, for all $n,T\in\mathbb N$,
\ben
\item [(i)]
$\max_{1\le j\le q}n^{-2} \E\l[ \max_{\vert h\vert \le \mathcal B_T} 
\l\vert
{\mu}_j(\wh f_y^n(\theta_h))-{\mu}_j(f_y^n(\theta_h))
\r\vert^2
\r]\le C\alpha_T$;
\item [(ii)]
$\max_{1\le j\le q} \max_{\vert h\vert \le \mathcal B_T} 
\l\vert
{\mu}_j( f_y^n(\theta_h))-{\mu}_j( f_\chi^n(\theta_h))
\r\vert \le C_\xi$;
\item [(iii)]
$\max_{1\le j\le q} n^{-2} \E\l[\max_{\vert h\vert \le \mathcal B_T} 
\l\vert
{\mu}_j(\wh f_y^n(\theta_h))-{\mu}_j(f_\chi^n(\theta_h))
\r\vert^2
\r]\le C_1\max(\alpha_T,n^{-1})$;
\een
where
$C$ and $\alpha_T$ are defined in Lemma \ref{lem:WZ}, $C_\xi$ is defined in Lemma \ref{lem:Cxi}, and $C_1$ is defined in Lemma \ref{lem:spectra}.
\end{lemma}

\begin{proof}
The proof follows from Weyl's inequality \citep[Theorem 1]{merikoski2004inequalities} plus:
 Lemma \ref{lem:spectra}(i) for part (i),
  Lemma \ref{lem:spectra}(iii) for part (ii), and
  Lemma \ref{lem:spectra}(iv) for part (iii).
\end{proof}

\begin{lemma}\label{lem:rowevec}
Under Assumptions A\ref{A: stat}, A\ref{ass:newGDFM}, A\ref{ass:diveval}, as $n\to\infty$,
\[
\max_{1\le i\le n}\max_{\vert h\vert \le \mathcal B_T} \l\Vert
s_i'\sqrt n  V_n^*(\theta_h)
\r\Vert=\mathcal O(1),
\]
where $s_i$ is the $i$th element of the canonical basis of $\mathbb R^{n\times 1}$ and $V_n(\theta_h)$ is the $q\times n$ matrix having as rows the normalized eigenvectors of ${f}_\chi^n(\theta_h)$ corresponding to its $q$ non-zero eigenvalues.
\end{lemma}

\begin{proof}
Let $v_{ji}(\theta_h)$
be the $(j,i)$th entry of $V_n(\theta_h)$, $i\in\mathbb N$, $j=1,\ldots, q$,
\begin{align}
\max_{1\le i\le n}\max_{\vert h\vert \le \mathcal B_T} [f_\chi^n(\theta_h)]_{ii} = \max_{1\le i\le n}\max_{\vert h\vert \le \mathcal B_T}\sum_{j=1}^q \mu_j(f_\chi^n(\theta_h))
\l\vert
v_{ji}(\theta_h)
\r\vert^2<\infty,\label{eq:evecevalchi}
\end{align}
indeed, by Assumption A\ref{ass:newGDFM}
\begin{align}
[f_\chi^n(\theta_h)]_{ii}&=\frac 1{2\pi}\sum_{l=-\infty}^\infty  e^{-\iota l \theta_h} 
\E [\chi_{it} \chi_{i,t-l}']\nonumber\\
&\le \sum_{l=-\infty}^\infty \l\vert \E [\chi_{it} \chi_{i,t-l}']\r\vert\nonumber\\
&\le \sum_{l=-\infty}^\infty \sum_{j,j'=0}^\infty \sum_{\ell,\ell'=1}^q \l\vert K_{i,\ell}(j)
K_{i,\ell'}(j') 
\E[\varepsilon_{\ell,t-j}\varepsilon_{\ell',t-l-j'} ]\r\vert 
\nonumber\\
&\le \sum_{l=-\infty}^\infty \sum_{j=0}^\infty \sum_{\ell=1}^q \l\vert K_{i,\ell}(j)
K_{i,\ell}(j-l) \r\vert 
\E[\varepsilon_{\ell,t-j}^2 ]
\nonumber\\
&\le \sum_{l=-\infty}^\infty (\rho^\chi)^{-l}  \sum_{j=0}^\infty (\rho^\chi)^{2j} \sum_{\ell=1}^q (A_{i,\ell}^\chi)^2\nonumber\\
&\le q(A^\chi)^2 \frac{1}{1-(\rho^\chi)^2}  \frac{1+\rho^\chi}{1-\rho^\chi}\nonumber\\
&\le q(A^\chi)^2 \frac{1}{(1-\rho^\chi)^2}, \label{eq:finitespectra}
\end{align}
moreover, $[f_\chi^n(\theta_h)]_{ii}\ge 0$ (see \citealp[Corollary 4.3.2, p.120]{brockwell2009time}).

Now, since, by Assumption A\ref{ass:diveval}, for all $j = 1,\ldots,q$, we must have
\[
 \lim_{n\to\infty} \max_{\vert h\vert \le \mathcal B_T} \frac{\mu_j\left(f_\chi^n(\theta_h)\right) }{n}\ge \max_{\vert h\vert \le \mathcal B_T}\alpha_j(\theta_h) >0
\]
it follows that from \eqref{eq:evecevalchi} we must have
\[
\lim_{n\to\infty} \max_{\vert h\vert \le \mathcal B_T}  n \l\vert
v_{ ij}(\theta_h)
\r\vert^2\le C_2<\infty
\]
with $C_2$ independent of $i$ and $j$. Therefore,
\[
 \lim_{n\to\infty}\max_{1\le i\le n}  \max_{\vert h\vert \le \mathcal B_T}  n \sum_{j=1}^q \l\vert
v_{ ij}(\theta_h)
\r\vert^2= \lim_{n\to\infty}\max_{1\le i\le n}  \max_{\vert h\vert \le \mathcal B_T}  n \Vert s_i' V_n^*(\theta_h) \Vert^2
\le q C_2<\infty.
\]
\end{proof}

\begin{lemma}\label{lem:DK}
Under Assumptions A\ref{A: stat}, A\ref{ass:newGDFM}, A\ref{ass:diveval}, A\ref{ass:kernel}, as $n,T\to\infty$, 
\ben
\item [(i)] $\min(\alpha_T^{-1},n^2) \E\l[\max_{\vert h\vert\le \mathcal B_T}\l\Vert V_n(\theta_h)- \mathcal R \widehat W_n(\theta_h)\r\Vert^2\r]=\mathcal O(1)$;
\item [(ii)] $\min(\alpha_T^{-1},n)\max_{1\le i\le n}  \E\l[\max_{\vert h\vert \le \mathcal B_T} \l\Vert
s_i'\sqrt n\l(V_n^*(\theta_h)\mathcal R- \widehat W_n^*(\theta_h)\r)
\r\Vert^2\r]=\mathcal O(1)$;
\een
where $\mathcal R$ is a $q \times q$ orthogonal matrix and
$s_i$ is the $i$th element of the canonical basis of $\mathbb R^{n\times 1}$.
\end{lemma}

\begin{proof}
For part (i), from Davis-Kahan theorem
\citet[Theorem 2]{yu2015useful}, for all $n\in\mathbb N$ there exists a $C_3\in(0,\infty)$, independent of $n$ and $T$, such that
\begin{align}\label{eq:DK0}
 \l\Vert V_n(\theta_h)- \mathcal R \widehat W_n(\theta_h)\r\Vert &\le  C_3 \frac{\l\Vert
\wh{f}^n_y(\theta_h)-f^n_\chi(\theta_h)
\r\Vert}{
\mu_q(f^n_\chi(\theta_h))-\mu_{q+1}(f^n_\chi(\theta_h))
}, \quad  \theta_h = \frac{\pi h}{\mathcal B_T}, \quad \vert h\vert\le \mathcal B_T.
\end{align}
Therefore, since  Assumption A\ref{ass:diveval} implies
\[
\min_{\vert h\vert\le \mathcal B_T}\lim_{n\to\infty} \frac{\mu_q(f^n_\chi(\theta_h))}n\ge \min_{\vert h\vert\le \mathcal B_T} \alpha_q(\theta_h)>0,
\]
and since $\mu_{q+1}(f^n_\chi(\theta_h))=0$ for $n>q$ for all $\theta_h$, by Lemma \ref{lem:spectra}(iv) and \eqref{eq:DK0},
as $n,T\to\infty$, 
\beq\nn
\min(\alpha_T^{-1},n^2) \E\l[\max_{\vert h\vert\le \mathcal B_T}\l\Vert V_n(\theta_h)- \mathcal R \widehat W_n(\theta_h)\r\Vert^2\r]=\mathcal O(1).
\eeq

For part (ii), from \eqref{eq:DK0} and Assumption A\ref{ass:diveval} there exists a $C_4\in(0,\infty)$, independent of $n$, such that
\begin{align}
\sum_{i=1}^n \l\Vert
s_i'\sqrt n\l(V_n^*(\theta_h)\mathcal R- \widehat W_n^*(\theta_h)\r)
\r\Vert^2 &= \l\Vert
\sqrt n\l(V_n^*(\theta_h)\mathcal R- \widehat W_n^*(\theta_h)\r)
\r\Vert^2_F\nn\\
&\le \sqrt q \l\Vert
\sqrt n\l(V_n^*(\theta_h)\mathcal R- \widehat W_n^*(\theta_h)\r)
\r\Vert^2  \nn\\
&\le n\sqrt q C_3 \frac{\l\Vert
\wh{f}^n_y(\theta_h)-f^n_\chi(\theta_h)
\r\Vert^2}{
\{\mu_q(f^n_\chi(\theta_h))-\mu_{q+1}(f^n_\chi(\theta_h))\}^2
}\nn\\
&\le C_4 n^{-1} \l\Vert
\wh{f}^n_y(\theta_h)-f^n_\chi(\theta_h)
\r\Vert^2_F\nn\\
&=C_4 n^{-1}\sum_{i=1}^n \l\Vert s_i'\l\{ \wh{f}^n_y(\theta_h)-f^n_\chi(\theta_h)\r\}\r\Vert^2,\nn
\end{align}
Hence, as $n\to\infty$,
$$
\max_{1\le i\le n}  \E\l[\max_{\vert h\vert \le \mathcal B_T} \l\Vert
s_i'\sqrt n\l(V_n^*(\theta_h)\mathcal R- \widehat W_n^*(\theta_h)\r)
\r\Vert^2\r]
$$ 
is of order no greater than
$$
\max_{1\le i\le n}n^{-1}  \E\l[\max_{\vert h\vert \le \mathcal B_T}
 \l\Vert s_i'\l\{ \wh{f}^n_y(\theta_h)-f^n_\chi(\theta_h)\r\}\r\Vert^2
\r].
$$
Therefore, by Lemma \ref{lem:spectra}(v), as $n,T\to\infty$,
\[
\min(\alpha_T^{-1},n)\max_{1\le i\le n}  \E\l[\max_{\vert h\vert \le \mathcal B_T} \l\Vert
s_i'\sqrt n\l(V_n^*(\theta_h)\mathcal R- \widehat W_n^*(\theta_h)\r)
\r\Vert^2\r]=\mathcal O(1).
\]
\end{proof}

\begin{lemma}\label{lem:evecevec}
Under Assumptions A\ref{A: stat}, A\ref{ass:newGDFM}, A\ref{ass:diveval}, A\ref{ass:kernel}, as $n,T\to\infty$, 
\ben
\item [(i)] 
$\min(\alpha_T^{-1},n^2)
\E\l[\max_{\vert h\vert \le \mathcal B_T}
\l\Vert V_n^*(\theta_h)V_n(\theta_h)-
 \widehat W_n^* (\theta_h)\widehat W_n(\theta_h)
\r\Vert^2
\r]=\mathcal O(1);
$
\item [(ii)]
$\min(\alpha_T^{-1},n)
\max_{1\le i\le n}
\E\l[\max_{\vert h\vert \le \mathcal B_T}
\l\Vert
s_i'\sqrt n \l\{V_n^*(\theta_h)V_n(\theta_h)-
 \widehat W_n^* (\theta_h)\widehat W_n(\theta_h)\r\}
\r\Vert^2
\r]=\mathcal O(1);
$
\een
where $s_i$ is the $i$th element of the canonical basis of $\mathbb R^{n\times 1}$.
\end{lemma}

\begin{proof}
By the $C_r$-inequality with $r=2$, we have the decomposition
%
\begin{align}
\l\Vert V_n^*(\theta_h)V_n(\theta_h)-
 \widehat W_n^* (\theta_h)\widehat W_n(\theta_h)
\r\Vert^2 \le&\, 
 2 \l\Vert V_n^*(\theta_h)\,\l( V_n(\theta_h)- \mathcal R \widehat W_n(\theta_h)\r)\r \Vert^2\nn\\
&+2  \l\Vert \l(V_n^*(\theta_h)\mathcal R- \widehat W_n^*(\theta_h)\r ) \widehat W_n(\theta_h)\r\Vert^2\nn\\
\le &\,2 \l\Vert V_n(\theta_h)- \mathcal R \widehat W_n(\theta_h) \r\Vert^2+ 2\l\Vert V_n^*(\theta_h)\mathcal R- \widehat W_n^*(\theta_h)\r \Vert^2,\label{eq:DK2}
\end{align}
since eigenvectors are normalized so have norm one. From Lemma \ref{lem:DK}(i) and \eqref{eq:DK2} we prove part (i).

Similarly, we have  the decomposition
\begin{align}
\l\Vert s_i'\sqrt n \l\{V_n^*(\theta_h)V_n(\theta_h)-
 \widehat W_n^* (\theta_h)\widehat W_n(\theta_h)\r\}\r\Vert^2\le&\, 2\l\Vert s_i'\sqrt n V_n^*(\theta_h) \l(V_n(\theta_h)- \mathcal R \widehat W_n(\theta_h \r)\r\Vert^2\nn\\
 &+2\l\Vert s_i'\sqrt n\l(V_n^*(\theta_h)\mathcal R- \widehat W_n^*(\theta_h)\r)   \widehat W_n(\theta_h)\r\Vert^2\nn\\
 \le&\, 2\l\Vert s_i'\sqrt n V_n^*(\theta_h)\r\Vert^2 
 \l\Vert V_n(\theta_h)- \mathcal R \widehat W_n(\theta_h \r\Vert^2
 \nn\\
 &+ 2 \l\Vert s_i'\sqrt n\l(V_n^*(\theta_h)\mathcal R- \widehat W_n^*(\theta_h)\r) \r\Vert^2\nn\\
 =&\, a_{ni}^{(1)}(\theta_h)+a_{ni}^{(2)}(\theta_h), \; \text{say,}\nn
\end{align}
since eigenvectors are normalized so have norm one. 
Then, from Lemmas \ref{lem:rowevec} and \ref{lem:DK}, it follows that, as $n,T\to\infty$, (notice that $\Vert s_i'\sqrt n V_n^*(\theta_h)\Vert $ is deterministic)
\[
\min(\alpha_T^{-1},n^2)
\max_{1\le i\le n}
\E\l[\max_{\vert h\vert \le \mathcal B_T}  
a_{ni}^{(1)}(\theta_h)
\r]=\mathcal O(1),
\]
and, by Lemma \ref{lem:DK}(ii),  as $n,T\to\infty$,
\[
\min(\alpha_T^{-1},n)
\max_{1\le i\le n}
 \E\l[\max_{\vert h\vert \le \mathcal B_T}a_{ni}^{(2)}(\theta_h)
\r]=\mathcal O(1).
\]
\end{proof}

\begin{lemma}\label{lem:idioz}
Let $z_t^n = \underline a_n(L)\xi_t^n$ and $\Gamma_z^n=\E[z_t^nz_t^{n'}]$. Under Assumptions A\ref{A: stat}, A\ref{ass:newGDFM}, A\ref{ass:ARGDFM}, for all $n\in\mathbb N$ there exists a $B_z\in(0,\infty)$ independent of $n$ such that
$\mu_1(\Gamma_z^n)\le B_z$.
\end{lemma}

\begin{proof}
The proof follows from Lemma \ref{lem:Cxi} and \citet[Proposition B.1 of the supplementary material]{barigozzi2024fnets}.
\end{proof}

\begin{lemma}\label{lem:chispectral}
Under Assumptions A\ref{A: stat}, A\ref{ass:newGDFM}, A\ref{ass:diveval}, A\ref{ass:kernel}, as $n,T\to\infty$,
\[
\min(\alpha_T^{-1}, n) \max_{1\le i,j\le n}\E
\l[\max_{\vert h\vert \le \mathcal B_T} \l\vert
[\widehat{f}_{\chi}^n(\theta_h)]_{ij}
-[{f}_{\chi}^n(\theta_h)]_{ij}
\r\vert^2
\r]=\mathcal O(1).
\]
\end{lemma}

\begin{proof}
The proof follows from Lemmas \ref{lem:weyl}(iii) and \ref{lem:DK}(ii), noticing that
\[
[\widehat{f}_{\chi}^n(\theta_h)]_{ij} =  s_i' \widehat W_n^*(\theta_h)\widehat M_n(\theta_h)\widehat W_n(\theta_h) s_j
\]
and
\[
[{f}_{\chi}^n(\theta_h)]_{ij} = s_i' V_n^*(\theta_h) M_n(\theta_h)V_n(\theta_h) s_j,
\]
where $M_n(\theta_h)$ is the $q\times q$ diagonal matrix containing $\mu_j(f_\chi^n(\theta_h))$, $j=1,\ldots, q$, and $s_i$ is the $i$th element of the canonical basis of $\mathbb R^{n\times 1}$.
See also \citet[Proposition F.15 of the supplementary material]{barigozzi2024fnets} for convergence in probability.
\end{proof}

\begin{lemma}\label{lem:covbasic}
Under Assumptions A\ref{A: stat}, A\ref{ass:newGDFM}, for all $n,T\in\mathbb N$, 
\ben
\item [(i)] there exists a $C^*\in(0,\infty)$ independent of $n$ and $T$, such that\\
$
\max_{1\le,i,j\le n} \E
\l[\max_{\vert \ell\vert \le \bar p} \l\vert
[\widehat{\Gamma}_{y}^n(\ell)]_{ij}
-[{\Gamma}_{y}^n(\ell)]_{ij}
\r\vert^2
\r]\le C^*T^{-1};
$
\item [(ii)] 
$
\max_{1\le i,j\le n}\max_{\vert \ell\vert\le \bar p}
\l\vert
[\widehat{\Gamma}_{y}^n(\ell)]_{ij}
-[{\Gamma}_{y}^n(\ell)]_{ij}
\r\vert^2
=O_P(\beta_{n,T});
$
\een
where
\[
\beta_{n,T}=\max\l(
\frac{n^{2/\nu}\log^3 n}{T^{1-2/\nu}}
,\sqrt{\frac{\log n}T} \r),
\]
and $\nu\ge 4$ is defined in Assumption A\ref{ass:newGDFM}.
\end{lemma}

\begin{proof}
The proof of part (i) follows directly from Assumption \ref{ass:newGDFM}. In particular, from linearity of the processes $(\chi_{it} : t\in\mathbb Z)$ and  $(\xi_{it} : t\in\mathbb Z)$, summability of the coefficients of associated linear filters along over lags and columns, and finite 4th moments of the associated innovations, imply finite summable 4th order cumulants of the processes and hence of $(y_{it} : t\in\mathbb Z)$. The latter is a necessary and sufficient condition for the lemma to hold (see \citealp[pp. 209-211]{hannan}).

For part (ii) we refer to \citet[Lemma F.5]{barigozzi2024fnets}
who derive the result from \citet{zhang2021convergence}. 
\end{proof}

\begin{lemma}\label{lem:IFTcov}
Under Assumptions A\ref{A: stat}, A\ref{ass:newGDFM}, A\ref{ass:diveval}, A\ref{ass:kernel}, as $n,T\to\infty$,
\ben
\item [(i)] 
$\min(\alpha_T^{-1}, n) \max_{1\le i,j\le n}\E
\l[\max_{\vert \ell \vert \le \bar p}  \l\vert 
[ \widehat \Gamma_\chi^n(\ell)]_{ij} -  [\Gamma_\chi^n(\ell)]_{ij}
\r \vert^2
\r]=\mathcal O(1);
$
\item [(ii)] $\min(\beta_{n,T}^{-1}, \alpha_T^{-1/2}, \sqrt n) \max_{1\le i,j\le n} \max_{\vert \ell \vert \le \bar p}  \l\vert 
[ \widehat \Gamma_\chi^n(\ell)]_{ij} -  [\Gamma_\chi^n(\ell)]_{ij}
\r \vert = \mathcal O_P(1).
$
\een
\end{lemma}

\begin{proof}
For any $1\le i,j\le n$, and fixed $0<\ell\le\bar p$
 \begin{align}
 \l\vert 
[ \widehat \Gamma_\chi^n(\ell)]_{ij} -  [\Gamma_\chi^n(\ell)]_{ij}
\r \vert =&\, \l\vert  \frac {2\pi}{2B_T+1}\sum_{ h=-\mathcal B_T}^{\mathcal B_T} e^{\iota \ell \theta_h} [\wh f_{n}^\chi(\theta_h)]_{ij} - \int_{-\pi}^\pi e^{\iota \ell \theta} [f_{n}^\chi(\theta)]_{ij}\mathrm d\theta  \r\vert\nn\\
\le&\, \frac {2\pi}{2B_T+1}\sum_{ h=-\mathcal B_T}^{\mathcal B_T} \l\vert e^{\iota \ell \theta_h}[\wh f_{n}^\chi(\theta_h)]_{ij} - e^{\iota \ell \theta_h}[f_{n}^\chi(\theta_h)]_{ij}\r\vert\nn\\
&+ \frac {2\pi}{2B_T+1}\l\vert \sum_{ h=-\mathcal B_T}^{\mathcal B_T} 
e^{\iota \ell \theta_h}[f_{n}^\chi(\theta_h)]_{ij}-\int_{-\pi}^\pi
e^{\iota \ell \theta}[f_{n}^\chi(\theta)]_{ij}\mathrm d\theta\r\vert \nn\\
\le&\, \frac {2\pi}{2B_T+1}\sum_{ h=-\mathcal B_T}^{\mathcal B_T} \l\vert[\wh f_{n}^\chi(\theta_h)]_{ij} - [f_{n}^\chi(\theta_h)]_{ij}\r\vert \, \vert e^{\iota \ell \theta_h}\vert\nn\\
&+ \frac {2\pi}{2B_T+1}\sum_{ h=-\mathcal B_T}^{\mathcal B_T} \max_{\theta_{h-1}\le \theta\le \theta_h}\l\vert
e^{\iota \ell \theta_h}[f_{n}^\chi(\theta_h)]_{ij}
-
e^{\iota \ell \theta}[f_{n}^\chi(\theta)]_{ij}
\r\vert\nn\\
\le&\, \frac {2\pi}{2B_T+1}\sum_{ h=-\mathcal B_T}^{\mathcal B_T} \l\vert[\wh f_{n}^\chi(\theta_h)]_{ij} - [f_{n}^\chi(\theta_h)]_{ij}\r\vert \nn\\
&+ \frac {2\pi}{2B_T+1}\sum_{ h=-\mathcal B_T}^{\mathcal B_T} 
\max_{\theta_{h-1}\le \theta\le \theta_h}\l\vert 
[f_{n}^\chi(\theta_h)]_{ij}-
[f_{n}^\chi(\theta)]_{ij}
\r\vert\nn\\
&+ \frac {2\pi}{2B_T+1}\sum_{ h=-\mathcal B_T}^{\mathcal B_T} \max_{\theta_{h-1}\le \theta\le \theta_h}\l\vert
e^{\iota \ell \theta_h}
-
e^{\iota \ell \theta}
\r\vert.\nn
 \end{align}
 and by using the $C_r$-inequality with $r=3$ 
 \begin{align}
 \l\vert 
[ \widehat \Gamma_\chi^n(\ell)]_{ij} -  [\Gamma_\chi^n(\ell)]_{ij}
\r \vert ^2 \le&\, 3  \l\{ \frac {2\pi}{2B_T+1}\sum_{ h=-\mathcal B_T}^{\mathcal B_T} \l\vert[\wh f_{n}^\chi(\theta_h)]_{ij} - [f_{n}^\chi(\theta_h)]_{ij}\r\vert\r\}^2\nn\\
 &+3\l\{
  \frac {2\pi}{2B_T+1}\sum_{ h=-\mathcal B_T}^{\mathcal B_T} 
\max_{\theta_{h-1}\le \theta\le \theta_h}\l\vert 
[f_{n}^\chi(\theta_h)]_{ij}-
[f_{n}^\chi(\theta)]_{ij}
\r\vert
 \r\}^2\nn\\
 &+3\l\{
 \frac {2\pi}{2B_T+1}\sum_{ h=-\mathcal B_T}^{\mathcal B_T} \max_{\theta_{h-1}\le \theta\le \theta_h}\l\vert
e^{\iota \ell \theta_h}
-
e^{\iota \ell \theta}
\r\vert
 \r\}^2\nn\\
 =&\, I_{ij}+II_{ij}+III, \;\text{say.}
\end{align}
  Then, by Cauchy-Schwarz inequality
  \begin{align}
  I_{ij}&\le \frac{3(2\pi)^2}{2B_T+1}\sum_{ h=-\mathcal B_T}^{\mathcal B_T} \l\vert[\wh f_{n}^\chi(\theta_h)]_{ij} - [f_{n}^\chi(\theta_h)]_{ij}\r\Vert^2\le 3(2\pi)^2\max_{\vert h\vert\le \mathcal B_T } \l\vert[\wh f_{n}^\chi(\theta_h)]_{ij} - [f_{n}^\chi(\theta_h)]_{ij}\r\vert^2,\nn
  \end{align}
  so, by Lemma \ref{lem:chispectral}, as $n,T\to\infty$,
  \[
 \min(\alpha_T^{-1},n) \max_{1\le i,j\le n}\E[I_{ij}] =\mathcal O(1).
  \]
  Moreover, the function  $\theta\mapsto f_\chi^n(\theta)$ is of bounded variation (see \citealp[Proposition 2]{forni2017dynamic}) and the function $\theta\mapsto g_{n}(\theta)$ and $\theta\mapsto e^{\iota \ell \theta}$ is also of bounded variation.  Hence $\max_{1\le i,j\le n} II_{ij}=O(\mathcal B_T^{-2})$ and $III=O(\mathcal B_T^{-2})$  and are dominated when using the Bartlett kernel for which $\vartheta=1$.
  
For part (ii) we refer to \citet[Proposition F.16 of the supplementary material]{barigozzi2024fnets}.
\end{proof}

\begin{lemma}\label{lem:VARparam}
Let $B_\chi^{(h)} = (\Gamma_\chi^{(h)'}(\ell) : 1\le \ell\le p_h)$ so that $A^{(h)}=B_\chi^{(h)}(C_\chi^{(h)})^{-1}$, where $C_\chi^{(h)}$ is defined in Assumption A\ref{ass:ARGDFM}.
Denote the matrix collecting all the coefficients of ${\underline a}^{[h]}(L)$ defined in Assumption A\ref{ass:ARGDFM} as $A_n=(A^{(h)} : h=1,\ldots, m )$ which is $n\times n\bar p$.
Let  $\wh B_\chi^{(h)}$ and $\wh C_\chi^{(h)}$
be the estimators of $\wh B_\chi^{(h)}$ and $\wh C_\chi^{(h)}$ computed using the entries of $\wh \Gamma_\chi^{(h)}(\ell)$. And finally, let $\wh A^{(h)}=\wh B_\chi^{(h)}(\wh C_\chi^{(h)})^{-1}$ and $\wh A_n$ be estimator containing all the coefficients of $\widehat {\underline a}^{[h]}(L)$. Under Assumptions A\ref{A: stat}, A\ref{ass:newGDFM}, A\ref{ass:diveval}, A\ref{ass:kernel}, A\ref{ass:ARGDFM}, as $n,T\to\infty$,
\ben
\item [(i)] $\min( \beta_{n,T}^{-1}, \alpha_T^{-1/2}, \sqrt n) n^{-1/2} \l\Vert \wh{A}_n -A_n\r\Vert_F = \mathcal O_P(1)$;
\item [(ii)] $\min( \beta_{n,T}^{-1}, \alpha_T^{-1/2},\sqrt n) \max_{1\le i\le n} \l\Vert s_i'(\wh{A}_n -A_n)\r\Vert = \mathcal O_P(1)$.
\een

\end{lemma}

\begin{proof}
First of all
\begin{align}
 \frac 1n \l\Vert \wh{A}_n -A_n\r\Vert^2_F&= \frac 1n \sum_{h=1}^m \l\Vert \wh{A}^{(h)} -A^{(h)}\r\Vert^2_F\le \frac{m}{n}\max_{1\le h\le m} \l\Vert \wh{A}^{(h)} -A^{(h)}\r\Vert^2_F=(q+1) \max_{1\le h\le m} \l\Vert \wh{A}^{(h)} -A^{(h)}\r\Vert^2_F.\nn
\end{align}
Consider the decomposition
\begin{align}
\l\Vert \wh{A}^{(h)} -A^{(h)}\r\Vert_F=&\,\l\Vert \wh B_\chi^{(h)}(\wh C_\chi^{(h)})^{-1} - B_\chi^{(h)}( C_\chi^{(h)})^{-1}
\r\Vert_F\nn\\
\le&\, \l\Vert \wh B_\chi^{(h)} - B_\chi^{(h)}
\r\Vert_F \l\Vert  (C_\chi^{(h)})^{-1}\r\Vert_F
+\l\Vert (\wh C_\chi^{(h)})^{-1} - ( C_\chi^{(h)})^{-1}
\r\Vert_F \l\Vert B_\chi^{(h)}\r\Vert_F \nn\\
&+\l\Vert
\wh B_\chi^{(h)} - B_\chi^{(h)}
\r\Vert_F
\l\Vert
 (\wh C_\chi^{(h)})^{-1} - ( C_\chi^{(h)})^{-1}
\r\Vert_F.
\nn
\end{align}
Then, since
\begin{align}
 \l\Vert \wh B_\chi^{(h)} - B_\chi^{(h)}
\r\Vert_F^2 =&\, \sum_{\ell=1}^{\bar p}\sum_{i,i'=1}^{q+1} \l([\wh\Gamma_\chi^{(h)}(\ell)]_{ii'}-[\Gamma_\chi^{(h)}(\ell)]_{ii'}\r)^2\le  \bar p (q+1)^2\max_{1\le \ell\le \bar p}\max_{1\le i,i\le q+1} \l([\wh\Gamma_\chi^{(h)}(\ell)]_{ii'}-[\Gamma_\chi^{(h)}(\ell)]_{ii'}\r)^2.\nn
\end{align}
by Lemma \ref{lem:IFTcov}(ii)
\[
\min(\beta_{n,T}^{-1}, \alpha_T^{-1/2}, \sqrt n) \max_{1\le h\le m} \l\Vert \wh B_\chi^{(h)} - B_\chi^{(h)}
\r\Vert_F = \mathcal O_P(1).
\]
Similarly,
\begin{align}
\min(\beta_{n,T}^{-1}, \alpha_T^{-1/2},\sqrt n)  \max_{1\le h\le m} \l\Vert \wh C_\chi^{(h)} - C_\chi^{(h)}
\r\Vert_F =\mathcal O_P(1).\nn
\end{align}
Last, notice that, by Assumption A\ref{ass:newGDFM} 
\begin{align}
\sum_{\ell=1}^{\bar p}\sum_{i,i'=1}^{q+1} \l\vert [\Gamma_\chi^{(h)}(\ell)]_{ii'}\r\vert \le&\, \bar p (q+1) \max_{1\le \ell\le \bar p} \max_{1\le i,i'\le q+1} \sum_{j,j'=0}^\infty\sum_{h,h'=1}^{q+1} \l\vert K_{i,h}(j)K_{i',h'}(j')\E\l[\varepsilon_{t-\ell-j,h}\varepsilon_{t-j',h'}\r]\r\vert \nn\\
\le&\, \bar p (q+1) \max_{1\le \ell\le \bar p} \max_{1\le i,i'\le q+1} \sum_{j=0}^\infty\sum_{h=1}^{q+1} \l\vert K_{i,h}(j)\r\vert \,\l\vert K_{i',h}(j+\ell)\r\vert \E\l[\varepsilon_{t-\ell-j,h}^2\r]\nn\\
\le&\, \bar p (q+1) \max_{1\le \ell\le \bar p} \max_{1\le i,i'\le q+1} \sum_{j=0}^\infty\sum_{h=1}^{q+1} A_{i,h}^\chi A_{i',h}^\chi 
(\rho^\chi)^{2j} (\rho^\chi)^{\ell}\nn\\
\le&\, \frac{\bar p (q+1)(A^\chi)^2 }{1-(\rho^\chi)^{2}}. \nn
\end{align}
Therefore,
\[
\l\Vert B_\chi^{(h)}\r\Vert_F \le\sum_{\ell=1}^{\bar p}\sum_{i,i'=1}^{q+1} \l\vert [\Gamma_\chi^{(h)}(\ell)]_{ii'}\r\vert = \mathcal O(1).
\]
And similarly, we have $\Vert C_\chi^{(h)}\Vert_F=\mathcal O(1)$ and $\Vert (C_\chi^{(h)})^{-1}\Vert_F=\mathcal O(1)$ by Assumption A\ref{ass:ARGDFM}, which, since the entries of $(C_\chi^{(h)})^{-1}$ are rational functions of the entries of $C_\chi^{(h)}$, implies
 \begin{align}
\min(\beta_{n,T}^{-1}, \alpha_T^{-1/2},\sqrt n) \l\Vert (\wh C_\chi^{(h)})^{-1} - (C_\chi^{(h)})^{-1}
\r\Vert_F =\mathcal O_P(1).\nn
\end{align}
For part (ii) we refer to \citet[Proposition B.2 of the supplementary material]{barigozzi2024fnets} which uses Lemma \ref{lem:IFTcov}(ii).
\end{proof}

\begin{lemma}\label{lem:phipsiz}
Under Assumptions A\ref{A: stat}, A\ref{ass:newGDFM}, A\ref{ass:diveval}, A\ref{ass:kernel}, A\ref{ass:ARGDFM}, as $n,T\to\infty$,
\ben
\item [(i)]
$
\min( \beta_{n,T}^{-1}, \alpha_T^{-1/2}, \sqrt n) n^{-1} \l\Vert \widehat{\Gamma}^n_{\widehat\psi}-{\Gamma}^n_{\psi}\r\Vert =\mathcal O_P(1);
$
\item [(ii)]
$
\min( \beta_{n,T}^{-1}, \alpha_T^{-1/2}, \sqrt n) \max_{1\le i\le n} n^{-1/2} \l\Vert s_i'\l\{\widehat{\Gamma}^n_{\widehat\psi}-{\Gamma}^n_{\psi}\r\}\r\Vert =\mathcal O_P(1);
$
\item [(iii)]
$
\l\Vert {\Gamma}^n_{\psi}-{\Gamma}^n_{\phi}\r\Vert \le B_z;
$
\item [(iv)]
$
\min( \beta_{n,T}^{-1}, \alpha_T^{-1/2}, \sqrt n) n^{-1} \l\Vert \widehat{\Gamma}^n_{\widehat\psi}-{\Gamma}^n_{\phi}\r\Vert =\mathcal O_P(1);
$
\item [(v)]
$
\min( \beta_{n,T}^{-1}, \alpha_T^{-1/2}, \sqrt n) \max_{1\le i\le n} n^{-1/2}\l\Vert s_i'\l\{\widehat{\Gamma}^n_{\widehat\psi}-{\Gamma}^n_{\phi}\r\}\r\Vert =\mathcal O_P(1);
$

\een
where $s_i$ is the $i$th element of the canonical basis of $\mathbb R^{n\times 1}$ and $B_z$ is defined in Lemma \ref{lem:idioz}.
\end{lemma}

\begin{proof}
For simplicity of notation and wlog hereafter we set $p_h=1$ for all $h=1,\ldots,m$. Then, we can write
\begin{align}
\psi_t^n &= (I_n- A_n L) y_t^n =  (I_n- A_n L) \chi_t^n +  (I_n- A_n L) \xi_t^n = \phi_t^n+z_t^n,\quad t=2,\ldots, T,\nn\\
\widehat \psi_t^n &= (I_n-\widehat A_n L) y_t^n ,\quad t=2,\ldots, T.
\end{align}
It follows that:
\[
{\Gamma}^n_{\psi} = \E\l[\psi_t^n\psi_t^{n'} \r] = \E\l[ (y_t^n- A_n y_{t-1}^n) (y_t^n- A_n y_{t-1}^n)'  \r] = \Gamma_y^n + A_n \Gamma_y^n A_n' - A_n\Gamma_y^{n}(-1)-\Gamma_y^{n}(1)A_n',
\]
and similarly,
\[
\wh {\Gamma}^n_{\wh \psi} = \wh \Gamma_y^n + \wh A_n\wh \Gamma_y^n\wh A_n' -\wh A_n\wh\Gamma_y^{n}(-1)-\wh\Gamma_y^{n}(1)\wh A_n'.
\]
Then, part (i) follows directly from Lemmas \ref{lem:covbasic}(ii) and \ref{lem:VARparam}(i) and using the same arguments used for Lemma \ref{lem:spectra}(i), while part (ii) follows from  Lemmas \ref{lem:covbasic}(ii) and \ref{lem:VARparam}(ii) and using the same arguments used for Lemma \ref{lem:spectra}(ii). See also \citet[Proposition F.22]{barigozzi2024fnets}.

For part (iii), because of Lemma \ref{lem:idioz}
$$
\l\Vert {\Gamma}^n_{\psi}-{\Gamma}^n_{\phi}\r\Vert \le\l \Vert \Gamma_z^n\r\Vert \le B_z.
$$

Parts (iv) and (v) follow from parts (i)-(iii) and (ii)-(iii), respectively and using the same arguments used for Lemma \ref{lem:spectra}(iv) and  \ref{lem:spectra}(v).
\end{proof}

\begin{lemma}\label{lem:rowevecPHI}
Let  $Q_n=Q_n(\Gamma_{\phi}^n)$ be the $q\times n$ matrix having as rows the normalized eigenvector of  $\Gamma_{\phi}^n=\E[\phi_t^n\phi_t^{n'}]$ corresponding to its $q$ largest eigenvalues.
Under Assumptions A\ref{A: stat}, A\ref{ass:ARGDFM}, as $n\to\infty$,
\[
\max_{1\le i\le n} \l\Vert
s_i'\sqrt n  Q_n'
\r\Vert=\mathcal O(1),
\]
where $s_i$ is the $i$th element of the canonical basis of $\mathbb R^{n\times 1}$.
\end{lemma}

\begin{proof}
First, notice that because of Assumption A\ref{ass:ARGDFM}, $\lim_{n\to\infty} n^{-1}R_n'R_n$ has as eigenvalues the eigenvalues of $\lim_{n\to\infty} n^{-1}\Gamma_{\phi}^n$ and these are all finite and positive since $\lim_{n\to\infty} n^{-1}R_n'R_n$ is finite and positive definite. Then the proof follows the same arguments used for Lemma \ref{lem:rowevec}. Indeed, the variances $[\Gamma_\chi^n]_{ii}$ are given by
\[
[\Gamma_\chi^n]_{ii} =\int_{-\pi}^\pi [f_\chi^n(\theta)]_{ii}\mathrm d\theta 
\]
and so are finite and positive for all $i$ because of \eqref{eq:finitespectra}.
\end{proof}

\begin{lemma}\label{lem:DKPHI}
Under Assumptions A\ref{A: stat}, A\ref{ass:newGDFM}, A\ref{ass:diveval}, A\ref{ass:kernel}, A\ref{ass:ARGDFM}, as $n,T\to\infty$, 
\ben
\item [(i)] $\min(\beta_{n,T}^{-1}, \alpha_T^{-1/2},\sqrt n) \l\Vert Q_n- \mathcal R \widehat Q_n \r\Vert=\mathcal O_P(1)$;
\item [(ii)] $\min(\beta_{n,T}^{-1}, \alpha_T^{-1/2},\sqrt n)\max_{1\le i\le n}  \l\Vert
s_i'\sqrt n\l(Q_n'\mathcal R- \widehat Q_n'\r)
\r\Vert=\mathcal O_P(1)$;
\een
where $\mathcal R$ is a $q \times q$ orthogonal matrix and
$s_i$ is the $i$th element of the canonical basis of $\mathbb R^{n\times 1}$.
\end{lemma}

\begin{proof}
The proof follows the same arguments used for Lemma \ref{lem:DK} based on Davis-Kahan theorem \citet[Theorem 2]{yu2015useful}, but using Lemma \ref{lem:rowevecPHI} jointly with either Lemma \ref{lem:phipsiz}(iv) for part (i) or Lemma \ref{lem:phipsiz}(v) for part (ii).
\end{proof}

\begin{lemma}\label{lem:evecevecPHI}
Under Assumptions A\ref{A: stat}, A\ref{ass:newGDFM}, A\ref{ass:diveval}, A\ref{ass:kernel}, A\ref{ass:ARGDFM}, as $n,T\to\infty$, 
\ben
\item [(i)] 
$\min(\beta_{n,T}^{-1}, \alpha_T^{-1/2},\sqrt n)
\l\Vert 
Q_n'Q_n-
 \widehat Q_n'\widehat Q_n
\r\Vert=\mathcal O_P(1);
$
\item [(ii)]
$\min(\beta_{n,T}^{-1}, \alpha_T^{-1/2},\sqrt n)
\max_{1\le i\le n}
\l\Vert
s_i'\sqrt n \l\{
Q_n'Q_n-
 \widehat Q_n'\widehat Q_n
 \r\}
\r\Vert=\mathcal O_P(1);
$
\een
where $s_i$ is the $i$th element of the canonical basis of $\mathbb R^{n\times 1}$.
\end{lemma}

\begin{proof}
The proof follows the same arguments used for Lemma \ref{lem:evecevec} but using Lemma \ref{lem:DKPHI}(i) for part (i) and \ref{lem:rowevecPHI} and \ref{lem:DKPHI}(ii) for part (ii).
\end{proof}

\begin{lemma}\label{lem:phihat}
Under Assumptions A\ref{A: stat}, A\ref{ass:newGDFM}, A\ref{ass:diveval}, A\ref{ass:kernel}, A\ref{ass:ARGDFM}, as $n,T\to\infty$, 
\ben
\item [(i)] for any fixed $t=\bar p,\ldots, T$,
$\min(\beta_{n,T}^{-1}, \alpha_T^{-1/2},\sqrt n)n^{-1/2}{\l\Vert \wh\phi_t^n -\phi_t^n\r\Vert} = \mathcal O_P(1)$;
\item [(ii)] for any fixed $t=\bar p,\ldots, T$ and $i=1,\ldots, n$,
$\min(\beta_{n,T}^{-1}, \alpha_T^{-1/2},\sqrt n) \l\Vert \wh\phi_{it} -\phi_{it}\r\Vert = \mathcal O_P(1).$
\een
\end{lemma}
\begin{proof}
 By definition we have $\phi_t^n = Q_n'Q_n \phi_t^n$, then
\beq\label{eq:psizz}
\frac 1{\sqrt n} \l\Vert \phi_t^n -Q_n'Q_n\psi_t^n\r\Vert = \frac 1{\sqrt n} \l\Vert Q_n'Q_n z_t^n\r\Vert \le \frac 1{\sqrt n}\l \Vert Q_n z_t^n\r\Vert = \mathcal O_{P}\l(\frac 1{\sqrt n}\r),
\eeq
since, because of Lemma \ref{lem:idioz} and by the multiplicative Weyl's inequality \citep[Theorem 7]{merikoski2004inequalities},
\begin{align}
\frac 1n \E\l[ \Vert Q_n z_t^n\Vert^2\r]&
=\frac 1n\E\l[z_t^{n\prime}Q_n'Q_n z_t^n\r]
=\frac 1n \mu_1\l(\Gamma_z^n Q_n'Q_n\r)
\le \frac 1n \mu_1\l(\Gamma_z^n\r) \mu_1\l(Q_n'Q_n\r)
=\frac 1n\mu_1\l(\Gamma_z^n\r)\le \frac{C_z}n,\nn
\end{align}
for some $C_z\in(0,\infty)$ independent of $n$, and where we also used the fact that $\mu_1(Q_n'Q_n) =\mu_1(Q_nQ_n')=\mu_1(I_q)=1$.
Then, from \eqref{eq:psizz},
\begin{align}\label{eq:phihatphi}
\frac 1{\sqrt n}\l\Vert\wh\phi_t^n -\phi_t^n\r\Vert
 &\le
\frac 1{\sqrt n} \l\Vert \wh Q_n'\wh Q_n \wh{\psi}_t^n -  Q_n' Q_n {\psi}_t^n\r\Vert + \mathcal O_{P}\l(\frac 1{\sqrt n}\r)\nn\\
&\le \l\Vert \wh Q_n'\wh Q_n-Q_n' Q_n\r\Vert \frac1 {\sqrt n}\l\Vert{\psi}_t^n\r\Vert + \l\Vert Q_n' Q_n\r\Vert \frac 1{\sqrt n}\l\Vert  \wh{\psi}_t^n- {\psi}_t^n\r\Vert+ \mathcal O_{P}\l(\frac 1{\sqrt n}\r).
\end{align}
For simplicity of notation and wlog hereafter we set $p_h=1$ for all $h=1,\ldots,m$.
Therefore,
\begin{align}
\frac{1 }{\sqrt n}\l\Vert \wh{\psi}_t^n- {\psi}_t^n\r\Vert&= 
\frac 1{\sqrt n} \l\Vert (\wh A_n -A_n) y_{t-1}^n\r\Vert \le  
 \sqrt{\frac 1{ n} \sum_{h=1}^m  \l\Vert (\wh A^{(h)} -A^{(h)}) y_{t-1}^{(h)}\r\Vert^2}
 \nn\\
 &\le  \sqrt{ \max_{1\le h\le m} \l\Vert (\wh A^{(h)} -A^{(h)}) \r\Vert^2_F \frac 1{ n} \sum_{h=1}^m\l\Vert y_{t-1}^{(h)}\r\Vert^2}\nn\\
  &\le  \max_{1\le h\le m} \l\Vert (\wh A^{(h)} -A^{(h)}) \r\Vert_F \sqrt{ \frac 1{ n} \sum_{h=1}^m\l\Vert y_{t-1}^{(h)}\r\Vert^2}\nn\\
  &=  \max_{1\le h\le m} \l\Vert (\wh A^{(h)} -A^{(h)}) \r\Vert_F  \frac 1{\sqrt{ n}} \l\Vert y_{t-1}^{n}\r\Vert\nn\\
    &=  \max_{1\le h\le m} \l\Vert (\wh A^{(h)} -A^{(h)}) \r\Vert_F \l\{ \frac 1{\sqrt{ n}} \l\Vert \chi_{t-1}^{n}\r\Vert+ \frac 1{\sqrt{ n}} \l\Vert \xi_{t-1}^{n}\r\Vert\r\}.\nn
\end{align}
And, from Lemma \ref{lem:VARparam}(ii), as $n,T\to\infty$
\beq\label{eq:AhatAhunif}
\min( \beta_{n,T}^{-1}, \alpha_T^{-1/2}, \sqrt n) \max_{1\le h\le m} \l\Vert \wh A^{(h)} -A^{(h)}\r\Vert_F^2=\mathcal O_P(1).
\eeq
Moreover, by Assumption \ref{ass:newGDFM},
\begin{align}
\E\l[\l\Vert \chi_{t-1}^{n}\r\Vert^2\r] &=
\E\l[\sum_{i=1}^{n} \chi_{i,t-1}^2\r]\nn\\
& = 
\E\l[ \sum_{i=1}^{n} \sum_{\ell,\ell'=0}^\infty\sum_{j,j'=1}^q K_{i,j}(\ell)K_{i,j'}(\ell') u_{j,t-1-\ell}u_{j',t-1-\ell'}\r]\nn\\
& = n\max_{1\le i \le n} \sum_{\ell,\ell'=0}^\infty\sum_{j,j'=1}^q K_{i,j}(\ell)K_{i,j'}(\ell') \E\l[u_{j,t-1-\ell}u_{j',t-1-\ell'} \r]\nn\\
& =n\max_{1\le i \le n} \sum_{\ell=0}^\infty\sum_{j=1}^q \l\vert K_{i,j}(\ell)\r\vert^2 \E\l[u_{j,t-1-\ell}^2 \r]\nn\\
&\le n\max_{1\le i \le n}   \sum_{j=1}^q(A_{i,j}^\chi)^2 \sum_{\ell=0}^\infty (\rho^\chi)^{2\ell}\nn\\
&\le n\max_{1\le i \le n}  \sum_{j=1}^q(A_{i,j}^\chi)^2 \frac1{1- (\rho^\chi)^{2}}\nn\\
&\le  \frac{n (A_\chi)^2}{1- (\rho^\chi)^{2}},\label{eq:maxchi2}
\end{align}
and similarly
\begin{align}
\E\l[\l\Vert \xi_{t-1}^{n}\r\Vert^2\r]&\le  \frac{n (A_\xi)^2}{1- (\rho^\xi)^{2}}.\nn
\end{align}
Thus, by \eqref{eq:maxchi2}, Chebychev's inequality and \eqref{eq:AhatAhunif}, as $n,T\to\infty$, we have
\[
\min( \beta_{n,T}^{-1}, \alpha_T^{-1/2},  \sqrt n)\frac 1{\sqrt n}\l\Vert \wh{\psi}_t^n- {\psi}_t^n\r\Vert
\]
and using also Lemma \ref{lem:evecevecPHI}(i) in \eqref{eq:phihatphi} we prove  part (i).

The proof of part (ii) is the same but using  Lemma \ref{lem:evecevecPHI}(ii) instead of  Lemma \ref{lem:evecevecPHI}(i).
\end{proof}

\begin{lemma}\label{lem:Ycovmat}
Under Assumptions A\ref{A: stat}, A\ref{ass:newGDFM}, A\ref{ass:divevalC}, for all $n,T\in\mathbb N$,
\ben
\item [(i)] 
$
n^{-2}\E\l[ \l\Vert \widehat{\Gamma}^n_{y}-{\Gamma}^n_{y}\r\Vert^2\r] \le C^* T^{-1},
$;
\item [(ii)]
$
\max_{1\le i\le n} n^{-1} \E\l[\l\Vert s_i'\l\{\widehat{\Gamma}^n_{y}-{\Gamma}^n_{y}\r\}\r\Vert^2\r] \le C^* T^{-1},
$;
\item [(iii)]
$
\l\Vert {\Gamma}^n_{y}-{\Gamma}^n_{C}\r\Vert \le C_e ;
$
\een
where $s_i$ is the $i$th element of the canonical basis of $\mathbb R^{n\times 1}$,  $C^*$ is defined in Lemma \ref{lem:covbasic}(i), and $C_e$ is defined in Assumption A\ref{ass:divevalC}.
Moreover, for all $n,T\in\mathbb N$ there exists $C_1^*\in(0,\infty)$ independent of $n$ and $T$, such that
\ben
\item [(iv)]
$
n^{-2} \E\l[\l\Vert \widehat{\Gamma}^n_{y}-{\Gamma}^n_{C}\r\Vert^2\r] \le C_1^*\max(T^{-1}, n^{-2});
$
\item [(v)]
$
\max_{1\le i\le n} n^{-1}\E\l[\l\Vert s_i'\l\{\widehat{\Gamma}^n_{y}-{\Gamma}^n_{C}\r\}\r\Vert^2\r] =C_1^*\max(T^{-1}, n^{-1}).
$
\een
\end{lemma}

\begin{proof}
The proof follows directly from Lemma \ref{lem:covbasic}(i) and Assumption A\ref{ass:divevalC} and using the same arguments used for Lemma \ref{lem:spectra}.
\end{proof}

\begin{lemma}\label{lem:rowevecC}
Under Assumptions A\ref{A: stat}, A\ref{ass:newGDFM}, A\ref{ass:divevalC}, as $n\to\infty$,
\[
\max_{1\le i\le n} \l\Vert
s_i'\sqrt n  \Omega_n'
\r\Vert=\mathcal O(1),
\]
where $s_i$ is the $i$th element of the canonical basis of $\mathbb R^{n\times 1}$ and $\Omega_n$ is the $r\times n$ matrix having as rows the normalized eigenvectors of $\Gamma_C^n$ corresponding to its $r$ non-zero eigenvalues.
\end{lemma}

\begin{proof}
First, notice that because of Assumption A\ref{ass:divevalC}, $\lim_{n\to\infty} n^{-1}\Lambda_n'\Lambda_n$ has as eigenvalues the eigenvalues of $\lim_{n\to\infty} n^{-1}\Gamma_{C}^n$ and these are all finite and positive since $\lim_{n\to\infty} n^{-1}\Lambda_n'\Lambda_n$ is finite and positive definite. Then the proof follows the same arguments used for Lemma \ref{lem:rowevec}. Indeed, 
%
%
%
by Assumption A\ref{ass:divevalC}
\begin{align}
[\Gamma_C^n]_{ii}&=\E [\chi_{it}^2] = \Lambda_i\Lambda_i' =\l\Vert \Lambda_i\Lambda_i'\r\Vert \le C_\Lambda^2<\infty,\nn
\end{align}
moreover, $[\Gamma_C^n]_{ii}\ge 0$.
\end{proof}

\begin{lemma}\label{lem:DKC}
Under Assumptions A\ref{A: stat}, A\ref{ass:newGDFM}, A\ref{ass:divevalC}, as $n,T\to\infty$, 
\ben
\item [(i)] $\min(T,n^2) \E\l[\l\Vert \Omega_n- \mathcal R \widehat \Pi_n\r\Vert^2\r]=\mathcal O(1)$;
\item [(ii)] $\min(T,n)\max_{1\le i\le n}  \E\l[ \l\Vert
s_i'\sqrt n\l(\Omega_n'\mathcal R- \widehat \Pi_n'\r)
\r\Vert^2\r]=\mathcal O(1)$;
\een
where $\mathcal R$ is a $r \times r$ orthogonal matrix and
$s_i$ is the $i$th element of the canonical basis of $\mathbb R^{n\times 1}$.
\end{lemma}

\begin{proof}
The proof follows the same arguments used for Lemma \ref{lem:DK} based on Davis-Kahan theorem \citet[Theorem 2]{yu2015useful}, but using Lemma \ref{lem:Ycovmat}(iv) for part (i) and \ref{lem:Ycovmat}(v) for part (ii).
\end{proof}

\begin{lemma}\label{lem:evecevecC}
Under Assumptions A\ref{A: stat}, A\ref{ass:newGDFM}, A\ref{ass:divevalC}, as $n,T\to\infty$, 
\ben
\item [(i)] 
$\min(T,n^2)
\E\l[
\l\Vert \Omega_n'\Omega_n-
 \widehat \Pi_n' \widehat \Pi_n
\r\Vert^2
\r]=\mathcal O(1);
$
\item [(ii)]
$\min(T,n)
\max_{1\le i\le n}
\E\l[
\l\Vert
s_i'\sqrt n \l\{\Omega_n'\Omega_n-
 \widehat \Pi_n' \widehat \Pi_n\r\}
\r\Vert^2
\r]=\mathcal O(1);
$
\een
where $s_i$ is the $i$th element of the canonical basis of $\mathbb R^{n\times 1}$.
\end{lemma}

\begin{proof}
The proof follows the same arguments used for Lemma \ref{lem:evecevec} but using Lemma \ref{lem:DKC}(i) for part (i) and \ref{lem:rowevecC} and \ref{lem:DKC}(ii) for part (ii).
\end{proof}

\begin{lemma}\label{lem:CHIcovmat}
Under Assumptions A\ref{A: stat}, A\ref{ass:newGDFM}, A\ref{ass:diveval}, A\ref{ass:kernel},   A\ref{ass:divevalC}, as $n,T\to\infty$,
\ben
\item [(i)] 
$
\min(\alpha_T^{-1}, n) n^{-2}\E\l[ \l\Vert \widehat{\Gamma}^n_{\chi}-{\Gamma}^n_{\chi}\r\Vert^2\r] = \mathcal O(1);
$
\item [(ii)]
$
\min(\alpha_T^{-1},  n) \max_{1\le i\le n} n^{-1} \E\l[\l\Vert s_i'\l\{\widehat{\Gamma}^n_{\chi}-{\Gamma}^n_{\chi}\r\}\r\Vert^2\r] = \mathcal O(1);
$
\een
where $s_i$ is the $i$th element of the canonical basis of $\mathbb R^{n\times 1}$. Moreover, for all $n\in\mathbb N$ there exists a $C'\in(0,\infty)$ independent of $n$, such that
\ben
\item [(iii)]
$
\l\Vert {\Gamma}^n_{\chi}-{\Gamma}^n_{C}\r\Vert \le C';
$
\een
and, as $n,T\to\infty$,
\ben
\item [(iv)]
$
\min(\alpha_T^{-1}, n) n^{-2}\E\l[ \l\Vert \widehat{\Gamma}^n_{\chi}-{\Gamma}^n_{C}\r\Vert^2\r] = \mathcal O(1);
$
\item [(v)]
$
\min(\alpha_T^{-1},  n) \max_{1\le i\le n} n^{-1} \E\l[\l\Vert s_i'\l\{\widehat{\Gamma}^n_{\chi}-{\Gamma}^n_{C}\r\}\r\Vert^2\r] = \mathcal O(1).
$
\een
\end{lemma}

\begin{proof}
The proof follows directly from Lemma \ref{lem:IFTcov}(i) and Assumption A\ref{ass:divevalC} and using the same arguments used for Lemma \ref{lem:spectra}. In particular, for part (iii), we have
\begin{align}
n^{-1}
\l\Vert
{\Gamma}^n_\chi-{\Gamma}^n_C
\r\Vert
&= n^{-1}
\l\Vert{\Gamma}^n_{e^\chi}
\r\Vert= n^{-1}
\l\Vert{\Gamma}^n_{e}-{\Gamma}^n_{\xi}
\r\Vert\le  n^{-1}
\l\Vert{\Gamma}^n_{e}\r\Vert + n^{-1}
\l\Vert{\Gamma}^n_{\xi}\r\Vert\le n^{-1} C_e + n^{-1} 2\pi C_\xi,\nn
\end{align}
because of  Assumption A\ref{ass:divevalC}, Lemma \ref{lem:Cxi}, with $C_e$ and $C_\xi$ independent of $n$, so that $C'=C_e+2\pi C_\xi$, and since
\begin{align}
\l\Vert{\Gamma}^n_{\xi}\r\Vert =\l\Vert \int_{-\pi}^\pi f^n_\xi(\theta)\mathrm d\theta
\r\Vert\le  \int_{-\pi}^\pi \l\Vert f^n_\xi(\theta)\r\Vert\mathrm d\theta \le 2\pi \sup_{\theta\in[-\pi,\pi]} \mu_1(f_\xi^n(\theta)).\nn
\end{align}
\end{proof}

\begin{lemma}\label{lem:DKCHI}
Under Assumptions A\ref{A: stat}, A\ref{ass:newGDFM}, A\ref{ass:diveval}, A\ref{ass:kernel},  A\ref{ass:divevalC}, as $n,T\to\infty$,
\ben
\item [(i)] $\min(\alpha_T^{-1},n) \E\l[\l\Vert \Omega_n- \mathcal R \widehat P_n\r\Vert^2\r]=\mathcal O(1)$;
\item [(ii)] $\min(\alpha_T^{-1}, n)\max_{1\le i\le n}  \E\l[ \l\Vert
s_i'\sqrt n\l(\Omega_n'\mathcal R- \widehat P_n'\r)
\r\Vert^2\r]=\mathcal O(1)$;
\een
where $\mathcal R$ is a $r \times r$ orthogonal matrix and
$s_i$ is the $i$th element of the canonical basis of $\mathbb R^n{\times 1}$.
\end{lemma}

\begin{proof}
The proof follows the same arguments used for Lemma \ref{lem:DK} based on Davis-Kahan theorem \citet[Theorem 2]{yu2015useful}, but using Lemma \ref{lem:CHIcovmat}(iv) for part (i) and  \ref{lem:CHIcovmat}(v) for part (ii).
\end{proof}

\begin{lemma}\label{lem:evecevecCHI}
Under Assumptions A\ref{A: stat}, A\ref{ass:newGDFM}, A\ref{ass:diveval}, A\ref{ass:kernel},  A\ref{ass:divevalC}, as $n,T\to\infty$,
\ben
\item [(i)] 
$\min(\alpha_T^{-1},n)
\E\l[
\l\Vert \Omega_n'\Omega_n-
 \widehat P_n' \widehat P_n
\r\Vert^2
\r]=\mathcal O(1);
$
\item [(ii)]
$\min(\alpha_T^{-1},n)
\max_{1\le i\le n}
\E\l[
\l\Vert
s_i'\sqrt n \l\{\Omega_n'\Omega_n-
 \widehat P_n' \widehat P_n\r\}
\r\Vert^2
\r]=\mathcal O(1);
$
\een
where $s_i$ is the $i$th element of the canonical basis of $\mathbb R^{n\times 1}$.
\end{lemma}

\begin{proof}
The proof follows the same arguments used for Lemma \ref{lem:evecevec} but using Lemma \ref{lem:DKCHI}(i) for part (i) and \ref{lem:rowevecC} and \ref{lem:DKCHI}(ii) for part (ii).
\end{proof}

\subsection{Proof of Proposition \ref{prop1}}\label{prop1proof}

\paragraph{Part I.a.} Recall that $g_{n}(\theta) = V_n^* (\theta)V_n(\theta)$,   $\theta\in[-\pi,\pi]$, has Fourier coefficients
\[
G_{n}(\ell)= \int_{-\pi}^\pi e^{\iota \ell \theta} g_{n}(\theta), \quad \ell\in\mathbb Z.
\]
It follows that,
\[
\chi_t^n = \sum_{\ell=-\infty}^\infty G_{n}(\ell) \chi_{t-\ell}^n =\sum_{\ell=-\infty}^\infty G_{n}(\ell) y_{t-\ell}^n,
\]
because of Assumption A\ref{ass:newGDFM} of orthogonality at all leads and lags implying $\E [\chi_{it} \xi_{\ell s}]=0$ for all $i,\ell \in\mathbb N$ and $t,s\in\mathbb Z$.
Then, let 
 $
 \mathcal D_T=\{\ell\in\mathbb Z : \ell < - \mathcal M_T\text { or }  > \mathcal M_T\},
$
 and for any $\mathcal M_T\le t\le T-\mathcal M_T$,
 consider the decomposition
 \begin{align}
  \wh{\chi}_t^n -\chi_t^n &=\sum_{\ell=-\mathcal M_T}^{\mathcal M_T} \l(\widehat D_n(\ell) -G_{n}(\ell)\r)y_{t-\ell}^n + \sum_{\ell\in \mathcal D_T} G_{n}(\ell) y_{t-\ell}^n= a_t +\widetilde a_t, \;\text{say.}\nn
 \end{align} 
 Start with $a_t$. We have,
 \begin{align}
\Vert a_t\Vert &\le \l\Vert \sum_{\ell=-\mathcal M_T}^{\mathcal M_T} \l \{ \widehat D_n(\ell) -G_{n}(\ell)\r\}\chi_{t-\ell}^n\r\Vert + 
 \l\Vert \sum_{\ell=-\mathcal M_T}^{\mathcal M_T} \l \{ \widehat D_n(\ell) -G_{n}(\ell)\r\}\xi_{t-\ell}^n\r\Vert= a_{1t} + a_{2t}, \;\text{say.}\nn
 \end{align}
 Then, by Cauchy-Schwarz inequality
 \beq
 \frac {\E[a_{1t}]}{\sqrt n}\le  \sum_{\ell=-\mathcal M_T}^{\mathcal M_T}\E\l[\l\Vert \l \{ \widehat D_n(\ell) -G_{n}(\ell)\r\}\frac{\chi_{t-\ell}^n}{\sqrt n}\r\Vert\r]\le \sum_{\ell=-\mathcal M_T}^{\mathcal M_T}
 \l\{
 \E\l[
 \l\Vert  
 \widehat D_n(\ell) -G_{n}(\ell)
 \r\Vert^2
  \r]
 \r\}^{1/2} 
 \l\{
 \E\l[
 \frac{\Vert\chi_{t-\ell}^n\Vert^2}{n}
 \r]
 \r\}^{1/2},\label{eq:a1t}
 \eeq
and
\beq
 \frac {\E[a_{2t}]}{\sqrt n}\le  \sum_{\ell=-\mathcal M_T}^{\mathcal M_T}\l\Vert \l \{ \widehat D_n(\ell) -G_{n}(\ell)\r\}\frac{\xi_{t-\ell}^n}{\sqrt n}\r\Vert\le \sum_{\ell=-\mathcal M_T}^{\mathcal M_T}
 \l\{
 \E\l[
 \l\Vert  
 \widehat D_n(\ell) -G_{n}(\ell)
 \r\Vert^2
  \r]
 \r\}^{1/2} 
 \l\{
 \E\l[
 \frac{\Vert\xi_{t-\ell}^n\Vert^2}{n}
 \r]
 \r\}^{1/2}.\label{eq:a2t}
 \eeq
 Now, 
 \begin{align}
 \Vert \widehat D_n(\ell) -G_{n}(\ell)\Vert =&\, \l\Vert  \frac {2\pi}{2B_T+1}\sum_{ h=-\mathcal B_T}^{\mathcal B_T} e^{\iota \ell \theta_h} \wh d_{n}(\theta_h) - \int_{-\pi}^\pi e^{\iota \ell \theta}g_{n}(\theta)\mathrm d\theta  \r\Vert\nn\\
\le&\, \frac {2\pi}{2B_T+1}\sum_{ h=-\mathcal B_T}^{\mathcal B_T} \l\Vert e^{\iota \ell \theta_h}\wh d_{n}(\theta_h) - e^{\iota \ell \theta_h}g_{n}(\theta_h)\r\Vert\nn\\
&+ \frac {2\pi}{2B_T+1}\l\Vert \sum_{ h=-\mathcal B_T}^{\mathcal B_T} 
e^{\iota \ell \theta_h}g_{n}(\theta_h)-\int_{-\pi}^\pi
e^{\iota \ell \theta}g_{n}(\theta)\mathrm d\theta\r\Vert \nn\\
\le&\, \frac {2\pi}{2B_T+1}\sum_{ h=-\mathcal B_T}^{\mathcal B_T} \l\Vert\wh d_{n}(\theta_h) - g_{n}(\theta_h)\r\Vert\, \vert e^{\iota \ell \theta_h}\vert\nn\\
&+ \frac {2\pi}{2B_T+1}\sum_{ h=-\mathcal B_T}^{\mathcal B_T} \max_{\theta_{h-1}\le \theta\le \theta_h}\l\Vert
e^{\iota \ell \theta_h}g_{n}(\theta_h)
-
e^{\iota \ell \theta}g_{n}(\theta)
\r\Vert\nn\\
\le&\, \frac {2\pi}{2B_T+1}\sum_{ h=-\mathcal B_T}^{\mathcal B_T} \l\Vert\wh d_{n}(\theta_h) - g_{n}(\theta_h)\r\Vert \nn\\
&+ \frac {2\pi}{2B_T+1}\sum_{ h=-\mathcal B_T}^{\mathcal B_T} 
\max_{\theta_{h-1}\le \theta\le \theta_h}\l\Vert 
g_{n}(\theta_h)-
g_{n}(\theta)
\r\Vert\nn\\
&+ \frac {2\pi}{2B_T+1}\sum_{ h=-\mathcal B_T}^{\mathcal B_T} \max_{\theta_{h-1}\le \theta\le \theta_h}\l\vert
e^{\iota \ell \theta_h}
-
e^{\iota \ell \theta}
\r\vert.\nn
 \end{align}
 Therefore,  by the $C_r$-inequality with $r=3$
 \begin{align}
\l \Vert \widehat D_n(\ell) -G_{n}(\ell)\r\Vert^2 \le&\, 3  \l\{ \frac {2\pi}{2B_T+1}\sum_{ h=-\mathcal B_T}^{\mathcal B_T} \l\Vert\wh d_{n}(\theta_h) - g_{n}(\theta_h)\r\Vert\r\}^2\nn\\
 &+3\l\{
  \frac {2\pi}{2B_T+1}\sum_{ h=-\mathcal B_T}^{\mathcal B_T} 
\max_{\theta_{h-1}\le \theta\le \theta_h}\l\Vert 
g_{n}(\theta_h)-
g_{n}(\theta)
\r\Vert
 \r\}^2\nn\\
 &+3\l\{
 \frac {2\pi}{2B_T+1}\sum_{ h=-\mathcal B_T}^{\mathcal B_T} \max_{\theta_{h-1}\le \theta\le \theta_h}\l\vert
e^{\iota \ell \theta_h}
-
e^{\iota \ell \theta}
\r\vert
 \r\}^2\nn\\
 =&\, I+II+III, \;\text{say.}
  \end{align}
  Then, by Cauchy-Schwarz inequality
  \begin{align}
  I&\le \frac{3(2\pi)^2}{2B_T+1}\sum_{ h=-\mathcal B_T}^{\mathcal B_T} \l\Vert\wh d_{n}(\theta_h) - g_{n}(\theta_h)\r\Vert^2\le 3(2\pi)^2\max_{\vert h\vert\le \mathcal B_T } \l\Vert\wh d_{n}(\theta_h) - g_{n}(\theta_h)\r\Vert^2,\nn
  \end{align}
  so, by Lemma \ref{lem:evecevec}(i), as $n,T\to\infty$,
  \[
 \min(\alpha_T^{-1},n^2) \E[I] = \min(\alpha_T^{-1},n^2)
\E\l[\max_{\vert h\vert \le \mathcal B_T}
\l\Vert V_n^*(\theta_h)V_n(\theta_h)-
 \widehat W_n^* (\theta_h)\widehat W_n(\theta_h)
\r\Vert^2
\r]=\mathcal O(1).
  \]
  Moreover, the function $\theta\mapsto g_{n}(\theta)$ is of bounded variation since $\theta\mapsto f_\chi^n(\theta)$ is of bounded variation (see \citealp[Proposition 2]{forni2017dynamic}) and eigenvectors are continuous functions of the spectral density matrix. The function $\theta\mapsto g_{n}(\theta)$ and $\theta\mapsto e^{\iota \ell \theta}$ is also of bounded variation.
  Hence $II=O(\mathcal B_T^{-2})$ and $III=O(\mathcal B_T^{-2})$ and are dominated when using the Bartlett kernel for which $\vartheta=1$. It follows that, as $n,T\to\infty$,
  \beq
 \min(\alpha_T^{-1},n^2)  \max_{\vert \ell\vert\le \mathcal M_T}  \E\l[\l\Vert \widehat D_n(\ell) -G_{n}(\ell)\r\Vert^2\r] =\mathcal O(1).\label{eq:DnIFT}
  \eeq
Furthermore, by Assumption A\ref{ass:newGDFM}
\begin{align}
\frac 1n \E \l[\l\Vert
\chi_{t-\ell}^n\r\Vert^2
\r] &\le \frac 1n \E \l[\l\Vert
\sum_{j=0}^\infty 
K_n(j) \varepsilon_{t-\ell-j}
\r\Vert^2_F
\r]  \nn\\
&=\frac 1n \sum_{j,j'=0}^\infty \text{trace}\l\{
K_n(j)
\E\l[
\varepsilon_{t-\ell-j}
\varepsilon_{t-\ell-j'}'\r]
[K_n(j')]'
\r\}
=\frac 1n \sum_{j=0}^\infty \text{trace}\l\{
K_n(j)
[K_n(j)]'
\r\}\nn\\
&=\frac 1n\sum_{j=0}^\infty\sum_{\ell=1}^q\sum_{i=1}^n \vert K_{i,\ell}(j)\vert^2\le \frac 1n\sum_{j=0}^\infty\sum_{\ell=1}^q\sum_{i=1}^n (A_{i,\ell}^\chi)^2 (\rho^\chi)^{2j}\nn\\
&\le q \max_{1\le i\le n}\max_{1\le \ell\le q} (A_{i,\ell}^\chi)^2 \sum_{j=0}^\infty(\rho^\chi)^{2j}\le  \frac{q(A^\chi)^2}{1-(\rho^\chi)^2}.\label{eq:chibound}
\end{align}
Substituting \eqref{eq:chibound} and \eqref{eq:DnIFT} into \eqref{eq:a1t}, we have
\[
 \min(\alpha_T^{-1/2},n)  \frac {\E[a_{1t}]}{\mathcal M_T \sqrt n} = \mathcal O(1).
\]
Following exactly the same reasoning as the one leading to \eqref{eq:chibound} we also have
\[
\frac 1n \E \l[\l\Vert
\xi_{t-\ell}^n\r\Vert^2
\r] \le \frac{(A^\xi)^2}{1-(\rho^\xi)^2},
\]
which once substituted into \eqref{eq:a1t}, together with \eqref{eq:DnIFT} implies
\[
 \min(\alpha_T^{-1/2},n)  \frac {\E[a_{2t}]}{\mathcal M_T \sqrt n} = \mathcal O(1).
\]
Last,
\[
\Vert \widetilde a_t\Vert \le \l\Vert \sum_{\ell\in \mathcal D_T} G_{n}(\ell)\chi_{t-\ell}^n\r\Vert + 
 \l\Vert \sum_{\ell\in \mathcal D_T} G_{n}(\ell) \xi_{t-\ell}^n\r\Vert= \widetilde a_{1t} +\widetilde a_{2t}, \;\text{say.}\nn
\]
Then,  there exists a $C_5\in(0,\infty)$,  such that
\begin{align}
 \frac {\E[\widetilde a_{1t}]}{\sqrt n}\le  \sum_{\ell\in \mathcal D_T}\l\Vert G_{n}(\ell) \frac{\chi_{t-\ell}^n}{\sqrt n}\r\Vert&\le 
 \sum_{\ell\in \mathcal D_T}
 \l\Vert  
G_{n}(\ell)
 \r\Vert
 \l\{
 \E\l[
 \frac{\Vert\chi_{t-\ell}^n\Vert^2}{n}
 \r]
 \r\}^{1/2}\nn\\
 &\le   \mathcal C
 \l\Vert  
G_{n}(0)
 \r\Vert
 \sum_{\ell\in \mathcal D_T} 
 \frac 1{(1+\varphi)^{  |\ell| }}
 \l\{
\frac{q(A^\chi)^2}{1-(\rho^\chi)^2}
 \r\}^{1/2}\nn\\
  &\le 
C_5 
\l\{\sum_{\ell=-\infty}^{\mathcal M_T} 
 \frac 1{(1+\varphi)^{ |\ell| }}+\sum_{\ell=T-\mathcal M_T}^{\infty} 
 \frac 1{(1+\varphi)^{ \ell }}\r\}
\nn\\
  &\le  
 \frac{ 2C_5(1+\varphi)^{ 1- \mathcal M_T }}{\varphi},\nn
 \end{align}
 because of \eqref{eq:chibound}, Assumption A\ref{eq:decay}, and since 
 \[
  \l\Vert  
G_{n}(0)
 \r\Vert
 \le  \int_{-\pi}^\pi \Vert g_{n}(\theta)\Vert \le 
  \int_{-\pi}^\pi \Vert
 V_n^* (\theta)\Vert^2\le 2\pi,
 \] 
 because eigenvectors are normalized. The same reasoning holds for $\widetilde a_{2t}$.
 Therefore, as $n,T\to\infty$
 \begin{align}
(1+\varphi)^{\mathcal M_T} \frac{ \E\l[\l\Vert
\widetilde a_{t}
 \r\Vert
 \r]}{
\sqrt n} =\mathcal O(1).\nn
 \end{align}
It follows that, by Assumption A\ref{ass:kernel}, as $n,T\to\infty$,
\begin{align}
 \min(\alpha_T^{-1/2}\mathcal M_T^{-1},nM_T^{-1},(1+\varphi)^{\mathcal M_T}) \frac{ \E\l[\l\Vert
  \wh{\chi}_t^n -\chi_t^n
 \r\Vert 
 \r]}{\sqrt n}&= \mathcal O(1).\nn
 \end{align}
 and since by Assumption A\ref{ass:kernel} $\mathcal M_T\asymp\log T$, then, $(1+\varphi)^{\mathcal M_T}\asymp T^{\log (1+\varphi)}$ and
 \[
 \min(\alpha_T^{-1/2}\mathcal M_T^{-1},n\mathcal M_T^{-1}, (1+\varphi)^{\mathcal M_T})\asymp \min(\alpha_T^{-1/2},n)(\log T)^{-1}.
 \]
The proof of part (i) follows from Markov's inequality.

For part (ii) the proof is the same but we need to use Lemma \ref{lem:evecevec}(ii) instead of Lemma \ref{lem:evecevec}(i).



\paragraph{Part I.b.}  For simplicity of notation and wlog hereafter we set $p_h=1$ for all $h=1,\ldots,m$.
 By definition 
\[
\chi_t^n = (I_n- A_nL)^{-1}  {\phi}_t^n 
\]
and
\[
\wh\chi_t^n = (I_n-\wh A_nL)^{-1} \wh Q_n'\wh Q_n \wh{\psi}_t^n = (I_n-\wh A_nL)^{-1}\wh{\phi}_t^n \nn\\
\]
Let 
 $
 \mathcal D_T=\{\ell\in\mathbb Z : \ell  > \mathcal K_T\},
$
 and consider the decomposition for $\mathcal K_T \le t\le T$
 \begin{align}
  \wh{\chi}_t^n -\chi_t^n &=\sum_{\ell=0}^{\mathcal K_T} \l(\wh A_n^\ell \widehat \phi_{t-\ell}^n - A_n^{\ell} \phi_{t-\ell}^n\r) + \sum_{\ell\in \mathcal D_T} A_n^{\ell} \phi_{t-\ell}^n\nn\\
  &= b_t +\widetilde b_t, \;\text{say.}\nn
 \end{align} 
  Start with $b_t$. We have,
 \begin{align}
 b_t &= \sum_{\ell=0}^{\mathcal K_T}  \l \{ \wh A_n^{\ell}- A_n^{\ell} \r\}\phi_{t-\ell}^{n} + 
\sum_{\ell=0}^{\mathcal K_T}   A_n^{\ell} \l\{\wh\phi_{t-\ell}^{n}-\phi_{t-\ell}^{n}\r\} 
+ \sum_{\ell=0}^{\mathcal K_T}    \l \{ \wh A_n^{\ell} - A_n^{\ell} \r\} \l\{\wh\phi_{t-\ell}^{n}-\phi_{t-\ell}^{n}\r\} \nn\\
&= b_{1t} + b_{2t}+ b_{3t}, \;\text{say.}\nn
 \end{align}
 Now, 
 \begin{align}
 \frac 1{\sqrt n}\Vert b_{1t}\Vert &\le \sum_{\ell=0}^{\mathcal K_T}   \frac 1{\sqrt n} \l\Vert \l \{ \wh A_n^{\ell}- A_n^{\ell} \r\}\phi_{t-\ell}^{n} \r\Vert\nn\\
 &  \le\sum_{\ell=0}^{\mathcal K_T}   \sqrt{\frac 1{ n}\sum_{h=1}^m \l\Vert \l \{ \wh A^{(h)\ell}- A^{(h)\ell} \r\}\phi_{t-\ell}^{(h)} \r\Vert^2}\nn\\
  &  \le \sum_{\ell=0}^{\mathcal K_T} \max_{1\le h\le m} \l \Vert \wh A^{(h)\ell}- A^{(h)\ell} \r\Vert_F 
  \sqrt{\frac 1{ n}\sum_{h=1}^m \l\Vert \phi_{t-\ell}^{(h)} \r\Vert^2}\nn\\
  &  \le (\mathcal K_T+1) \max_{0\le \ell\le \mathcal K_T} \max_{1\le h\le m} \l \Vert \wh A^{(h)\ell}- A^{(h)\ell} \r\Vert_F  \sum_{\ell=0}^{\mathcal K_T}
\frac 1{  \sqrt n} \l\Vert \phi_{t-\ell}^{n} \r\Vert.\nn
 \end{align}
 Notice that 
 \begin{align}
 \Vert \wh A^{(h)\ell} - A^{(h)\ell} \Vert_F&= \Vert  \wh A^{(h)} (\wh A^{(h)\ell-1} - A^{(h)\ell-1}) + A^{(h)\ell-1} (\wh A^{(h)}-A^{(h)})  \Vert_F \nn\\
 &\le  \Vert  \wh A^{(h)}\Vert\,\Vert \wh A^{(h)\ell-1} - A^{(h)\ell-1}\Vert +\Vert A^{(h)\ell-1}\Vert\,\Vert\wh A^{(h)}-A^{(h)}  \Vert_F 
 \end{align}
Therefore, by induction over $\ell$ we have from Lemma \ref{lem:VARparam}(ii), as $n,T\to\infty$,
 \begin{align}\label{eq:boundAell}
\min( \beta_{n,T}^{-1}, \alpha_T^{-1/2}, \sqrt n) \max_{0\le\ell\le\mathcal K_T} \max_{1\le h\le m}\l\Vert \wh A^{(h)\ell} - A^{(h)\ell} \r\Vert_F =\mathcal O_P(1).
 \end{align}
 since $\max_{1\le h\le m} \Vert A^{(h)}\Vert <1$, because of Assumption A\ref{ass:ARGDFM}, and thus $\max_{1\le h\le m}\Vert A^{(h)\ell}\Vert \le\max_{1\le h\le m} \Vert A^{(h)\ell-1}\Vert<1$ and also $\max_{1\le h\le m} \Vert  \wh A^{(h)}\Vert<1$ by Lemma \ref{lem:VARparam}(ii).
 Moreover, for any fixed fixed $0\le \ell\le\mathcal K_T$
 \begin{align}
 \frac 1{  \sqrt n} \l\Vert \phi_{t-\ell}^{n} \r\Vert &= \frac 1{  \sqrt n} \l\Vert (I_n- A_nL) \chi_{t-\ell}^{n} \r\Vert\le  \frac 1{  \sqrt n}\l\Vert\chi_{t-\ell}^{n}\r\Vert + \frac 1{\sqrt n} \l\Vert A_n \chi_{t-\ell}^{n} \r\Vert\nn\\
 &\le  \frac 1{  \sqrt n}\l\Vert\chi_{t-\ell}^{n}\r\Vert + \sqrt{ \frac 1{ n}\sum_{h=1}^m \l\Vert A^{(h)} \chi_{t-\ell}^{(h)} \r\Vert^2}\nn\\
 &\le  \frac 1{  \sqrt n}\l\Vert\chi_{t-\ell}^{n}\r\Vert + \max_{1\le h\le m} 
\Vert A^{(h)}\Vert   \frac 1{\sqrt n}\l \Vert \chi_{t-\ell}^{n} \r\Vert=\mathcal O_P(1),
\label{eq:phiOp1}
 \end{align}
because, $\max_{1\le h\le m} \Vert A^{(h)}\Vert <1$ by Assumption A\ref{ass:ARGDFM} and $\frac 1{\sqrt n}\l \Vert \chi_{t-\ell}^{n} \r\Vert=\mathcal O_P(1)$ as shown in \eqref{eq:maxchi2}.
Furthermore, again by \eqref{eq:maxchi2} and Chebychev's inequality for any $\epsilon>0$
\[
\mathrm P\l( \frac 1{\sqrt n}\l \Vert \chi_{t-\ell}^{n} \r\Vert >\epsilon \r) \le \frac{\E[ \frac 1{ n}\l \Vert \chi_{t-\ell}^{n} \r\Vert^2]}{\epsilon^2}\le \frac{(A_\chi)^2}{1- (\rho^\chi)^{2}{\epsilon^2}}
\]
Thus, by Bonferroni inequality, 
\begin{align}
\mathrm P\l( \max_{0\le \ell\le\mathcal K_T}\frac 1{\sqrt n}\l \Vert \chi_{t-\ell}^{n} \r\Vert >\epsilon \r) &\le \sum_{\ell=0}^{\mathcal K_T} \mathrm P\l( \frac 1{\sqrt n}\l \Vert \chi_{t-\ell}^{n} \r\Vert >\epsilon \r)\le (\mathcal K_T+1)\frac{(A_\chi)^2}{1- (\rho^\chi)^{2}{\epsilon^2}}.
\end{align}
Therefore,
\[
\max_{0\le \ell\le\mathcal K_T}\frac 1{\sqrt {n \mathcal K_T}} \l \Vert \chi_{t-\ell}^{n} \r\Vert = \mathcal O_p(1),
\]
which implies
\begin{align}
\min( \beta_{n,T}^{-1}, \alpha_T^{-1/2}, \sqrt n) \frac {\Vert b_{1t}\Vert}{\sqrt {n}\mathcal K_T^{3/2}}  =\mathcal O_P(1).
\end{align}
For $b_{2t}$ we can use the same reasoning and Lemma \ref{lem:phihat}(i),
while $b_{3t}$ is dominated by $b_{1t}$ and $b_{2t}$. So
\begin{align}
\min( \beta_{n,T}^{-1}, \alpha_T^{-1/2}, \sqrt n) \frac {\Vert b_{t}\Vert}{\sqrt {n}\mathcal K_T^{3/2}}  =\mathcal O_P(1).
\end{align}
Finally, by Assumption A\ref{ass:ARGDFM} there exists $\varrho\in(0,\infty)$ such that $\max_{1\le h\le m} \l\Vert A^{(h)}\r\Vert <\frac 1{1+\varrho}$, therefore
\begin{align}
\frac{\Vert \widetilde b_t\Vert }{\sqrt n} &\le \sum_{\ell\in \mathcal D_T} \sqrt{\frac 1{ n} \sum_{h=1}^m \l\Vert A^{(h)\ell} \phi_{t-\ell}^{(h)}\r\Vert^2}\nn\\
&\le\sum_{\ell\in \mathcal D_T} \max_{1\le h\le m} \l\Vert A^{(h)\ell}\r\Vert  \frac 1{\sqrt n} \l\Vert \phi_{t-\ell}^{n}\r\Vert\nn\\
&\le  \sum_{\ell\in \mathcal D_T} \max_{1\le h\le m} \l\Vert A^{(h)}\r\Vert^{\ell}  \frac 1{\sqrt n} \l\Vert \phi_{t-\ell}^{n}\r\Vert\nn\\
&\le \max_{1\le h\le m} \sum_{\ell= \mathcal K_T+1}^\infty  \l\Vert A^{(h)}\r\Vert^{\ell}  \frac 1{\sqrt n} \l\Vert \phi_{t-\ell}^{n}\r\Vert\nn\\
&= \max_{1\le h\le m} \frac{ (1+ \varrho)^{1-\mathcal K_T} }{\varrho} 
\frac 1{\sqrt n} \l\Vert \phi_{t-\ell}^{n}\r\Vert\nn
\end{align}
Then, as $n,T\to\infty$,  by \eqref{eq:phiOp1}
\begin{align}
(1+\varrho)^{\mathcal K_T} \frac{\Vert \widetilde b_t\Vert }{\sqrt n} =\mathcal O_P(1)
\end{align}
It follows that, as $n,T\to\infty$,
\[
 \min(\beta_{n,T}^{-1}\mathcal K_T^{-3/2},\alpha_T^{-1/2} \mathcal K_T^{-3/2}, \sqrt n\mathcal K_T^{-3/2},(1+\varrho)^{\mathcal K_T}) \frac{ \E\l[\l\Vert
  \wh{\chi}_t^n -\chi_t^n
 \r\Vert 
 \r]}{\sqrt n}= \mathcal O(1),\nn
\]
 and since by Assumption A\ref{ass:kernel} $\mathcal K_T\asymp\log T$, then, $(1+\varrho)^{\mathcal K_T}\asymp T^{\log(1+ \varrho)}$ and, since $\varrho\in(0,1)$
 \[
\min(\beta_{n,T}^{-1}\mathcal K_T^{-3/2},\alpha_T^{-1/2} \mathcal K_T^{-3/2}, \sqrt n\mathcal K_T^{-3/2},(1+\varrho)^{\mathcal K_T})\asymp \min(\beta_{n,T}^{-1},\alpha_T^{-1/2} , \sqrt n)  (\log T)^{-3/2}.
 \]

For part (iv) the proof is the same but using only one row of $\wh A_n$ and thus no sum over the $m$ blocks is taken and, thus, no rescaling by $\sqrt n$ is needed. In this case we use Lemma \ref{lem:phihat}(ii) instead of Lemma \ref{lem:phihat}(i). \hfill \rule{.7em}{.7em}

\subsection{Proof of Proposition \ref{prop3}}\label{prop3proof}
\paragraph{Part I.a-II.a.} 
First, by the multiplicative Weyl's inequality \citep[Theorem 7]{merikoski2004inequalities}, 
\begin{align}
\frac 1{ n}\E\l[  \l\Vert \Omega_n e_{t}^{\chi,n}\r\Vert^2\r] &= \frac 1n
\E\l[e_{t}^{\chi,n'}\Omega_n'\Omega_n e_{t}^{\chi,n}\r]
=\frac 1 n \mu_1(\Gamma_{e^\chi}^n \Omega_n'\Omega_n)\nn\\
&\le\frac 1 n \mu_1(\Gamma_{e^\chi}^n) \mu_1(\Omega_n'\Omega_n)=\frac 1 n \mu_1(\Gamma_{e^\chi}^n)\le \frac{C_0}n, \label{eq:OOGWCC}
\end{align}
for some $C_0\in(0,\infty)$ independent of $n$, and 
where we also used the fact that $\mu_1(\Omega_n'\Omega_n)=\mu_1(\Omega_n\Omega_n')=\mu_1(I_r) =1$ and that $(e_{it}^\chi)$ is statically idiosyncratic, hence, it satisfies Theorem \ref{thm: charact stat idiosyncratic}.

Then, notice that $C_t^n = \Omega_n' \Omega_n  C_t^n$, by definition, so, since by Theorem \ref{thm: relation of r-SFS and q-DFS}, 
$\chi_t^n = C_t^n+e_t^{\chi,n}$, by \eqref{eq:OOGWCC},
\begin{align}
    \frac 1{\sqrt n} \l\Vert C_t^n-\Omega_n' \Omega_n  \chi_t^n \r\Vert &=  \frac 1{\sqrt n}\l\Vert \Omega_n' \Omega_n e_{t}^{\chi,n}\r\Vert \le \frac 1{\sqrt n} \l\Vert \Omega_n e_{t}^{\chi,n}\r\Vert =\mathcal O_P\l(\frac 1{\sqrt n}\r).\label{eq:COOchi}
\end{align}
Thus, by \eqref{eq:COOchi},
 \begin{align}
\frac 1{\sqrt n}\l\Vert \wh C_t^n - C_t^n\r\Vert &= \frac 1{\sqrt n}\l\Vert\wh\Pi_n' \wh\Pi_n  \wh \chi_t^n  -\Omega_n' \Omega_n\chi_t^n\r\Vert+\mathcal O_P\l(\frac 1{\sqrt n}\r)\nn\\ 
 &\le  \l\Vert \wh\Pi_n' \wh\Pi_n  -\Omega_n' \Omega_n\r\Vert  \frac 1{\sqrt n}\l\Vert \chi_t^n\r\Vert + \l\Vert\wh\Pi_n' \wh\Pi_n\r\Vert  \frac 1{\sqrt n} \l\Vert \wh \chi_t^n-\chi_t^n\r\Vert +\mathcal O_P\l(\frac 1{\sqrt n}\r),\nn
 \end{align}
and the proof of part (i) follows directly from Lemma \ref{lem:evecevecC}(i), Proposition \ref{prop1}(i), and since, $n^{-1/2} 
\Vert \chi_t^n\Vert=\mathcal O_P(1)$ because of \eqref{eq:maxchi2} in the proof of Lemma \ref{lem:phihat}, and  $\Vert \wh\Pi_n' \wh\Pi_n\Vert \le \Vert \wh\Pi_n\Vert^2=1$ (as eigenvectors are normalized).

For part (ii) the proof is the same as part (i),  but \eqref{eq:COOchi} is replaced by
\begin{equation}\label{eq:COOchirow}
   \l\Vert C_{it}- s_i'\Omega_n' \Omega_n \chi_t^n\r\Vert =
   \l\Vert s_i'\Omega_n' \Omega_n e_t^{\chi,n}\r\Vert \le
  \l\Vert s_i'  \sqrt n \Omega_n'\r\Vert \frac 1{\sqrt n} \l\Vert \Omega_n e_{t}^{\chi,n}\r\Vert =\mathcal O_P\l(\frac 1{\sqrt n}\r),
\end{equation}
because of  Lemma \ref{lem:rowevecC}.
And then we need to use \eqref{eq:COOchirow} together with Lemma \ref{lem:evecevecC}(ii) instead of Lemma \ref{lem:evecevecC}(i) and Proposition \ref{prop1}(ii) instead of Proposition \ref{prop1}(i).

\paragraph{Part I.a-II.b} The proof is the same as for part I.a-II.a, but using Lemma \ref{lem:evecevecCHI} instead of Lemma \ref{lem:evecevecC}.

\paragraph{Part I.b-II.a.} For parts (iii) and (iv) the proof is the same as parts (i) and (ii) in part I.a-II.a, respectively, but we need to use Proposition \ref{prop1}(iii) and \ref{prop1}(iv) instead of Proposition \ref{prop1}(i) and \ref{prop1}(ii).

\paragraph{Part I.b-II.b} The proof is the same as for part I.b-II.a, but using Lemma \ref{lem:evecevecCHI} instead of Lemma \ref{lem:evecevecC}.
\hfill \rule{.7em}{.7em}

\subsection{Proof of Theorem \ref{thm: weakCC}}\label{weakCCproof}

\paragraph{Part I.a-II-III} For any fixed $t=\mathcal M_T+1,\ldots, T-\mathcal M_T-1$ and any given $i=1,\ldots, n$,
from Propositions \ref{prop1}(i) and \ref{prop3}(i), when using the Barlett kernel $\vartheta=1$, and since $\nu\ge 4$ by Assumption \ref{ass:newGDFM},
\begin{align*}
    \vert \wh{e}_{it}^{\,\chi}-e_{it}^\chi \vert \le 
\vert \wh{\chi}_{it}-\chi_{it} \vert+
\vert \wh{C}_{it}-C_{it} \vert&=  \mathcal O_P\left(
\max\l(
\frac{\mathcal B_T}{T^{1-2/\nu}},
\sqrt{\frac{\mathcal B_T\log \mathcal B_T}T},
\frac 1{\mathcal B_T},
\frac 1{\sqrt n}
\r)
\right) \log T\nonumber\\
&=  \mathcal O_P\left(
\max\l(
\sqrt{\frac{\mathcal B_T\log \mathcal B_T}T},
\frac 1{\mathcal B_T},
\frac 1{\sqrt n}
\r)
\right) \log T.\nonumber
\end{align*}

\paragraph{Part I.b-II-III} For any fixed $t=\mathcal K_T+1,\ldots, T$ and any given $i=1,\ldots, n$,
from Propositions \ref{prop1}(iii) and \ref{prop3}(iii), when using the Barlett kernel $\vartheta=1$, and assuming $\nu>6$, and $n=\mathcal O(T^\gamma)$ for some $\gamma\in(0,\infty)$,
\begin{align*}
    \vert \wh{e}_{it}^{\,\chi}-e_{it}^\chi \vert \le 
\vert \wh{\chi}_{it}-\chi_{it} \vert+
\vert \wh{C}_{it}-C_{it} \vert&=  \mathcal O_P\left(
\max\l(
\frac{n^{2/\nu}\log^3 n}{T^{1-2/\nu}}
,\sqrt{\frac {\log n}T},
\frac{\mathcal B_T}{T^{1-2/\nu}},
\sqrt{\frac{\mathcal B_T\log \mathcal B_T}T},
\frac 1{\mathcal B_T},
\frac 1{\sqrt n}
\r)
\right) \log T\nonumber\\
&=  \mathcal O_P\left(
\max\l(
\sqrt{\frac{\mathcal B_T\log \mathcal B_T}T},
\frac 1{\mathcal B_T},
\frac 1{\sqrt n}
\r)
\right) \log T.\nonumber\qquad\qquad\qquad\qquad\qquad  \rule{.7em}{.7em}
\end{align*}

\subsection{Proof of Theorem \ref{thm: weakfactest}}\label{weakfactproof}

Recall that by Theorem \ref{thm: relation of r-SFS and q-DFS}, ${\Gamma}_{e^\chi}^n=\Gamma^{ n}_\chi - \Gamma^{ n}_C$ and note that $\Gamma_C^n=\Omega_n'\Omega_n \Gamma^{ n}_C \Omega_n'\Omega_n$, by definition. Let $s_i$ be the $i$th element of the canonical basis of $\mathbb R^{n\times 1}$. Then, for all $i,j=1,\ldots,n$,
\begin{align}
\l\vert [\Gamma_C^n]_{ij}-s_i'\Omega_n' \Omega_n  \Gamma_\chi^n\Omega_n' \Omega_n s_j \r\vert 
    &=  \l\vert s_i'  \sqrt n\Omega_n' \Omega_n \frac{\Gamma_{e^\chi}^{n}}n \Omega_n' \Omega_n\sqrt n\, s_j \r\vert\nn\\
    &\le  \l\vert s_i'  \sqrt n\Omega_n' \Omega_n\r\Vert^2 \, \l\Vert \frac{\Gamma_{e^\chi}^{n}}n\r\Vert=\mathcal  O\l(\frac 1n\r).\label{eq:COOchicov}
\end{align} 
because of Lemma \ref{lem:rowevecC} and since, by Theorem \ref{thm: relation of r-SFS and q-DFS}, $(e_{it}^\chi)$ is statically idiosyncratic, hence, it satisfies Theorem \ref{thm: charact stat idiosyncratic}.
Then, 
from \eqref{eq:GammahatWCCSI}, and given $ \widehat\Gamma^{ n}_\chi$ as estimated in \eqref{eq:GammahatChiSI}, 
by Lemmas \ref{lem:IFTcov}(i)  and \ref{lem:evecevecCHI}(ii), and \eqref{eq:COOchicov},
\begin{align}
\l\vert [\wh{\Gamma}_{e^\chi}^n]_{ij}-[{\Gamma}_{e^\chi}^n]_{ij}\r\vert
=&\,
 \l\vert s_i'\l\{ \widehat\Gamma^{ n}_\chi - \widehat P_n'\widehat P_n \widehat\Gamma^{ n}_\chi\widehat P_n'\widehat P_n\r\}s_j-s_i'
\l\{\Gamma^{ n}_\chi - \Gamma^{ n}_C\r\}s_j \r\vert\nn\\
\le&\,  \l\vert[\widehat\Gamma^{ n}_\chi]_{ij} -[\Gamma^{ n}_\chi]_{ij} \r\vert +\l\vert
s_i' \widehat P_n'\widehat P_n \widehat\Gamma^{ n}_\chi\widehat P_n'\widehat P_n s_j
-[\Gamma_C^n]_{ij}
\r\vert\nn\\
\le&\,  \l\vert[\widehat\Gamma^{ n}_\chi]_{ij} -[\Gamma^{ n}_\chi]_{ij} \r\vert +\l\vert
s_i'\sqrt n \widehat P_n'\widehat P_n \frac{\widehat\Gamma^{ n}_\chi}n \widehat P_n'\widehat P_n\sqrt n\, s_j
-s_i'\sqrt n \Omega_n'\Omega_n \frac{\Gamma^{ n}_\chi} n \Omega_n'\Omega_n \sqrt n\, s_j
\r\vert \nn\\
&+ \l\vert [\Gamma^{ n}_C]_{ij}-s_i' \Omega_n'\Omega_n \Gamma^{ n}_\chi \Omega_n'\Omega_n s_j\r\vert
\nn\\
=&\, \mathcal O_P\l(\max\l(\sqrt{\frac{\mathcal B_T\log \mathcal B_T}{T}},\frac 1{\mathcal B_T},\frac 1{\sqrt n}\r)\r)+\mathcal O\l(\frac 1n\r),\label{eq:crucialeq}
\end{align}
where we used the fact that  $\nu\ge 4$ by Assumption A\ref{ass:newGDFM}, and we are using a Bartlett kernel, so $\vartheta=1$.
From \eqref{eq:crucialeq} it follows that for any fixed $\bar n\in\mathbb N$, i.e., $\bar n$ is independent of $n$, such that $\bar n\ge r_w$,
\beq
\l\Vert \wh{\Gamma}_{e^\chi}^{\bar n} - {\Gamma}_{e^\chi}^{\bar n}\r\Vert  \le \l\{ \sum_{i=1}^{\bar n} \sum_{j=1}^{\bar n} 
\l\vert [\wh{\Gamma}_{e^\chi}^n]_{ij}-[{\Gamma}_{e^\chi}^n]_{ij}\r\vert^2
\r\}^{1/2} 
=\mathcal O_P\l(\max\l(\sqrt{\frac{\mathcal B_T\log \mathcal B_T}{T}},\frac 1{\mathcal B_T},\frac 1{\sqrt n}\r)\r).\label{eq:consistGammaWCC}
\eeq

Now, from \eqref{eq:consistGammaWCC} and by Weyl's inequality \citep[Theorem 1]{merikoski2004inequalities}:
\beq
\l\Vert  \wh{\mathcal M}-\mathcal M\r \Vert \le \l\Vert \wh{\Gamma}_{e^\chi}^{\bar n} - {\Gamma}_{e^\chi}^{\bar n}\r\Vert=\mathcal O_P\l(\max\l(\sqrt{\frac{\mathcal B_T\log \mathcal B_T}{T}},\frac 1{\mathcal B_T},\frac 1{\sqrt n}\r)\r).\label{eq:consistGammaWCCeval}
\eeq
Furthermore, for all $j=2,\ldots, r_w-1$,
\beq\label{eq:seqeval}
0<\mu_{r_w}(\Gamma^{\bar n}_{e^\chi})\le\mu_{j+1}(\Gamma^{\bar n}_{e^\chi})< \mu_j(\Gamma^{\bar n}_{e^\chi})<\mu_{j-1}(\Gamma^{\bar n}_{e^\chi})\le \mu_1(\Gamma^{\bar n}_{e^\chi})\le \sup_{n\in\mathbb N} \mu_1(\Gamma^{n}_{e^\chi})<\infty,
\eeq
since eigenvalues are distinct by Assumption A\ref{A weakfactor ID}; $(e_{it}^\chi)$ is statically idiosyncratic by Theorem \ref{thm: relation of r-SFS and q-DFS} and, therefore, we can apply Theorem \ref{thm: charact stat idiosyncratic}; $\Gamma^{\bar n}_{e^\chi}$ has rank $r_w$ because of  Theorem \ref{thm: weakfactors}, which holds for any fixed $n\in\mathbb N$; and $r_w>0$ by Assumption A\ref{A weakfactor ID}. By letting $\iota_j$ be the $j$th element of the canonical basis of $\mathbb R^{r_w\times 1}$, from \eqref{eq:consistGammaWCC} and \eqref{eq:seqeval}, and by Davis-Kahan theorem \citep[Corollary 1]{yu2015useful}, for all $j=1,\ldots, r_w$, 
\begin{align}
\l\Vert
 \iota_j' \wh{\mathcal W}_{\bar n}\pm \iota_j'{\mathcal W}_{\bar n}
\r\Vert
&\le C_0 \frac{\l\Vert \wh{\Gamma}_{e^\chi}^{\bar n} - {\Gamma}_{e^\chi}^{\bar n}\r\Vert}{
\min
\l\{
\mu_{j-1}({\Gamma}_{e^\chi}^{\bar n})-
\mu_{j}({\Gamma}_{e^\chi}^{\bar n}),
\mu_{j}({\Gamma}_{e^\chi}^{\bar n})-
\mu_{j+1}({\Gamma}_{e^\chi}^{\bar n})
\r\}
}\nn\\
&=\mathcal O_P\l(\max\l(\sqrt{\frac{\mathcal B_T\log \mathcal B_T}{T}},\frac 1{\mathcal B_T},\frac 1{\sqrt n}\r)\r),\label{eq:consistGammaWCCevec}
\end{align}
where, by definition, $\mu_{0}({\Gamma}_{e^\chi}^{\bar n})=\infty$ and 
$\mu_{r_w+1}({\Gamma}_{e^\chi}^{\bar n})=0$. From, \eqref{eq:consistGammaWCCevec} it follows that
\beq\label{eq:consistGammaWCCevecALL}
\l\Vert
\wh{\mathcal W}_{\bar n}- \mathcal S{\mathcal W}_{\bar n}
\r\Vert=\mathcal O_P\l(\max\l(\sqrt{\frac{\mathcal B_T\log \mathcal B_T}{T}},\frac 1{\mathcal B_T},\frac 1{\sqrt n}\r)\r),
\eeq
where $\mathcal S$ is a $r_w\times r_w$ diagonal matrix with entries $\pm 1$.

Finally, by Assumption A\ref{A weakfactor ID}, for any fixed $t=\mathcal M_T+1,\ldots, T-\mathcal M_T$ (if we compute $\wh e_t^{\, \chi,\bar n}$ using $\wh{\chi}_t^{\bar n}$ as in part I.a)
or $t=\mathcal K_T+1,\ldots, T$ (if we compute $\wh e_t^{\,\chi,\bar n}$ using $\wh{\chi}_t^{\bar n}$ as in part I.b),
\begin{align}
\l\Vert\wh F_t^{w}-\mathcal S F_t^w\r\Vert =&\, 
\l\Vert 
\wh{\mathcal M}^{-1/2}\wh{\mathcal W}_{\bar n}\, \wh e_t^{\,\chi,\bar n}-
{\mathcal M}^{-1/2}\mathcal S{\mathcal W}_{\bar n}\,  e_t^{\,\chi,\bar n}
\r\Vert 
\nn\\
\le &\,
\l\Vert
\wh{\mathcal M}^{-1/2}-{\mathcal M}^{-1/2}
\r\Vert\,
\l\Vert
\mathcal S{\mathcal W}_{\bar n}
\r\Vert\,
\l\Vert
e_t^{\,\chi,\bar n}
\r\Vert
+
\l\Vert
{\mathcal M}^{-1/2}
\r\Vert\,
\l\Vert
\wh{\mathcal W}_{\bar n}-\mathcal S{\mathcal W}_{\bar n}
\r\Vert\,
\l\Vert
e_t^{\,\chi,\bar n}
\r\Vert\nn\\
&
+
\l\Vert
{\mathcal M}^{-1/2}
\r\Vert\,
\l\Vert
\mathcal S{\mathcal W}_{\bar n}
\r\Vert\,
\l\Vert
\wh e_t^{\,\chi,\bar n}
-e_t^{\,\chi,\bar n}
\r\Vert
+
\l\Vert
\wh{\mathcal M}^{-1/2}-{\mathcal M}^{-1/2}
\r\Vert\,
\l\Vert
\wh{\mathcal W}_{\bar n}-\mathcal S{\mathcal W}_{\bar n}
\r\Vert\,
\l\Vert
e_t^{\,\chi,\bar n}
\r\Vert\nn\\
&+
\l\Vert
\wh{\mathcal M}^{-1/2}-{\mathcal M}^{-1/2}
\r\Vert\,
\l\Vert
\mathcal S{\mathcal W}_{\bar n}
\r\Vert\,
\l\Vert
\wh e_t^{\,\chi,\bar n}
-e_t^{\,\chi,\bar n}
\r\Vert
+
\l\Vert
{\mathcal M}^{-1/2}
\r\Vert\,
\l\Vert
\wh{\mathcal W}_{\bar n}-\mathcal S{\mathcal W}_{\bar n}
\r\Vert\,
\l\Vert
\wh e_t^{\,\chi,\bar n}
-e_t^{\,\chi,\bar n}
\r\Vert
\nn\\
&+ 
\l\Vert
\wh{\mathcal M}^{-1/2}-{\mathcal M}^{-1/2}
\r\Vert\,
\l\Vert
\wh{\mathcal W}_{\bar n}-\mathcal S{\mathcal W}_{\bar n}
\r\Vert\,
\l\Vert
\wh e_t^{\,\chi,\bar n}
-e_t^{\,\chi,\bar n}
\r\Vert\nn\\
=&\,\mathcal O_P\l(\log T\max\l(\sqrt{\frac{\mathcal B_T\log \mathcal B_T}{T}},\frac 1{\mathcal B_T},\frac 1{\sqrt n}\r)\r),
\end{align}
because of \eqref{eq:consistGammaWCCeval} (and the continuous mapping theorem), \eqref{eq:consistGammaWCCevecALL}, and Theorem \ref{thm: weakCC}, and since
$\l\Vert{\mathcal M}^{-1/2}\r\Vert=1/\sqrt{\mu_{r_w}(\Gamma_{e^\chi}^{\bar n})} < \infty$ by \eqref{eq:seqeval}, 
$\l\Vert\mathcal S{\mathcal W}_{\bar n}\r\Vert=1$ since eigenvectors are normalized regardless of their sign,
and $\l\Vert
e_t^{\,\chi,\bar n}
\r\Vert=\mathcal O_p(1)$. Indeed, $\bar n$ is fixed and by the $C_r$-inequality with $r=2$, 
\begin{align}
\E\l[\l\Vert
e_t^{\,\chi,\bar n}
\r\Vert^2\r]
\le &\,
2\l\{\E\l[\l\Vert
\chi_t^{\bar n}
\r\Vert^2\r]+
\E\l[\l\Vert
C_t^{\bar n}
\r\Vert^2\r]\r\} \le \frac{2\bar n (A_\chi)^2}{1- (\rho^\chi)^{2}} +
2\sum_{i=1}^{\bar n} \Lambda_i\E[F_{t}F_{t}']\Lambda_i'
\le2\bar n \l\{ \frac{ (A_\chi)^2}{1- (\rho^\chi)^{2}} + C_\Lambda^2\r\},\nn
\end{align}
because of \eqref{eq:maxchi2} in the proof of Lemma \ref{lem:phihat}, and Assumption A\ref{ass:divevalC}.
\hfill \rule{.7em}{.7em}

\clearpage
\FloatBarrier
\section{Additional simulation results}\label{app: simres}

\begin{figure}[h!]
    \centering
	\begin{tabular}{cc}
    \includegraphics[width=.4\textwidth]{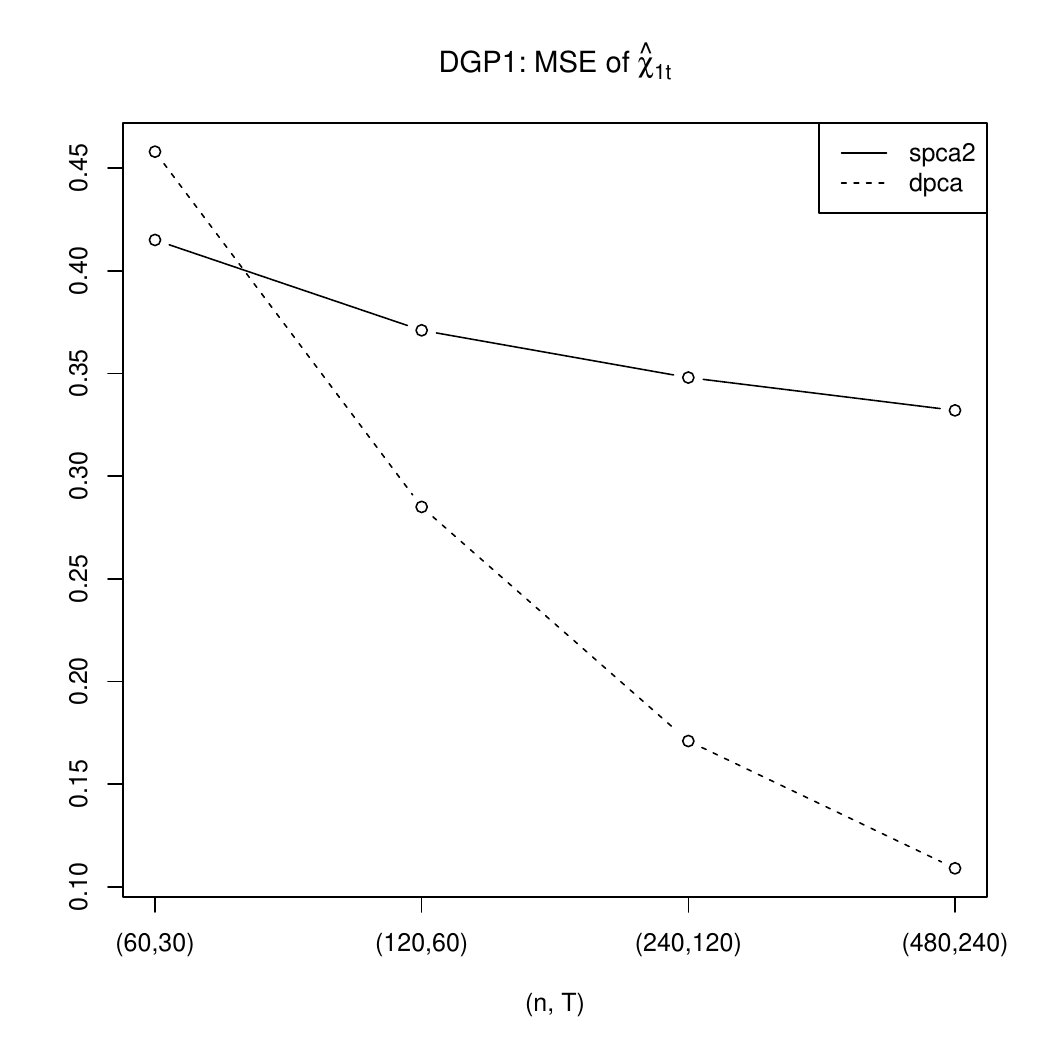}&
    \includegraphics[width=.4\textwidth]{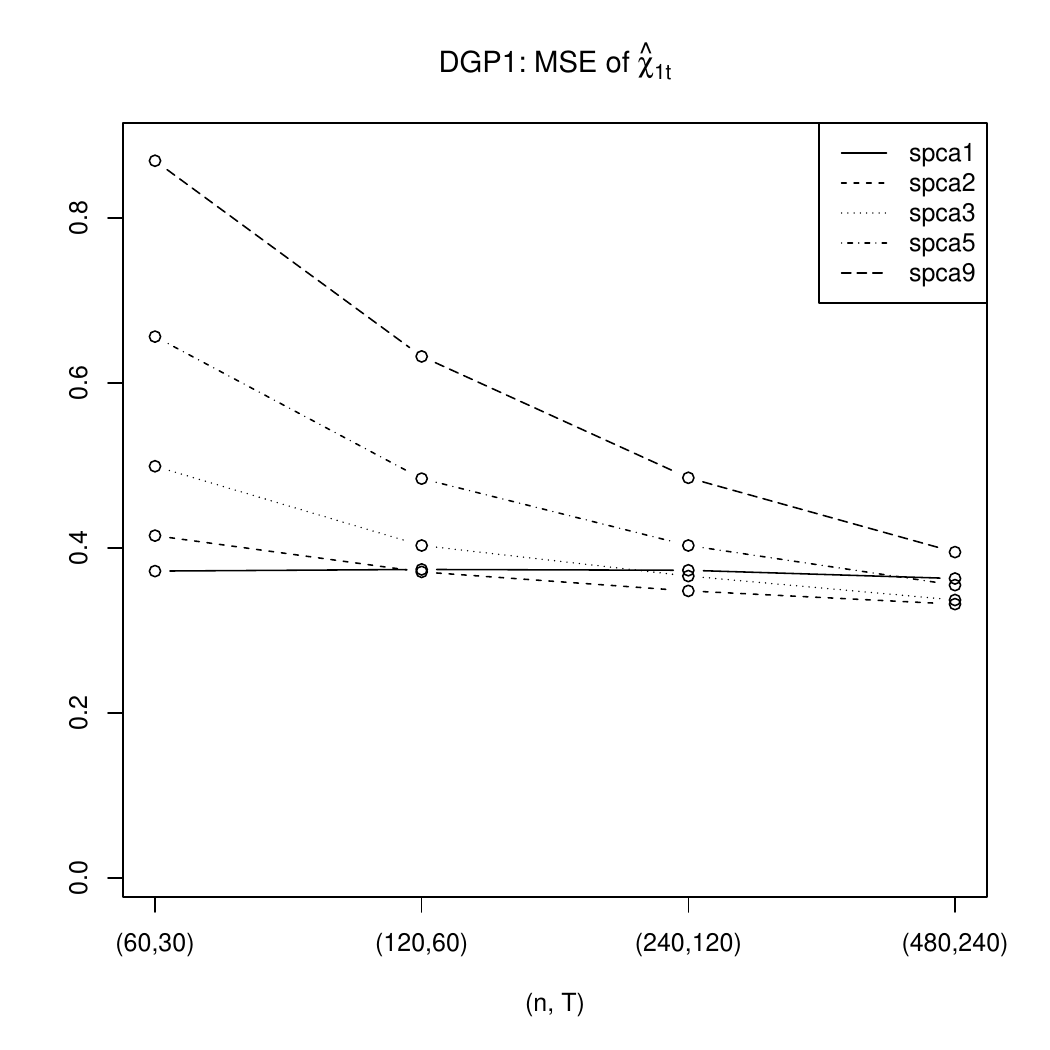}
    \end{tabular}
    \caption{ 
    \footnotesize DGP1. 
    Mean Squared Error of $\widehat \chi_{1t}$ over 500 replications, when $n>T$. \texttt{spcar}: estimation with static PCA with \texttt{r} $=1,2,3,5,9$, \texttt{dpca}: estimation by PCA with $q = 1$.}
    \label{fig: inconsistency of weak factors T < n, 0505}
\end{figure}

\begin{figure}[h!]
    \centering
	\begin{tabular}{cc}
    \includegraphics[width=.4\textwidth]{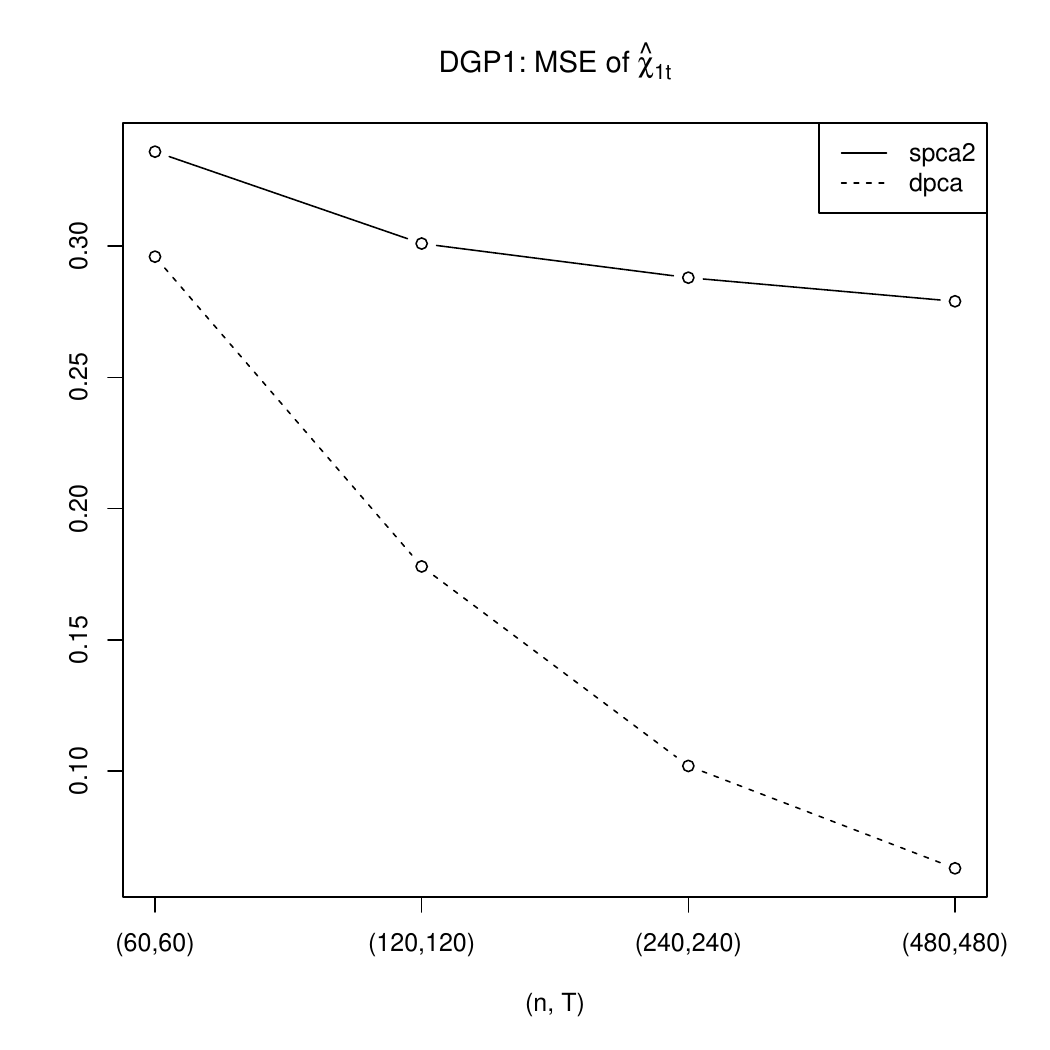}&
    \includegraphics[width=.4\textwidth]{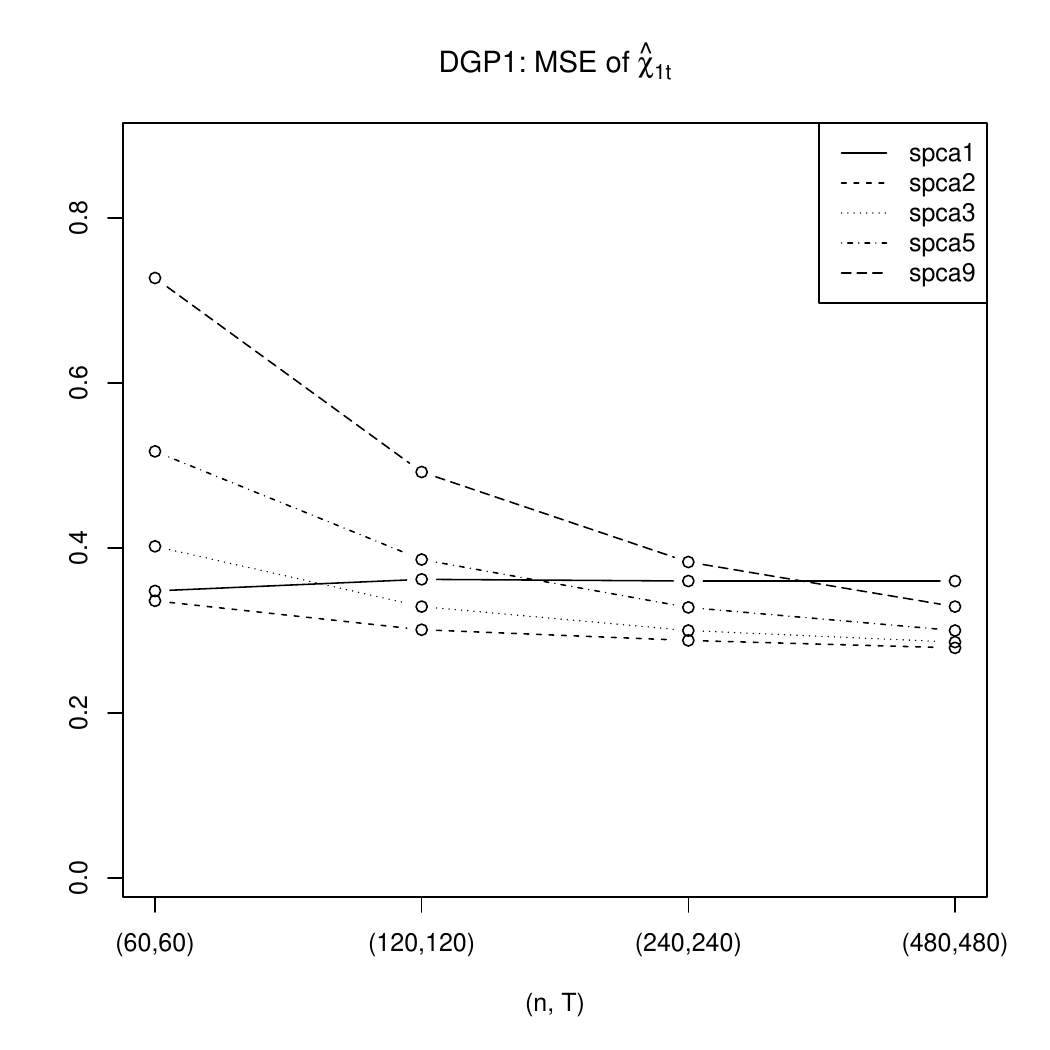}
    \end{tabular}
    \caption{ 
    \footnotesize DGP1. 
    Mean Squared Error of $\widehat \chi_{1t}$ over 500 replications, when $n=T$. \texttt{spcar}: estimation with static PCA with \texttt{r} $=1,2,3,5,9$, \texttt{dpca}: estimation by PCA with $q = 1$.}
    \label{fig: inconsistency of weak factors T=n, 0505}
\end{figure}


\begin{table}[ht]
\centering
\begin{tabular}{l|llll}
\toprule
\multicolumn{5}{c}{\textbf{MSE, Canonical Decomposition Estimates}} \\
\midrule
\hline
$(n,T)$ & (60,30) & (120,60) & (240,120) & (480,240) \\ 
\\[-1.8ex]\hline 
\hline \\[-1.8ex]

$\widehat{\chi}^{Ia}_{1t}$ 
& 0.487 (0.409) & 0.29 (0.191) & 0.174 (0.114) & 0.102 (0.059) \\[0.2em]

$\widehat{\chi}^{Ib}_{1t}$ 
& 0.496 (1.899) & 0.206 (0.173) & 0.105 (0.093) & 0.06 (0.044) \\[0.2em]

\hline \\[-0.8em]

$\widehat{C}^{Ia, IIa}_{1t}$ 
& 0.246 (0.239) & 0.135 (0.116) & 0.07 (0.07) & 0.039 (0.036) \\[0.2em]

$\widehat{C}^{Ia, IIb}_{1t}$ 
& 0.269 (0.266) & 0.146 (0.13) & 0.073 (0.074) & 0.04 (0.037) \\[0.2em]

$\widehat{C}^{Ib, IIa}_{1t}$ 
& 0.131 (0.193) & 0.051 (0.062) & 0.022 (0.026) & 0.012 (0.015) \\[0.2em]

$\widehat{C}^{Ib, IIb}_{1t}$ 
& 0.162 (0.24) & 0.065 (0.085) & 0.024 (0.03) & 0.013 (0.017) \\[0.2em]


\hline \\[-0.8em]

$\widehat{e}^{\chi, Ia, IIa}_{1t}$ 
& 0.217 (0.105) & 0.14 (0.063) & 0.1 (0.039) & 0.064 (0.022) \\[0.2em]

$\widehat{e}^{\chi, Ia, IIb}_{1t}$ 
& 0.21 (0.095) & 0.141 (0.059) & 0.1 (0.037) & 0.065 (0.022) \\[0.2em]

$\widehat{e}^{\chi, Ib, IIa}_{1t}$ 
& 0.434 (1.697) & 0.191 (0.205) & 0.101 (0.103) & 0.055 (0.043) \\[0.2em]

$\widehat{e}^{\chi, Ib, IIb}_{1t}$ 
& 0.441 (1.702) & 0.181 (0.192) & 0.093 (0.095) & 0.051 (0.041) \\[0.2em]



\hline \hline
\end{tabular}\\[3pt]
\caption{
\footnotesize 
Mean Squared Error and standard deviation (in parentheses) evaluated over $B = 500$ replications, when $n>T$.
}
\label{tbl: MSE canon decomp 005 ngT}
\end{table}
\begin{table}[ht]
\centering
\begin{tabular}{l|llll}
\toprule
\multicolumn{5}{c}{\textbf{MSE, Canonical Decomposition Estimates}} \\
\midrule
\hline
$(n,T)$ & (60,60) & (120,120) & (240,240) & (480,480) \\ 
\\[-1.8ex]\hline 
\hline \\[-1.8ex]

$\widehat{\chi}^{Ia}_{1t}$ 
& 0.299 (0.216) & 0.168 (0.099) & 0.104 (0.06) & 0.063 (0.03) \\[0.2em]

$\widehat{\chi}^{Ib}_{1t}$ 
& 0.244 (0.697) & 0.106 (0.074) & 0.061 (0.045) & 0.033 (0.023) \\[0.2em]

\hline \\[-0.8em]

$\widehat{C}^{Ia, IIa}_{1t}$ 
& 0.147 (0.129) & 0.073 (0.061) & 0.042 (0.037) & 0.02 (0.018) \\[0.2em]

$\widehat{C}^{Ia, IIb}_{1t}$ 
& 0.152 (0.138) & 0.075 (0.064) & 0.043 (0.038) & 0.021 (0.018) \\[0.2em]

$\widehat{C}^{Ib, IIa}_{1t}$ 
& 0.061 (0.059) & 0.027 (0.029) & 0.014 (0.016) & 0.007 (0.008) \\[0.2em]

$\widehat{C}^{Ib, IIb}_{1t}$ 
& 0.068 (0.081) & 0.029 (0.032) & 0.015 (0.016) & 0.007 (0.008) \\[0.2em]


\hline \\[-0.8em]

$\widehat{e}^{\chi, Ia, IIa}_{1t}$ 
& 0.158 (0.069) & 0.098 (0.036) & 0.065 (0.021) & 0.047 (0.014) \\[0.2em]

$\widehat{e}^{\chi, Ia, IIb}_{1t}$ 
& 0.153 (0.063) & 0.1 (0.035) & 0.065 (0.021) & 0.047 (0.014) \\[0.2em]

$\widehat{e}^{\chi, Ib, IIa}_{1t}$ 
& 0.239 (0.658) & 0.104 (0.084) & 0.058 (0.044) & 0.032 (0.023) \\[0.2em]

$\widehat{e}^{\chi, Ib, IIb}_{1t}$ 
& 0.222 (0.659) & 0.096 (0.078) & 0.053 (0.041) & 0.031 (0.022) \\[0.2em]



\hline \hline
\end{tabular}\\[3pt]
\caption{
\footnotesize 
Mean Squared Error and standard deviation (in parentheses) evaluated over $B = 500$ replications, when $n=T$.
}
\label{tbl: MSE canon decomp 005  neqT}
\end{table}
%

%
%
%
%
%
\clearpage
\section{Additional empirical results}\label{app:EVfig}

\begin{figure}[h!]
    \centering
    \includegraphics[width=.8\textwidth]{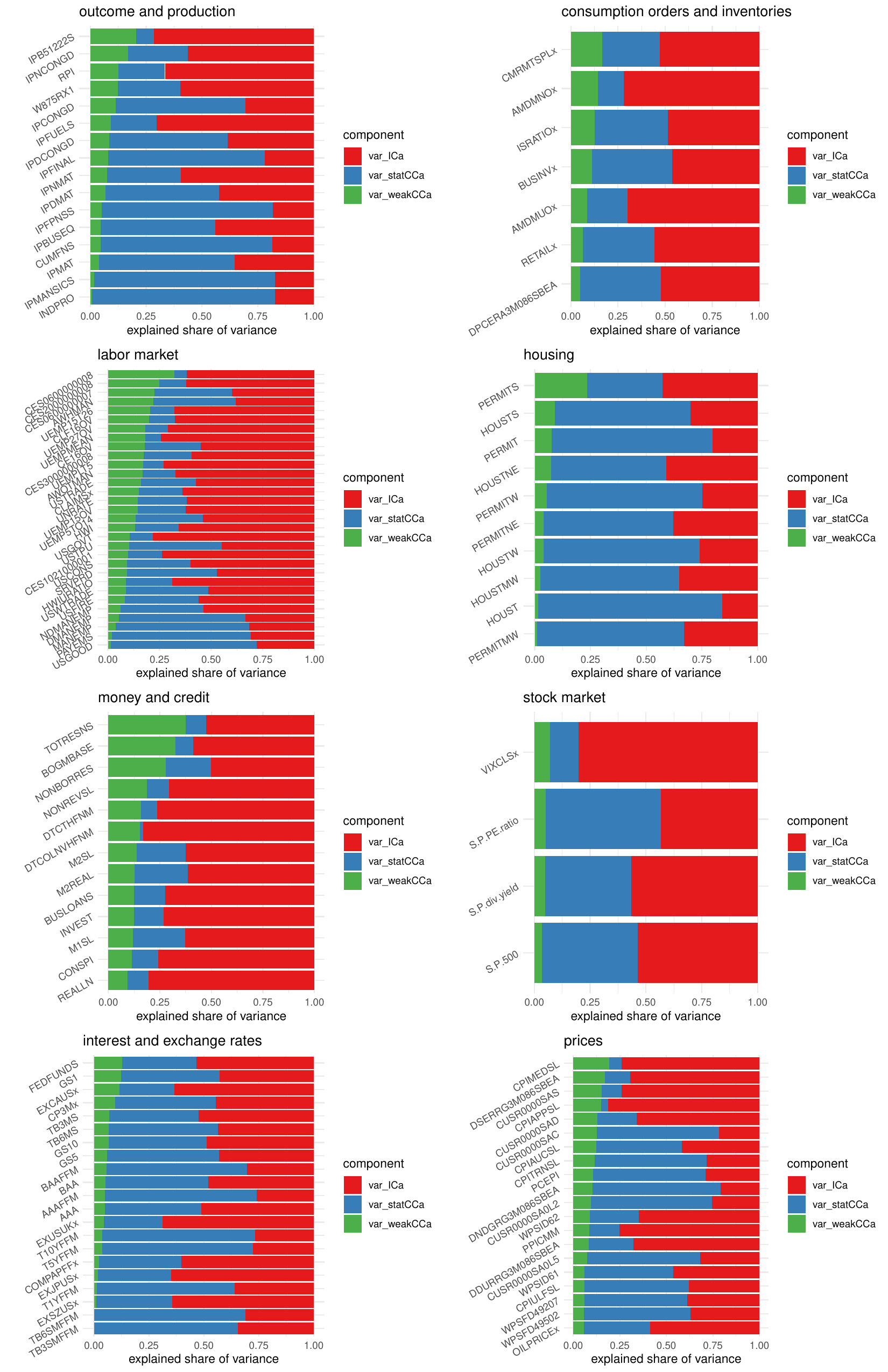}
    \caption{\footnotesize Share of variance explained by each component per variable with $q = 4$ and $r=8$. Estimates are obtained by using part II.a when estimating $C_{it}$. Here \texttt{var\_statCCa} $= EV_i^C$ (given in \eqref{eq:EVCa}),  \texttt{var\_weakCCa} $=EV_i^{e^\chi}$ (given in \eqref{eq:EVWC}).
     Last, \texttt{var\_ICa} is the variance explained by the dynamic idiosyncratic component, which is given by $EV_i^\xi=1-EV_i^C-EV_i^{e^\chi}$.}
    \label{fig: share_of_var_r8_method_a}
\end{figure}

\begin{figure}[h!]
    \centering
    \includegraphics[width=.8\textwidth]{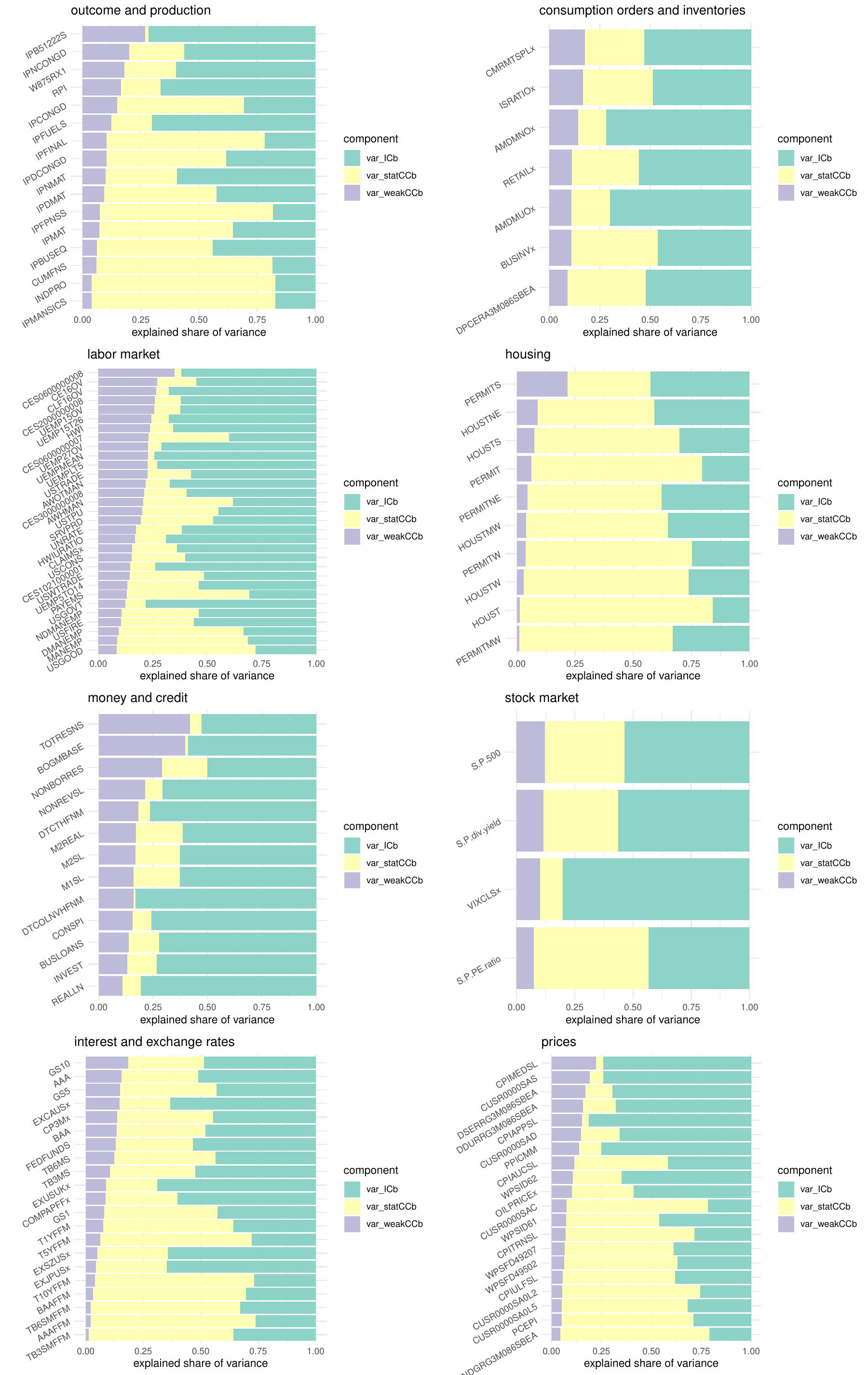}
    \caption{\footnotesize Share of variance explained by each component per variable with $q = 4$ and $r=6$. Estimates are obtained by using part II.b when estimating $C_{it}$. Here \texttt{var\_statCCb} $= EV_i^C$ (given in \eqref{eq:EVCa}),  \texttt{var\_weakCCb} $=EV_i^{e^\chi}$ (given in \eqref{eq:EVWC}).
     Last, \texttt{var\_ICb} is the variance explained by the dynamic idiosyncratic component, which is given by $EV_i^\xi=1-EV_i^C-EV_i^{e^\chi}$.} 
    \label{fig: share of var r6}
\end{figure}

\begin{figure}[!ht]
    \centering
    \includegraphics[width=.8\textwidth]{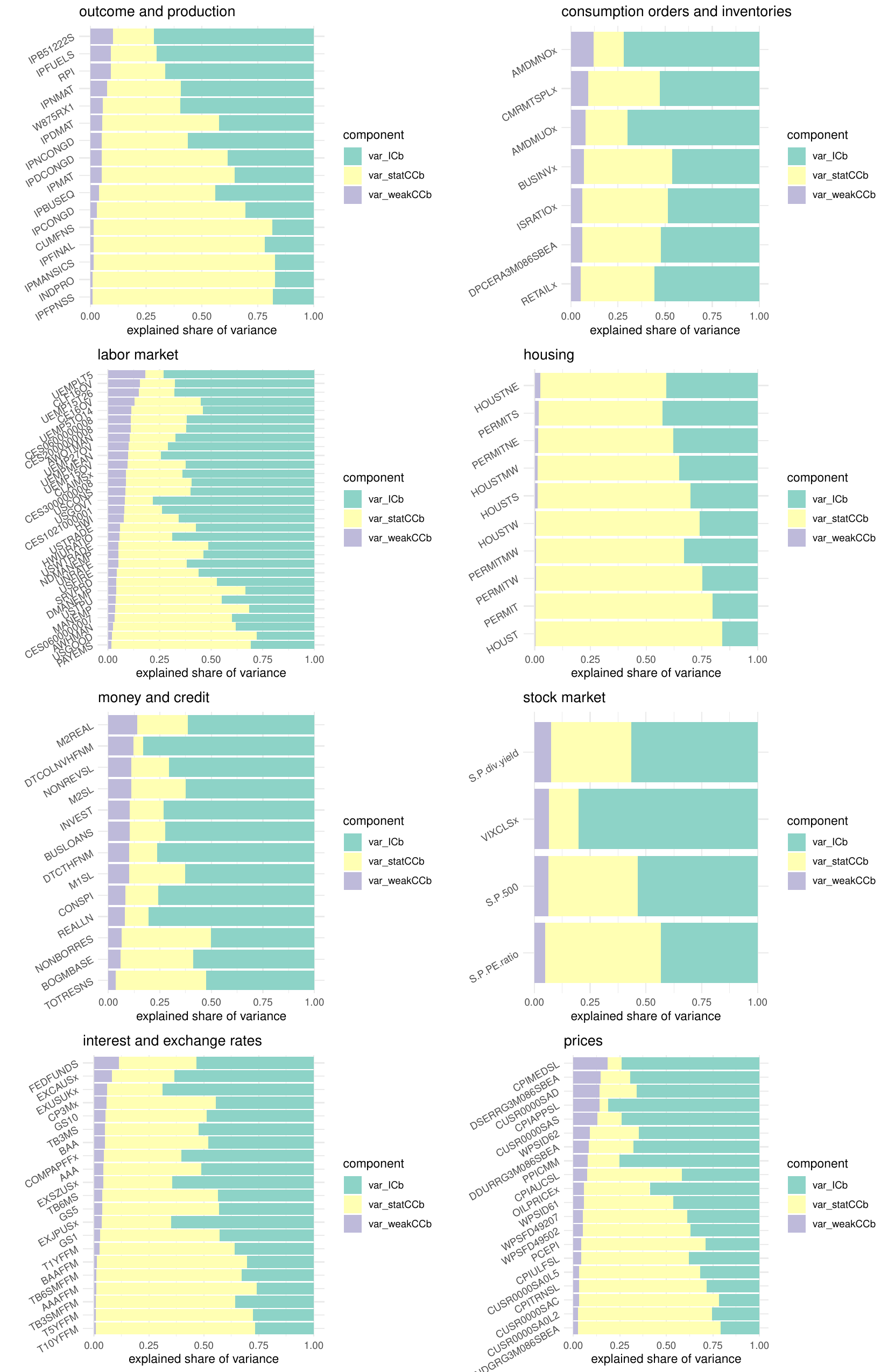}
    \caption{\footnotesize Share of variance explained by each component per variable with $q = 4$ and $r=12$. Estimates are obtained by using part II.b when estimating $C_{it}$. Here \texttt{var\_statCCb} $= EV_i^C$ (given in \eqref{eq:EVCa}),  \texttt{var\_weakCCb} $=EV_i^{e^\chi}$ (given in \eqref{eq:EVWC}).
     Last, \texttt{var\_ICb} is the variance explained by the dynamic idiosyncratic component, which is given by $EV_i^\xi=1-EV_i^C-EV_i^{e^\chi}$.} 
    \label{fig: share of var r12}
\end{figure}
\end{document}